\documentclass[11pt,a4paper]{article}
\usepackage{graphicx} 
\usepackage{subfig}
\usepackage{amsmath} 
\usepackage{amssymb}
\usepackage{lineno}
\usepackage{mathrsfs}
\usepackage{array}
\usepackage{lmodern}
\usepackage{sectsty}
\usepackage[toc,page]{appendix}
\usepackage{braket}
\usepackage{setspace}
\usepackage{titlesec}

\newcommand\secspace{1.2cm}

\usepackage{amsmath}

\begin{document}
\begin{center}

{\huge A Mini Review of some Dark Matter/BSM Physics and a Bit More}\\
\vspace{1cm}
{Shmuel Nussinov, Tel Aviv University}
\end{center}
\vspace{1cm}
\section*{Abstract}
There is a  vast literature on Dark Matter (DM) with many reviews of specific topics only a small fraction of which will be mentioned. I will not attempt - for the most part - to guess which among the many suggested DM variants and attendant Beyond Standard Models (BSM) physics is not yet proven to be wrong. If, following Sherlock Holmes, the true DM is to be found by "Eliminating all the impossible" the way there will be long and torturous. Rather, the choice of topics is guided by the highly subjective criterion of my own familiarity with, or beauty of, the DM model, the BSM physics underlying it, the experimental methods suggested to search for it, and general approaches for testing its soundness. I also mention models/scenarios/ideas which already were or are likely to be soon excluded if the arguments suggesting them or their inconsistency are interesting.\par
\noindent I start with a very brief review of cosmology which underlies much of DM research and some relevant General Relativity (GR). I next discuss Self Interacting Dark Matter (SIDM) models and upper bounds on the mass $M(X)$ of point-like, symmetric DM. This is followed up by some general aspects of DM detection and  directional/temporal variations. I dwell on efforts to explain the ratio of the contributions to the (critical) density of DM and baryonic matter and conclude this part  with a speculation connecting this ratio with the existence of three fermionic families. I discuss DM models tied with BSM physics scenarios including Primordial Black Holes (PBH’s), new physics in the neutrino sector, ultra-light DM and axions. I next mention various "Gems" - beautiful ideas, constructs and suggested experimental set-ups that arose in or were borrowed into the DM field. Another repeating motif are scenarios where most DM searches to-date naturally fail for a variety of reasons culminating with cases where it is impossible to experimentally find non-gravitational evidence for the DM.   \par\noindent
The possible existence of other communicative civilizations appears in the "Bit more" part in the last sections. It is motivated by BSM “Quirks” which may offer a new method for communication across our galaxy but could entails a superstrong self interacting DM.

\bfseries
\tableofcontents
\normalfont




\newpage

\section{\hspace{\secspace}Introduction: our motivation \& philosophy and what we select to present}
Two basic assumptions underlie DM research:
\begin{enumerate}
 \item[I] DM exists and is the single strongest argument for new Beyond the Standard Model (BSM) physics 
 \item[II] Discovering and understanding DM can guide our search for BSM physics
\end{enumerate}

The arguments  for DM are widely described and well known. Early hints of missing light versus gravitating mass were noted by Fritz Zwicky \cite{Rotverschiebung} and indicated by the flat rotation curves - the graphs of v(r)-velocities of hydrogen atoms or of stars, as a function of distance from the center of the corresponding galaxy measured by Vera Rubin and others \cite{rubinVC}. These were later complemented by evidence from the Cosmic Microwave Background (CMB) and "structural" data pertaining to measurement of the spatial distribution of galaxies and from the abundance of light nuclei produced in the Big Bang Nucleosynthesis (BBN). Gravitational lensing probes mass distributions in general and in and around pairs of recently collided galaxies such as in the famous "Bullet cluster" galaxies. Jointly all of these also ruled out DM which is made of baryons in various forms ranging in size from "atoms" to golf ball to "MACHOs" - Massive Compact Halo Objects - such as brown dwarfs, old neutron stars or black holes formed by collapse of ordinary compact baryonic stars. Thus DM is likely to entail completely new particles/fields and interactions.\par\noindent
Cosmological data from CMB and from structure and supernovae research instated the present "$\Lambda$ CDM Paradigm" where, in addition to dark matter and a smaller fraction of ordinary baryonic matter, we have also a bigger, enigmatic contribution of "Dark energy" or a Cosmological Constant mentioned in ref. \cite{Zeldovich:1968ehl} (D.E or C.C) - the origin of which is largely a mystery. The attention of the particle community still remains largely focused on dark matter particles or fields that were motivated by many theoretical ideas and are being searched over a huge range of masses and other DM properties.\par\noindent
The discovery of astrophysical data motivating DM happened roughly at the time when it was realized that a TeV scale supersymmetry (SUSY) may resolve the "Hierarchy Problem” of the low, $\approx$ 200 GeV scale of weak interactions relative to the natural ($m_{Planck} \approx 10^{19}$ GeV) cutoff -  and potentially also other problems in high energy physics. If R-parity, relating SM particles to their SUSY counterparts, is conserved, then the Lightest Susy Partner (LSP) is stable, making a most natural DM candidate. As briefly sketched in Sec IV, the "Freeze Out" relic abundance of TeV mass LSP's which have ordinary weak interactions, "miraculously" provides the observed contribution of DM to the total, critical, energy density. Such SUSY particles have substantial (and calculable!) rates of production and very specific decay chains terminating with LSP’s at the LHC (Large Hadron Collider). Many LSP Weakly Interacting Massive Particle (WIMP) candidates for making up the DM and in particular the DM in galactic halos have a large, spin independent, coherent scattering on heavy nuclei \cite{DMC} helping their detection \cite{Drukier:1984}. Also the annihilation of DM particles in the galactic halo into SM particles allows "indirect searches" of LSP DM. \par 
In the fifties and early sixties of the last century, working in theoretical particle physics required mastering  analytic functions and group theory. The first were used to investigate Lorentz invariant scattering amplitudes of strongly interacting particles. Unlike for elementary point-like electron, muon, neutrino and photon, a "bootstrap” approach was adopted where all strongly interacting particles (a.k.a, Hadrons) were bound/resonant states of each other in a self-consistent way. It explained the systematic of resonances and high energy behaviors of scattering amplitudes by "Regge" poles in the complex angular momentum  of the partial wave amplitudes. David Horn \& Christoph Schmidt suggested a "dual” description of scattering amplitudes by "Regge poles" in the t channel \textit{or} by resonances in the s channel. This and similar developments by Silvio Fubini, Daniele Amati and collaborators lead to the Veneziano and Virasoro amplitudes and to open and closed string theory.
The second approach focused on the internal symmetries generating multiplets of particles of similar masses and interactions. The octet version of Gel-mann and Ne'eman of the SU(3) flavor group led to quarks and to QCD -the gauge field theory describing hadrons. Bigger, albeit more strongly broken, symmetry groups, culminated in Grand Unified Theories (GUTs). In such theories all known gauge interactions: $SU(3)_{color}$ and electro-weak $SU(2)\times U(1)$ Grand-Unify into GUTs Gauge groups at a high energy scale. Both approaches are incorporated in relativistic local field theories.  We will encounter examples of each of the above two approaches. \par The three pronged search: in LHC, directly, using cryogenic underground detectors, and "Indirectly" by looking for the products of LSP annihilation using large telescopes/satellites and the underlying theory, dominated DM research for several decades. While some LSP variants remain viable DM candidates, the failure to-date of these searches to discover TeV SUSY particles has changed the field. Research is no longer guided by one, well defined, theoretical frameworks such as LSP SUSY WIMPs or DM made of axions\footnote{
The very light axions may accompany the resolution of the "Strong CP Problem", namely the need to suppress the CP violating term $\theta \tilde{G}_{\mu\nu} \ G^{\mu\nu}$ with $\tilde{G}_{\mu\nu}$ the $E \leftrightarrow B$ dual of the gluon field, suggested in \cite{CPviolation}.}. \par 
 The research field of DM owes much of its charm to its difference from the SM of particle physics, where experiments during the last forty years did not find convincing evidence for BSM physics. The future here is shrouded in mystery as any one of the many DM models (or combination thereof, or none) may turn out to be correct. It is worthwhile to be familiar with as many DM alternatives as possible and many practitioners work in multiple DM fields. 
\par \noindent 
Suggestions (and even performance!) of novel experiments often using condensed matter and Atomic/Molecular (AMO) physics to construct sensitive DM detectors were often made by HEP (High Energy Physics) theoreticians. Early examples are the suggestion of using resonant cavities to discover axion conversion into photons  \cite{Sikivie1983} and - the suggestion of searching for massive, extra neutrinos via kinks in $\beta$ decay spectra \cite{Shrock1980}. Conversly the experimentalist Andzej Drukier suggested with Kathy Freese \& David Spergel the annual modulations of WIMPs signals 
\par \noindent
That this requires broad if superficial knowledge lured me to this field. The following nicely illustrates this. A young postdoc gave a seminar on one aspect of one DM model. After her talk, she asked me what my research focus is. I answered "lately it is DM". She followed up with "What type of DM?" - "Most of them" I said and her eyes glazed, having concluded that I am a charlatan. I believe however that specialization is appropriate in mature fields such as astrophysics or condensed matter but not for DM, which does not have a single non-gravitational evidence.\par \noindent
 DM research may remind one of the wild west where each researcher/research group hastily makes new suggestions, staking out as many new territories as they can. However, as workers keep reminding their competitors, some basic rules apply. These include unitarity, causality, the need for an effective field theory description that allows separation of the relevant scales and consistency with experimental limits, cosmology and stellar dynamics.\par \noindent 
 Many alternative BSM theoretical DM frameworks besides SUSY and axion physics are presently being followed. The dark sector can be rather complex - a "Hidden Valley" separated from the SM by potential barrier "Mountains" (As noted, in particular in \cite{ZurekStrassler}. DM could have several components, reside in extra dimensions of various types and sizes, be a new light vector or Pseudo - Nambu-Goldstone, Pseudo-scalar boson (PNBG) associated with (un)broken symmetries in the dark sector, or could be made of topological objects: monopoles, cosmic strings, (reviewed by \cite{AlexanderVilenkin} (see also work on cosmic (E.M) superconducting strings by Eduard Witten \cite{WITTENEDUARD}), be Coleman's Q balls \cite{SidneyQ} etc. \footnote{Q balls are coherent collections of bosons sharing a conserved global quantum number and having an effective potential $V(\phi)$ with a local minimum at non zero $\phi$. Non-Relativistically this can reflect overall attraction of the bosons at low temperatures.} \par 
 
BSM physics and new types or new properties of dark and/or of ordinary matter were often suggested by anomalies - such as an apparent excess in a DM detector or other experiments testing the SM \footnote{
$^*$Such an anomaly has occurred in the discovery of the first electron neutrino by C.L. Cowan and Fred Reines \cite{neutrino}. An underestimate of the reactor neutrino flux and an upward fluctuation of counts suggested that weak interaction failed to predict the large rate of the neutrino interactions. This and repeated wrong claims by Reines of neutrino oscillations in short, high energy neutrino beams, caused the lapse of 40 years between the discovery and its Nobel recognition}.
A partial list of anomalies includes the apparent proton decay events seen by Kamiokande and IBM groups, tachyonic neutrinos suggested by several Tritium end point spectrum measurements ,the "high y” anomaly in neutrino scattering (y is the fraction of energy carried by the produced lepton), the $17 KeV$ neutrino found as a kink in the tritium decay spectrum, the $\sim eV$ majorana neutrinos suggested in searches for neutrinoless double $\beta$ decay, the apparent mixings with a $4^{th}$ sterile neutrino of order eV mass suggested in the Los Alamos neutrino experiment (Liquid Scintillator Neutrino Detector LSND) and by experiments done near nuclear reactors (the Mini/Micro Boon experiments - designed to test this effect - tended to exclude it, but found new anomalies of their own), the "fifth interaction" a bit weaker than gravity and of O(Km) range suggested by "Reanalysis of the Eotvos Experiment" \cite{Fischbach1985}, the apparent violation of the "GZK" (Greisen–Zatsepin–Kuzmin) upper bound on cosmic ray energies\footnote{ 
Kenneth Greisen \cite{greisen} and G.T. Zatsepin \& V.A. Kuzmin \cite{ZK} noted that upon traveling $\sim$100 megaparsecs in the CMB background of $\sim$400 $\text{photons}/cm^3$ of few tenths of milli electron volt energy, cosmic rays protons of energy higher than $\sim10^{20}$ eV collide with CMB photons forming the $\Delta$(1238) resonance which decays to a pion and a lower energy nucleon. This cuts-off the CR energy at $10^{20}$ eV. An analog cutoff of $\gamma$ energies is implied by electron-positron pair production off star light in the galaxy.
Near threshold where: $$ 2 E(\gamma) E(\text{starlight})=4m_{e}^2$$
$\gamma\gamma-> e^+ e^-$ has a cross-section of $$\sigma_{\text{ann}}( s\sim 4m_{e}^2) \sim \sigma ( \text{Thompson}) \sim 6 .10^{-25} cm ^2$$ with eV starlight energies. This prevents photons of energy $E \gg TeV$ from crossing the galaxy}, the mono-jet events in UA(1) and UA(2) CERN experiments suggesting the discovery of many SM and SUSY particles, the Attic, Pamela, and Alpha Magnetic Spectrometer (AMS) groups finding an excess of low energy anti-protons and positrons in 
 cosmic rays, the Colar mine HE muon cluster, the Integral detector anomaly suggesting a new $U^{\prime}(1)$ vector particle coupled to MeV DM, the gap in the Ice-cube spectrum, possible indication of very heavy D.M. particles that decay or annihilate into neutrines, the 3.5 KeV X-ray suggesting new physics in the neutrino sector and the "magnetic" modes in the pattern of the CMB polarization seen by the BICEP collaboration suggesting a tensor part in the density fluctuations due to gravitational waves emitted in a high scale primordial inflation.\par\noindent
More recently we had some indications of a 700 GeV Di-photon resonance in the ATLAS detector at the LHC, the Ryken anomaly in a rare $K_L$ decay mode, the inconsistency between Lamb shifts in ordinary and muonic Hydrogen, the upward moving UHE neutrinos seen in one of the ANITA balloon flights, the overly cooled "21 Cm'' line suggested by the EDGE collaboration, the excess of low energy O(KeV) events in the Xenon 1T underground experiment, the neutron life-time anomaly, the muon g-2 apparent disagreement with calculation of the hadronic contributions, the apparent violation of lepton universality in weak decays involving $2^{nd}$ \& $3^{rd}$ generation fermions, the slight discrepant abundance of lithium, the Hubble "Tension" between determinations of the Hubble "Constant" using CMB and near type 1-a supernovae and many more apparent inconsistencies between different cosmological/astrophysical measurements and between such data and the $\Lambda$ CDM "standard" cosmology (See \cite{snowmass}). \par\noindent
Most anomalies disappeared with more data, reanalysis and/or better experiments. A careful analysis explained the CERN "Mono-Jet” events as misidentified $\tau$ lepton decays, more data at the highest CR energies tend to reinstate the GZK bound, a Berkeley experimental group ruled out the 17 KeV neutrino\footnote{The leader of this group, the late Stuart Freedman, said that "given a sufficient budget any anomaly can be ruled out". A better rendering of Archimedes saying that "given one stable point to put my lever on I can move the earth" is that "One precise experiment (or one solid theoretical result) can kill thousand wrong theories"}, the muon Lamb shift anomaly was due to overconfidence in electron atomic co-data, the ANITA UHE earth traversing events have not recurred in more recent balloon flights and the NUSTAR and XRISM X ray satellite did not find the 3.5 KeV line suggested by data of the XNM satellite, the 750 GeV diphoton state was not seen in the CMS experiment at the LHC, and with more data, it faded also in ATLAS. Recently the anomalies in $B \rightarrow K ^{*} \rightarrow e$ versus $K^{*} \rightarrow \mu$ decays disappeared. Finally, it has been suggested by Wendy Freedman and collaborators \cite{Freedman_2021} that different methods using special stelar objects yield different values of the Hubble constant and may alleviate the Hubble Tension \cite{Freedman2024}. This and other works, using the difference of different gravitationally lensed quasar images weakens the "Hubble tension" anomaly. We encounter later some more anomalies most of which may soon disappear.\par\noindent 
Theoretically studying such anomalies, and the new BSM scenarios they suggest, enriched the field. Also, the "Solar Neutrino Anomaly" - the paucity of the high energy solar "Boron ($\beta$ decay) Neutrinos" expected from \hspace{0.1cm} $p \hspace{0.1cm} + \hspace{0.1cm} ^7Li \hspace{0.1cm} \rightarrow \hspace{0.1cm} ^8B + \hspace{0.1cm} \gamma$ \hspace{0.1cm} reactions - found by Ray Davies and championed by John Bahchall, was the first indication for massive/mixing neutrinos. Equal credit goes to the atmospheric neutrino "anomaly" - the deficit of events induced by $\mu$ neutrino interactions relative to $e$ type neutrino events. These and D.M. are the only solid BSM physics we have to date. \par\noindent 
Efforts to exclude/incorporate/explain anomalies are often referred to as "Ambulance Chasing". Sheldon L.Glashow, Vernon Berger, Rabindra Nath Mohapatra, Savas Dimopoulos, Maxim Pospelov, Katerin Zurek and many other prominent physicists raised such "Ambulance chasing" to a level of an art and many physics including Phd works were inspired by these. Often careful comparison with other data casts doubt on the validity of the "Anomaly". Thus Andrew Cohen and Sheldon Glashow discounted Opera's Claim of faster than light neutrinos by invoking "Cherenkov" radiation of tachyonic neutrino pairs \cite{superluminal}. This most sensational anomaly was finally resolved by tightening a loose contact in the experimental set-up. \par\noindent 
 The physics community is rather lenient in its attitude to "wrong" theoretical papers. When foraying ahead into the unknown, theorists suggest many models. Models which do not conflict with data or basic principles and are not mathematically flawed, are well received - even though later we find that nature adopted just one or none of these suggestions. GUT, Susy, LR symmetric models, Mirror, Twin Higgs, Axions, Dark photons and most DM models may be of this type. In designing experiments aiming to test theoretical suggestions, experimentalists tend to follow the theoreticians with the best track record. This paid off accidentally when groups attempting to detect proton decay predicted in GUT's (Grand Unified Theories) built large underground water Cherenkov detectors. They did not find proton decay but detected atmospheric neutrinos and neutrinos from the 1987a supernova to which we will return later.\footnote{GUTs were suggested by \cite{PattyAbdus} and the specific SU (5) version by \cite{PhysRevLett32}. The latter predicted proton decay, the unification of all gauge couplings, charge quantization, reasonable values of $sin^2(\theta_W)$ and $m(b)/{m(\tau)}$.
 Also the renormalization group (further described in Appendix H) flows of the electric, weak and QCD couplings tend to unify them at a scale ($\sim 10^{15} GeV$) \cite{UGT}.}
The large investments required for doing dedicated experiments do not allow checking all the new anomalies and/or theoretical suggestions which often last only for a short period.\footnote{When asked "what could be the lifetime of the 17 KeV neutrino?" a question relevant to the astrophysical implication of this neutrino "discovered" at the time in several experiments starting with \cite{Simpson:1985xc}) the great experimental physicist Maurice Goldhaber replied "No more than a year" and he was right!} \par\noindent
Experimentalists claiming important discoveries which turn out to be wrong are, however, treated most harshly. Indeed, new, unexpected experimental results, if correct, can suggest  whole new theories and motivate larger/better and more expensive experiments aiming to exclude or confirm them. A famous case is that of Joe Webber, a pioneer in Laser research and Gravitational Wave (GW) detection. His wrong claim of discovering GW made it to the Encyclopedia Britannica, to the chagrin of most physicists. The Israeli astrophysicist Dror Sadeh from TAU (Tel Aviv University) installed sensitive seismographs in caves near Eilat. Finding signals with a period close to that of a pulsar, he claimed that it was a GW. It was way stronger than expected and may have originated from near-by spy submarines in the red Sea. Indeed among  $\sim 150$ observations of GW‘s events at LIGO and VIRGO, not a single event with a pulsar frequency was found, implying very tiny time varying quadrupole moments of pulsars .\par\noindent 
 Fear from such a backlash made Sam Ting force his group to keep secret the finding of a striking narrow peak in the invariant mass of $\mu^+ \mu^-$ pairs produced in a fixed target experiment at Brookhaven. This peak, also hinted at, but initially unresolved, in the SLAC $e^+ e^-$ collider, was the all important J/$\psi$ particle made of $\bar{c}c$ new quarks.\par\noindent 
 Some suggestions of new models, new analyses or new experiments were "Dead on Arrival D.o.A" as experiments and/or theories excluded them already at their inception. To reduce the noise in the field these have to be taken back ASAP, following the example of Sidney Coleman: learning that a recent paper of his is flawed he wrote to the $\sim 600$ recipients of Harvard preprints "my Paper xxx is wrong, throw it in the trash can!". \par\noindent 
Many different DM types have been suggested. A partial list includes: cold/hot DM particles which were N.R/Relativistic when they decoupled from radiation, asymmetric/symmetric DM, Bosonic-Fermionic DM (with Bosonic DM including field -like coherent DM), elementary/composite DM, iDM - inelastic (excitable) DM, Nugget/Solitonic or more generally DM made of a (very) large collection of particles often with nuclear density, self interacting DM (SIDM), Dissipative DM, Unstable DM - which decays on super-hubble time scales, H-L"Heavy"-”Light" ($m<GeV$) DM, VH - Very heavy "Wimp-zillas" of masses $\geq100$TeV, RDM - (presently) relativistic DM, WEI-MUMps and ST DM - Weakly \& Medium and Strongly interacting (with SM particles) DM, Fuzzy/Feeble ultra-light DM with galactic size de-Broglie wavelengths, and ultra weakly coupled particles, FO DM - which was in thermal equilibrium and "Froze Out" to the correct $\Omega(DM) \sim 0.25$, FIDM "Frozen -In" in out of equilibrium processes, SGL -MUL models with a single or multi-DM component, SI/SD - with spin (in)dependent interactions with SM particles, ST DM made of stellar mass objects, CSDM - Cosmic String DM, D.M. models based on a new partially broken U(1) symmetry with non- zero mass dark photons kinetically mixing with the SM photons or an unbroken U(1) with ensuing mili-charged DM, CON/NCO DM models which are in conventional cosmology with radiation \& matter and cosmological constant dominations occurring in this order or DM within Non Conventional cosmologies (such as cosmologies with early matter domination era which transited back into the regular radiation dominated epoch by a decay of some unstable heavy DM into light particles), D.M. which underwent various phase transitions, co-annihilating \& resonant DM, DYDM - Dynamical DM with a conformal like continuum spectrum suggested in ref. \cite{dienes2022differentnonminimaldarksectors}), DM made of "Q balls" and many more. Only a small fraction of this wealth will appear in this review. \par\noindent 
To fully specify a particular DM requires:
\begin{enumerate}
 \item[1] The BSM scenario in which it could arise
 \item[2] The specific dominant characterizing property that it is required to have
 \item[3] The experimental methods that could  be used to discover (or to exclude) it
\end{enumerate}
Most combinations do not correspond to a viable DM candidate and arguments why some are consistent and others are not, make up much of DM research and of this review.
\par
DM models and BSM physics in general, were strongly constrained by demanding their consistency with SM gauge invariance, renormalizability and freedom from triangular anomalies. In effective Lagrangians, terms of dimensions $d>4$ are suppressed by $(E/{\Lambda})^{d-4}$ with E the low energy of interest and $\Lambda$ the scale of the corresponding "U.V complete" renormalizable theory which yields the Effective Lagrangian at low energies. The Effective Lagrangian approaches are becoming increasingly popular. The freedom in constructing new DM models is much larger within the effective Lagrangian framework. Thus, the double insertion of weak Isospin  $I(W) = \frac{1}{2}$ Higgs fields or their V.E.V.'s (Vacuum Expectation Values) generate the Weinberg $\Delta(L)=2$ term where $L$ is the Lepton number, leading to Majorana neutrino masses. Such masses were achieved in a particular, renormalizable field theory, by the "Majoron", a Goldstone boson associated with spontaneous lepton number violation having Yukawa couplings to the light, left-handed neutrinos. \par\noindent 
 The volume and quality of data ,which can weigh in on DM/BSM physics, is increasing at a very fast rate. This includes the bounds on coherent nuclear cross-section of WIMP DM obtained using large underground liquid XENON detectors. For masses $M(X) \sim 50 GeV$ the impressive upper limits approach $\sigma(X-N)= 10^{-48} cm^2$ - (N=n,p a neutron or a proton) close to the "neutrino floor" - the irreducible low background of solar neutrinos and geothermal anti-neutrinos. Lighter DM is being searched for via its interaction with electrons, molecules and various new microscopic/ mesoscopic forms of ordinary matter.\par
 Surveys of large and "small" scale structures and many body/hydrodynamical simulations keep improving. These and precise measurements of the CMB lead to strong limits on hot components of DM such as the upper bound of 0.12 eV on the sum of the masses of the three neutrinos. \par
Further improved tests of the close-to-perfect black body spectrum of the CMB at both the Raleigh-Jeans IR and the short wavelength (Wien) tails, limit photonic decays of light and somewhat heavier (Pseudo) scalars or of dark photon into three photons. Precise measurements of the CMB anisotropies at ever decreasing angles - higher angular momenta - are being performed or planned, yielding critical information on cosmology and on structure formation. Many measurements of the "21 cm” line are ongoing or being planned. They seem to exclude the EDGE anomaly which may have suggested extra, efficient cooling mechanisms at the time of "Cosmic Dawn” ($z \approx 17$). \par\noindent The "line imaging technique" of looking in a particular direction and inferring the "depth” - namely the redshift z from molecular/atomic lines, can help in the study of the earliest stars and galaxies \cite{Creque_Sarbinowski2018}. \par\noindent
Gravitational lensing keeps yielding information on the distribution of gravitating matter and helps find evidence for DM (in particular DM with limited self-interactions) in "colliding” galaxies in galaxy clusters. Micro and Femto Gravitational lensing are used to look for extrasolar planets. The ”Haze” expected in such experiments from certain mass B.H.s and various types of "Micro-haloes” may, as possibly indicated in recent ALMA observations, exclude such BHs and microhaloes of $\sim$ earth mass \footnote{ Gravitational lensing is due to the bending of light originating in a far source by intermediate mass(es) along the line of sight. Unlike lensing in a microscope or telescope the multiple/enhanced/distorted images of the far source are used mainly to study the "lense" namely - the intervening (Baryonic and Dark) matter}, The GAIA and future LSST projects of astrometric measurements of $\sim$ a billion stars will yield a high precision phase space distribution of halo + galaxy baryonic and dark matter. It also vastly increases our knowledge and understanding of various stellar types such as White dwarfs, Neutron stars, and B.H.s which are in binaries and emit radiation. \par
The improving measurements of cosmological abundance of light nuclei limit DM and radiation which could influence primordial nucleosynthesis at temperature of ~ 0.1-1 MeV and spoil the largely successful prediction of Big Bang Nucleosynthesis (BBN). In particular, these led to the upper bound on $\Delta N_{eff}\le 0.2$ additional light D.o.F. These D.o.F could be 0.2 extra neutrinos, or the equivalent amount of light dark photons, light BSM gluons or axions and other light scalar/vector particles. All of the above mentioned particles can affect the rate of expansion at the time of BBN.\par
The careful observation and analysis of the evolution of all stars: the sun, red giants, white dwarfs, supernovae and neutron stars/pulsars of various ages, limits DM/BSM scenarios (G. Raffelt in \cite{Fundamental}). This is achieved by considering the heating up of the stars due to accretion of DM, or of the excessive cooling of various stellar objects by volume emission of sufficiently light and weakly interacting DM/ BSM motivated particles. Particularly intriguing is the generation of energetic neutrinos in the solar or in the Earth's core by the annihilation of heavy DM collected therein. Also very heavy DM particles ("Wimp-Zillas") can be captured in neutron stars, fall to the in centers, and form black holes which, in turn, "eat up" all old neutron stars. \par 
Measurements at various high intensity fixed target accelerators or at colliders, done beyond thick shielding or at beam dumps, search for long lived particles which could be DM candidates. Of particular note is the proposed FASER detector "Forward Search ExpeRiment at the LHC" \cite{Feng2017} and \cite{Feng_2023} to search for non-Wimp lighter DM via distant detectors placed in the direction of one of the beams colliding in the CMS detector at the LHC.\par
The Webb telescope has already collected much data at wavelengths of a meter and above. The highly redshifted "21 cm" line found indicates that star/galaxy formation may have started earlier than anticipated. \par 
Detailed analyses of the CMB may reveal non-Gaussian fluctuations. This and also G.W's emitted early on from "Bubbles" in some first order phase transitions may lead to a better picture of the very early universe as a "Cosmological Collider" \cite{arkani} a modern reincarnation of Yakov Zeldovich's "The early universe as a hot laboratory" \cite{zeldovitch} or \cite{Sunyaev:1970er} 
\par\noindent 

Many other optical UV/X ray and $\gamma$ detectors carried on satellites are presently operational or planned. They will find or limit putative annihilation of symmetric DM, the direction of which should be correlated with regions of increased DM density. There are many radio telescopes, radio arrays and large multi-ton water Cherenkov facilities underground like Super-K or spread out at high altitude like HAWC (High Altitude Water Cherenkov) . All of these will help find the origin of cosmic rays of various energies and of Gamma Ray Bursts (G.R.B). Recent studies of Fast Radio Bursts (F.R.B) lasting a few milliseconds tend to correlate them with magnetars - namely pulsars with very strong magnetic fields. UHE neutrino telescopes like the enhanced Ice-cube and the Askarian array at the south pole will continue probing cosmological neutrinos\footnote{$^*$ Gurgen Askarian suggested that an underground, almost horizontal extensive shower develops in a grazing collision of an UHE neutrino or cosmic ray, the preferential absorption of positrons generates a negatively charged residual propagating object that emits strong radio Cherenkov radiation. It is particularly relevant for the Antarctic ice sheet which is transparent to much of this radiation. Askarian's free spirit was tolerated by the soviet regime due to his invention of tiny electronic devices later used in constructing spying bugs}.
\par
The multi-messenger approach where gravity waves and excess counts in many of the detector types are time and direction correlated was remarkably successful in the detection and measurements done on the binary neutron star merger at a distance of 400 million light years. In particular, it verified the equality of the speed of light and of G.W’s that arrived within a few seconds \footnote{This also limits the difference between gravitationally induced delays of the photons and G waves from a distant galaxy upon leaving the parent galaxy and  entering our milky way. Recent observations greatly improve previous upper bounds on the difference of the velocities and gravitationally induced time delays of neutrinos and photons inferred from the few hour delay between the neutrino and em signal in supernova 1987a, Also the sharp features seen in the cosmological GRB’s (Gamma ray Bursts) strongly limit such deviations \cite{Amelino-Camelia1997}}.\par \noindent

Most improved optical, I.R, radio, $\gamma$ ray earthbound or space telescopes, many CMB, X ray and cosmic ray detecting satellites and GW detectors, originated in the astronomical community. Their serving also as DM searches is an unexpected bonus.\par\noindent
Finally, the great advances in the precision of atomic clocks (and the specter of a $10^5$ more accurate clock based on nuclear isomers looming), quantum cantilevers, and various other aspects of atomic/molecular and new forms of matter such as Graphene, allow new classes of table top experiments. These experiments can look for short (Micron) range deviation from Newtonian gravity, put better bounds yet on the neutrality of matter and search for milli-charged DM. Axions $\rightarrow$ photon Conversion can be most sensitively searched in resonant cavities such as described in  \cite{ADMX_Collaboration}. Axion, dilaton, and ultra light dark photon fields can show up in measurements of small transient changes of physical constants such as charge, magnetic moments or Newton's $G_N$. Interference of cold atoms falling through a 100 meter deep vertical pipe testing various aspects of gravity and fifth interactions was considered. Related ideas for GW detection use Atomic Interferometry (MAGIS) or superconductors.\par \noindent While many individuals are involved, Savas Dimopoulos was particularly influential directing towards this research many outstanding students including Asimina Arvanitaki, Peter Graham, Surjeet Rajendran, Ken Van Tilburg, Masha Bakhtiar, Asher Berlin, Junwu Huang and many more. "A Search for Variations of Fundamental Constants using Atomic Fountain Clocks" \cite{marion2003} is an early work in this field. 
  \par\noindent
 The detection of massive DM particles by using the tiny gravitationally induced oscillations during their passage, induced in a micro-pendulums array (White paper\cite{carney}) or of axion clouds with a global net of sensitive magnetometers and of DM chunks with a network of accelerometers were discussed.\par\noindent
We cannot describe all the many experimental methods. However in Sections XI-XIII we recall several suggestions of using stellar observations and temporal and/or directional information to enhance the sensitivity of DM searches and later in sections XXI we list few other "Gem" ideas for detecting or limiting certain types of DM. Two special classes of DM will be discussed in sec. XIV-XX:
\begin{itemize}
\item[A)]DM types which were especially designed to be detectable by an appropriate experimental technique such as the "Self-destructing DM," DM clustering on earth size scales, DM that was accelerated to relativistic energies, "Resonant DM" and more.

\item[B)]DM types made of sterile right handed neutrinos and of Primordial Black Holes( PBH’s) - both being well defined and understood objects but finding credible scenarios for producing their correct relic density is challenging.
\end{itemize}

\noindent 
Sec. II is a short excursion into cosmology. It mentions the initial strong inflation - the big bang and the various stages of standard evolution leading to BBN and CMB and terminating with the present expansion. Cosmology provides among others the $\Delta Neff \leq 0.2$ limit on the number of light D.o.F (Degrees of Freedom) of DM which could affect BBN and CMB. Finally we mention present inflation and the problem of the tiny cosmological constant (CC) and the tiny present acceleration.\par \noindent 
Section VII and VIII describe models attempting to explain the $r\sim 5$ ratio of dark and baryonic matter contributions to the critical cosmological energy density and the fermion mass hierarchy problem is described in Section IX. A rather unconventional approach which may relate the ratio r to the existence of three fermionic families is mentioned in section XXV. \par\noindent
 \noindent
The last sections present the "bit more" in the title. 
In sec. XXV we consider a most strongly self-interacting DM which can arise in the BSM "Quirk" scenario with TeV mass "Quirks" which carry both ordinary color and a new color'. An extreme version with a very low new color scale $\Lambda$’ is of particular interest. Despite the tiny production rate of $\sim$ one in $10^{14}$ pp collisions in LHC-like colliders, extracting a handful of Quirks produced and stuck in neighboring rocks from the $\sim 10^{34}$ atoms therein may be feasible. As shown in sections XXV and XXVI and associated appendixes,  this scenario may allow a novel method for communication over galactic distances. The novel communication technique may complement the SETI search for intelligent civilizations by radio waves which we discuss in section XXIV. If the Quirks are ultra heavy or absent, then the extension of the SM to include the Quirky gauge interactions can be non-refutable nor confirmable. Along with the DM model in Section XXVII, with completely neutral non-interacting Planck mass particles and the multiverse concept, these may provide crude analogs of Goedel's undecidability in physics. The anthropic principle (AP) is briefly revisited in Sec XXVIII.\par\noindent
Footnotes explaining various technical terms are often included. To avoid early crowding of such footnotes they often appear in later chapters. Also, to avoid extra length, some references appear only in the text. F.N$^{*}$ indicates footnotes of some historic interest. Relying often on second hand information these footnotes may be inaccurate. F.N$^+$ refer to the few new suggestions made and F.N${^{++}}$ to footnotes which are highly speculative and barely connected with the main text yet of some interest. In discussing DM/BSM issues I try to provide intuitive understanding.  \par
Georg Raffelt book on "Stars as laboratories for H.E. physics" \cite{Fundamental}, the books by Rabindrana Nat Mohapatra \& Palash B. Pal \cite{Mohapatra:1998rq} and by John N. Bahcall \cite{science1996review} have been extensively used. The Springer Theses book "Beyond the BSM Cocktail" by Yann  Gutternoier along with the $\sim 20$ expert summaries in the PDG (Particle Data Group) of various SM, BSM/Dark matter, astro-particle physics and cosmology issues and many references therein are very helpful.
A monumental review of D.M. by Marco Cirelli, Alessandro Strumia \& Jure Zupan appeared recently \cite{cirelli2024darkmatter} has an extremely broad scope and exhaustive referencing.

\section{\hspace{\secspace}An introduction to cosmology}
\textbf{in which we most briefly introduce this vast discipline which reflects on DM.}\par\vspace{5pt}
\hspace{1cm}

Einstein deserved Nobel prizes for special and general relativity. He also predicted Bose - Einstein condensates and gravitational waves ($GW$). The Maser and Laser use his “A\&B coefficients” for spontaneous and induced emission, quantum entanglement followed from the EPR (A. Einstein, B. Podolsky and N. Rosen) paper \cite{epr} and the recent discovery of the accelerated expansion connects with his cosmological constant \footnote{$^*$ Amusingly, the prize he did get for the photoelectric effect was not obvious at the time.The astronomer Royal Sir James Jeans ridiculed photons in the fourth, 1925 edition of his book "The Dynamical Theory of Gases" Dover publication by noting that a diffraction picture of a very far star appears despite having at any given time at most one photon originating from this star in his telescope and "Clearly a photon is not able to interfere with itself!". Little did he know that this ability \textit{defines} quantum mechanics!.  Having the frequencies of the resonantly absorbed radiation match those of the matter oscillators is an alternative explanation - just as in Planck's derivation of blackbody spectrum. In hindsight, photons and light, the first case of particle-wave duality, inspired  QM and are one of Einstein's most important discoveries.}.\par\noindent   
In a giant leap of faith Einstein applied the equation of GR: 
\begin{equation}
    \label{GR1}
    G_{\mu \nu}-G g_{\mu \nu}=G_N \hspace{0.2cm} T_{\mu \nu}
\end{equation}
which explained the anomaly in the precession of Mercury's orbit, to the whole universe. $G_N$ is Newton's constant, $G_{\mu \nu}$ is Riemann’s curvature tensor, $ G= G_{\mu \nu} g^{\mu \nu}$ it’s trace, $T_{\mu \nu}$ is the energy momentum tensor. The metric $g_{\mu \nu}(x,t)$ fixes the line element $(ds)^2= g_{\mu \nu} dx^{\mu} dx^{\nu}$ and other relevant quantities are constructed from products of $g_{\mu \nu}$ and its derivatives. The upper and lower tensor indices dictate behavior under local Lorentz transformations\footnote{$^*$The many indices in GR underlie a Feynman story. Arriving a day late to attend a GR conference he forgot which of the three universities in the Raleigh triangle hosts the meeting and asked the cab driver at the airport to go where people mumbling $\mu\nu, \mu\nu$ went the day before}.
Ordinary partial derivatives $ \partial / \partial(x_{\mu})$ are not
 invariant under local reparameterization transformations. However the result that the rotation angle of a line element upon "parallel transport" around a closed orbit equals half the integral of the Gaussian curvature over the enclosed surface is invariant. Thus traversing the equator of a two dimensional sphere enclosing half of the area ($2 \pi$) of the sphere we return to the starting point from the opposite $-\pi$ rotated-direction. This holds also for the "Covariant derivative" which accounts-like in internal gauge theories-for the changes induced by local curvature in the parallel transport of the quantity in question. \par \noindent 
In popular lingo it is stated that massive bodies (and energy in general) "curve space". According to Fermat's principle, light rays follow the geodesics on the curved manifold. The world line of massive test particles minimizes another ,modified world line action. As our focus here is on the big-bang we largely skip these issues. \par
The uniform, static universe that Einstein hoped for is unstable against gravitational collapse  To stabilize it he introduced a "Cosmological Constant" (CC) term $\Lambda \hspace{0.1cm} g_{(\mu\nu)} (x,t)$. When confronted with evidence for the Hubble expansion and a time-dependent cosmology he called the cosmological constant "My gravest error". Present evidence for an accelerating universe and the strong arguments for an early "inflatonary" period of a fast expansion, suggest an additional "Dark energy" (D.E.) component. Unlike ordinary radiation/relativistic particles or cold baryonic/DM with Eq. of state: $ $pressure$ =p=\frac{\rho}{3}$ or p$\sim$0, the E.o.S of dark energy is the unusual p=-$\rho$. The simplest "dark energy" scenario adds a positive CC term to the energy momentum tensor on the R.H.S of the basic GR equation transforming Einstein's error into yet another brilliant prediction \footnote{The tiny positive CC corresponding to the present day expanding De-Sitter space seems irrelevant to fundamental particle physics. Yet the remarkable ADS-CFT (Anti-de Sitter Space-Conformal Field Theory) duality of Juan Maldacena, a central idea in modern day theory - which will not be discussed further here - favors an ADS space corresponding to a negative cosmological constant which readily accommodates local supersymmetry and conformal field theories.}.

The Cosmological FLRW (Friedman, Lemaitre , Robertson ,Walker) time dependent metrics:  
\begin{equation}
    \label{GR2}
ds^2= dt^2 -a(t)^2 [\frac{dr^2}{1-kr^2} + r^2( d\theta^2 +\sin^2\theta d\phi^2 )]         
\end{equation}
describe for k = +1, -1 or 0 an open, closed or critical expanding universe, and a(t) is the time dependent, homogenous and isotropic scale factor.

 Using the above metric to calculate the curvature tensor in Eq \ref{GR1} yields for a homogeneous energy momentum tensor $T(\mu \nu)$ the evolution of the scale facor a(t) : 
\begin{equation}
    \label{GR3}
    \begin{aligned}
 H^2 +\frac{k}{a^2} = \frac{\rho}{3m(PL)^2}, \hspace{5mm} with \hspace{3mm}\ m(PL)^2 =\frac{1}{8\pi G_N}
 \end{aligned}
\end{equation}
where $\rho$ is the energy density and H = the Hubble “constant” is defined as [da/{dt}]/a. Also  
 \begin{equation}
    \label{GR4}
        \frac{d^2(a)/{dt^2}}{a^2}=\frac{\rho-3p} {6m(PL)^2}     
\end{equation}
The conservation of the momentum energy tensor $\partial_ \mu T^{\mu \nu} =0$ leads to:
 \begin{equation}
    \label{GR5}
        \frac{d\rho}{dt} = - 3H(\rho+3p)            
\end{equation}                         
 Present observations suggest a flat universe with k=0 and a corresponding critical density
\begin{equation}
    \label{GR6}
        \rho(c) =\frac{3(m(PL)^2}{H^2}           
\end{equation}
\noindent
 Using the present H=70 (Km/Sec)/megaparsec, \hspace{0.2mm}yields $\rho(c)$ $\sim$ 5 Kev.$cm^{-3}$. Allowing for k=1 or -1 leads to
\begin{equation}
    \label{GR7}
 \Omega=  \rho/{\rho(c)}=\rho/3m(\text{Planck})^2 H^2 = 1 \pm k/(aH)^2.        
\end{equation}
\noindent
$\Omega(X)$ refers to the fractional contribution of the specific component X = DM, X= Baryons or X= radiation to the critical density $\Omega= 1$ corresponding to a flat universe which inflation naturally provides.\footnote{$ \rho$(crit) is also the minimal average density $M/R^3$ of “autonomous” regions such as galaxies or clusters thereof which do not participate in the Hubble expansion. The inward gravitational acceleration  $G_N M/R^2=M/R^2 (m(PL)^2$ of a test mass at the edge of such a structure then exceeds the opposite effect:
$ d/{dt} [ v(Hubble] = R H^2 $  where $v(Hubble)=d(a)/dt=aH$} The Hubble constant which specifies the fractional change  of the scale factor H=[da/dt]/a  controlling the growth of the redshift factor z  or of the relative velocities of galaxies with their mutual  distance, is often written as  H= h. [100(km)/sec ]/ {Kilo-Parsec}\par
The energy density consists of a radiation + matter (baryonic and dark) and presently a DE or CC  like- contribution which is $\sim$3 times that of matter. The time dependence of H, $\rho$ and other quantities during the various cosmological phases are fixed by the EoS  of  the component of $\rho$ which dominates at that time. \par 
The other hallmarks of the big bang besides the Hubble expansion, the CMB and BBN were suggested by Lemaitre and by Gamow.\footnote{$^*$ Gamow’s popular science books were most influential. His unique sense of humor made him add the name of Hans Bethe between Alpher \& Gamow on a paper on stellar nuclear reactions.} The abundance of Helium and other light elements could not be explained by nuclear reactions in stars. The adiabatic Hubble expansion cools by the $(1+z)^{-1}$ "Red-shift" factor the initial "fire-ball" where the light elements were produced and which became the present CMB - a feature emphasized by Robert Dicke and John Peebles. \par
The radiation dominated era (RDE) extends over a large range of red-shifts terminating when matter and radiation energy densities become equal at $z\sim 10^4$ and while non-trivial is the simplest cosmological era. During most of this era  $T\gg m(i)$ and $T\gg U(int)$ with U the potential energy due to interactions between the particles.
The homogeneous energy momentum tensor $T(\mu\nu)$ is diagonal with $T_{00}= \rho$ and $T_{11}=T_{22}=T_{33} = p/3$. The EoS $\rho=3p$ of free relativistic particles reflects E=P, the equality of their momentum and energy. It implies that the trace of $T(\mu,\nu)$ vanishes.\footnote{Emmy Noether’s theorem and the definition of the energy-momentum tensor as $\delta{L}/{\delta(g(\mu,nu)}= T(\mu,\nu)$ with L the Lagrangian then allows inferring that a traceless $T(\mu,\nu)$ (with $\rho=3p)$ implies invariance under ($x(\mu) \rightarrow \lambda x(\mu)$) scaling. This allows S matrix elements to depend only on ratios of and angles between momenta. In Lorentz invariant field theories this leads to a more extended conformal invariance which includes inversions and fixes not only the form of the two point propagator but also of the three point vertex function.} \par
The $SU(3)\times SU(2)\times U(1)$ SM field theory has, apart from the scalar Higgs sector, no mass terms making it scale invariant and conformal for temperatures/energies higher than $ <H> \sim m(\text{Higgs})$ $\sim$ breaking scale of the E.W part. Much higher masses of order $10^{15}$ GeV in say GUT are irrelevant as all the massive particles are very short lived. Stable particles are the exception and as shown in section IV,  their relic “Freeze out”, abundance is fixed by the $X-\bar{X}$ annihilation cross section.\par
The measured CMB frequency/energy spectrum is remarkably close to an ideal Planck distribution for $E(k) =\sqrt{m^2+k^2}$  appropriate for massive and massless bosons:
\begin{equation}
    \label{GR8}
f_B(k)= {\frac {k^3}{E(k)}}\frac{ e^{\frac {-E(k)}{T}}}{1-  e^{\frac {-E(k)}{T}}} \hspace{4mm}\ \text{for photon}\ \hspace{4mm}  E(k)=k=\omega(k)
\end{equation}
with present CMB temperature T(photon now)=T(z=1)= 2.78$^o$ Kelvin. $\sim2.5 \, \, 10^{-4} eV$ \par
Beside the Bosonic photons we have today the Fermi-Dirac distributed neutrinos:  
\begin{equation}
    \label{GR9}
{f_{FD}(k)} = {\frac {k^3}{E(k)}} \frac  {e^{\frac  {-E(k)}{T}}}{ 1+ e^{\frac {-E(k)}{T}}} \hspace{0.3cm}  \text{with} \hspace{0.3cm} E(k) \sim k  
\end{equation}
As noted below, the latter is at a lower temperature of $\sim 1.9^o$ K. The energy and number densities of the CMB photons and of the neutrinos obtain by integrating the distributions: 
\begin{equation}
    \label{GR10}
    \begin{split}
        \rho(CMB) =\frac{\pi^2T^4}{15}  T(\gamma)^4  \hspace{1cm} n(CMB)= 4\zeta(3) T(\gamma)^3, \\  
n(\nu_i) = \frac{7}{2}\zeta(3)  T(\nu)^3  \ \hspace{5mm} 
   \text{with} \hspace{5mm}\ \zeta(s) = \Sigma_n n^{-s}
    \end{split}                                                   
\end{equation}

 Running the Hubble expansion backwards in time, the contraction blue shifts the photons (and any other relativistic particles) to higher temperatures, i.e higher energies and higher number densities. In the most common scenarios this ensures that all known SM particles which did not decay fast enough started in thermal equilibrium.\par
The slow, adiabatic, forward evolution guarantees that the equilibrium form above will persist along with the Boltzmann $\exp^-E(i)/T$ factor and, comoving entropy which for the relativistic particles is $n_B +7/8  n_F$, is conserved. Thus, the decoupling and Freeze-out of a massive ($m(i)> T$) species X increases the temperature of the remaining DoF which were in equilibrium with X and are in equilibrium with each other by a factor of:
\begin{subequations}
    \label{GR11}
    \begin{align}
       \bigg( \frac{N(D.o.F)-2}{{N(D.o.F)}} \bigg) ^{1/ 3}\ for\ X=B=\text{Boson}\\
        \bigg( \frac{N(DoF) -2(7/8)}{{N(D.o.F)} }\bigg)^{1/3}\ for\ X=F=\text{fermion} 
    \end{align}                                                   
\end{subequations}
 Where N(DoF) is the number of DoF of the coupled thermal particles prior to the decoupling of X ( with X included). We focused on the thermal heat part which only mildly changed. The remaining  un-annihilated massive X which eventually becomes in many models the dominant matter component is discussed at length in Sec V. Electron - positron annihilation dumps $\delta(N) =4 \times (7/8 )$\footnote{The 7/8 times fewer DoF of fermions versus bosons obtains by denoting  $E/T =k/T =x$, and expanding $ (1- e^{-x})^{-1} = \Sigma_n e^{-nx}$; $ (1+e^{-x})^{-1} = \Sigma_n (-1)^n e^{-nx}$ and integrating $\int k^2 dk f(k) \rightarrow \Sigma \int_0^\infty x^2 dx e^{-nx}= \Sigma 1/{n^3}$ with the sum over even n missing for fermions we get the  $1- 1/ 8  =7/ 8$  factor.} helicity states into the photon radiation of N(DoF) = 2 helicities and Eq~\ref{GR11}b implies:
\begin{equation}
    \label{GR12}
\frac{T(\gamma)_{\text{Final}}}{T(\gamma)_{\text{initial} }}=[(4. (7/8) +2)/2]^{1 /3}=(11/4) ^{1 /3}     
\end{equation}
 The initial photon and electron/positron temperatures are the same as those of the neutrinos. Since at the time of $e^- e^+$ annihilation the neutrinos are largely decoupled from the rest of the radiation, the electron and positron endow their entropy to the photons only. The $(11/4)^{1/3}$ ratio of temperatures of the CMB and Cosmic Neutrino Background (CNB) then follows.\par
The neutrino or any species X is in chemical equilibrium if the rate of inelastic reactions, say  $\nu+\bar{\nu} \rightarrow  e^+  e^-$ which by dimensional arguments is:
\begin{equation}
    \label{GR13}
     \Gamma(\text{Reaction}) = \text{Constant} \cdot \ G(\text{Fermi})^2 T^5
\end{equation}
exceeds the (Volume) expansion rate 
\begin{equation}
    \label{GR14}
        \frac{\frac{dV}{dt}}{V}=\frac{\frac{3d(a)}{dt}}{a}=3H
\end{equation}
 Neglecting small temperature anisotropies of the CMB and attendant density fluctuations and using (Eq.\ref{GR3}) and (Eq.\ref{GR6}) the condition for maintaining chemical equilibrium for the neutrinos becomes:
\begin{equation}
    \label{GR15}
  T^3 \gtrsim = (G_F)^{-2} m(PL)^{-1}  \ or\ T > MeV 
\end{equation}
 The mass difference $\Delta(m^2_{2-3})$ found from atmospheric neutrino oscillations, along with the \[ \sum_{i=1}^{3} m (\nu _i) \leq 0.12 ev\] Planck satelite constaint implies that the heaviest neutrino has a mass
\begin{equation}
    \label{GR16}
           m(3) \leqq
           0.12/2\ eV=0.06\ eV                  
\end{equation}
When this mass exceeds the temperature T, the non-relativistic $\nu(3)$ particle has energy $\sim m(\nu) +k^2/{2m}$, and contributes as any matter component $ m T^3$ to the energy density $\rho$. However, the number density $n(\nu)$ is that of the \textit{massless} spectrum at T $\sim$ MeV adiabatically expanded, cooled and diluted to the present day values \footnote{
Eq \ref{GR16} obtains for the inverted $\nu$ mass hierarchy when the two heavier neutrino mass eigenstates m(3) and m(2) are split by the tiny $\Delta(m^2)$ Solar.  As long as gravitational interaction energies are smaller than the kinetic energies, the temperature of these particles decrease as $ z^2$ -rather than z. Discovering the CNB would be a great event, yet its much inferior angular/energy resolution prevents it from adding much cosmological information beyond what the CMB provided}. In standard cosmology the CMB and CNB  subsume the complete initial Entropy of particles that were relativistic, decoupled and for $M \neq 0$ "froze out" when the temperature dropped below their mass. Late decay or annihilation of massive particles happening when these particles are \textit{not} in thermal equilibrium, inject a lot of energy at a relatively low temperature, thereby enhancing the entropy of the radiation dominated universe. Such decays/annihilations do not occur in the SM, but were invoked to resuscitate otherwise excluded BSM scenarios or DM types. The eventual thermalization erases foot-prints of such events in the photon distribution. This also applies to the SM phase transitions (PTs) -the EW breaking at $T \sim v \sim 200 GeV$ and that of QCD confinement at $T\sim \Lambda(QCD) \sim 150 \text{MeV} $ which are \textit{not} first order PTs where bubbles form and may leave gravitational radiation footprints.\par
Coalescence of bubbles formed in first order phase transitions occuring in many BSM and inflatonary model variants could be a potential source of GWs.
 Apart from lensing effects, GWs freely traverse cosmological distances and the long period GWs recently reported by the Nanograv project and expected to be seen by space laser interferometry, may have -in part, an early universe origin.\par
The apparent “magnetic” vortex-like patterns in the polarization of the CMB claimed by the BICEP group suggested such primordial G.Ws. It generated great excitement which soon faded when the effect was traced to interstellar dust.\\
While radiation dominates, the following holds:
\begin{equation}
    \label{GR17}                                             
    H=\frac{\frac{da}{dt}}{a} \sim \rho^{1 /2} \sim T^2 \sim a^{-2}
\end{equation}
Integrating  $d(t)/{da} \sim a$ yields:
 \begin{equation}
    \label{GR18}         
    t (in\ RDE) \sim a^2 \sim T^{-2}
\end{equation}
With $\rho (X) \sim T^3$ in the matter dominated era a similar argument yields 
\begin{equation}
    \label{GR19}  
    t (in\ MDE) \sim a^{3/2} \sim T^{-3/ 2} 
\end{equation}
Used in Eq.\ref{GR5}, Eq.\ref{GR19} indeed yields p=0 appropriate for non-relativistic particles almost at rest.\par
The third pillar of the big-bang cosmology is  the BBN of light elements: Helium, Deuterium, Lithium, Beryllium and Boron. These are also produced and/or further processed in stars and finding the pre-stelar early universe precise abundances is non trivial.\footnote{Most of the solar luminosity and solar neutrinos are generated in the "proton cycle" where $4p + 2e^-  \rightarrow 2\  ^4He^{++} + 2 \nu_e$. Burning hydrogen into Helium enhances the He/H ratio as we go deeper into the sun. Furthermore, our sun is made of reprocessed matter which contains some He of stellar origin. Studies of young stars at high z and the “Forest” of absorption "Lyman alpha" lines of atomic Hydrogen/Helium at various redshifts, yields He/H and other light element abundances at earlier times which are relevant for BBN.}.\par
BBN starts when the CMB and CNB temperature are $T\sim MeV$ at a corresponding time of order 1 Sec, and terminates around $t= 10^3 Sec \sim \tau (neutron)$ -the neutron's lifetime. The process $e^- + p \rightarrow \nu_e + n$  produces neutrons that are incorporated into deuterons and Helium nuclei which then serve as a gateway to heavier elements. The abundances are affected by the decoupling of the $\nu_e$ at $T\sim MeV$, and the annihilation of electron-positron pairs around the same temperature. This reduces the electrons density to be equal to that of the protons as required by the overall charge neutrality. The successful prediction of BBN depends on the correct expansion rate at BBN (which is enhanced by any additional light DoFs) \footnote{$^*$The SM prediction of three active (left handed) neutrinos lighter than $m(Z)/2$ was confirmed at $\sim 1 \%$ level by the Z boson decay width at LEP. Gary Steigman and David Schramm argued much earlier for three light neutrinos using BBN in ref. \cite{SCHRAMM1979}} \par
Predicting the various nuclear abundances requires elaborate numerical studies which in the spirit of this review we do not discuss. Among the inputs into the calculations of BBN and CMB studies, the well measured  $\eta= n(B)$ /{entropy} $\sim 6 .10^{-10}$ which is fixed by some stage of “Baryogenesis” is the least understood \footnote{The complex computations of BBN are still easier than a complete, accurate description of the nuclear reactions in the sun. The sub-KeV solar temperatures require careful measurements or calculations of rates of reactions impeded by Coulomb barriers such as $p+ Be \rightarrow ^8B+\gamma$ producing energetic neutrinos by the Boron $\beta$ decays. Doing this and calculating heavy elements production in White dwarfs and Red Giant  with final further processing and ejection into the galaxy by supernova explosions are triumphs of astrophysics}. \par
While Inflation was suggested earlier \cite{Brout} and \cite{kasanas} its later more complete form is due to Alan Guth \cite{guth}, and subsequent researchers\footnote{ $^*$ A. H. Guth and S. H. H. Tye\cite{guth_tye} suggested inflation to avoid the over-abundance of ultra heavy GUT monopoles generated after Spontaneous Symmetry Breaking (SSB) of GUT (John Preskill \cite{Preskill}) These monopoles of mass $M(Gut)/\alpha \gg M(Gut)$ cannot be thermally produced after the GUT breaking phase transition at temperature $T \sim M(Gut)$- Or more generally in perturbative physics at any energy \cite{drukier} Yet, the Kibble mechanism \cite{kibble} utilizing the hedge-hog pattern of correlated internal SU(2) DoF of the Higgs field, causing the SSB, and the spatial direction to points where the Higgs vev vanishes, allows efficient creation of monopoles at these points.}. 
More generally, Inflation is motivated by asking: "How far backwards in time (to some t(min) and corresponding temperature $T(max)$) can we extrapolate the RDE (Radiation Dominated Era)"? \par
Jointly Quantum Mechanics and GR suggest a minimal length $l_{\text{Planck}} \sim 10^{-33}\ cm$ and a corresponding maximal temperature $T(\text{Max})= m_{\text{Planck}} \sim 1/l_{\text{Planck}} \\  \sim 10^{19}Gev$. The QM momentum-length uncertainty relation $\delta(P) \delta(x) \geqq \bar{h}/2$ suggests that the above E(Planck)=P(Planck) is the maximum energy/momentum attainable by an elementary “point -like particle”. Indeed, otherwise we could generate wave packets smaller than  l(Planck). This fails in string theory where a system with energy exceeding the square root of the string tension  $\sigma(\text{string})^{\frac{1}{2}} \sim \mu$ is excited to higher energy, more extended states in the string tower. In general as more energy is pumped into a system the average energy of the particles therein namely the temperature increases. However if the number of available states increases faster than $\exp^{E/{T(\text{lim})}}$, then the temperature stops rising beyond a limiting temperature T(lim) as inputting extra energy  produces the many almost degenerate states rather than raising the energies/ temperature of existing particles\footnote{$^{++}$The number of states, at energies up to $m$ in string models grow as  $\exp{\sqrt{m^2/{\mu}^2}}$. 
This follows from the number P(N) of ways that a large integer N (which in the string case is $N= m^2/{\mu}^2$) can be partitioned into a sum of integers provided by Srinvanatan Ramanujan. The relevant qualitative aspect of the answer, namely that $P(N) \sim \exp{(c\sqrt{N})}$ with a constant c, can be intuitively motivated: P(N) is the number of ways that N can be written as $N=\Sigma {n(i)k(i)}$. To qualitatively understand this, consider a Young tableaux made of a total number of squares = $N$ and  with $n(i)$ rows of length $k(i)$ corresponding to the above partition. Alternatively, we can use the 90 degree rotated Tableaux where $N=\Sigma {n'(i)k'(i)}$  with $n'(i)$ the number of Columns of length $k'(i)$ in the original tableaux. As is often the case, the maximum, namely most partitions correspond to almost symmetric tableaux with roughly equal number of rows and columns $\sim \sqrt{N}$ as shown in fig. 1. Consecutive rows and consecutive columns in the Young tableaux are of non-increasing lengths.
The number of partitions or tableau then is roughly the same as the number of random walks across the $N^{1/2} \times N^{1/2}$ square where only down or right directed steps are allowed which is $\sim\ 2^{\sqrt{N}}$}. 
 
In the 1980’s when the discussion of Inflation started, GUTs  were in vogue. Following the early authors we assume first that $T(\text{max}) \sim M(\text{GUT}) \sim 10^{15}\ GeV$. This choice dramatizes the following difficulties which persist for much lower T(max). Since $T=T(\text{CMB}) \sim 1/a$ with a the scale factor: 
\begin{equation}
    \label{GR20}  
 a(min) =a(\text{now}) \frac{T(\text{now})}{T(\text{max})} \sim 10^{28} cm \frac{2.10^{-4} eV}{ 10^{15}\ GeV}\sim 1 cm .
\end{equation}
Using the connection between time and temperature during the radiation dominated era 
\begin{equation}
    \label{GR21}  
t(\text{min}) = t(\text{BBN}) (\frac{T(\text{BBN})}{T(\text{max})})^2= 1 sec (\frac {MeV}{10^{15} GeV})^2 = 10^{-36} sec
\end{equation}

The corresponding causal horizon defined as the distance light travels during this time $c(t(\text{min}))= 3.10^{-26}$ cm is vastly smaller than the above cm size of the early universe. This raises the first  “Horizon” problem: how can a region  with $\sim ({10^{26}})^3=10^{78}$ causally disconnected parts have a CMB of essentially uniform temperature? 

Eq \ref{GR7} raises another problem. Using $H \sim 1/t$ we find that to keep k=0, namely a flat universe, $\rho$ has to be exactly $\rho(\text{critical})$ with a precision of $(t/a)^2 \sim 10^{-52}$!
Inflation can solve both these "Horizon" and "Flatness" problems. To see this, let us assume that early on cosmological evolution is dominated by a cosmological constant-like contribution to $\rho$ with \textit{constant} energy density $\rho(\Lambda)\sim \Lambda^2$. The basic Eq.\ref{GR3}  with k=0 then reads:
\begin{equation}
    \label{GR22}  
H=t^{-1}da/{dt}=d(\text{log}(a))/{dt}=[\rho/{m(\text{Planck})^2}]^{1/2}=\Lambda/{m(\text{Planck})}=\text{const}
\end{equation}
This implies an exponential "Inflatonary" growth during a slow roll of the $\phi$ inflaton field over a rather flat section where $V(\phi)$ is almost constant as in fig.\ref{fig:02}
\begin{figure} [h]
\begin{center}
 \includegraphics[width=0.4\textwidth]{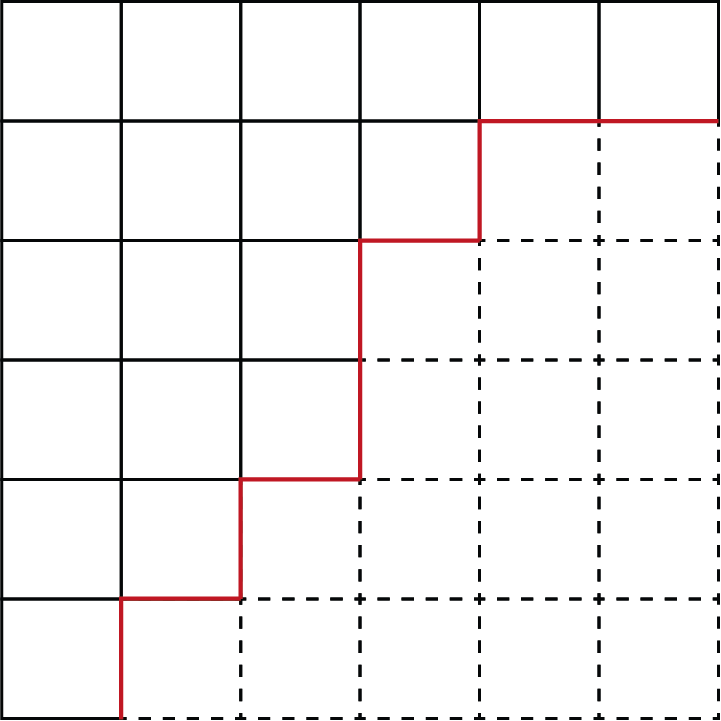}
    \caption{Yang tableaux made of n squares arranged in adjacent s rows of non- increasing lengths are partitions of n  expressed as a sum of lengths of the rows, or alternatively as the sum of the lengths of the Columns.} 
\label{fig:01}
\end{center}
\end{figure}
allows 60 e-fold expansion. This then suffices to close the above $10^{26}$ gap and stretch a single horizon size region to the desired a $\sim$ cm. It also smooths out any initial "Wrinkles" and dilutes away any residual massive particles and the above mentioned GUT monopoles in particular \footnote{In addition to the point like - zero dimensional magnetic monopoles associated with the breakdown of a non-abelian gauge symmetry group G to $G'\times U(1)$ we can have 1 and 2 dimensional topological remnants. These remnants are cosmic strings and domain walls associated with broken U(1) and Z(N) symmetries respectively. If the latter straddle the complete horizon, then the corresponding energy densities dilute upon expansion as $a^{-2}$ and $a^{-1}$.For a large enough surface tension, the energy of one defect $ R^2 \cdot \sigma '$ may exceed the total energy in the observable universe of $\sim 10^{80} GeV$ as would a single string of planckian string tension.} Our choice of $T(\text{max})\sim M(\text{GUT})$ was arbitrary as there may be no GUT symmetry. A landmark in the possible “Energy Desert” extending between v(E.W.)$\sim$ 200 GeV  and m(Planck) is the Peccei Quinn scale relevant for axion physics of $10^8 - 10^{12} GeV$ \footnote{In the standard E.W. Model, $\sim$ a TeV Higgs bosson would  imply a large $\lambda \sim 10$ coupling in the quartic Higgs term in the lagrangian which upon SSB yields $m^2_{\text{Higgs}} \sim \lambda v^2$.
As noted by Roger Dashen \& Herbert Neuberger \cite{Dashen} the large scalar $H^4$ self coupling may result in a theory which incurably diverges. In turn this may render the theory "trivial" (A technical term implying that at the  fundamental, point-like level, the divergent theory requires that the "bare" coupling be set to 0). This is avoided if there is new physics at the TeV scale recquird. The observed light Higgs at 125 GeV corresponds to a reasonable $\lambda H^4$  coupling $(\lambda\sim \frac{1}{4})$ avoids a divergence and allows a "desert" with no new physics all the way to very high energies.}. \par \singlespacing

That inflation may reflect a “false vacuum” of energy higher than that of the true vacuum during a first order phase transition, helped its acceptance. It was soon realized that  collisions of bubbles of the true vacuum typical of first order PT’s lead to a chaotic universe and versions avoiding this were suggested by Andreas Albrecht \& Paul J. Steinhardt \cite{albrecht}, and by Andrei Linde \cite{linde}. Most models require a new scalar field, the Inflanton and later variants have additional scalar(s). The graph of the potential $V(\phi)(t)$  includes an extended almost flat section as in fig.\ref{fig:02}. During this period when the almost constant $V(\phi) $ slowly rolls off towards the minimum of $V(\phi)$, the system is in a “Wrong” vacuum with an effective cosmological constant and energy density $\rho=V(\phi) \sim V(\phi(0))$. Jointly  $\rho \sim V(\phi(0))$ and the duration of the roll-over fix the extent of inflation which should exceed 60 e-folds. After that the Inflaton field oscillates for a while around the minimum of V where the curvature of the  potential is the inflaton's mass 
\begin{figure} [h]
\begin{center}
 \includegraphics[width=0.5\textwidth]{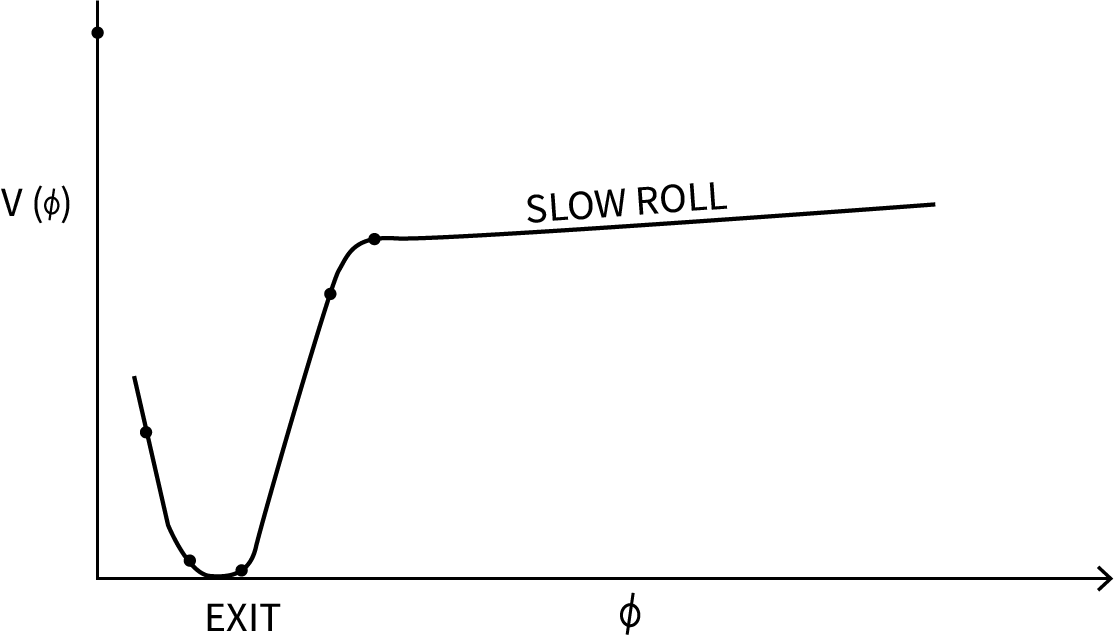}
    \caption{The potential $V(\phi)$ of the inflaton field $\phi$ exhibiting the almost flat "slow roll" section and the exit oscillatory decay section further down}
\label{fig:02}
\end{center}
\end{figure}

The inflaton then decays into SM (or other) particles. The need of having a "graceful exit" from inflation into the radiation dominated era or to some transient matter dominated period, is an important constraint. In principle $V(\phi(0)) ^{1/4}$, m(Dilaton) and T(exit) are independent mass parameters and very low (down to $\sim 10 MeV$) exit temperatures have been considered by some authors. While the idea of inflation as that of starting with a charge symmetric universe and generating the observed baryon asymmetry are extremely appealing, the BSM models realizing them are unwieldy and we do not describe them.\par
The decoupling of baryons from the CMB at the $ep \rightarrow \text{Hydrogen}$ recombination at temperature of $\sim 0.15 eV$ or redshift of  $z\sim 10^3$  and the angular anisotropy of the CMB are all important exceptions to the complete erasure in the  RDE of any previous information. The precise measurements allowed finding angle dependent temperature fluctuations of $\Delta(T)/ T \sim 10^{-5}$. That these angular fluctuations originate from different tiny quantum fluctuations of the inflaton field at different regions of space which could delay or accelerate the conclusion of inflation and therefore slightly enhance or reduce the amount of stretching may turn out to be the single most striking aspect of inflation.\par
At the early stages of inflation and throughout the slow roll toward its minimum 
the inflaton field $\phi$ is treated  as a classical (almost free) field. However beside the coherent $n \gg 1$ quanta in its various modes we also have "Vacuum fluctuations" of $\delta n = \pm \frac{1}{2}$  which, as expected for any free field, are Gaussian.\par
The resulting perturbations of the metric persist all the way to recombination when the CMB stops scattering from free electrons. The Sachs Wolf effect -namely the climbing out of the slightly deeper (or less deep) grav. potential wells at regions with positive (negative)  $\delta(\rho)$ -causes slight red or blue shifting of the photon energy generating the $\delta(T)/T \sim 10^{-5}$ fluctuations observed in the CMB. These obviously retain their shape during the free adiabatic expansion from decoupling until today when they have been beautifully mapped by The WMAP and by PLANCK collaborations \footnote{Some non-gaussianity (if suggested by the data) could manifest deviation from a free Inflaton field due to possible collisions of the inflanton or other heavy particles.}
Expanding the measured angular distribution in spherical harmonics, the distribution of the fluctuation power in the different l values
\begin{figure} [h]
\begin{center}
 \includegraphics[width=0.7\textwidth]{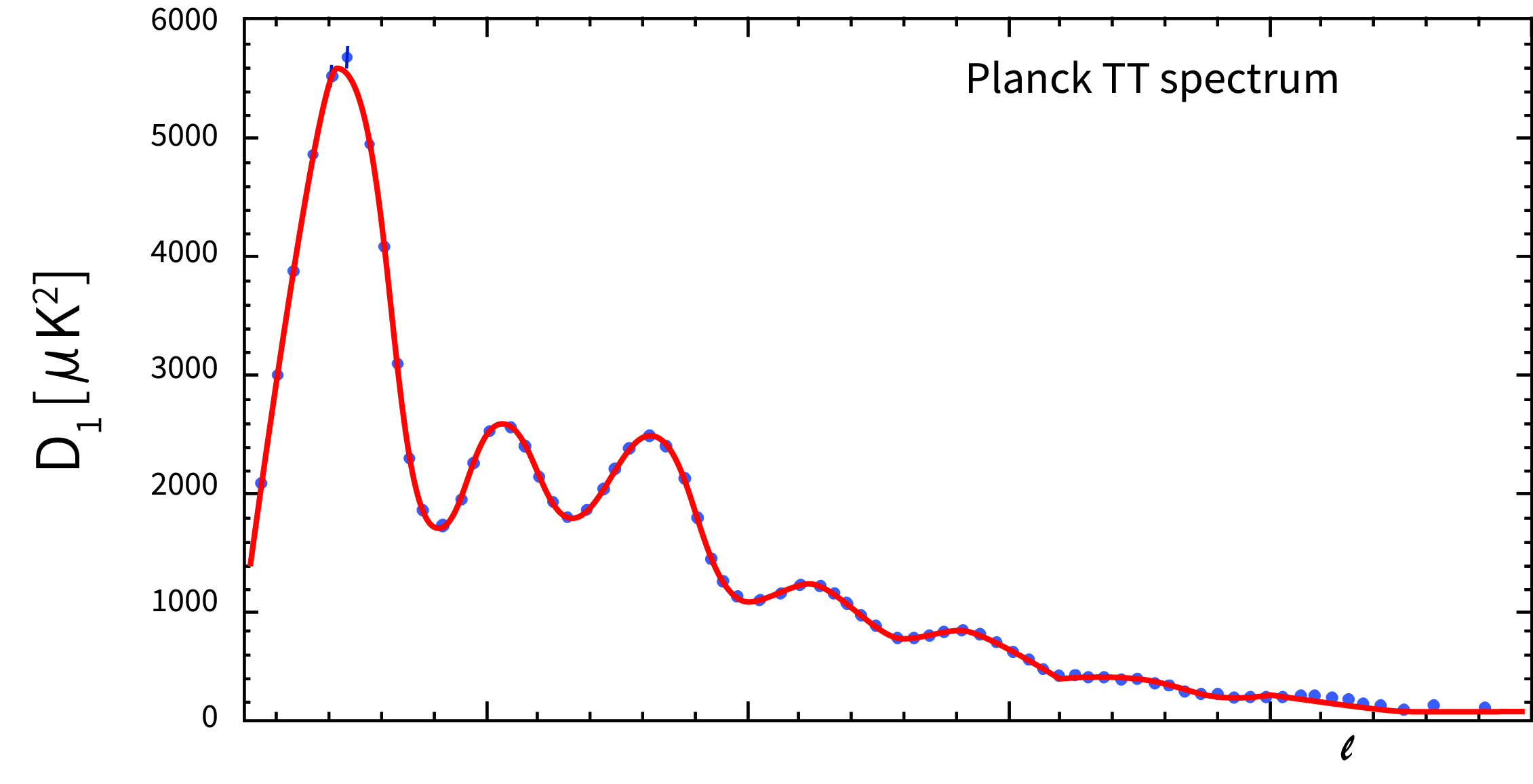}
    \caption{A picture of the power spectrum of the measured angular oscillations of the CMB as a function of l the angular momentum.}
\label{fig:03}
\end{center}
\end{figure}
is very informative. Many key cosmological parameters can be inferred from this graph shown in fig. 3.\par
\begin{enumerate}
    \item[1] When divided by the distance to the surface of the last scatter l becomes k, the magnitude of a (transverse) momentum. The 1/k decrease of the envelope of the graph (Known as the Harrison Zeldovich behaviour) reflects the conformal invariance of the early field theory. An intuitive understanding of the larger fluctuations at lower k (larger wavelength modes) is that those enter the horizon later and therefore are more extended.
    \item[2] Most oscillations are Baryonic Acoustic Oscillations (BAO) originating as waves in the plasma of electrons \& protons prior to recombination. The increased (Landau/ Silk) damping at shorter wavelengths is due to diffusion from hotter to colder regions of the plasma which is faster for smaller regions. The spacing and magnitude of the oscillations depend on and help pin down $\eta(B)$ and $\Omega(\text{Matter})$
    \item[3] The overall fit indicate a Hubble constant of h=0.65
    \item[4] If the CMB fluctuations trace Quantum fluctuation of an Inflaton field, then in the SM and many BSM extensions, the large scale structure seeded by them are universal, adiabatic and Gaussian. Finding "Isocurvature" fluctuations deviating from this would be of great intrerest.
    \item[5] The small pure number $\delta(T)/T \sim 10^{-5}$ is jointly fixed by the small ratio $n^{-1}$ of the quantum oscillation and coherent classical part in the dominant modes of $\phi$ and other parameters such as the value and slope of the potential $V(\phi)$
    \item[6] The very low l values seem to deviate from the general pattern and may contain as yet not fully understood information.
\end{enumerate}
Most aspects of the $\Lambda$ CDM cosmology described above (see also review in ref. \cite{Frieman_2008}) are likely to survive. Still we recall outstanding challenges including the recent JWST (Webb Space Telescope) measurements in the infrared suggesting unexpectedly very young galaxies/stars at high redshift up to z=20.\par
The early cosmic dawn due to brightly shining large stars can reflect a “Top weighted” Initial Mass Function (IMF) i.e a mass distribution of early stars with strong preference for more massive stars as compared with the present IMF. This in turn requires enhanced baryonic interactions or mutual DM  interactions in the early universe. Measurements of millions of redshifts by the DESI collaboration \cite{DesyCola2024} using in part Lyman alpha absorption forests of distant quasar light-offers an independent reconstruction of the BAO’s due to both baryonic and DM redshifts. Its preliminary results challenge a time independent cosmological constant.

\section {\hspace{\secspace} Self-interacting DM (SIDM)}
\textbf{In which we mention the complex problem of galactic and sub-galactic structures in the frame-work of particle-like D.M.  We use it to motivate SIDM yet limit DM -DM scattering. We mention IDM (inelastic) DDM (Dissipative) and mirror DM models.}

\hspace{1cm}

We first address self-interacting DM of the general particle form. Such interactions were suggested by the analysis of galactic and sub-galactic structures. It is an area of DM research where new relevant data from Gaia, from the Webb Space telescope from Gravitational lensing and from many other sources keep flowing in. Hopefully this will put  DM on firmer footing by excluding alternative theories and help clarify the type of DM required.\par
The density contrasts considered here are far bigger than the initial small “adiabatic” fluctuations manifesting in the $\delta(T)/T \sim 10^{-5}$ directional variations  in the CMB  discussed in the previous section. Rather we are in the truly nonlinear regime with $\delta(\rho)/{\rho} \geq 1$. The long range nature of gravity poses formidable difficulties leading to instabilities in the numerical simulations and in the real world \cite{Mace} \cite{palubski2024}. We can have dramatic purely gravitational effects when sufficiently dense systems "heat-up". Kinetic energy of stars, gas or DM particles can then be transferred via collisions from the denser, inner regions to the outer cooler regions of the galaxy, thereby allowing the central region to collapse even more, a phenomenon known as the "Gravothermal collapse”. This and a "Linden- Bell Statistics" \cite{Lynden-Bell1966} were introduced because the Bose-Einstein, Fermi-Dirac and Boltzmann statistics face difficulties in systems with long range interactions and no clear separation of extensive and intensive quantities.\par \singlespacing
Self-interacting DM was motivated in \cite{Spergel_2000} by measured galactic/halo structure which, at that time, conflicted with the many body simulations of weakly interacting cold DM Particles with no baryon back reaction. 
A Universal density profile was computed from hierarchical clustering  \cite{Navarro_1997}. The resulting "NFW" profile manifested "Spiked" rather than the gentler "Core"-like enhancement near $r=0$, and predicted excess of power on small scales such as many more satellite galaxies of the Milky than the few known then \footnote{The nice review of SIDM by Sean Tulin and Hai-Bo Yu (2017), has much of what follows, which is addressed also in Snow mass white papers \cite{snowmass} and in ref. \cite{Butler:2023glv}}. \par
It was noted by D. Spergel \& P. Steinhardt that a DM-DM elastic scattering cross-section of:
\begin{equation}
    \label{sidm1}
        \sigma(X-X)  \sim \frac{\text{Barn}} {\left( \frac{M(X)}{GeV} \right) }           
\end{equation} 
may resolve such difficulties. For a "local" DM mass density $\sim 0.4 GeV cm^{-3}$ and the typical Virial $3. 10^7 cm Sec^{-1}$ velocity, a DM particle then suffers over a galactic lifetime several collisions. A collision of a fast DM particle falling in from the outer part  of the halo will then disrupt structures such as a cusp forming in the center of the halo where the slower DM particles have a higher density. Recent higher statistics simulations resolving smaller fluctuations and, in some cases also accounting for the back-reaction of baryons on DM, suggest that the original simulations may have been misleading \footnote{Large scale computer aided calculations and/or simulations are vital in many areas including the proof of the four color theorem in mathematics. Without some intuitive understanding it is however difficult to assess the reliability of such calculations. Thus QCD calculations of the contribution of the hadronic vacuum polarization (HVP) to the anomalous muon's magnetic moment using the measurements of $e^+ e^-  \rightarrow \pi\pi$ pairs produced near threshold, suggested initially that the recent measurement of $(g-2)\mu$ significantly deviates from the SM predictions and is the harbinger of new BSM physics/DM scenarios. However lattice calculations give widely different results. Also when used in dispersive calculations, recent measurements of the cross-section of $e^+ e^-  \rightarrow \pi\pi$ near threshold yield HVP values closer to what the measured $(g-2)$ requires.}.

The above remark, the additional satellite galaxies which keep being discovered and the upper bounds on $XX$ cross sections restrict the allowed self interacting DM models. The SIDM has to be consistent with the bounds from the Bullet (and other) cluster data which is viewed as the most direct evidence for dark matter. Two neighboring galaxies in the cluster are separating, leaving in the space between them an excess of hot gas with enhanced X ray emission - as expected from a prior collision of the two galaxies. On the other hand, gravitational lensing data indicates that unlike ICM and ISM (Intra Cluster and Inter Stellar Medium) baryonic gasses, most dark matter has freely sailed through. 
This would not happen if the $X-X$ collision cross section is large enough and can cause multiple DM DM collisions. Escape velocities from smaller structures are lower, and cross-sections which rise at lower velocities help explain some features of dwarf galaxies or other small structures. Rutherford-like scattering mediated by a dark photon has a cross-section scaling as $v^{-4}$-  the fastest rise at low velocities allowed within $S$ matrix theory. A nonzero mass $m(\gamma')$ of the dark photon serves as an infra-red cut-off for $k(\gamma')\rightarrow 0$. \par 
The Kinetic $ \epsilon F_{\mu\nu} F'^{\mu\nu}$ mixing between our photon (or more exactly the $U(1)_Y$ part thereof) with another "dark U’(1) vector meson" can be generated via new heavy particles which carry both charges as shown in fig. 4
\begin{figure} [h]
\begin{center}
 \includegraphics[width=0.5\textwidth]{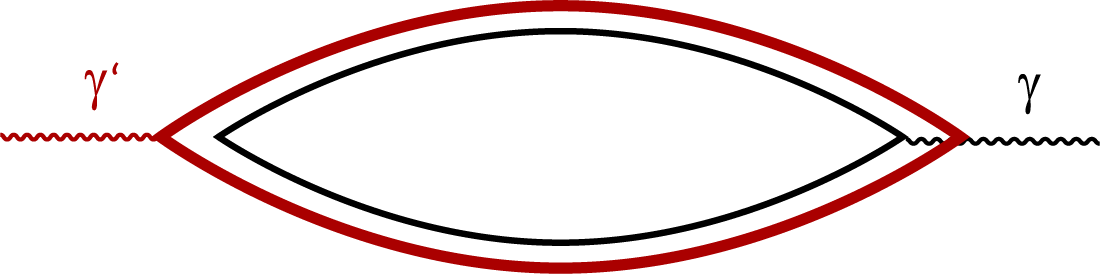}
    \caption{The kinetic mixing of a Dark and ordinary photon induced by a loop where particles carrying both dark (black line) and ordinary (red line) charges $e’$ and $e$ respectively, circulate.}
\label{fig:04}
\end{center}
\end{figure}
and the resulting "vector portal" is often used to connect the two sectors and generate DM- baryon  or DM electron interactions. 
On dimensional grounds the diagram of fig. 4 contributes 
$\epsilon \sim$ Log ($\Lambda$) but various extra selection rules imposed by an appropriate underlying theory can dramatically decrease $\epsilon$. \par
For very light vector or scalar mediators of spin independent interactions, the total elastic scattering cross-section, which is proportional to  $g^4/{m^2}$ with m and g the mass and coupling of the mediator to the DM (X) particles, can be very large. This reflects the huge  $[1/{\theta}]^4$ forward (small scattering angle) peak occurring when $M(X). v(X) \gg m$. The  acceleration-slowing down of DM particles by the mutual interactions are better described by the "transport" cross - section
\begin{equation}
    \label{sidm2}
\sigma(X-X)_{\text{transport}} = \int{ d\Omega (1-\cos\theta) \frac{d\sigma}{d \Omega} }    
\end{equation}

The overall velocity pattern is not affected by a maximal $180^o$ backward scattering of two identical SIDM particles which simply exchanges the momenta of the two colliding particles. A better definition of the transport cross-section in this case is given by
\begin{equation}
    \label{sidm2}
\sigma(X-X)_{\text{transport}} = \int{  d\Omega \hspace{0.15cm} sin^2 (\theta)  \frac{d\sigma}{d \Omega} }    
\end{equation}

At small angles $1-cos(\theta)$ and $sin^2 \theta$ behave as ${(\theta)^2}$. With $d\Omega = 2\pi.d(cos(\theta) \sim d (\theta^2)$  the transport cross-section is integrable up to logarithmic factors. In galaxy clusters the velocities are 2-3 times larger than those in galactic halos and the resulting more weakly interacting SIDM  there is consistent with the observed bounds. Point-like/contact interactions due to exchange of heavy mediators yield Isotropic $XX$ scattering cross-sections $\sigma \sim \sigma(\text{Transport}) \sim [M(X)v(X)]^{2}$  which $\it{\text{rise}}$ with energy. \par
The exchange of a light mediator generates a long range $V=\pm g^2 \frac{e^{-mr}}{r}$ potential which for asymmetric DM is repulsive/attractive for vector/scalar mediators respectively. Very light scalars generating attractive forces of long ranges  could aid collapse of asymmetric heavy DM  to form heavy dense nuggets and even BH’s- a possibility that we further discuss later.\footnote{    
$^{++}$ For SIDM with mutual elastic scattering much higher than the “bench-mark"  of Eq \ref{sidm1} above, the mean free path (mfp)  for DM-DM collisions becomes smaller than the distance over which the local gravity varies by say $\sim 0. 1\%$. A fluid -like streaming motion where all DM particles within a mfp move coherently with similar accelerations integrating over time into similar velocities is then expected. Hydrodynamical simulation typically used for the gas in the ISM can then be used for such DM as well. For a very large  $\sigma(X-X)/{M(X)}$, then the m.f.p. for X-X collisions  becomes shorter than earth radius $R(E) \sim 10^9$ cm and the ordered collective motion of DM adapts to that of near-by compact objects such as the sun and the earth as suggested by the E.P. (Equivalence Principle), DM will then move along with them at minimal relative velocities. Direct searches relying on nuclear recoil due to collisions with DM become ineffective. The interpretation of the colliding galaxies data from the Bullet and other clusters as a proof for DM and their use to exclude X-X cross sections higher than the above  "benchmark" would have to be reconsidered and possibly could be accounted for in the ultra strong SIDM framework. However the fact that the DM spatial distribution may then tend to follow that of the galactic disc and will not provide the spherically symmetric extended halo which the rotation curves seem to require- may exclude this possibility.} \par
Inelastic SIDM of a \textit{very} specific form has been conceived almost seventy years ago, before evidence for DM was found. As noted in \cite{pomeran} it appears in the "Mirror model" suggested by Tsung Dao Lee \& Chen Ning Yang to correct the "Blemish" of parity violation due to the left handed charged weak interaction. This correction is achieved by having for every known particle p and interaction I, a matching, mirror particle p' interaction I' with the left handed weak interactions "mirrored" into equal, right handed interactions. A particular mirror models will be discussed in Appendix A. A related "Tween Higgs" concept was suggested in \cite{arnik}.\par
The L-R symmetry is restored at high energies in a model suggested by Rabindra Nath Mohapatra \& Goran Senjanovich \cite{senjanovic}. It includes an extra set of  $SU(2)_R \times U(1)$ coupling to the right handed SM  fermions and gauge bossons which obtain TeV masses generated by a new rich Higgs sector.\par
A very different solution to the cusp vs core and excessive small scale structure has been suggested in \cite{WayneHu}. (See also \cite{LamHui}). It calls for extremely light $(M(X) \sim 10^{-22} eV)$ “Fuzzy” DM component. The very long wavelength of FDM provides a quantum outward pressure resisting inward collapse. FDM particles in the halo may then be in the same self consistent ground state. This fully justifies a classical treatment and as in case of light axions, this hopefully provides the pre-recombination potential wells needed to start structure formation. High precision tests using coherent fields and E.P. (Equivalence Principle) tests tend to limit this scenario.
A recent search strategy was suggested in \cite{blum2024axionh0graphyhuntingultralightdark}. Also efforts to find for such long wave classical DM the distribution of halo mass, the analog of the NFW profile for CDM, by Tomer Vollanski, Salvatore Botaro and Guiseppe Rossi are ongoing.
\par
Following the main motif of the present work we ask: “What features is the SIDM likely to share with other types of DM and what other characteristics it should not have?” Specifically for SIDM models which are largely symmetric with roughly equal amounts of X and $\bar{X}$ we should verify that:
\begin{enumerate}
    \item[a]The residual "density" of X and $\bar{X}$ which remains after Hubble expansion sufficiently dilutes the number densities of X and $\bar{X}$ (and all other particles which were in thermal equilibrium) has the correct $\Omega(X)$ when X is the dominant component of CDM.
    \item[b] The present rate of X $\bar{X}$ annihilation into light SM particles in the galaxy and the galactic halo is not excessive. Even small annihilation (and or decay) rates into S.M. photons, leptons or pions can be excluded by the “indirect” searches, by distortions of the CMB spectrum and by breaking down light nuclei conflicting with the correct BBN predicted abundances.
\end{enumerate}
         
SIDM models satisfying a\&b have been proposed by Yonit Hochberg, Erik Kuflik, Tomer Volansky and Jay Wacker \cite{Yonit} and by other authors. SIDM can also be the charge symmetric stable, lightest particle in the DM sector \cite{yonit2}.
In our sector efficient annihilations left \textit{no} observable residual primordial anti-protons or positrons. An excess of positrons or low energy anti-protons and even more so of anti-deuterons in the AMS (Anti Matter Search) or similar projects flying a large magnet, if inconsistent with the expectation from reactions of cosmic-rays and the interstellar medium, would then suggest that they originated in the decay or annihilation of DM.\par
 
We next mention two SIDM "Relatives": IDM and DDM. IDM, Inelastic DM, where the $X(0)$ DM particle is accompanied by one (or several) nearby higher states $X'(i)$ is readily excitable in DM - nucleus/electron collisions (see ref. \cite{David}). It had an important impact on the field and was incorporated in a "Grand unified" model of DM \cite{NimaArka} where the DM X(0) and a nearby X’(0) are the Majorana DM split by $(10 -10^3)$ MeV. In this model DM scattering $(Z,A) +X(0) \rightarrow (Z,A) +X'(0)$ is mediated by a  dark photon $\gamma '$. More massive $\gamma$’s are free from many experimental/ astrophysical constraints on lighter dark photons. Its decay could then help explain the apparent positron $\gamma$ and antiproton excess in the Pamella and Attic and AMS experiments. This model along with most WIMP type models is by now excluded.\par
In the second related form of DDM - Dissipative DM, the DM couples to light particles in the dark sector. These particles are emitted when an excited DM state decays or when DM particles collide thereby facilitating dissipation of kinetic energy. If they are pseudo-scalars (pseudo) Nambu Goldstone bosons,then their derivative couplings reduce their emission rates. In the following we assume mass-less U(1) vector particles with D.M couplings similar to those of our photon to ordinary electrons or protons. The DDM would then tend to form disc-like structures-the lowest energy configuration permitted by angular momentum conservation. Ref. \cite{Fan2013} envisioned a sub-dominant component of DDM with mass  $M(X) \gg m(p)$, that forms a thin disc inside, and parallel to our disc. Motion transverse to the disc periodically brings the solar system and its asteroid rich Kuiper belt/Oort cloud inside this thin disc. This kicks  out asteroids which may hit earth causing catastrophic extinction events spaced by a common period \footnote{Unlike for non interacting halo DM of various masses for which the equivalence principle implies the same velocities, the velocities of the heavier DDM and thickness d of the discs decrease with their mass when the DDM is in thermal equilibrium The analog of the atmospheric Boltzmann distribution implies that $d \sim 1/{M(X}$. This dramatically manifests in the tiny aspect ratio of Saturn's rings which are  made of pebbles of normal matter. While the assumed additional dark disc in our galaxy is thinner than ours, its  upper and lower surfaces are fuzzy, making the entrance and exit by our solar system "adiabatic"-lasting longer than the period of the motion of the asteroids in the Kuiper belt and rendering disruption events unlikely. \par 
Ref. \cite{geller} suggested heavy dark "protons" and far lighter dark "electrons" that are in thermal equilibrium. This pushes up the velocity of the dark electrons to be way larger than the escape velocity from the galaxy, with the Coulomb like attraction of the dark proton keeping them there, with greatly enhanced detectability.}. \par
The $10^8-10^9$ Year formation time of galaxies is dictated by the rate of EM dissipation via bremsstrahlung which scales with $\alpha (em) \sim 1/137$. Analog formation of dark Discs at a rate  $\sim \alpha '$ requires that $\alpha' \sim \alpha$ and (the stricter $\alpha'= \alpha$ is imposed in Mirror/Twinn Higgs models). 
\par
If the CDM is the mirror neutron in a broken mirror model and the baryon numbers in ordinary and mirror sectors are equal and opposite due to N(B)-N(B') conservation, then the ratio r $\sim 5$ between the CDM and ordinary baryonic matter contributions to the cosmological energy density fixes the mass of the CDM particles to be $M(X) \sim 5 m(n)$. Such nucleon' masses are readily generated by changing the VEVs and/or Yukawa couplings of the mirror Higgs relative to the ones in our sector. A further minimal tweaking of the small Yukawa mirror Higgs couplings can reverse the mass ordering of the up and down, first generation quarks, making  $m(d') < m(u')$ so that $m(n')$ is smaller than $m(p')$. With mirror electromagnetic self energy contributing (see, André Walker-Loud \cite{walker2019} - and W. N. Cottingham  \cite{Cottingham1963}), a substantial $\Delta(m') = m(p')-m(n')$ of order 7 MeV can arise, leading to a neutral  non dissipative $n'$ DM. Appendix A provides further variants of this DM model. In general diversifications developed for many DM types explain the huge volume of research in the DM field. Note that if $p'$ with $m(p') \sim 5m(p)$ was the lightest mirror particle and along with $e'$ constituted a dissipative $H'$ DM, then it would form mirror disc five times thinner and five times heavier than our disc in the middle of the latter which is clearly ruled out. More generally, the cosmological expansion at the time of BBN is affected by all energy density sources and the doubling of the number of neutrinos and photons is a problem which suggests a lower background temperature$ T' < T$  in the mirror sector. \par
In concluding this section we briefly return to the opening motif. A new problem that keeps arising with more precise observations of faint and dwarf-galaxies via special Dragon-fly multi-lense telescopes and of more distant galaxies is their very large variability. In particular we encounter (dwarf) galaxies where DM completely dominates and the opposite case of galaxies and halos (almost) completely devoid of DM. This adds another dimension to the well known diverse -elliptical, spiral and bar-like types. The extra diversity may reflect  the pattern of mergers forming the galaxy in question. Indicators of the variability - though most likely not its source, are the massive black holes in the galactic centers of masses between a few million to almost 10 billion solar masses \footnote{ Analysis of recent astrometric data suggests that our galaxy merged with another rather large secondary galaxy leading to the "Gaya Sausage" like pattern of high gas/stellar velocity in our neighborhood. The large variability of galaxies may bear on the existence of extraterrestrial intelligent life to which we return in the last  sections of this work. Specifically in order to develop intelligent life we may need not only planets similar to the rare and very special earth. The candidate planet  should also belong to a special - though less rare-  class of galaxies similar to the milky-way}.\par 
\noindent
Two main features characterize structure formation in the present $\Lambda$-CDM Paradigm are:
\begin{enumerate}
    \item[i] Matter domination over radiation occurs  before recombination of protons and neutrons into neutral Hydrogen atoms at  $z \sim 1600 $ . This allows for a period of a faster than logarithmic growth of the $\delta ( \rho )/ \rho $ perturbation in the CDM to commence prior to that in the baryonic sector. The pregalactic structures formed by DM then serve as potential wells for attracting the baryonic content of the galaxies.
    \item[ii] The formation of structures in CDM has a down-up hierarchical pattern: smaller structures such as micro/mini haloes and dwarf galaxies form first merging later into larger \& heavier structures. This merging is  ongoing, manifesting in our galaxy where recent merge left a "stream" extending to our own solar system.
\end{enumerate}
The basic features of $\Lambda$(CDM) cosmology were confirmed by most observations on structures and by the pattern of Baryon acoustic Oscillations (BAO) in the CMB fluctuation spectrum. As noted in the introduction, the putative discovery of  large and luminous galaxies (made of low metallicity, massive, bright stars) at very high redshifts, some of which seem to be purely baryonic with no DM. is however puzzling and challenges  $\Lambda$(CDM). \par
Self-interactions can help DM accumulate in stars and in the sun. References \cite{Press1985CaptureBT}, \cite{osti_5939847} pointed out that accumulation in the sun of DM particles of mass of 5$ GeV < M(X)<10 GeV$ with mass density $\rho(X) \sim 10^{-12}\rho(\text{Nucleons}) \sim 10^{-12}  gr.cm^{-3}$, modifies some solar properties. Specifically the X particles conduct heat from the solar core thereby slightly lowering the temperature of the core. Due to the strong temperature dependence $\sim T^{18}$ of the rate of the fusion reaction generating the energetic $^8B$  neutrinos -this lowers the flux of these  neutrinos, the paucity of which in the Davis Home-Stake mine experiment was the "Solar neutrino anomaly". The minimal $ 5 GeV \hspace{2mm} $X mass avoids excessive "evaporation" of the assumed X particles after achieving the solar core temperature of  $T \sim 1 KeV$  and  an average rms  velocity $ v(X) = \sqrt{2T/M(X)}$ potentially exceeding the escape velocity.\par
The required X particles densities $n(X) \sim 10^{-11} cm^{-3}$ or total number $N(X) \sim 10^{44}$ in the sun are close to the maximal values achieved when every X particle hitting the sun is captured. The $\sigma(X-N)$ required for this greatly exceeds the maximal value presently dictated by the strict limits from the underground large liquid Xenon experiments. However a large XX cross section causes a fast non-linear growth leading to saturation where each infalling X particle scatters off X particles already bound in the solar gravitational field and eventually both are recaptured into the sun.

\section{\hspace{\secspace}The “Wimp Miracle” and the Greist Kamionkowski (GK) Unitarity Bound}
\textbf{We describe the dependence of the “Freeze Out”  temperature and  residual D.M. density on the $\bar{X}X$ annihilation cross section. It leads to the "Wimp Miracle" - the correct relic - freeze out density  for weakly interacting TeV DM. Also  $M(X)\leq 100 TeV$  follows from the “Unitarity Bound -on $\bar{X}X$ annihilaton cross-section for point-like and symmetric D.M }

\par

\hspace{1cm}

While excessive $\bar{X}X$ annihilations pose a problem on galactic scales, too small annihilation rates leave too large a co-moving DM relic cosmological mass density.\par
The co-moving density of Wimps or more generally of symmetric DM that was in thermal equilibrium decreases as the universe cools to temperatures below M(X) according to the Boltzmann factor $\exp^{-M(X)/T}$. Finally at a "Freezeout" temperature  $T(fo) \ll M(X)$, the WIMPs cease to be in chemical equilibrium with "Radiation" -the  light relativistic particles, and their comoving density becomes constant. Given standard or other cosmology and annihilation cross-section, the task of computing the resulting relic X particle density has been widely addressed. An early discussion in the context of heavy Dirac neutrinos is presented in ref. \cite{Lee:1977ua}. \par 
The present number density of the CDM particles n(X) is
\begin{equation}
    \label{W1}
    n(X)+n(\bar{X})=\rho(X)_{(\text{now})}/{M(X)}
\end{equation}
 where  the corresponding energy density is 
\begin{equation}
    \label{W2}
\rho(X)_{\text{now}} =  \rho(c) \Omega(X). h^2 ;  \quad	\quad \rho (\text{crit} )\sim 5 KeV/{cm^3}       
\end{equation}         

Here we used the critical density of Eq.(\ref{GR16}) in Sec II and our discussion of the radiation dominated era therein. As we go backward in time the temperature rises and the number density of X increases according to:
\begin{equation}
    \label{W3}
    n(X)(T) = n(X) (\text{now}) . \Big( \frac{T}{T (\text{now})} \Big)^3 
\end{equation} 
all the way up to temperature $T= T_{fo} (X) =T(f.o)= \frac{M(X)}{f}$. At earlier times and higher temperatures the X particles were in thermal equilibrium with the radiation. At T= T(fo) we have equal rates of annihilation
\begin{equation}
    \label{W4}
\Gamma(an)=  n(X)^{-1}d(n(X))/{d(t)}= (v_X).n(X) .\sigma(\text{ann}) 
\end{equation}
(with $v_X$, the X velocity) and volume V expansion
\begin{equation}
    \label{W5}
    \frac{1}{V} \frac{dV}{dt}=3 \frac{1}{a} \frac{da}{dt}=3H \ \ \quad \quad
\text{namely}:\ 
v_X.n(X) .\sigma(\text{ann}) =3H 
\end{equation}                  
 At  T=T(f.o), with the rad dominated cosmological density of N (D.o.F) -  H is given by  
\begin{equation}
    \label{W6}
H^2(fo)=\frac{\rho}{(3m(PL)^2)} = N(dof)\frac{\pi^2}{15} \frac{T(f.o)^4}{(3m(PL)^2)}
\end{equation}
substituting this value in Eq.(\ref{W5}) , using N(DoF) $\sim 80$, restoring factors of c and noting that careful calculations using Boltzmann equation allowing for variations between $T= M(X)$ and 
$T=T(f_o)$ and also for the backward reaction of WIMP pair production, consistently yield.  $\frac{M(X)}{T(f_o)} = f \sim 25 = \beta(f_o)^{-2}$ we find that:
\begin{equation}
    \label{W7}
    v(\sigma(\text{an}))_{T=T(\text{f.o})}  \sim 3. 10^{-27} cm^3/{\text{sec}} 
\end{equation}
 as the condition for obtaining the correct residual X DM density \footnote{The following suggests that f values of  $25 \pm 5$ are indeed reasonable. As T drops from T= M(X) to $T(f.o) = \frac{M(X)}{f}$, the Boltzmann factor $ e^{\frac{-M(X)}{T}}$ decreases by $1.4 \cdot 10^{-11}$. $200^{\pm 1}$.  Starting with the equilibrium value of X particles at T=M(X) $n(X)\sim n(\gamma)/40$ it goes down by a factor of $10^{-12} .200^{+/- 1}$ at T= T(f.o). ($n(\gamma)$ is the present CMB number density and 40 stands for $\sim$ for half the $\sim 80$  DoF s whose entropy is eventually channeled to the CMB). Recalling that $\rho(X) \sim 5 \rho (B)$ and $(n_B/(n_{\gamma})=\eta(B) = 6 \cdot 10^{-10}$, the required CDM mass roughly obtains. The SM N(D.o.F) $\sim 80$ when $T\sim 200 GeV$ is found by adding to the 2  photon helicities and 6 majorana neutrinos ,4x3x3 = 36 (Dirac) quarks of the three colors and three generations 16= 8x2 gluons of both helicities 12 = 4X3 Dirac leptons of the three generations,10 =3x3+1 massive $W^+ W^-$ and $Z^0$ Weak bosons and the scalar Higgs $H^0$ .}. \par
Interestingly this happens for WIMP DM which has ordinary weak interactions - such as many LSP candidates. \par 
The cross section for the process mediated by the t and s channel exchange diagrams in fig. 5a,b:
\begin{figure}[h]
  \centering
  \subfloat[]{\includegraphics[width=0.5\textwidth]{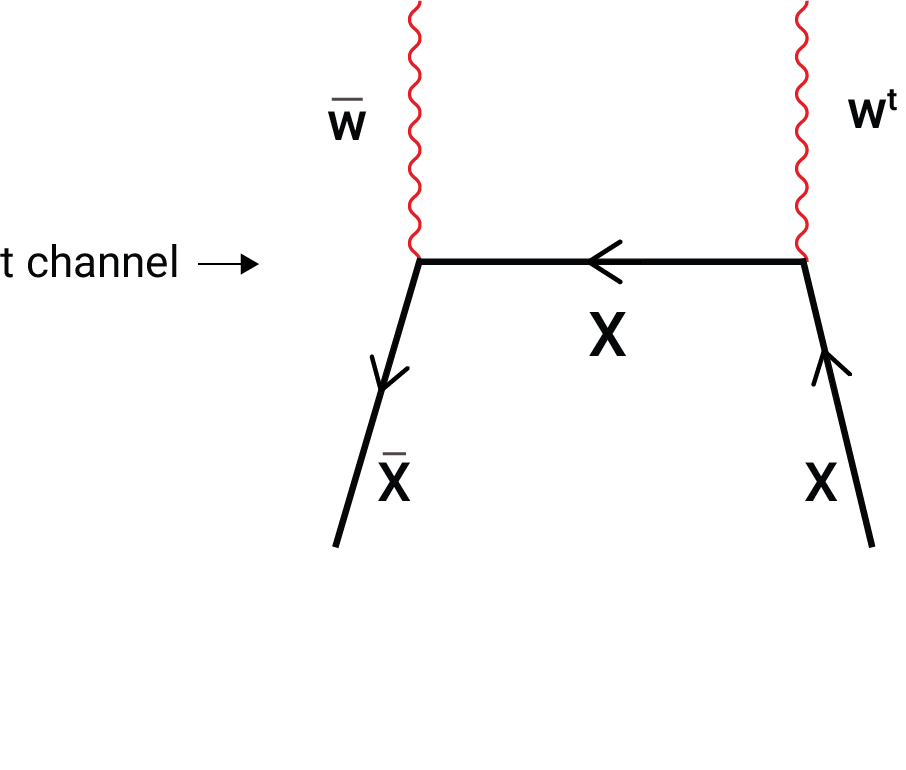}\label{fig:05a}}
 \hspace{3cm}
  \subfloat[]{\includegraphics[width=0.2\textwidth]{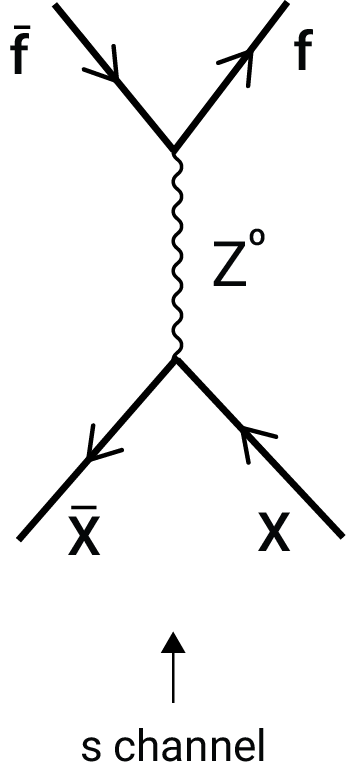}\label{fig:05b}}
  \caption{a-b Annihilation diagrams for X = LSP WIMP Dark Matter with ordinary weak interactions generated by t channel and s channel exchanges respectively. }
\end{figure}

is:
\begin{equation}
    \label{W8}
\sigma(\text{an})_{\text{Weak}} \sim \frac {\alpha(W)^2} {[M(X)^2 \hspace{0.15
cm} \beta(X)]}  \sim 4. 10^{-34} cm^2
\end{equation}
For $M(X) \sim TeV$, $\alpha(W) \sim 0.03 \hspace{0.2cm}\text{we find}$:  $v.\sigma \sim 2.4 10^{-27}$ - close to the value required- a result referred to as the "WIMP miracle" which further bolstered the belief that DM could be an LSP sharing weak interactions. \par
In Appendix C we derive the Unitarity upper bound of:
\begin{equation}
    \label{W9}
        \sigma (\bar{X}X) (\text{an}) \leq \frac{2\pi}{k^2} 
\end{equation}
 with k the momentum in the CMS Lorentz frame, on the annihilation cross section of point-like "elementary" DM X particles. Using $k^2 \sim M(X)$ E(X) with  Kinetic energy. 
 $E (X) \sim T$ we see that the maximal cross-section is $2\pi f/(\alpha_{(W)^2} \hspace{0.1mm
}$ $\sim 10^5$ times larger than $\sigma(\text{Weak})$. This allows a maximal M(X) values 315 times higher than that for  M(X) = TeV of LSP WIMPs, namely: 
\begin{equation}
    \label{W10}
              M(X) \leq 315 ~TeV  
\end{equation}
which is referred to as the GK or the Unitarity bound. Updating the cosmological $\Omega(X)$ and $h^2$ to 0.2 and 1/2 respectively, instead of the values of 1 used in the original paper \cite{Griest1989} by Greist and Kamionokowski tightens the bound by a factor of $10^{-1/2}$ so that:

\begin{equation}
    \label{W11}
    M(X) \leq 100 ~ TeV
\end{equation}  
In invoking the unitarity bound of Eq.(\ref{W9}) it is stated, often without detailed explanation, that point-like elementary X particles can annihilate only via the l=0 S wave. Appendix C clarifies this correct intuitive result.  \par
Interactions endow  “elementary particles” with “structures”- such as the photon cloud around an electron or the Z / W cloud around an elementary DM X particle carrying the SM weak charges. An “elementary particle” is defined as having  no structure beyond such clouds implied by its fundamental interaction\footnote{The probability of having the Bosonic cloud is $P \sim \alpha=g(Y)^2/{4\pi}$ where $g(Y)$ is the "Yukawa" dimensionless coupling of the boson to the particle of interest. It becomes P=1 when $\alpha \sim 1$, blurring the distinction between elementary and composite X. Strongly coupled gauge theories tend to confine which, as elaborated in Sec V, evade the unitarity bounds}. 
For larger boson masses $\mu$ the cloud size shrinks as $R \sim 1/{\mu}$. If the momentum of the colliding particles k is much smaller than $\mu$, the mass of the exchanged particle, then only the lowest, zero angular momentum, partial wave amplitude in the direct $X -\bar{X}$ “ s channel” survives since all higher partial waves $a_l(k)$ decrease as $(k/{\mu})^{2l}$. The $a_l(k)$ are the coefficients in the Legendre Polynomials  expansion of the scattering amplitude 
$A( X\bar{X} \rightarrow X\bar{X})$ :
\begin{equation}
    \label{W12}
    \text{A(s,cos}(\theta))= \sum_{l}{(2l+1)a_l(k)P_l(\text{cos}\theta))} 
\end{equation}  
with k and $\theta$ the relative momentum and the scattering angle in the center mass frame \footnote{The fact that S wave dominance at threshold of $ X-X or X-Nucleus$ scattering due to massive mediators implies essentially isotropic scattering is a recurring motif in the following}.
Finally the contribution of the S wave to the  annihilation channels, is limited by the unitarity bound 
\begin{equation}
    \label{W13}
    \sigma(l=0) \leq \frac{2\pi}{k^2}.
\end{equation}  
In Appendix C we prove all the above using the original S matrix approach.\par
Here we elaborate the main physics ingredient of the proof - that the minimal mass
 $\mu=M(Y)$ exchanged in $X \bar{X} \rightarrow x \bar{x}$ (or $x' \bar{x}'$) where $ x \bar{x} (\text{or} \hspace{0.15cm}x' \bar{x'}) $ belong in the S.M (or in the BSM) sector, satisfies:  
\begin{equation}
    \label{W14}
                M(Y) \geq M(X). 
\end{equation}
The stability of X requires that it is the lightest particle in the new sector carrying some new conserved quantum number. Even if this new symmetry is "softly" broken (as in the S.M where the Higgs VeV induces masses of the EW bosons breaking down the $SU(2)_L \times U(1)$ symmetry to $U(1)_{e.m}$), the vertices in Feynman diagrams are inherited from the original Lagrangian and respect the symmetry. Thus as indicated in fig. 6a,b in the annihilation diagrams the new conserved charge Q(X) is either carried out by the new $x'-\bar{x'}$ pair of final particles or "loops" back via the exchanged Y particle. 
\begin{figure}[h]
  \centering
  \subfloat[]{\includegraphics[width=0.36\textwidth]{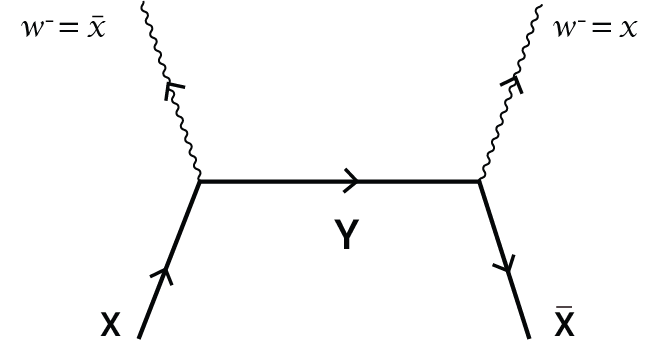}\label{fig:06a}}
 \hspace{2.6cm}
  \subfloat[]{\includegraphics[width=0.28\textwidth]{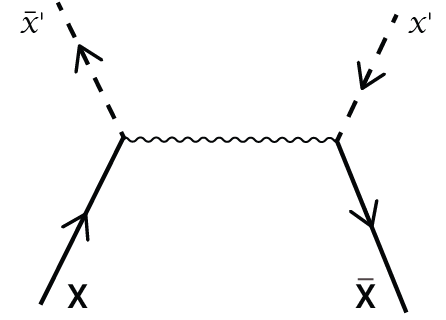}\label{fig:06b}}
  \caption{a-b Feynman diagrams for $X-\bar{X} \rightarrow x\bar{x}$ or $x’\bar{x’}$ with $x$ or  $x’$, light, SM, or dark particles showing the conserved new charge which loops back or is being transmitted to the final state. For an assumed fermionic LSP X the outgoing lines in 6-a are S.M. bosons and in 6-b the exchanged horizontal line corresponds to a bosson and the outgoing upper final state particle $x’$ and $\bar{x’}$ are fermions.}
\end{figure}

Since by assumption X is the lightest particle carrying the new charge we have in the case \ref{fig:06b} $M(X) < m(x')$ so that the process $X \bar{X} \rightarrow x' \bar{x'}$ is kinematically forbidden at $X \bar{X}$ threshold. In the case shown in fig. \ref{fig:06a} Y carries the Q(X) charge and hence the condition  $M(Y) >M(X)$  of Eq.(\ref{W14}) above is satisfied.\par

If the X particle participates in the ordinary weak interactions then there is also the s channel annihilation via the $Z^0$ exchange into $f\bar{f}$ where f refers to any SM quark and/or lepton or $W^+ W^-$. The annihilation is via S wave so that the GK bound in its original or improved new version applies. Alternatively these could be gauge particles associated with the new dark sector - which we assume are all much lighter than M(X). If (some of) the gauge symmetries in the dark sector are softly broken then we have also to the new 'dark' Higgs particle(s). In this case a small multiplet of Y particles exhaust the complete set which can be exchanged in the annihilation diagram. We cannot have gauge couplings $g^2$  bigger than $8\pi$ as the theory then is non-perturbative and the Feynman diagrammatic considerations implicitly used in our discussion become inoperative.

\section{\hspace{\secspace}Relaxing  the GK bound}
 \textbf{We describe how composite DM can evade the G.K bound and the collisions of such objects with various types of stars. We mention the Sommerfeld Enhancement (S.E.) of annihilation which also weakens the G.K  bound.}
\par
\hspace{1cm}

The G.K. bound can be relaxed or completely evaded by modifying the usual Big - Bang scenario. Thus a late, out of equilibrium, decay of some heavy particle(s) dilutes the relic comoving DM density. Alternatively the dark matter particles were never in thermal equilibrium and are largely absent in the early universe. The present required amount was "frozen in" at a later stage \cite{hall} as was suggested for DM made of R.H (Right Handed) neutrinos. Here we focus on modified annihilation scenarios.\par
The G.K. bound may be relaxed at the elementary DM particles level in several ways.
It was suggested in ref. \cite{kramer} that adding just one extra dark particle Y which  decays into SM particles say $Y \rightarrow sm\ + \ sm$ \ and with gX $Y^3$ interaction can  evade the bound. At  temperatures $T < M(X)$  the population of $X$ and $\bar{X}$ is Boltzman suppressed by  $\exp^{- (M(X)/T)}$ to the point that a too small annihilation cross-section leads to an early decoupling from the chemical equilibrium, and to excessive relic X particles. However X particles disappear via the $Y + X \rightarrow Y+Y$ reaction. For the specific choice of $M(Y)$:
 \begin{equation}
     \label{gk1}
     M(X)  >  M(Y) > \frac{M(X)}{3}
 \end{equation}
- the Y abundance is much larger, by a factor of up to   $\sim \exp(2M(X)/(3T))$ than that of X. The $Y+X \rightarrow YY$ reaction is correspondingly faster than the direct $X \bar X$ annihilation, and this can yield the desired  $X$ and ($\bar X$) relic density. The Judicious mass choice of Eq. (\ref{gk1}) also forbids the  direct on shell decay of X via $X \rightarrow Y\ Y\ Y$ which would prevent the required  longevity of $X$: $\tau(X) \geq t_{(Hubble)} $. The latter decay, achieved via the tree diagram for $X\rightarrow\ (\text{sm})\ (\text{sm})\ (\text{sm})\ (\text{sm})\ (\text{sm})\ (\text{sm})$ proceeding via 3 off-shell Y particles is suppressed by an extra $g^6$ with g the small coupling $g=g\ (Y(sm) (sm))$ (and the by the milder effect of the small 6 body phase space). A related earlier work appeared in ref. \cite{Smirnov2019}. \par     
Velocities of the DM particles in haloes are $\beta \sim 10^{-3}$ and their extra kinetic energy is $m(X)\beta^2 \sim 10^{-6}M(X)$. A putative new elementary particle with a mass $M$ very close to $2M(X)$ mediating the annihilation could then damatically enhance the annihilation. In early universe annihilations, the freezeout temperature and the kinetic energy of the relative motion are T(fo) $\sim M/ 25$ diluting the effect of any new finely tuned state coupled to $X \bar{X}$.
\par
Many different $Y$ particles of masses $M(Y_n) > M(X)$ can be exchanged and many new dark sector particles can be produced in the annihilation process. The resulting total annihilation rate could in principle be much larger and relax the GK bounds. "Towers" of exchanged/produced particles arise in string theory models though cancellations between contributions of various exchanges provide even lower cross-sections. \par 
Composite DM particles were suggested by John March-Russell, David. E. Kaplan, Gordon Krajniak, Kathryn Zurek, Maxim Pospelov and others. These composites readily evade the unitarity bound if they form after freeze-out at temperatures lower than $T_{fo}$. If the original lower mass, elementary, constituent particles satisfy the GK upper bound, then their early, efficient annihilation can yield the correct relic DM abundance. The composites which constitute the present DM- do not have to satisfy the GK bound and may be much heavier. They can be atomic, nuclear, grain-like and even macroscopic nuggets. In many cases the composite can have rather large annihilation cross-sections. Thus Dark Hydrogen and antiHydrogen-like atoms rearrange into into $e' \bar{e'}$ and $\bar{p'}  p'$ bound pairs (that are guaranteed to annihilate later) with very large rearrangement cross-sections: $\sigma \cong a(\text{Bohr}')'^2$ which for the case of ordinary hydrogen is $\sim 5. 10^{10}\ m(N)^{-2}$. This clearly enhances the prospects of indirect detection but as the atomic composites form  late in the cosmological evolution they do not modify the residual CDM density. \par
To form composites in the first place we should have a SIDM. Typically the $X$ and $\bar X$ annihilate before the formation of clusters down to the excess of say the $X$ particles. The clusters then will be asymmetric DM made of $X$ particles only.
The mass of moderately bound composites made of $\mathscr{N}$ dark matter X particles increases linearly with $\mathscr{N}$  yet it's  geometric size and mutual scattering cross section increases only as ($\mathscr{N}^{2/3}$) and very heavy nuggets become collisionless.\par
Such a geometric $\sigma$ grain-grain $ \sim \pi R(\text{grain})^2$  cross section is obtained when the scattering of the individual DM  particles is large enough so that $$\frac{\sigma(XX) \mathscr{N}}{R(\text{grain})^2} >>1$$ and there is significant mutual shadowing. \par
Searches (as in ref. \cite{Acharyya2023}) via UHE $\gamma$ rays from annihilation of composites of $X$ and of $\bar{X}$ DM particles can fail to put upper bounds on such ultra heavy composite WIMPs. The point is that in a grain-grain collision, the SM particles are produced by the short distance high energy annihilations of $X$ particle from one grain and $\bar X$ from the other grain and $\textit{cannot}$ generate monochromatic photons of energies beyond those achieved in individual $\bar{X}  X$ annihilations. \par Further it is quite possible that the rate of these annihilations will be suppressed by the fragmentation of the colliding grains due to the heat and pressure generated by the emitted light Dark bosons. Since some such bosons are needed to bind the grain in the first place, their emission in the annihilations is obligatory.\par
If masses in the dark sector are bigger than those in the SM sector say by $E$ for the heavy "baryons" and by $e$ for the light "electrons", then the mass densities inside grains of dark atoms $\sim m(B) m(e)^3$ are enhanced by $E\ e^3$ in the dark sector so that even a modest $e\sim E \sim 10^4$ results in dark grains which are $10^{16}$ times denser than S.M. grains.  \par
If Nuggets consisting of DM $X$ particles  with  mass $M(\text{Nugget}) = \mathscr{N}.M(X)$ make up the "Local"  DM density of $\rho \sim 0.4 GeV/{cm^3}$, then their number density $n_{\text{Nuggets}}$ is $\rho/(\mathscr{N}.M(X))=\rho/(M(\text{Nugget}))$. The earth will on average be hit by Nuggets of weight $M(\text{Nugget}) \sim  10^{12} gr$  once every thousand years. If an asteroid of such a mass and velocity $\text{V}_{(\text{Vir})} = 300 \frac{\text{km}}{{\text{sec}}}$ hits earth it would release upon impact its kinetic energy
\begin{equation}
     \label{gk2}
     W(\text{asteroid}) = M(\text{asteroid}) \frac{v^2}{2} \sim  10^{27} erg  
 \end{equation}
with disastrous consequences. However the radius of a nugget with the same mass but with nuclear internal density $\rho \sim 10^{15} gr (cm)^{-3}$ is only $\sim 0.1 cm$. While traversing the earth it displaces $\sim 10^8 gr$ of  earth's material residing inside the cylinder of $0.1 cm$ radius along the $\sim 10^9 cm$ path of the nugget through the earth -a tiny  fraction of the  mass of the DM nugget. If this displaced matter is ejected with the same velocity of $v \sim 300 \text{Km}/ \text{sec}$ then the energy deposition is $10^{-4}$ smaller than in the case of an asteroid (Eq. \ref{gk2}). This tiny nugget will punch through the earth and keep sailing. Also with  most of the energy deposition occurring deep in earth the observed effects may be quite moderate. 
\par
Nuggets making up galactic D.M. which are $10^3$ times lighter are $10^3$ times more numerous and will hit earth once a year. For the same nuclear density they will be one tenth in diameter and consequently will replace $1\%$ as much- namely $10^{6} gr$ of earth material. Still the special pattern of extended albeit rather weak earthquakes generated may be detectable (\cite{derujula} and \cite{herrin}).
The motion of the nugget inside earth is highly supersonic. The resulting sonic shock wave can enhance the energy transferred and the visibility of the impact. When emerging from earth the nugget is followed by a trail of hot matter seen at night by satellites with optical and infra-red capabilities and the coincidence with the above mentioned mini-earth-quake will help its discovery.\par
In order to maintain charge neutrality some nuggets also include electrons which along with their counterpart heavier charged partners make the nugget charge neutral but can cause strong ionization. When this happens in clouds it would manifest as lightning which -unlike ordinary lightning - do not follow a jagged but rather a straight and long line as in: "A Straight Lightning Bolt?!" \cite{Starkman2022}. \par
This effect was noted earlier in the context of experimental signatures of supersymmetric dark matter Q balls \cite{Kusenko1997}. \par \singlespacing
As noted early on by Bernard Carr \& S.W. Hawking \cite{Carr:1974nx}, primordial Black Holes (PBH's) may be DM. The experimental manifestations and associated bounds on such PBH’s and the mechanisms for their production in the early universe were discussed in detail in a recent white paper (Simeon et-al). Here and in section XIV below we note the miniscule effect of collisions between various stars and PBH’s due to the remarkably high density of the latter. Thus for a $10^{16} gr$ PBH the (Schwarzschild) radius is only $10^{-12} cm$. Consequently PBHs in the mass range of $10^{16} - 10^{22} gr$ - presently allowed for PBH s which make up most of DM, will almost freely sail through earth or any star including neutron stars. This holds even though the  rate of acretting/compressing/heating matter in the vicinity of the straight track through the stellar object  is enhanced  (see discussion below of the Bondi- Lyttleton accretion) by $(c/v)^4$  where $v \sim v(\text{Virial}) \sim 10^{-3}$ \footnote{Apart from BHs, neutron stars have the highest roughly nuclear density and the highest escape  velocity $v\sim c/2$. It has been suggested by Marcos M. Flores \& Alexander Kusenko \cite{Flores_2021} that very slow halo PBHs (of velocity $\sim 3 Km/s$) can lose a fraction of $10^{-10}$ of the kinetic energy while traversing the Neutron star, become gravitationally bound, eventually be captured in the NS, migrate to its center, keep accreting and growing and eventually convert it to a BH. The resulting unexpected very light ($\sim$ solar mass) BH’s would be quite striking.
Yet the small phase space $(v/ (v_{virrial}))^{-3} \simeq 10^{-8}$ factor and the requirement that the BH be a part of a binary with a luminous partner to indicate its existence, strongly reduce the prospects of discovering this effect.}.
\par
Forming DM clusters resolves the other potential difficulty with too strongly self interacting (SIDM). Thus in our S.M. sector Hydrogen and Helium are most strongly interacting with $\sigma[\text{(Hy) -(Hy)}] \sim \text{Angstrom} ^2 =10^{-16}cm^2 = 10^8   \hspace{1mm}\text{barns}$ exceeding by $10^8$ the bullet cluster upper bound of $\sigma( D-D)/m(D) \leq \text{barns}/GeV \sim cm^2/gr$.
On the other hand for our sun the $\sigma/M$ ratio is $2.10^{-11} cm^2/gr. -a$ reduction as compared with the unclustered case by $\sim 10^{19}=(10^{57})^{1/3}$. In general for closed packed DM particles where their size inside the star or cluster is the same as their original size, the reduction is by ${(\mathscr{N})^{-\frac{1}{3}}}$.
Thus the stars which contain most of the galactic baryons are collisionless.\par
Dark stars made of DM have been suggested by K. Freese awhile ago and recently in ref. \cite{ilie2023} and by others as in \cite{Gemmell2023}.
To avoid discovery by micro-lensing these stars need to be light, reminiscent of the small rock buckling under the pressure of growing trees that Saint Exupery's little prince is standing on as in fig. 7. \par 
\begin{figure}[!h]
\begin{center}
 \includegraphics[width=0.5\textwidth]{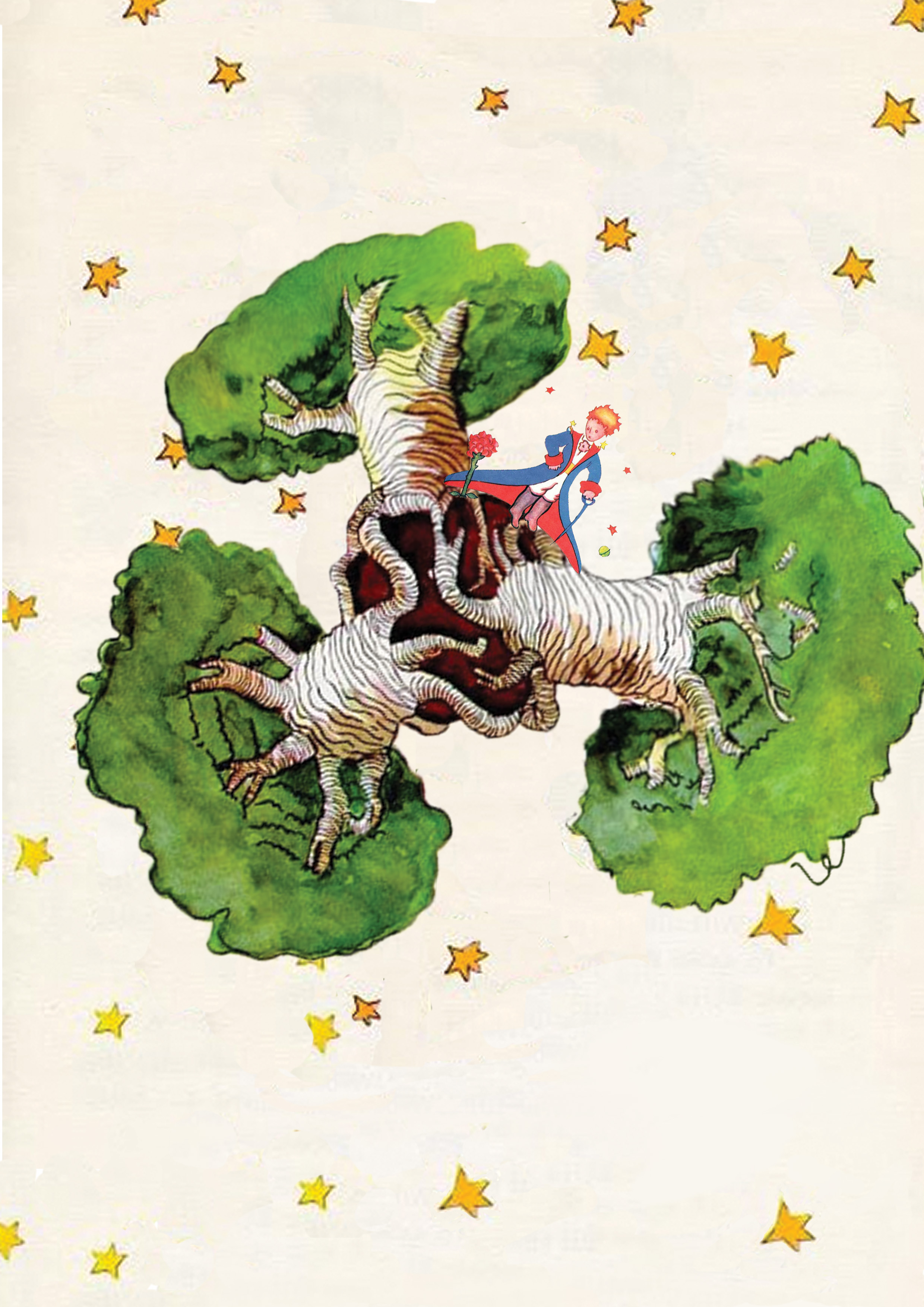}
    \caption{From Saint Exupery "Little prince" Showing the L.P on a small rock in space}
\label{fig:07}
\end{center}
\end{figure}
If made of ordinary or similar mirror matter the gravity on this $R \sim 10$ meter star is $~ 10^{-6} g$ making the L.P float into space \footnote{$^{++}$The little prince picture can be more realistic if the rock material is very heavy D.M. generating sufficient gravity}. \par
That some planets in extrasolar systems are dark-manifesting in the radial doppler modulation but being transparent to ordinary light will not cause partial eclipsing and be discovered also via the transit method, is unlikely. Even a 1 PPM (Parts Per Million) contamination of the dark stars by ordinary baryonic matter will make them opaque to optical light. Also the mixed planetary systems arising around the same parent star may be as rare as births of multiple twins some of which are black and some white.
If macroscopic DM has substantial interactions with ordinary matter it may be efficiently detected as suggested in \cite{Dhakal:2022rwn}. 
DM clusters on other scales such as DM grains and /or extended clouds may have unique footprints of multiple interactions in underground detectors or time coincident interactions in detectors located on different continents. We will consider this in more detail later. \par
The unitarity bound is weakened when there is a relatively strong, long range Coulomb/ gravitational -like attraction between $\bar{X}$ and $X$. If only DM particles arriving with impact parameter $b < r_0$ can be captured/annihilated on an object of size $r_0$, then the extra attractive force enhances the flux of DM particles reaching such distances by a factor of $\mathscr{E} =[v_{r0}/{v_{inf}}]^2$ .Here $v_{inf}$ is the initial $X$ velocity-at "infinity" - $b$ is the corresponding impact parameter and $v_{r_0}$ the velocity achieved upon reaching $r_0$ which by energy conservation is  
$v^2_{r_0}=v_{inf}^2 + v^2_{\text{escape}} (r_0). \hspace{1mm}\hspace{0.5mm} \text{with} \hspace{0.5mm}  v_\text{escape}(r_0) $ the escape velocity from  $r_0$.
Angular momentum conservation \newline
$L_i=M(X)v_{inf}.b =L_f= M(X)V{r_0}.r_0$ - implies that particles with initial velocity at "infinity" $v_{inf}$ arriving with impact parameters smaller than b will all be “funneled” into the object of radius $r_0$ arriving there with velocities  $v^2_{r_0}$. This $[b/{r(0)}]^2 = [v(\text{escape})/v(\infty)]^2$ enhancement is familiar and relevant for accretion onto neutron stars with escape velocity of $v_e  \sim 0.5 c$ which is $\sim 500$ times larger than the typical halo/galactic virial initial velocity of $10^{-3} c$. The last expression assumes a classical picture valid for large angular momenta.\par  
The analog effect in the quantum case, (S.E.) It reflect the sum over the diagrams with multiple soft photon exchange and also has --- $v^2(\text{escape})/v_p^2$  (the "Sommerfeld enhancement"). The escape velocity for the Coulombic potential - the average velocity in the 1S $X \bar{X}$ bound state, is $c \hspace{0.1cm} \alpha'$) \footnote{$^*$ $\alpha= e^2/{\bar{h}c} \sim 1/{137}$ was introduced a century ago by Arnold Sommerfeld  in order to explain the "fine structure" namely, the departure of Hydrogen atomic energy levels from the NR Bohr-atom prediction of $E{n,l} =-.Ry/{n^2} \quad \text{by} \hspace{0.15cm}  \beta^2= (\alpha/n)^2$ relativistic correction to the kinetic energy. $\alpha$ is the most important dimensionless constant in physics. Presently it is known to better than 0.1 PpB (Parts per Billion) thanks to the precise measurements of the g-2 anomalous deviation of the electrons magnetic moment from that of a point-like free Dirac particle, combined with the computation of the theoretically expected value by the life-time work of Toichiro Kinoshita. Precise values are independently provided by measurement of the quantum Hall effect (See e.g. "Integral quantum Hall effect for nonspecialists" \cite{Yennie:1987sa})}. Exchanging $Z^0$ can generate significant S.E. for very heavy DM transforming under ordinary weak interactions.  \par
The SE can also enhance the annihilation rate of DM made of neutral  $X^0$  so long as $X^0$ is part of a multiplet say ($X^-, X^0 , X^+$)  with mass splittings smaller than $M(X)/f$, the freeze-out temperature. In this "co- annihilation" scenario the neutral and charge components of the multiplet are for some time in thermal equilibrium so that the faster annihilating $X^+  +  X^-$ keep being replenished by the reaction: $X^0 +X^0 \rightarrow X^- + X ^+$. This co-annihilation mechanism is particularly important for neutral Majorana LSP’s such as the bino $\tilde{B^{0}}$, the annihilation of which happens only via p waves and therefore, unlike for the charged wino, is suppressed at lower energies.
Also DM belonging in higher $SU(2)_W$ representations with corresponding stronger $Z^0$ and $W$ couplings was considered. For a recent work see "The quintuplet annihilation spectrum" \cite{Baumgart2023}. Earlier work was done in ref. \cite{Cirelli:2005uq}. The SE has been extencively studied and we move further discussion to appendix D.\par
If the accreted particles are mutually strongly interacting and dissipative, then the angular momentum of the system can be radiated away and the centrifugal barrier becomes inoperative. This results in a much faster "Bondi Litleton" accretion rate $\sim v(e)^4/v^3$ as all particles arriving at a distance $r$ from a body of mass $M$ with escape velocity $v(e)=[G_{\text{Newton}} M/r]^{1/2}$ can eventually be accreted. For neutron stars $v(e)\sim c/2$ and the the accretion is indeed rather fast. This applies to SIDM particles whereas magnetic fields strongly limit the accretion onto neutron and other compact  stars of Hydrogen and Helium in the Inter Stellar Medium (ISM).

\section{\hspace{\secspace}Evading the unitarity bound in confining gauge theories and the effect of resonances on $\bar{X}X$  annihilations}
\subsection{\hspace{\secspace}DM Scenarios in confining theories}
The above discussion radically changes if the CDM $X$ particles are confined at temperatures below T(con) $\sim \Lambda'$ by a non-abelian gauge interaction. S matrix bounds are expected to fail in such theories as there are no asymptotic “in” and “out” plane wave states of the $X$ particles which define the $S$ matrix. Also in the analytic $S$ matrix all potentials are superposition of Yukawa potentials of different ranges:
\begin{equation}
    \label{6_1.1}
    V(r) = \int d\mu^2\rho(\mu^2)  \frac{e^{- \mu r}}{r}.
\end{equation}
For particle antiparticle scattering $\rho(\mu^2)$ is positive and all the potentials in the superposition are attractive. A linear potential expected in confining theories obtains when the spectral function $\rho(\mu^2)$ is the non positive derivative of the $\delta$ function and the momentum space propagator of the exchanged particle is $\sim 1/{k^4}$. This  corresponds to higher derivative field theories often beset by Ghosts . It is still instructive to see how the evasion of bounds happens.\par

For simplicity we assume an $SU(N)'$ group where $\Lambda'$ -the analog of $\Lambda$(QCD)-  is the scale of the theory. We assume that the DM particles $X$, just  like our quarks, are in the fundamental ,$N'$ representation and that $\bar{X} X$ pairs are confined. This confinement induces a very efficient $\bar{X}-X$ annihilation which leaves only the asymmetric excess part of the DM population so that the GK bound for \textit{symmetric} CDM becomes irrelevant.\par
Let  an independent (Higgs-like) mechanism provide masses $M(X_i)$ to the DM particles. We discuss separately the cases where:
\begin{itemize}
    \item[a]$\Lambda'$ -the (confinement) scale of the dark gauge group is (much) larger than $M(X_i)$ 
    \item[b]$\Lambda'$ is (much) smaller than all $M(X_i)$ -the masses of the new heavy "Quarks” and:
    \item[c]$\Lambda' \sim M(X_i)$- similar to the case of the SM where $\Lambda$(QCD) $\sim 200 MeV$ is intermediate between the light u,d, s and the heavy c,b,t quarks.
\end{itemize}
It is important to emphasize that scales far higher or far lower than the “Quark” masses can be naturally achieved with no fine tuning. The scales, are extremely sensitive to $N'$ of the confining $SU'(N')$, to the fermionic content and to the values of the gauge coupling at a high scale. This manifests in the dimensional transmutation relation.  
\begin{equation}
    \label{6_1.2}
    \Lambda' \sim \exp({\frac{- 8 \pi} {{\beta_0 g'^2 N'}}})
\end{equation}
where $\beta_0$ is the coefficient of the lowest order term in the perturbative expansion of the $\beta$ function of the theory (the $\beta$ function relating the couplings in the theory at different scales. \par
In case a) the confinement and spontaneous chiral symmetry breaking phase transitions occur at temperature(s) $T'  \sim \Lambda'$ and all $X_i=Q'_i$ (with i a flavor index i= 1, 2, ...,  N(F) are incorporated into pseudo Goldstone "dark" pions of masses:
\begin{equation}
    \label{6_1.3}
 m(\pi(i,\bar{j}) \sim [\Lambda' (m'_i+m'_j)]^{1/2}
\end{equation}
These "pions" are much lighter than other $Q'_i$ composites and glue’-balls of masses$\sim \Lambda'$ so that the latter quickly decay into the pions. The same holds for the $(N'_F)^2$ t’h  $SU(N'_F)$ singlet Goldstone boson - the analog of the ninth $SU(3)_F$  Goldstone boson in our SM sector - which is heavy thanks to non-perturbative "gluonic" effects.\par
If the $Q'_i$ share the ordinary weak interactions or have a new, dark weak interaction analog, then the dark pions decay:
\begin{equation}
    \label{6_1.4}
    \pi'_{i,\bar{j}} \rightarrow e^-+\nu_e\ (or\ e'^-+\nu_{e'})
\end{equation} 
providing that  $m_e \hspace{0.15cm} \text{and} \hspace{0.15cm} m_{e'} \hspace{0.15cm} \text{are both smaller than} \hspace{0.15cm} m(\pi'_{i,\bar{j}})$.\par
The analog of the QCD inequalities (reviewed in \cite{nussinov}) between masses of (pseudo) Goldstone dark pions of various "flavors" 
$$ 
2m(\pi'[(i,\bar j)] - m(\pi'[(i,\bar i)] -m(\pi'[(j,\bar i)])= \delta(m) >0  $$

allows the reaction
$$
\pi'[(i,\bar j)] +\pi'[(j,\bar i)]\rightarrow \pi'[(i,\bar i)] + \pi'[(j,\bar j)]  $$

\par 
Once the temperature T' in the DM sector falls below $\delta(m)$ this rearrangement of mixed flavor pions into diagonal flavor pions which may annihilate, becomes energetically favored.\par
If the CDM particles $X'_i$ have no SM nor dark electro-weak (or E.M. like) interactions, then the lightest pion which remain after the extensive annihilation at confinement may serve as symmetric DM of mass exceeding the GK bound, as noted in ref. \cite{De_Luca_2018} and in ref. \cite{Geller2018} \par
Case b) appeared in "split Susy" with Gluino LSP and heavy squarks suggested in ref. \cite{arkani2005supersymmetric} and in the Quirk model of Markus Luty. The early, perturbative stage of the annihilation of the heavy Quirks or squarks, say: 
$$
Q'_j+\bar{Q}'_j \rightarrow 2\ gluons\ ( or\ 2\ gluon's) 
$$
already leaves less than the relic co-moving freeze out density required for CDM.
For temperatures lower than $T\sim \Lambda'$, the $SU'(N')$ confinement kicks in and $SU'(N)$ strings form between the near-by $\bar{Q'_j}$ and $Q'_k$. 
 Despite the small string tension $\sim \Lambda'^2$ this provides a constant force which keeps pulling the heavy $Q'_j$ and $\bar{Q'_k}$ 
 towards each other so that eventually they form $Q'_i \bar{Q}'_j$ bound states. 
 The bound states are Coulomb-like, with bindings: $ B.E. \sim \mathscr{\mu} \alpha'^2/{n^2}$.
 They quickly cascade via gluon emission to the 1S ground state and the  diagonal $ Q'_j \bar{Q}'_j$ states annihilate into gluon or gluon pairs.\par
The general 
QCD inequalities between 
masses of pseudoscalar mesons of different flavors 
mentioned above, transform in the present N.R. case into inequalities between the bindings: 
$$
 B.E(\Pi'_{jj}) + B.E(\Pi'_{kk}) \geq 2 B.E (\Pi'_{jk})
$$
With the Coulombic binding given by $B.E(\bar{Q'}_j Q'_k) = \alpha'^2 \mathscr{\mu}_{jk}$ the last equation becomes the trivial algebraic relation $\mathscr{\mu}_{kj} \leq (M_k + M_j)/4$ ,with $\mathscr{\mu}_{kj}$ the reduced mass of the $Q'_j$ and anti-$Q'_k$. It  ensures that at all times the flavor diagonal states are populated (at least) as much as the non-diagonal states, leading eventually to the complete annihilation of all $Q'_i$ and leaving only the asymmetric excess in the form of baryon-like $N'$ Quirks' states. Quirks which are a specific example of case b, are discussed in detail towards the end of this review, We therefore turn to:  \par 
Case c) It is closest to our QCD where we encounter special features stemming from the lightness - on the scale of $\Lambda(\text{QCD}) \sim 200 MeV$, of the u,d  and to a lesser extent of the s quark: At the confinement phase transition at temperature $T_{P.T} \sim \Lambda$ all relevant length scales - the average distance between neighboring quarks or gluons, and the forming $\bar{q} -q$ and $\text{gg}$ states have size $\sim 1/T \hspace{0.2mm}\sim 1/\Lambda$. 
The $\bar{q}-q$ 
pairs which are spatially close to each other readily form the confined meson states which  completely decay into photons, electrons, neutrinos and muons. 
For our $SU(3)_c$ there is also  the formation of baryons and antibaryons by triplets of $q_i$ coupled to color singlets. 
Thanks to the very efficient annihilations of the remaining protons and anti- proton only the small asymmetric excess $\eta(B).n(\gamma)$ - of  the baryons survives.\par

\subsection{\hspace{\secspace}Resonance effects in general}
Returning to the original general theme the following question naturally arises: "Can we enhance the $\bar{X}-X$ annihilations (and evade the GK bounds) by postulating a new vector or a scalar particle $R$ to be exchanged in the s channel of mass:
\begin{equation}
    \label{6_2.1}
    M(R) = 2M(X) +\delta(m)\ \ ;\ |\delta(m)| \ll M(X) 
\end{equation}
For positive $\delta(m)$, $R$ corresponds to a resonance - a complex pole at $W =M(R) + i\Gamma/2$ where $W$ is the center mass energy and $\Gamma$, the total width of the resonance, is the sum of the "elastic" $R \rightarrow X\bar{X}$ width and the widths $\Gamma_i$ for $R$ decay into the final ($\bar{i} i$) states:
\[
\Gamma = \Gamma(el) +\Sigma_i {\Gamma_i}
\]
Unitarity limits the magnitude of each partial wave amplitude by $a_l(W) \leq 1$. The maximal contribution of any partial wave to the total cross-section then is:
$$(2l + 1)\pi/{k^2} = \sigma_l(max)$$
This maximal value is achieved only at center mass energy $W = W(R) = M(R)$, the peak of the the Breit Wigner distribution corresponding to the resonance R where the partial wave contribution to any $\bar{i} i$ annihilation channel is:
\[
\sigma_{Max}(X\bar{X} \rightarrow i\bar{i}) =(2l +1)\frac{\pi}{k^2} \frac{\Gamma(el)\Gamma(i)}{[(W -M(R))^2+(\Gamma/2)^2]} 
\]
The annihilation cross-section into any $i \bar{i}$  final state is proportional to the product of the decay widths\[\Gamma(el)= \Gamma(R \rightarrow X\bar{X})\ and\  \Gamma(i) =\Gamma \left( R \rightarrow (i\bar{i}) \right)
\]
The total width $\Gamma$ is the sum of the elastic and inelastic widths
\[
\Gamma(in)= \sum_{i} {\Gamma(i)}
\]
so that the total annihilation rate is:
\begin{equation}
    \label{6_2.2}
    c(2l+1)\frac{4\pi}{k^2} \frac{\Gamma(el)\Gamma(in)}{[(W-W(R))^2+(\Gamma/2)^2]}
\end{equation}
Since the sum of elastic and inelastic is the total width $\Gamma$ the product of $\Gamma(el)$ and $\Gamma(in)$ in the numerator of the last equation is maximized when
$\Gamma(el) = \Gamma(in) = \Gamma/2$

In equilibrium at temperature $T$, the annihilation rate is given by an average weighted by the Boltzmann factor.
\[
Rate \sim c\pi (2l+1) \Gamma(el) \Gamma(in) \int  \frac{dk}{(W-W(R))^2 + \Gamma^{\frac{2}{4}}} \hspace{0.2cm} \frac{{e^-{(W/T)}}}{Z}
\]
where we used $d^3k=4\pi k^2 dk$ and canceled the $k^2$ factor with the $1/{k^2}$ in the expression for the cross section. Assuming  that the masses of the final SM or dark sector $x \hspace{0.1cm} \text{or} \hspace{0.1cm} x'$ particles into which the $\bar{X} X$ annihilate are $\textit{much}$ lighter than $M(X)$ the latter particles are relativistic so that the final velocity, $c$ appears in front. In the above we use the non-relativistic expressions for the heavy DM(X) particles:
\[
W=k^2/{M(X)},\ W(R) =\delta(m)\ 
\]
\hspace{3cm} and
\[
Z =4\pi \int dk\ k^2 \hspace{0.3cm} e^- \hspace{0.05cm}{\frac{k^2}{M(x)T}}
\]
For a near threshold $\bar{X} X$  bound state/resonance a large elastic decay width to the  \textit{initial} $\bar{X} X$ state is expected. However the strong coupling to SM   $X\bar{X} \hspace{0.1cm} \rightarrow  \hspace{0.1cm} x \bar{x} $ states or to a light $x'\bar{x'}$ state dark sector pair, in the \textit{final} state, needed in order to generate a large $\Gamma(annihilation)$ is unlikely when $R$ is an $X \bar{X}$ composite: We have to consider the $t$ channel exchanges responsible for $X \rightarrow x\ (or\ x')$ and $\bar{X} \rightarrow \bar{x}\ (or\ \bar{x}')$ transition - precisely what led to the unitarity bound in the first place.\par
Conversely consider the annihilation of the near threshold DM $X \bar{X}$ pair  into $Y \bar{Y}$ where the $\bar{Y} Y$ \textit{final} state interactions generate the resonance at $2M(X) +\delta(m)$. One would then expect an enhanced $X \bar{X}$ annihilation at the corresponding (kinetic) energy $W=W(R)= \delta (m)$. However if the initial primary interaction coupling $X \bar{X}$ into $Y \bar{Y}$ and/or the resonance $R$ is weak, then we have a corresponding small elastic width of $R$ which appears in the rate equation above and reduces the effect.\par
This reflects the "Final state interaction theorem" that in the absence of further important annihilation final states beyond $Y \bar{Y}$, the strong interaction in the final state  simply introduces a "Final state" phase but does not change the rate which is fixed by the initial weak interaction. This can be generalized to a more complex multi-channel case  where the phase $\exp(i\delta) \sim S^{1/2}$ is replaced by the \textit{unitary} matrix $U(0,\infty)$. A familiar incarnation of that theorem occurred in QCD where the confining interactions after an initial perturbative $e^+ e^-  \rightarrow \bar{q}q$ (or $\bar{Q}-Q)$ process, dramatically change the character of the observed final state but do \textit{not} change the overall interaction rate beyond some calculable radiative correction effects.\par
We next consider the case where R is a new elementary particle. A field theory theorem excludes \textit{interacting} fields (which $R$ must be to in order to mediate annihilations) with spin $ > 1$ so that the $R$ particle is a vector gauge particle or a scalar "Higgs" like particle. The fine tuning required for having $M(R)$ near $2M(X)$ is not the main difficulty of this scenario. If the $X\bar{X}$ annihilate into SM particles, then $R$ is the vector (or scalar) “portal” connecting the dark and SM sector. The vector portal can be only via abelian $U(1)$ gauge. The special case of kinetic mixing of our $U(1)_{e,m}$ with a  dark $U'(1)$ with a light (or massless) dark photon has been extensively discussed but the portal can also be leptonic, baryonic and or iso-spin dependent. Unless we ensure sufficient neutrality of the new charge on length scales of order of the range $m^{-1}$ of the new force - its couplings $\sim \alpha'.\epsilon^2$ are strongly restricted. 
The discussion of the scalar "Higgs" portal (suggested by Wilczek) follows similar lines and will not be reproduced here.\par

\section{\hspace{\secspace}In which we recall baryon asymmetry generation and describe models which  attempt to explain why \* $r= \Omega(DM) / \Omega(baryons) \sim 5 $  }
The relic density in symmetric DM, fixed by the ($X-\bar{X}$) annihilation rate, seems to be unrelated to the baryonic mass density. The latter reflects the baryon asymmetry
\begin{equation}
    \label{7.1}
    \eta(B) \sim [n(B) -n(\bar{B})]/ {n(\gamma)} \sim 6.10^{-10} 
\end{equation}
We do not have a consensus on how the baryon asymmetry comes about. However, this asymmetry is likely to be generated by completely different processes from those controlling the relic density of symmetric DM. Consequently the ratio $r$ could be extremely large or extremely small. In the next four sections we will describe DM scenarios in which a ratio $r\sim 1$ may arise\footnote{
The necessary conditions for generating a baryon asymmetry starting with a charge symmetric universe have been spelled out by Andrei Sakharov in 1968 "Violation of CP Invariance, C asymmetry, and baryon asymmetry of the universe" \cite{Sakharov:1967dj}. They include beside the existence of baryon number violating interactions, an out of equilibrium set-up and time reversal violation. The CPT theorem states that in all local, Lorentz invariant, quantum field theories (QFT) the product of the three discrete transformations; $C$ (Charge conjugation); $P$ (Parity) and $T$ (Time reversal) is conserved. Hence CP violation is also needed. Grand unified theories (GUT) and many SUSY variants readily supply the first requirement and the CKM matrix of the SM has CP violation. The effect of a large CP violating phase there, is diluted by a tiny product of the mixings of the three pairs of quark generations (if  any one of these mixings vanish the $3\times 3$ CKM matrix reduces to a $2\times 2$ matrix where CP violation is impossible). This led to the failure of efforts to explain the baryon asymmetry within the standard model. 
The 't Hooft anomaly manifests in the effective interaction generated by instanton effects via the determinant prefactor involving quarks and leptons, the connection with the "Sphaleron" -the topological structure associated with the weak phase transition, and the possibility that baryogenesis is induced by an earlier leptogenesis further complicate the issue of Baryogenesis.
}.
\par
What then are the possible BSM approaches for explaining the DM to baryon ratio $r$?
One possibility is to give up the initial charge- symmetric universe, start with maximal baryon asymmetry and then dilute it by an out of equilibrium decay of a heavy long lived object. Since BH evaporation does not conserve the global baryonic number Bernoulli fluctuations can generate a  baryon excess $\delta(n(B))\sim N^{\frac{1}{2}}$, where $N \approx \frac{m(PBH)}{M_n} $ which will survive after the symmetric part annihilates.
The ratio of baryon excess to entropy generated by  a single  PBH  is $\delta N(B)/ Entropy = [N]^{-1/2}$. The correct value of $\eta \sim 6 .10^{-10}$ can obtain for small Primordial BH’s which evaporate much before BBN. Unfortunately the sign of $\delta(N(B))$ will fluctuate between the different PBHs .
To generate both the baryons and radiation in the SM, an appreciable  CP violating $\theta(QCD)$ is required at the time when the PBH’s can emit nucleons via Hawking radiation. While this can be achieved in certain Axion type models  ,the large mis-alignment of the axionic field may cause undesirable, late strong inflation. An early discussion appears in ref. \cite{Zeldovich:1976vw}. \par 
Barring the above alternatives we assume that generation of the DM and of baryon asymmetry are in fact  correlated . This can be in the framework of a symmetric DM or that of asymmetric DM. Cui Yanoue and Raman Sundrum \cite{cui} and others suggested that in symmetric WIMP (often SUSY inspired) models the baryon asymmetry results from a strongly B violating decay of a bosonic WIMP $H$ into DM and Baryon products which self annihilate but leave the required relic CDM and baryons.
\par
In the rest of the discussion we will consider r in the framework of asymmetric DM. That the baryon asymmetry arises albeit in a manner which is not fully understood as yet, suggests that a similar "Darko-genesis" mechanism generates a similar asymmetry in the dark sector.
This was considered in \cite{nussinov1985}, in the frame-work of the Techni-Color $SU(3)_{TC}$ confining gauge theory suggested by Leonard Susskind in ref. \cite{Susskind:1978ms} and by S. Weinberg in \cite{Weinberg:1977ma}. In this theory the Higgs particle arises as a Nambu-Goldstone Boson. The massless $H^{\pm}$ and $H^0$ are "eaten up" to yield the missing longitudinal polarization D.o.F of the massive $W^+$, $W^-$ and $Z^0$ boson -leaving an analog of the ordinary QCD  $\sigma$, a scalar  meson $S^0$ of $\sim TeV$ mass. Choosing $\Lambda_{TC}\sim 10^3  \Lambda_{QCD} $ yields the scale of the SM $SU(2)XU(1)$ breaking. The lightest stable Tecni-baryon of $\sim$ mass $10^3 m(N)$ could then be the DM. With somewhat smaller asymmetry in the techni-sector this yields $r \sim 20-40$ expected at that time before the $\Lambda CDM$ paradigm with large contributions to the energy density due to  a Cosmological constant or another source (referred to as “Dark energy”) was established (see \cite{Zurek2014} for a review of the asymmetric DM) \footnote{$^*$ For almost two decades SUSY and Technicolor vied to explain the Higgs/EW scale and the Fermion hierarchy problems. The original simple Technicolor model naturally yielded  the electro-weak scale as the $F(Technicolor) \sim \Lambda(Technicolor)$ analog of  $f_{\pi}\sim \Lambda(QCD)$. However its barely "Walking", (Slow Ren group running of couplings) extended Tecnicolor versions which aimed to explain also the Fermion mass hierarchy did not do as well. For Susy the explanation of the weak scale hierarchy problem turned into the prediction of the sub TeV breaking scale of SUSY and the Fermion Hierarchie was partially explained by a doubled  $[H(u), H(d)]$ Higgs sector where $H(u)/ H(d)$ couple to the upper/lower members of the weak isospin doublets .The LHC discovery of a “light” 125 GeV Higgs shot down Technicolor as no light Higgs is expected in this framework. For SUSY the fact that only one Higgs particle rather than five was discovered and its relatively “High” mass were a source of difficulty as detailed calculations implied that a single Higgs of$\sim 100 GeV$ mass allows for the desired Sub TeV SUSY breaking scale.}.
An approach somewhat similar to that using Technicolor was adopted in \cite{murgui}. This work is an example of UV complete DM model. Such models require specifying the under-lying Y.M. gauge interactions, the fermionic content and the bosonic part. This must include the $SU(3) \times SU(2)\times U(1)$, YM gauge interactions, the three fermion families, and the Higgs part of the SM Lagrangian, their analogues in the Dark sector and fields connecting the two sectors\footnote{
Inspired by Greek tragedies the authors classified the many fields in the model as main,
secondary or chorus characters.  After finding the spectrum of light states bound by the gauge interactions they verify that a correct cosmology obtains. As in Greek tragedies or many operas most characters  “die” - namely  most massive particles decay or annihilate leaving the desired amounts of baryons and dark baryons  without compromising (or even helping achieve) the correct CMB spectrum \& angular distribution, the correct BBN and the observed large/small scale structures.}. 

One needs to verify that the model is free  of possible "anomalies" which constrain various sums of gauge quantum numbers and/or powers thereoff. These were first discovered by Stephen Adler \cite{Brandt:1969gh} as divergences of triangular diagrams and by John Bell\& Roman Jackiw \cite{Bell:1969ts}. Later a deeper interpretation as quantum effects violating various symmetries which the classical system and its ground state have was found. Another issue is the stability of the preferred classical ground state against tunneling into undesirable vacua of lower energy. Also we may want to require that the theory is asymptotically free\footnote{Asymptotic freedom of non-abelian gauge theories means that at short distances or large momenta- the theory approaches its free field limit. This contrasts with any other field theory such as the $U(1)$ gauge theory of em, the $\lambda \phi^4$ and $\bar{\psi} \gamma(5) \psi$ Yukawa theory, all of which exhibit the opposite behavior namely of the coupling being naturally “Screened” by polarization charges as the distance at which the charge is probed increases and conversely blows up at short distances rendering these local field theories “ Trivial”. The unexpected opposite behavior in Non abelian gauge theory reflects the effect of the trilinear gauge coupling.\par For an appropriately limited  number of Fermions in the fundamental representation of the gauge group the "Ferromagnetic" type effect due to the self gauge coupling dominates and leads to this behavior. This asymptotic freedom was however not anticipated but rather discovered when calculating the  $\beta$ function of such theories by careful evaluation of vacuum polarization and vertex correction diagrams by Gerard Hooft and in \cite{Politzer1973fx} and \cite{Gross1973id}. The $\beta$ function controls the behavior of coupling constants which "Run" - decreasing or increasing with the log of the external momenta. This, the discovery of the chiral anomaly via the study of the triangular diagram for the 3 current VVA interaction by Adler and Sudakov’s discovery of the “double log” ($Log ^2$) behavior of highly off-shell high energy scattering amplitudes  illustrate the power of conventional Feynman diagrams. (Thus carefully calculated diagrams seem to be smarter than most physicists- present author included)  A novel , more efficient, diagrammatic approach, is being developed  by many physicists including, Zvi Bern,David Dunbar, Lance Dixon, David Kosower, E. Witten ,Freddy Cachazo and by Arkany-Hamed who believes that this approach will lead to deeper physics insights.}. \par
Another format for presenting BSM models uses the language of extra dimensions of R.S (Lisa Randall \& Raman Sundrum \cite{randall}), "Large" universal internal dimensions \cite{arkani2} or the "Fat brane - split fermions" Introduced by Keith Dienes Emilian Dudas and Anthony Gargenta Phys.Lett.B 429 (1998) 263-272 and used by Arkani-.Hamed \& Martin Schmaltz \cite{Arkani-Hamed2000}.
In the RS case one needs to specify the fields on the UV ``Planck'' brane and those residing on the low energy, infra-red (IR) brane. There may be additional branes in the fifth dimension interval between the above  two branes. The fifth dimension which provides the scaling of other dimensions and fields is appended to the scale-free ordinary $4 d$ field theories. The location of each brane indicates the scale of the physics associated with it playing a role similar to that of the  VEVs of  scalars in the usual field theoretic description. In particular the famous ‘Hierarchy problem “ is mapped to the distance between the IR and Planck branes which is exponentially enhanced by the warp factor in the metric on the fifth dimensional interval.
 \par
While the field-theoretic and the brane description largely overlap, the geometric brane description is quite appealing. The appropriate fall off of  effects from one brane to the other or towards the bulk where certain fields reside or the gaussian falling overlap of left and right parts of the wave functions of "split fermions" map the fermion and other mass Hierarchies into a moderate tuning of locations. Various "K.K.". excitations associated with non-zero (angular) momenta in the compact internal dimensions are new elements transcending low energy effective field theories. These appeared first in Theodore Kaluza(1921) \cite{KALUZA_2018} and "Quantum Theory and Five-Dimensional Theory of Relativity" by Oskar Klein \cite{ Klein:1926tv}
The non-decaying higher Kaluza Klein (KK) excitations in the universal large extra dimension scheme can be problematic (see ref. \cite{Hannestad2001}). This difficulty is further elaborated in appendix G. It is evaded in R.S schemes by having non - periodic boundary conditions.\par
In the Murgui-Zurek model  $r= \rho(DM)/{\rho(B)}$ is fixed by having the confining YM gauge theories  $SU(2)_{dark}$  in the dark sector and $SU(3)_{color}$ of the SM unify into $SU(5)$ at $\sim 10^9 GeV$. This yielded an $SU(2) _{Dark}$ scale $\Lambda(D)$ of about five times larger than $\Lambda(QCD)$ and a similar ratio of the masses of the dark neutron and our neutron. Along with similar "Baryo" and "Darko" genesis this can then produce the correct ratio of the cosmological densities of baryon and dark baryons. 
\par
In the more concrete and economical "Maryland model" \cite{bodas} DM is the $n'$ within a mirror sector with broken mirror symmetry. The $r \sim5$ comes about by having $\Lambda'\sim  5\Lambda(QCD)$ inducing $m(n') \sim 5(m(n)$. A Yanou Sundrum mechanism generates equal asymmetry both in our and the dark sector. Only a few percent increase in the $\alpha(QCD')$ relative to $\alpha (QCD)$ suffices to induce the desired ratio of scales. A similar ratio arises in the weak sector for $v'$ and $m(Higgs') \sim $   $5m(Higgs)$. This allows reheating of the SM via the out of equilibrium late decay of a "Reheaton" $R$  of mass $2m(Higgs') > m(R) > 2 m(Higgs)$ to two ordinary Higgs but not to two dark Higgs particles.  This makes $T > T'$, preventing the light $\gamma$’s and $\nu$’s from violating the $\Delta(n(eff) < 0.2$ Bound. Also care was taken to make $m(u') > m(d')$ and thus  $m(n') < m(p')$.\par
An earlier paper using a broken mirror model with m(n’)= 5 m(n) and where n(B)=n(B’) follows from a conservation law is ref. \cite{An2010}\par
While we have not described the above models in detail, the little we did, shows how complex DM models can be. This is why we do not dwell in great detail on any single BSM DM model. It is adviseable to have some hint as to the nature of DM before investing much effort in any particular model. Once we know the type of DM we will find the proper, hopefully simple and elegant, model in which this DM naturally arises. 

\section{\hspace{\secspace}Can DM made of quark nuggets or small quark composites explain r?}
\textbf{where we detail the advantages/difficulties of models of DM  made of quark nuggets and the extreme version where DM is the (us ds ud) Hexa= Sexa quark }\par

We next turn to special scenarios where $\rho(DM) \sim \rho(B)$ is achieved by having DM made of quarks- just like the ordinary baryons,  and \textit{no} new BSM fields are introduced.  This  allows eschewing their predictions rendering them very testable. Such models need to satisfy:
\begin{itemize}
    \item[1]The presence of such  CDM at BBN and at recombination when the abundance of light elements and the pattern of the CMB were fixed, does not affect their successful predictions, and
    \item[2]The scattering of this DM on baryons/electrons satisfies the direct search bounds.
\end{itemize} 
 We first address hadronic CDM in the form of 'Nugget' of nuclear density. Both demands are satisfied if the hadronic CDM is inside Nuggets which form early on - before BBN. Also with $\mathscr{N} \sim 10^{30}-10^{40}$ baryon-like DM particles per nugget and with nuclear density
 $n(internal) \sim 10^{39} cm^{-3}$ the collisions with and of nuggets are sufficiently rare. An early suggestion of such DM by E. Witten was based on two ingredients \cite{witten}
\begin{itemize}
\item[a] In  dense Strange Quark Matter (SQM) = $(uds)^{\mathscr{N}}$  made of equal large numbers of u d and s quarks the Fermi energy gain offsets the $m(s)-m(d)$ mass difference making it the most stable, true ground state of baryonic matter, and 
\item[b]Early on at temperatures $T \sim \Lambda(QCD)$ $\sim 200$ MeV a phase separation occurred segregating the SQM in massive nuggets which survive and constitute present day DM with no excessive baryonic density at the time of BBN and recombination. Most importantly the calculations suggested roughly similar overall baryonic masses in the Nuggets and in the unclustered, ordinary component. 
\end{itemize}
The absence of the lightest “Nugget”,namely a strong interaction stable $(uds)^2$  Hexa-quark, suggested that the strange quark  is a bit heavier than what was assumed by Witten, tending to destabilize SQM\footnote{
A Hexa-quark made of $ud$ $sd$ $su$ diquarks or a $\Lambda\Lambda$ bound state which is stable against fast strong decays into $\Lambda \Lambda$ was suggested by Robert Jaffe using the MIT "bag model". It was searched for and some evidence against the existence of a bound $\Lambda \Lambda$ state was found in sequential decays of doubly strange hypernuclei \cite{Farhi1984}. By now many other penta and tetra-quark bound states containing heavy b and c quarks  predicted by the “Naive” quark model were found. Remarkably precise predictions of their masses were made by Marek Karliner and the late Zvi (Harry) Lipkin and Jonathan Rosner by extrapolating from similar known quark structures. This is not readily applicable to the Hexa quark as the latter is the first example of a hadron that can be a bound state of the three $I=0\ S=0$  sd su and du di-quarks. Various lattice and other calculations did not find bound H states but did not conclusively rule it out. Hopefully this issue will be soon decided.} This along with the fact that the QCD phase transition is not a first order PT with no bubble formation and the likely evaporation of early forming small SQM nuggets heated by the intense flux of photons and neutrinos tend to rule out Witten’s fascinating SQM scenario.\footnote{The lower mass of the u quark is biasing the SQM nuggets to contain more u (and d) quarks than the heavier s quark, causing them to be electrically positively charged and therefore have an electron "halo" making the nugget more susceptible to heating by radiation.}\par
More speculative models e.g. by Ariel Zhitnitsky \cite{zhitnitsky}, try to explain the baryon asymmetry as well by having DM Nuggets made of baryons or of anti-baryons form in the early universe. If the CP violating  $\theta_{QCD}$ was initially $\sim 1$ and later relaxed into a tiny present value, then the formation of anti-baryonic Nuggets can be enhanced relative to those made of baryons leaving a matching excess of the observed un-clustered baryons. The relaxing of $\theta$ and the binding of the (anti) baryonic nuggets involve complex dynamics and in some of these models  there is also an appreciable Axion DM. A similar theme appears in ref. \cite{Kharzeev:2007tn}.  \par
We next consider more daring models with CDM of mass $M(X) \sim 1-2 GeV$ made of SM quarks. In these models satisfying requirement (1) and (2) are most challenging. Still we find these models intriguing.\par
In the model suggested and valiantly defended by Glennis Farrar \cite{farrar2018stablesexaquark}, DM consists of "Sexa-quarks"-S, made of tightly bound us, ds, and ud spin zero di-quarks. After the QCD phase transition the un-annihilated excess u.d. and s quarks get confined into the ordinary nucleons and into the Sexaquarks. Tuning the parameters of the model and the S mass in particular can yield a ratio of the density of the sexa-quarks i.e CDM density and that of the ordinary baryons of  $r \sim 5$ - (a possibility strongly contested by Mike Turner and Rocky Kolb in ref. \cite{kolb})\par
 The  Sexaquark  S  is a mythical reincarnation of the “normal” Hexaquark H whose apparent instability to $H \rightarrow \Lambda \Lambda$  was one of the reasons for abandoning Witten’s SQM. The mass of S is postulated to be close to the two nucleon threshold in order to avoid its weak decay. To avoid direct detection its size should be $\sim 0.2 Fermi$ -much smaller than that of ordinary baryons or mesons. Neither of these requirements can be provided by QCD -let alone the two jointly. Thus the interactions in the asymptotically free QCD at short distances are weak and cannot restrict the six quarks to a required small common volume of $(0.2 Fermi)^3$. The  GeV momenta of each quark implied by the uncertainty relation would then yield  $m(S) \sim 6 GeV \gg 2m(N) \sim 2 GeV$. Also both the Coulomb-like short distance and the long distance linear potentials between the $q_iq_j$ in the scalar di-quarks are half as strong as those between  $q_i\bar{q}_j$ in the corresponding pseudoscalar mesons. The size of each di-quark and a fortiori that of the three diquark composite, should then be larger than the size of the K meson implying normal hadronic nuclear cross-section of the sexa-quarks rather than the $\sim 20-200$ times smaller values of $\sim 0.1 -1 mb$ required.\par 
While the above casts doubts on the viability of Farrar’s model, the model raises the interesting question whether $O(GeV)$ halo DM particles with $\sim 0.1-1 mb$ cross-section on nucleons may have escaped detection. In Appendix F we recall that most such particles hitting earth are likely to be reflected after multiple collisions with atmospheric nuclei or otherwise stop before reaching the sensitive large detectors deep underground illustrating how reasonably strong $X N$ interactions can actually hinder the detection of certain DM types. \par
Neutral sexa-quarks can bind to heavy nuclei and form new isotopes the absence of which may argue against their existence. Still their reflection from earth, their small size weakening their nuclear interaction and disallowing bindings to Oxygen nuclei in ocean water and the proximity in mass  of high Z,A and new Z, A+2 isotopes mitigate potential difficulties of this scenario due to bounds on heavy isotopes (Glennys Farrar P.C). A careful, exhaustive, recent  critique - by Marian Moore and Tracy Slatyer "On the cosmology and terrestrial signals of sexa-quark dark matter" \cite{Moore2024} has largely debunked the model.

\section{\hspace{\secspace}The Fermionic Hierarchy and approaches to its resolution}
 \textbf{In which we describe the inter family Hierarchy of Fermion masses and some  approaches to its resolution.}
 \par
 \hspace{1cm}
Along with the Cosmological Constant (CC) and the hierarchy of Planck versus Higgs scale, the Fermion families constitute a most puzzling feature of the SM. A possible, albeit highly speculative, connection with “baryon” like DM prompted the following and the previous two sections. \par
The three families are often written in columns where fermions at the same vertical position  share \textit{identical} $SU(3)_c \times SU(2) _W \times U(1)$ em gauge interactions:
\begin{center}
\begin{tabular}{ m{2cm} m{2cm} m{2cm} }
 F(1) & F(2) & F(3) \\ 
 $u$ & $c$ & $t$ \\  
 $d$ & $s$ & $b$ \\
 $\nu(e)$ & $\nu(\mu)$ & $\nu(\tau)$ \\
 $e$ & $\mu$ & $\tau$
\end{tabular}
\end{center}

At present the $\sim 15$ dimensionless parameters: mass ratios and mixing of the Quarks and leptons of the various families are not explained\footnote{The masses of the Top quark and of the W,Z and Higgs boson are set by the single explicitly dimension-full parameter in the SM namely the VeV  $<H> =v$ of the  Higgs boson. The quark masses refer to the lagrangian masses induced by the Yukawa couplings to the Higgs rather than the “constituent” masses of the naive Quark model which include the $O(300)\ MeV$ mass induced by the confining strong interactions.}.  
The intergenerational “horizontal” hierarchy of masses of fermions in different generations, 
$$m_{top}>60\ m_{charm},\ m_{charm}>300\ m_{up} $$ 
$$m_{bottom} > 40\ m_{strange},\ m_{strange}>20\ m_{down} $$
\begin{equation}
    \label{9.1}
    m_{\tau}  >     17\  m_{\mu},\ m_{\mu}  >  200\   m_{e}
\end{equation}
is large . It can be summarized in a compact form:
\begin{equation}
    \label{9.2}
    m(3)  >    m (2)  >     m(1) 
\end{equation}
Where m(3) is a shorthand for the masses of the third and likewise m(2) and m(1) for the second and first generation.\par
The relation in Eq’s \ref{9.1} and \ref{9.2} seem to be a tautology: The top quark has identical gauge interactions as the charm and up quarks so that the only feature which distinguishes between them is their different masses and the same holds for $b$ $s$ and $d$ quarks and for $\tau$ $\mu$ and $e$ leptons. Changing this point of view underlies the new speculative "explanation" for $r=5$. in the next section.\footnote{A natural approach to  explaining this hierarchy is to assume that a small part of the large masses of the third generation fermions is  transferred  via a  “radiative mechanism” to the second generation and via a further iteration from the second to the first generation.} \par
The intergenerational mixing via the charge current ($W$ exchange) weak interaction - the V(1,2) element of the CKM (Nicola Cabbibo, Makoto Kobayashi \& Toshihide Maskawa) matrix conforms to the relation of Harald Fritzch \& Peter Minkowski\cite{fritzsch}:

\begin{equation}
    \label{9.3}
          V(1,2) \sim [m(d)/m(s)]^{1/2}  - [m(u)/m(c)]^{1/2}
\end{equation}

This suggested a perturbative mass leakage approach for explaining the intergenerational mass hierarchy. Thus let us assume that in some zeroth approximation the mass of the $d$ quark $m(d^0)$ vanishes but that of the $s$ quark $m(s^0)$ does not. If it is admixed with an amplitude $V(12)$ with the $s$ quark then the new "physical" state $\ket{d} = \ket{d^0} + V(ds) \ket{s^0} $ has a mass $m(d) \sim V(1,2)^2 m(s)$. This was suggested before the discovery of charm. The introduction of the second $[m(u)/m(c)]^{1/2}$ smaller correction term further improved the agreement with the measured value of the CKM entry. We note that the minus sign of the second "up" term relative to the first, "down" term, corresponds to the conjugate of the first matrix in the definition of the CKM matrix:
 $V(CKM) = C^+(up) C(down)$ where these unitary C matrices diagonalize the Higgs coupling to the up and down type quarks in the three generations  yielding a V(CKM) matrix that is necessarily unitary. Broader systematics and better understanding of flavor physics make eq. \eqref{9.3} a numerical curiosity. \par
 Efforts to explain the hierarchy using “radiative leakage’ of mass between neighboring families have been ongoing for $\sim 80$ years\footnote{$^{*}$ A serious obstacle facing such an approach is the very stringent 
experimental upper bounds on neutral flavor changing transitions. Thus an electromagnetic origin of the approximate relation between the muon and electron masses $m(e) = \frac{2}{3} \alpha m(\mu)$ considered by Dirac, Weinberg and Georgi \&Galshow, is hard to entertain when the upper limit on the branching ratio of radiative decay  $\Gamma( \mu\rightarrow e  + \gamma) \leq 10^{-12}\Gamma (\mu)=10^{-6}Sec^{-1}$ is far smaller than the value expected for a 100 MeV excited electron state.} 
The analog of inter family leptonic transitions in the quark sector - the flavor generation changing transitions as $s\rightarrow d$ or $b\rightarrow s$ via a second order two W exchange box diagrams such as for $K_{Long} \rightarrow \mu^+ \mu^-$ are also strongly suppressed. This was explained by the GIM mechanism "Weak Interactions with Lepton-Hadron Symmetry" S.L. Glashow J. Iliopoulos \& L. Maiani \cite{Glashow:1970gm}. 
It manifests via a cancellation between the two  Feynman Box diagrams contributing to the process with up and charm quark exchange. The underlying feature is the unitarity of the $3\times 3$ matrices C(up) and C(down) which diagonalize the $3\times 3$ hermitian matrices H(u) and H(d) describing the Higgs coupling to the up (or down) quarks-  thereby expressing the physical mass states as linear combination of the states participating in the weak interactions. 
The fact that charge current transitions were -by definition diagonal in the 3 dim family space preserves diagonality, up to the  calculable $m(c)^2 / {m(W)^2}$ correction was used to predict $m(c) \sim 1.5  GeV$! in the physical mass eigenstate basis\footnote{$^*$ Remarkably after the GIM mechanism paper was published and the discovery of charm, the electroweak model originally known as the "Weinberg Salam Model" became the celebrated GWS - Glashow Weinberg Salam" model- as it should have been all along.} \par
A simple approach to explaining the mass hierarchy is to have three Higgs fields H(1), H(2) and H(3) coupling to the first, second and third generation respectively with hierarchical vevs : 
\begin{equation}
    \label{9.4}
    \langle  H(3) \rangle  \hspace{2mm} \quad \gg \hspace{2mm} \langle H(2) \rangle  \hspace{2mm} \gg \quad \hspace{2mm}  \langle H(1) \rangle.
\end{equation}
This faces difficulties as experimentally only one Higgs particle at a mass of 125 GeV was discovered. On a more fundamental level the fact that the fermions of the first generation along with their gauge interactions are reproduced in two extra copies which seem to differ only by their heavier masses is highly puzzling. In particular there seems to be no obvious Anthropic advantage as the CP violation associated with the $3 \times 3$ CKM generational mixing matrix suggested in ref. \cite{Kobayashi:1973fv} is too weak to generate the correct excess $\Delta(B)=n(B) - n (\bar{B})$. The famous question of Isidor Isaac Rabi after the discovery of the muon "Who ordered this?" can be repeated within the more elaborate 3 families context.\par   

Returning to our main  theme -the 5:1 DM to baryonic matter ratio, it can be generated by using extra replicas of the SM residing on 5 different "Branes"
\footnote{$^*$Branes of various dimensionalities were introduced in the framework of string theory typically as a locus of string ends. Having "four branes" parallel to our SM slightly displaced in the extra dimension(s) can be used to introduce BSM physics and DM in particular, independent of the stringy motivation.}.   
A compulsory difference between the fermions in our brane and those on the other five branes which allows the latter to be "dark matter" is that the lightest, stable baryons on these five branes are neutral. As we saw above this is readily achieved by reversing the unusual sign of $m(u^0) - m(d^0)$ in our first brane so that 
\begin{equation}
    \label{9.5}
m(u^0_{K}) - m(d^0_{K}) = -  [m(u^0) - m(d^0)] \sim 2-5 MeV   
\end{equation}
for branes with index  $k = 2 -6$. The fermions in each brane have identical $SU(3)\times SU(2) \times U(1)$ gauge interactions and possibly further GUT or other Gauge interactions and identical Higgs sectors up to small corrections due to the flip of the tiny u-d mass difference. Hence we expect the same baryon excess in each of the branes with our $n(p) = n(e^-)$  baryon = lepton excess being matched by the $n'=\nu(e')$ excess in the other five branes yielding the ratio $r=5$  between the dark and  baryonic matter.\par

\section{\hspace{\secspace} A speculative connection between the three families and a particular SIDM}
The key to our further discussion of the Fermion problem is that we do \textit{not} view Eq.(\ref{9.2}) as a definition of the three generations. Rather we assume that the twelve entities in the three families  \#3 \#2 and \#1 $(t, b, \nu(\tau) ,\tau),\ (c, s, \nu(\mu) \mu )$ , and  $(u,d, \nu(e), e)$ are separate, well defined, fermionic fields.\par
In analogy with the motivation for the mirror models which realized the left right asymmetry by having in addition to the ordinary left sector also the sector with the right handed currents  we wish to restore a three family permutation symmetry by having all six different versions of the mass hierarchy of Eq.\ref{9.2} above, realized. Thus in addition to the particular mass hierarchy of Eq \ref{9.2} - we want to realize in nature five more possible mass orderings adding up with our sector  to the 6 = 3! permutations of the three families:   
\begin{subequations}
\begin{align}
\label{10.1}
             m(3)  >    m (2)  >     m(1) \\
             m(2)  >    m (3)  >     m(1) \\
             m(3)  >    m (1)  >     m(2) \\
             m(1)  >    m (2)  >     m(3)\\
             m(2)  >    m (1)  >     m(3)\\
             m(1)  >    m (3)  >     m(2) 
\end{align}
\end{subequations}
Not only the ordering of the masses is being permuted but so are  the actual masses. Thus, the top flavored quark has the measured mass $m(t) = 180 GeV$ in the first SM domain $D(a)$. However in domain $D(f)$ defined by eq(54f) above it has the mass of the corresponding quark in the middle family, namely the $\sim 1,4\ GeV$ charm quark mass as measured in our domain $D(a)$. The up-flavored quark in our first family which is the lightest among the upper members of the weak doublet of quarks  attains in the $D(f)$  domain the value of the corresponding entry in the third family, namely the 180 GeV mass of the $t$ quark in our domain. The charmed flavor fermion attains in  $D(f)$ the value of $m(u)$ (Which was upped by $\sim 3 MeV$ so as to be the heavier member in the Weak isospin quark doublet in the lightest ,first, family). This lower member in $D(f)$ domain has the “s quark'' flavor but with the mass $m(d)$ taken to be the original $m(u)$ mass in our domain of $\sim 3 MeV$.\par
We cannot however have all these possibilities realized in our full 3 dimensional space. In particular Eq(54-b) corresponds to a sector where just the two heavier families \# 2 and \#3 have been permuted but our own lightest \#1 family stays almost the same. The only change is that the neutron therein, which we will denote by $n'$, rather than the proton $p'$ is now the lightest baryon. This $n'$ of mass of $\sim 930 MeV$ has the usual $ddu$ quark content. 
\par
The novelty suggested here is that each of these six different permutations is realized in a\textit {separate} domain in our 3d space. This avoids the above serious difficulty and more generally reduces the problem of strong interaction of ordinary and dark matter. Also  the  proliferation of the light neutrinos as in the above simplistic six brane scenario is avoided.\par
We still have the desired six fold enhanced "matter" by the five extra neutral baryons each having a mass $m(n') \sim 5-8 MeV$ lighter than that of our neutron in the five domains $b-f$ which not only harbor the same gauge interactions but also have overall the same physical size and shape and therefore the same baryon asymmetry. We identify these n’s with the asymmetric CDM thereby explaining an $r$ value which indeed is quite close to 5.\par   
Just like the $udd$ in Domain (b), the $css$ flavored baryon in domains $D(c)$ and $D(d)$ and the $tbb$ flavored baryon in the remaining two domains $D(e)$ and $D(f)$ all will be neutral and will have a mass and interactions very similar to our neutron.  Hence unlike in our domain where recombination $p +e \rightarrow H$ must occur before the baryons decouple from the CMB and start clustering, the neutral neutron like DM is a good CDM clustering early and supplying the  potential wells which our baryonic matter later fall into and form stars and other structures on all scales, sub-galactic galactic and beyond. Only our S.M. domain has the ($p-e$) plasma this accounting for the ‘BAO’ like structures in the CMB power spectrum observed.\par
\begin{figure}[h]
\begin{center}
 \includegraphics[width=0.3\textwidth]{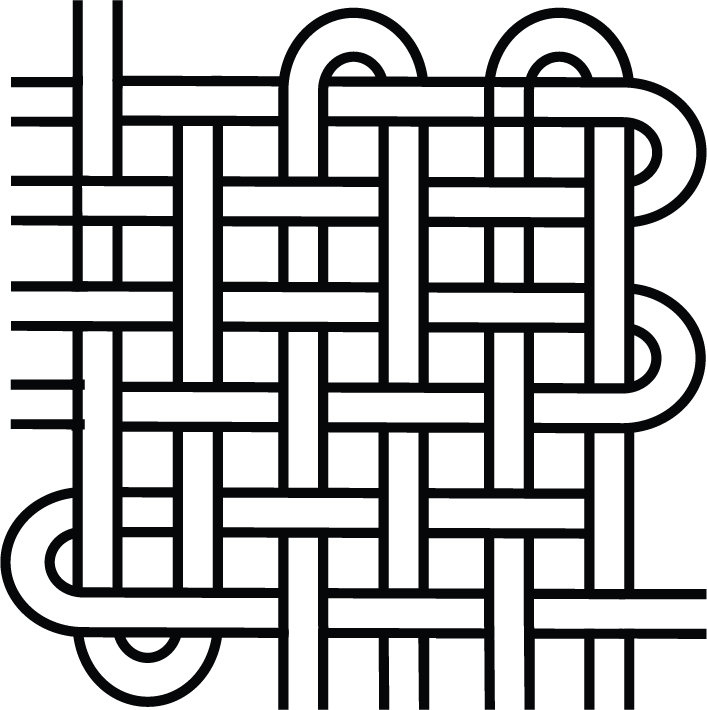}
    \caption{Tightly interwoven air duct systems}
\label{fig:08}
\end{center}
\end{figure}

In order to maintain some degree of isotropy and homogeneity the six different domains should be tightly interwoven in a manner similar to say 6 distinct convoluted systems of air-ducts is in fig. \ref{fig:08}
\par
This cannot happen in two space dimensions and $d=3$ is the lowest dimensionality allowing it.\par
One immediate important consequence of our unusual set-up is that the DM which moves around us with typical halo-galactic virial velocities of $v\sim 300 Km/Sec = 10^{-3} c$ \textit{cannot} reach any detector in our domain. Thus if the DM particle in question is the $n'$ in domain $c$ its true flavor content is one $c$ quark and two $s$ quarks. If it penetrates the domain wall and transfers to our domain it would become the neutral but heavy baryon of mass equal to that of the $css=\Omega_c \sim 2 GeV$ in our domain. Clearly the tiny kinetic energy of the DM particle $KE= \frac{1}{2} m(n')v^2 \sim 1/2 KeV$ is insufficient to allow crossing the $\sim GeV$ potential barrier separating domain $c$ and domain $a$. The mass difference barrier is minimal for $n'$ DM particles in domain $b$ but $m(n)-m(n') \sim 8 MeV$ is still insurmountable. This then would ‘explain” why we have not detected to date any DM despite the fact that $1/5$ of it  is made of the same quarks as our own nucleons and the remaining part of other neutral combinations of SM quarks.\par
 The See-Saw mechanism that generates the light neutrino masses in our domain operates in the other domains. Then the mass differences  $|m(i)-m(k)| \leq 0.1-1 eV$ of  neutrinos produced by weak decays of moving or stationary hadrons or nuclei are much lower than their energies and they can freely sail between the different domains suggestive of larger mixing angles of neutrinos relative to those of the quarks.\par
One last curiosity is that since $m(n) > m(n')$ with $n'$ the neutron analog making up the DM in domain $b$, our neutrons could occasionally cross the domain wall  into this domain  gaining the mass difference  $m(n) -m(n') \sim 6.2 -9.2 MeV$ as kinetic energy. As is well known quantum mechanics tends to hinder transitions from regions of high energy to regions of lower energy thereby reducing the rate of such meandering - which conceivably would account for the neutron lifetime anomaly by having some fraction of the neutrons in our domain $a$, disappear in the domain $b$  before the original neutron $\beta$ decays.\par
Even if the neutron lifetime anomaly persists, the above  “explanation”  faces difficulties. Thus consider a particular stable nucleus $(Z, N)$ with $N>Z$ of mass lower than that of its neighboring Isotopes. If it moves from  our domain to domain $b$ it keeps its quark content and has very similar nuclear bindings but each  $n'$ is lighter than the original neutron by $\sim 6-9 MeV$. The original nuclide will decay into $( Z, N-1, n')$ + a 6-9 MeV $\gamma$. \par
Much more serious difficulties are tied with the  new length scale $l$ characterizing our domains\footnote{$l$ is not related to the equal volumes of the separately connected and structurally similar, infinite domains each extending over a sixth of all space. Rather it is the typical width of the  intertwining "Air duct pipes" fig.\ref{fig:08} making up the domains which for simplicity we take to be similar to the distance separating these junctions.}.
Unfortunately no $l$ value is satisfactory. $l$ cannot  be astronomical as this conflicts with the  isotropy and homogeneity of space. Microscopic $l$ is unsatisfactory as well. Adopting the natural assumption that the fermions have the same gauge interactions irrespective of the domain where they reside we can have some interactions between the neutron’s in the other domain and our hadrons and electrons respectively -akin to what happens in Farrar’s model.\par
The above difficulties pale  in comparison with the following : Our protons cannot move along straight line trajectories if their energy does not exceed $2 m(t) \sim 400 GeV$ and likewise  electrons of energy below $m(\tau) \sim 2 GeV$ would reflect from the domain wall separating our domain $a$ and domain $f$. The reason is that upon traversal of this domain wall our $uud$ proton attains the mass of the $ttb$ heavy fermion and the electron attains the mass of the $\tau$ lepton. Phrased differently we will  fail to accelerate electrons or protons as after moving a distance l the accelerated particle will reflect from a domain wall!\par
To address these difficulties and preserve the $r=5,  DM/B$ ratio and spatial segregation of the DM particles we are then led to postulate that $l$ is the minimal  length $l \sim l(Planck)$. This can happen within frameworks where space time is emergent and at the shortest distances consists of six interpenetrating dynamical lattices with such a lattice spacing. One particular such framework  without the six domain option -was discussed by me in "Net for Toe" presented in Aharonov 80th birthday FestSchrift \cite{Struppa2013}. A new version containing also the six domains scenario is presently under preparation.

\section{\hspace{\secspace}Some basics of WIMP detection}
\textbf{In which we list some difficulties facing the direct detection of  particle-like  DM  and efforts to overcome them.}
\par
\hspace{1cm}
The following may help appreciate the difficulty of discovering D.M. particles which interact (super) weakly with ordinary nucleons. Thus consider a DM particle with mass $M(X) \sim 10 GeV$ on the verge of being discovered in the underground 10 tonne liquid Xenon detector. Let’s assume spin and isospin independent DM nucleon scattering with  DM -Nucleon cross-section of $10^{-45} cm^2$ which present measurements still allow for the above mass. For momentum transfers $q= p(DM) \sim 10 MeV$ the scattering from the $O(5 \text{Fermi})$ nucleus with $qR \sim 1/4$ is largely coherent and the isotropic, with $\sigma(X-\text{Xenon}) \sim A^2.\sigma(XN)\sim 3.10^{-41} cm^2$. The total number of expected DM collisions then is:  $$N_{\text{collisions}} \text{(total)} = N(Xe).\Phi(DM) \sigma(X - Xenon) =10 \hspace{0.1cm} \textnormal{events/year} $$ 
with $N(Xe)= 0.5 \ .10^{29}$ Xenon nuclei in 10 tonne of liquid Xenon and the DM flux $\Phi$, the product of the DM number density $n(DM) \sim 0.3 GeV/{M(DM)}cm^{-3}$ and their average velocity  $v\sim 300 Km/{sec}$.\par 
The  task of finding over a period of t $\sim$ 1 year, 10 events with deposited recoil energies of order $(m(DM) v_{virrial}^2)/2 \sim 5 KeV$ inside the big detectors is most demanding. That the large Xenon underground detectors have managed to exclude DM of mass $\sim 50 GeV$ and cross sections on nucleons as small as $\sigma \approx 10^{-48}cm^2$ is truly remarkable.\par
Despite their depth the detectors are inundated by muons. The  spallation neutrons they produce and a variety of other sources of radioactivity around and inside the detectors contribute to the background. The heroic efforts reducing the background by the many orders of magnitude required, done early on for germanium detectors which utilized also the pulse shape expected, were nicely described in \cite{avignone}. 
\par
The challenge of detecting tiny energy depositions arose first in efforts to detect solar neutrinos. Even for the $\sim 12 MeV$ Boron neutrinos the nuclear recoil energies are tiny. The methodology suggested for detecting these small recoil energies utilizes the very short paths of the heavily ionizing recoils which imply that the energy deposition is localized within a tiny region. If the material in these tiny regions is unstable, then the resulting tiny heat deposition can -just as in the case of bubble chambers- induce a phase transition which greatly amplifies the initial signal. An ingenious idea presented in ref. \cite{Drukier:1984} realizing this concept was to use superconducting micro spheres held in a rigid matrix permeated by a strong uniform magnetic field and kept at a temperature slightly below the (rather low) critical temperature $T_c$. The tiny $E(recoil)$ deposited heats up one sphere of  heat capacity reduced according to the $T_c^3$ scaling at low temperatures, by a tiny amount. This suffices however to stop it from superconducting, the previously excluded magnetic field rushes in and the changing flux is picked up by squids. \par
This original idea was surpassed by a simpler approach using much larger ultra cold, single crystal detectors (see ref. \cite{Fiorini:1983yj}). Appropriately doped pure crystals almost freely propagate scintillations and phonons and may offer directional information as we show later. A somewhat similar more recent suggestion is to utilize superconducting nanowires (of diameter d $\sim$ few nanometers), again maintained underground at a temperature slightly below  a very low  $T_c$ . The tiny amounts of heat transferred in a light DM collision then revert a short section from superconducting to normal dramatically decreasing the current in the wire. This has in particular been adopted for detecting dark photons from the sun where we can use regular lenses which in principle allow focusing of the converted photons magnifying the flux by up to $[T(\text{Sun})/{T \text{(Earth)}}]^4 \sim 10^{6}$. See e.g. ref. \cite{Chiles2022} \par
Recently A. Drukier suggested utilizing chemical/biological instabilities to amplify weak signals due to DM interactions. DNA based detectors were suggested in \cite{Drukier:2014rea} and searching "Foot-prints" left by DM interactions over some fractions of earth's lifetime in geologically stable rocks. \par
Another approach for detecting solar neutrinos using liquid Helium detectors championed by Robert Lanoue and collaborators some 40 years ago may be resurrected in efforts to detect light DM in the Herald project. The $T_L \sim 4^o Kelvin$ liquefaction temperature can be readily achieved and being the lowest liqufication point of all gasses it allows sedimentation of all chemical impurities. The absence of stable or long living radioactive isotopes is another great advantage. If the Helium is at a temperature slightly below $T_L$, say by $\delta(T) < 0.01 \text{Kelvin}$, then an energy deposition $\delta(E)$ vaporizes $\frac{10^9 {\delta(E)}}{\text{Kelvin}} $  Helium atoms. Since $\rho(\text{Liquid Helium}) \sim 0.1 gr (cm)^{-3}$ we find that for $\delta (E) \approx 10 \text{Kev}$ these atoms occupy initially a volume of $\sim 10^{10} (\text{Angstrom})^3$ which expands after evaporation forming a $10^{12} \text{Ansgstrom}^3 = \text{micron}^3$ hot bubble which will quickly buoy to the upper surface where it emits phonons (or rotons). The latter can be detected by sensitive thermistors and the pattern of emission can carry directional information. Due to the 30 fold lower density and $\sim 30$ times less gain from coherent scattering it would require a $\sim 1000$ times bigger liquid Helium detector as compared with a liquid Xenon detector to have the same number of interactions of the $>$ 100 GeV LSP particles. This may explain why in the LSP = WIMP era the Helium alternative was not followed up. Amusingly, the intermediate case of liquid Argon Time Projection Chamber (TPC) presently used in DUNE, the main detector in US long baseline neutrino effort, can serve also as a multi-purpose DM detector.\par  
Returning to present day WIMP detectors, the mandatory purification of the detector material is often  achieved by growing single crystals, a process that segregates chemical impurities. This  still leaves radioactive Isotopes. To distinguish the multiple interactions in the detector of a  $\beta$ electron or $\alpha$ particle, constituting radioactive and other backgrounds, from the spatially and temporally localized DM interactions -two signal types are used by most collaborations. These pairs of signals are chosen among the three possibilities of Ionization, fluorescence and phonon detection. Only the few “good” events where both signals indicate the same energy deposition \textit{and} are consistent with the spatial and temporal localization demanded, are used in the analysis. Having presumably detected some DM events we still need further indications for their DM source to help distinguish the rare DM events from background. We mention some strategies for achieving  this goal in Sec XIII

\section{\hspace{\secspace}A south-pole neutrino detector which  helped exclude “Classical” LSP WIMPs  and the  “Paradigm shift” to lighter DM
}
The beauty and grandeur of the idea \cite{silk} of using the whole sun or earth as DM detectors, which excluded many "classical" $(M(X)\sim 0.1-1 TeV)$ WIMP scenarios, behooves us to mention it, leaving  details to the excellent review in \cite{jungman} (which provides also a clear account of the "Wimp Miracle and SUSY DM" in general). It illustrates how well we can search for something in the rare cases when we know exactly what we are looking for. The  search strategy using particle and astro-physics proceeds in four stages:
 \begin{itemize}
     \item[1] DM particles interact occasionally with a solar/terrestrial nucleus (A,Z) at radius $r$. The condition that this will gravitationally capture the WIMP into a bound orbit around the sun/earth is that the fraction $fr \sim A.m(N)/M(X)$ of the X kinetic energy E(r) at the time of collision at a distance r from the center of the sun (earth):  
\begin{equation}
    \label{12.1}
E(r) =E(gained\ in\ infall) + E(initial)= M(X) (v_{\infty}^2 +v_{es}(r)^2)/2 
\end{equation}
transferred to the hit nucleus exceeds the Wimp’s initial energy $M(X)v_{\infty}^2/2$ implying a negative $E(final)_{\infty}$ i.e. a bound orbit. $v_{esc} (r)$ is the escape velocity from the internal point at r. For collision with Iron (A=56) nucleus $fr \sim 0.1$ if $M(X) \sim 0.5 TeV$. With $v(r)$ in earth being $20 Km/{sec}$ this condition becomes $v_{\infty} < f^{1/2} v_{\text{escape}}(r) \sim 2 Km/Sec$. Only a small fraction of slow DM particles  $[v(\infty)/v(\text{Virial}) ]^3$ can than be captured.\par
Thus despite the lighter composition of solar nuclei and the proximity of Earth's center, the sun with its very large (up to 1500 Km/Sec) escape velocities from internal points is a more effective capturer of DM than Earth. 
\item[2] The bound Wimps keep traversing the sun/earth colliding with nuclei therein, losing energy and migrating inward. This continues until they reach a kinetic energy $\sim KeV$ or $\sim eV$  corresponding to the temperature at the solar (or earth’s) center. They then settle into an “atmosphere” with a gaussian profile and size  $R=[kT(0)/({G_N.\rho(0) M(X)})] ^{1/2}$. Using $T=KeV$ and $\rho(0)\sim 150 ~gr/{cm^3}$  we find $R\sim 2.\ 10^{-4}$ of the solar radius. Once the total number of accreted WIMPs gets to a critical value a steady state is reached where the rate of mutual WIMP annihilation in the above region  is matched by the rate of accreting new WIMPs. Detailed calculation shows that it takes a fraction of the 5 BYR solar age to establish such a steady state.
\item[3]The annihilations lead, via various decay chains, to final photons or charged leptons which are absorbed and $e$ $\mu$ and $\tau$ neutrinos (and antineutrinos). Most neutrinos have energies lower than 100 Gev and the resulting small nuclear cross-sections allow them to freely escape the sun The energy spectrum of the neutrinos, like all other aspects of the previous stages can be exactly calculated in any specific SUSY model.
 \item[4]Enter the fourth, experimental part of the story. It is the giant Ice Cherenkov detector at the south pole "Ice Cube" which after installing all $160 \sim Km$  long strings each carrying $\sim 50$ photomultipliers, covered an effective volume of $Km^3$. When a muon neutrino interacts inside the detector or in near-by surrounding ice and rock, the direction of the resulting muon moving inside the detector can be measured with $\pm (1/ 2)^\circ$ accuracy and, in the above scenario, should point to the sun (or be in an upward direction pointing to the earth's center). Having calibrated the Ice cube detector with known atmospheric muon and neutrino fluxes the absence of the above signals kept restricting the allowed SUSY models. 
\end{itemize}
This, the tight bounds from the Xenon and other DM direct searches and the lack of evidence for SUSY at the LHC keep narrowing the scope of classic LSP Wimps. For many researchers this led to a paradigm shift where theoretical works and experiments focus on sub Gev DM and/or DM interacting mainly with electrons occupy center stage. The shift started gradually and at various places. Jonathan Feng \& Jason Kumar \cite{feng}, emphasized the possible WIMPless miracle, namely that the $\bar{X} X$ annihilation cross-section $\sim \frac{g^4}{M(X)^2}$ can track the value producing the correct relic density when $g^2$ is decreased by the same factor as $M(X)$. \par
Dark photons mediate and generate mutual DM interactions and via kinetic mixing with the SM photon : $\epsilon F^{\mu.\nu} F'_{\mu,\nu}$, also DM-SM interactions in many DM scenarios. Substantial regions in the $ln(M(X))- ln( \epsilon)$ plane have been excluded by fixed target and beam-dump experiments initiated in CEBAF and others accelarator labs (see ref. \cite{Bjorken:2009mm}) and by new direct detection experiments dedicated to light DM searches.\par
High A nuclei such as the Xenon Cesium or Argon and Silicon cannot be the targets in searches for light DM which interacts only with nucleons since the recoil energy $\sim \frac{M(X)^2 \hspace{0.1mm} v \hspace{0.1mm}(X)^2}{M(A,Z)}  \sim \frac{GeV.10^{-6}}{100}\sim 10eV$ for $M(X) \sim GeV\ A \sim 100$ and $v=10^{-3}c$ is tiny.\par
If the DM interacts with electrons and its kinetic energy $$M(X) v^2/2=(M(X)/{GeV})\ 5.10^{-7} GeV \sim 50 ev$$ exceeds the $\sim 9 eV$ ionization energy of Xenon then even DM with $M(X) ~\sim 0.2 GeV$ can be detected by looking for freed electrons (and ions). This simple observation allowed R. Essig, A. Manalaysay, J. Mardon, P. Sorensen, T. Volansky in "First direct detection limits on sub-GeV dark matter from Xenon10" \cite{Essig2012} to put useful upper bounds on the cross-section of such DM with electrons - an approach extensively followed up and improved on. Some of the strategies underlying nuclear WIMP detection described in the previous section can be translated over to the new regime and more generally applied to the new -electron/atomic/molecular/condensed matter excitations. The strong limits on DM of masses in the 20 -GeV - TeV mass range implied by the multi-ton underground cryogenic Xenon experiments suggested using also the "Migdal effect" of ionization via nuclear recoil shown in fig. 9 \footnote{$^{++}$ That Leptophilic DM (X) - scattering may generate Nuclear recoil by the inverse Migdal effect namely X interaction with electrons with the later columbically interacting with the nucleus as shown in fig. 10 is quite unlikely}. 

\begin{figure}[h]
\centering
\begin{minipage}{0.3\textwidth}
\captionsetup{justification=centering, aboveskip=20pt}
    \centering
    \includegraphics[width=\linewidth]{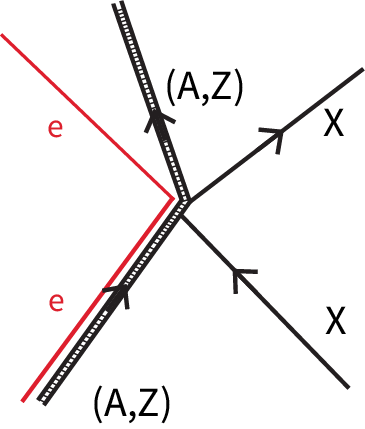}
    \label{fig:09}
    \caption{Feynman Diagram for the Migdal effect, namely, shaking off electrons by nuclear collisions of heavy Hadrophilic DM X}
\end{minipage}
\hspace{2.9cm}
\begin{minipage}{0.34\textwidth}
\captionsetup{justification=centering, aboveskip=29pt}
    \centering
    \includegraphics[width=\linewidth]{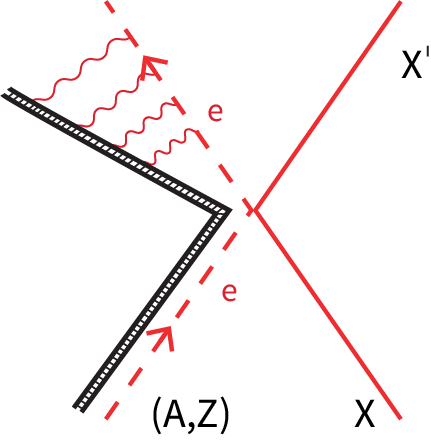}
     \label{fig:10}
    \caption{Feynman diagram for the less likely inverse effect in the case of leptophilic DM colliding first with electrons which subsequently generate nuclear recoil}
\end{minipage}
\end{figure}

The analysis of limits due the high energy ICECUBE neutrinos generated by DM capture and annihilation in the sun was extended to the leptophilic case in \cite{maity2023}. Free electrons do not contribute much to the capture of the DM particles due to the tiny fraction of the X energy $~ m(e)/ m(X) \sim 10^{-4}$ lost in X-e collisions. Also scattering on the predominant light elements in the sun is inefficient for slowing down D.M. X particles. \par  
Ref. \cite{Essig2012} above is an example of a sensitive search for a new type of DM (here, DM of masses $GeV> M(X) > MeV$ which interacts with electrons), carried out by reanalyzing  data from an existing experiment originally designed with a different goal in mind - searching for massive WIMPs with nuclear interactions in the present case.
\par
The G.W. detectors can search for special forms of D.M. The low mass - needed to guarantee a sufficient number of near fly-by at $ \leq 4km$ distance of DM chunks, cannot move by gravitationally pulling the LIGO mirrors by the smallest distance detectable -a fraction $h=10^{-22}$ of the total effective path length of the arms in the LIGO interferometer. However future planned G.W detectors in space like LISA may be sensitive to a range of DM masses. Another suggestion uses the ongoing “nano - gravity” project - monitoring the very precise rotation periods of many pulsars in an effort to see a common oscillating Doppler shift due to a passing GW or a passing DM such as a PBH  of mass $\sim 10^{22}- 10^{23} gr$.\par
A beautiful analysis attempting to detect clouds of DM made of ultra light dilatons was performed by the LIGO group. The classical scalar dilaton field $\phi$ in a cloud induces in many models, a tiny fractional change of the electric charges  
\begin{equation}
    \label{12.2}
           \epsilon = |\delta(q)/q|  \sim \phi(\vec{r},t)/ M 
\end{equation}
where the dilaton field $\phi$ oscillates in time with a frequency $\sim m$, the mass of $\phi$ and $M$ is the high scale of the new physics allowing this unusual behavior. This changes by $\epsilon$ the atomic radii/electron densities and therefore also the optical path of the laser beams inside the beam splitter- a center-piece of the LIGO detector. While the interfering beams traverse the LIGO arms and the beam splitter the same number of times the $\sim 10 cm$ beam splitter  is $\sim 10^{-4}$ times shorter than the Km long LIGO arm making the sensitivity to the above $\delta q/q = \epsilon  10^{-4}$ times smaller than the optimal $h \sim 10^{-21}$, namely
\begin{equation}
    \label{12.3}
        \epsilon \gtrsim 10^{-17} 
\end{equation} can be looked for.
Just as for a GW, the effect is transient lasting as long as the cloud overlaps LIGO.  The $\sim 50$ Hertz frequency where LIGO’s  sensitivity is maximal corresponds to a rather long wavelength $\sim 10^{8} cm\sim 1000 Km$. The size of the cloud is typically larger than that of the carrier wavelength. If it is big enough to simultaneously affect the LIGO and VIRGO GW detectors, then the background is reduced enhancing the prospect of detection. We note that the above wavelength is, up to a $2\pi$ factor, the range of the Yukawa potential generated by the Dilaton exchange  Since the force induced by dilaton exchange is not proportional to the gravitational attraction, it is constrained by recent high precision tests of the equivalence principle at distance $r$ of $1000 Km$ or more by the MICROSCOPE satellite experiment. \par 
We close this subsection with special purpose detectors aiming to detect DM particles of low masses. A prime example is SENSEI aiming to limit DM as light as 10 Mev interacting with electrons. Using  semiconductors where electron-hole creation occurs at significantly lower energies the detection threshold is reduced by an order of magnitude  and by repeated querying multiple times many individual CCDs it becomes sensitive to even single electron excitations.  For a recent status and summary see: "First Direct-Detection Results on sub-GeV Dark Matter from SENSEI at SNOLAB" \cite{sensei2023}.\par 
All the above notwithstanding, many physicists working at the LHC, or DM direct searches would argue for the viability of WIMPs in general and LSP DM in particular, paraphrasing Mark-Twain saying into  “Rumors about the  death of the (heavy) WIMPs were highly exaggerated”. Thus Dan Hooper presented in a recent cosmology-particle physics meeting at Saint Louis six arguments why indirect searches of WIMPs need not be affected by bounds on Direct Detection. The halo can yield an appreciable indirect detection signal particularly via an almost monochromatic $\gamma$ line. Interesting papers devoted specifically to the prospect of using this to discover Multi-TeV WIMP are: \cite{bottaro2023} and \cite{Bottaro_2022}. We will revisit indirect detection below.

\section{\hspace{\secspace}Directional, Temporal variations - A possible Key to Dark Matter Discovery}
 \textbf{In which we recall the expected anisotropy of the WIMP’s and the resulting annual variation which can be seen  in purely calorimertic detectors. We comment on DAMA-LIBRA, the only experiment claiming to have observed the modulation and on directional effects in crystals that may help detect axion-like particles.} 
 \par
 \hspace{1cm}
 
The annual modulation effect namely the periodic variation of massive DM signals suggested in \cite{Drukier1986tm} is presented next in three steps:

\begin{itemize}
    \item[a.] Assuming uniformity and isotropy of the halo DM, the rotation of the solar system with the rest of the galactic disc at $\sim 220 Km/Sec$ relative to the halo, exposes us to a "DM wind" coming from a specific sky direction near the Cygnus X(3) constellation. If we had directional DM detectors we would expect  to see about twice the flux and DM  with 4 times the average energy coming from this direction. While this strongly motivates directional DM detectors, most present detectors and the bigger ones in particular, are calorimetric and a more subtle approach is needed in order to utilize this directionality .
\item[b.]The $60^o$ angle between the galactic and the ecliptic planes makes the component of earth’s  velocity in the galactic plane equal $17 Km/{Sec}$. This component  is (anti)parallel to the WIMP wind in the ( summers) ,winters of the northern Hemisphere. Hence the “wind velocity ”  or flux  which our detectors see  is  $\pm 8\%$ stronger/weaker in the summer/winter and the kinetic energy is correspondingly $\pm 16\%$ bigger/smaller.

\item [c.]  The resulting annual variations can be seen in calorimetric detectors via a small periodic change of the rate of events observed  and of  the energies deposited.
\end{itemize}
Streams of (Dark and baryonic) matter at our location suggested by analysis of the results of the Gaya collaboration can affect the annual modulation. However the velocities of most streams are smaller than the galactic rotation velocity and the resulting change of the WIMP wind direction and magnitude are thus relatively small.\par
The annual modulation differently affects different  DM types and their prospective detectors.\par 
Cosmological neutrinos have temperatures of $T(\nu) \sim 1.9 Kelvin\sim 1/6000 eV$. This is lower than  putative ($0.1 eV$) neutrino masses implying NR velocities $v(\sim[2T/m]^{1/2}\sim c/16$ which however is much higher than the escape velocity from our galaxy of $600 Km/Sec$. Consequently there will be no accumulation of neutrinos in the halo. Annual modulation cannot help detecting such low mass Cosmological Neutrino Background (CNB) nor axion detection via resonant cavities. Also the ingenious idea of Stodolsky (1974) to use an asymmetric degenerate neutrino sea in the galaxy to induce macroscopic parity violation effects, is by now mute\footnote{It has been recently suggested in ref. \cite{Arvanitaki:2022oby} that the different index of refraction of electron neutrinos and antineutrinos in earth dramatically enhances (suppresses) the density of anti(neutrinos) near the earth surface. Unfortunately a more careful analysis in \cite{Gruzinov2024} largely refuted this truly fascinating suggestion. The "Wimp wind" effect while reduced for 0.1 eV neutrinos by  $ v(ViR) / v(\nu) \approx v(ViR) / 0.1c \approx  10^{-2}$ generates some anisotropy which may allow some residual effect}  \footnote{If B-L is conserved, then any excess of neutrinos over antineutrinos implies an equal excess of neutrons. The tiny baryon asymmetry of $\sim 6. 10^{-10}$ then forbids appreciable neutrino excess. It also forbids the $\Delta(L)=2$ mass term responsible for the seesaw mechanism explaining the smallness of neutrino masses. A stronger than gravity interaction stemming from Gauging the B-L as a conserved local $U(1)$ conflicts with precision tests of the equivalence principle or searches for shorter range "Fifth interactions". Extra $U'(1)$ long range interaction should therefore couple precisely in the same manner as does our electromagnetism allowing massive/massless mixing dark photon with Milli-Charged (MC) DM particles in the massless case.  Rabindra Nath Mohapatra \& Goran Senjanovic found that B-L is naturally conserved in their L-R framework. This B-L gauge symmetry is broken at a high scale, the B-L "photon" then becomes heavy and we can have a Majorana neutrino and–$\bar{n}n$ mixing suggested by Vadim Kuzmin and  by Robert Maharshak \&  R. Mohapatra .}.
\par
The Rutherford scattering of milli-charged DM particles off charged nuclei grows  as $v^{-4}$ with decreasing velocity, implying opposite phase annual variations-as compared with the case of interactions which grow with $v^2$. The opposite phase is expected also for Axion Like Particles (ALPs) originating from the core of the sun which is closer in the winters in the northern hemisphere.\par
While the recoiling Ion is always moving in the forward hemisphere relative to the incoming DM particle direction, its path is too short to be observable. An exception can be provided by very thin mono-layered devices such as the planned Graphene detectors.
Inside crystals the direction of the very short recoil may be important making. Detectors consisting of single crystals can --- directional information in search of solar axions. The the rate of conversion of a solar axion of $\sim KeV$ energy into an X ray photon in the strong electric fields of the the crystal Ions depends on the direction of the photon (and of it’s parent axion) relative to the crystal Bragg planes as the latter direction determines how long the coherent conversion can be maintained. This  beautiful idea - suggested in "Coherent production of light scalar particles in Bragg scattering" ref. \cite{Buchmuller:1989rb}, \cite{PASCHOS1994367}, was applied by Creswick et-al in \cite{Creswick:1997pg}. The analysis here is much more complicated than for the above annual modulation as we need to track the daily and yearly changing direction of the sun relative to the crystal planes during all the experimental run but it paid off by almost a  hundred fold gain in sensitivity.\par
Another direction related effect is the "Channeling" of the recoiling ions as they travel longer distances between the main, more widely separated, lattice planes and scintillate more in the process. Using in DM searches the daily modulations induced by channeling as the planes of the crystal rotate relative to the WIMP wind was first suggested in measurements of anisotropic scintillation efficiency for carbon recoils in a stilbene crystal \cite{Sekiya2003} and later in the context of the DAMA experiment in \cite{creswick}. The effect of channeling can be tested experimentally by neutron beams with tagged scattering events\footnote{While external ions shot into a crystal can directly enter optimal "Channels", an internal ion that the DM collided with causing it to recoil in the direction of a "channel", tends to be  blocked by the next ion along these directions. As noted in \cite{bozorgnia} this "Blocking" considerably reduces the effect of channeling.}.
\par
Annual modulations extending over 12 years were seen by the Dama and the Dama-Libra collaborations see e.g.
\cite{DAMA:2008jlt}
but not by any other group- which excluded the large  D-L effect by several orders of magnitude. The perseverance and ingenuity of the D-L collaboration manifesting, in particular, in the ultrapure optimally doped $NaI(Tl)$ crystals generated by alternating crystallizations and meltings are commendable. However their unwillingness to share this with other groups impeded similar set-ups and in particular a  D-L "Tween" at the south hemisphere which would allow zeroing out other seasonal variation. One such variation was discovered by the MACRO collaboration many years earlier in the same Gran-Sasso tunnel while conducting a search for magnetic monopols. It is that the expanding (contracting) warmer (colder) atmosphere in the summer (winter) increases  (decreases) the number of energetic pions decaying in the atmosphere into energetic penetrating muons which can generate spallation neutrons under-ground in the detector’s neighborhood. The D-L collaboration claims to exclude the effect of penetrating muons by verifying that their counts were not in coincidence with those in a shielding external muon counter\footnote{David Nygren, who developed the all important Time Projection Chamber (TPC) used in accelerator and non-accelerator detectors, noted that this may fail  for the following reason : The  Cesium Iodide crystal used by D-L can misfire via sudden release of strains induced by earlier interactions of energetic muons keeping the connection between the muon rate and annual modulations of the count rate.  Such flashes cannot happen in  liquid detectors as liquids cannot sustain strains. Also combining the ionization and fluorescence signals expected from a Wimp induced isolated nuclear recoil can reduce to acceptable levels the “Spontaneous flash signals”  in crystals but this was not done by the D-L collaboration. Such flashes may indicate excessive stresses in rocks and predict pending earthquakes.}.
 Most experimental details including time flags for individual events were not released by the D-L collaboration and by other experimental groups performing direct searches for DM. In some cases where the data are in the public domain and can be "Recast" for any analysis, eager theoreticians “discovered” production and decay of SUSY particles at the LHC and annihilating DM in denser spots of our halo in the Fermi-Lat satellite data. A joint approach revealing important features which escaped the separate analyses in the several large scale direct DM searches justifies risking such multiple DM “discoveries”.\par
Repeats of  D-L  are presently being done. The COSINE collaboration in Korea using the same procedure as DL found similar annual oscillations but with an opposite phase \cite{PhysRevLett.123.031302}. Both the D-L and COSINE results can be attributed to different annual background subtraction. While the D-L Saga is by now largely closed, we discuss in chapter XVIII a special type of a "resonant" DM inspired by it - a prime example of DM that cannot avoid being discovered.

\section{\hspace{\secspace}DM and Black Holes}
 \textbf{In which we digress on various lucky accidents and in particular  the unexpected large number of coalescing  massive binary BHs which greatly helped  LIGO to discover GW but cannot be a substantial part of DM}
 \vspace{0.2cm}  \par
"Lucky" circumstances often helped advance human civilization and science. This includes the serendipitous discoveries of radioactivity by Beckerel and of Penicillin by Fleming. More relevant to the following  are astronomical facts such as the presence of the massive Jupiter at an outer orbit allowing it to deflect and prevent many asteroids from hitting earth, our large moon slowing earth's rotation, stabilizing its axis of rotation and enhancing tides which helped life migrate from sea to land. Also a solar year of $\sim$ twelve lunar months inspired astronomical studies in order to correct- via a system of leap years- the discrepancy between the lunar and solar calendars. The equality of the angular size of the solar and lunar discs enabling total eclipses helped the ancient Greeks measure the distances to the moon and to the sun and Eddingtons mission to verify Einstein’s GR by measuring the bending of light during a complete solar eclipse.\par
Recent examples are the (approximate) equality of the distances of Japan's reactors from Kamland and  the “ Solar neutrino” (1.2) oscillation length and those in Daya bay China with the "atmospheric neutrino" (2,3) counterpart allowing  the discovery of $0\leq\theta (e3)$ entry in the PMNS neutrino mixing matrix. [(1,2) (2,3) denote pairing in the “Normal Hierarchy” case where $\nu(e)$ the neutrino associated with the electron - the lightest lepton, is mainly made of  the two lighter neutrino mass eigenstates $\nu(1)$ and $\nu(2)$].\par
The lucky circumstances considered here are the large number of merging binary BH’s with masses $\sim 10-60 M(Solar)$ which facilitated discovering gravitational waves. GW observations impact the special DM type we discuss next - namely DM made of BH’s of different masses. Such BH’s must be primordial (PBH’s) as many constraints, in particular the upper bounds on baryonic matter at the time of BBN and CMB decoupling, exclude the possibility that "ordinary" B.H's generated by the gravitational collapse of ordinary massive stellar cores constitute more than $1\%$ of DM\footnote{That sufficiently massive stellar objects gravitationally trap the emitted light “corpuscles” and are “black” was noted already in the 18th century. BH’s were introduced to modern physics by Carl Schwarzschild’s solution of Einstein’s Eq. -a static, spherically symmetric, diagonal metric
  $$ (ds)^2 = (1- \frac{r}{R}) dt^2 - (1-r/R)^{-1} (dr)^2 +r^2 [d\theta)^2+ \sin^2(\theta) (d\phi)^2] $$
where $g_{t,t}(r)$ ($g_{r,r}(r)$) vanish (diverge) at $r= R(Schwarzschild) =2GM/{c^2}$. BH’s are found in nature over a broad range of masses and their rotating and charged variants along with magnetic monopoles and cosmic strings are some of the most  fascinating structures in theoretical physics.}.  
There is ample evidence for BHs of $\sim 4-6$ solar masses. Most main sequence stars collapse at the end of their radiative life into White Dwarfs (W.D.’s), Neutron Stars (NS’s) or into BH’s depending on the final masses of  their cores. No NS with mass $M> 2.5  M(Sun)$ has been observed to date. The maximal core mass above which the nuclear equation of state (EoS) fails to prevent the collapse of the core into BH’s is not known exactly but is believed to be $\sim
2.5M(Sun)$. Absent reliable QCD calculations of the nuclear EoS at high densities (due to the "Sign problem" arising in the presence of the baryonic chemical potential) the above maximal mass and GW  expected in neutron star mergers cannot be reliably calculated. \par
Before the discovery of GW’s, NS mergers were thought to be the most likely source of GW’s  but only after several dozens of  binary BH mergers were recorded, was such an event discovered. This discovery was kept secret until the collected information by many types of observations led to this epitome of the “Multi-
messenger approach” which  allowed verifying that this was indeed a binary neutron star merger. Further spectral analysis suggested that such mergers may help ordinary type II core collapse supernovae in synthesizing the observed amounts of gold and other trans-Lanthanide elements.\par
As noted above the LIGO experiment benefitted from a bonanza of merging massive BH binaries. These included in particular few in the region above 30 solar masses which was supposed to be forbidden by the need for too massive, unstable progenitors.\par
The probability of detecting the coherent GW falls linearly rather than quadratically with increasing distance, which, with the few solar masses of energy emitted in GW in the heavy BH mergers, allowed their detection at cosmological distances.\par
Given their initial orbits every aspect of the BH’s mergers and of the resulting GW can be calculated. Their larger separation when merging starts:
$$
 r \geq R(SW)_1+R(SW)_2 \sim 3 Km.[M(BH)_1+M(BHB)_2]/M(Sun) \gtrsim 2R(NS)\sim 24 km \hspace{0.1cm}   
$$
increases by Kepler's third law the pre-collapse orbital period with BH mass according to 
 $t(p) \sim M(BH)^{-1}r^{3/2}\sim M(BH)^{1/2}$. The longer time during which the stronger "Chirping" GW pattern can be followed in the merger of the larger BH’s greatly helped the GW Discovery. 
 \par
Shortly after the discovery of the first 30 + 15 Solar mass binary BH merger it was suggested in \cite{bird} that such heavy BH’s make up DM in which case they have to be primordial (PBH’s). Unfortunately the "Dream Scenario" where Massive PBH’s which facilitated the discovery of GW ‘s (and thereby of these very BH’s!) are the long sought for DM - is not viable. It was realized that BH’s can readily form binaries by dynamical Friction (including the kicking away of nearby stars) - in the early universe with a dense population of massive stars and eventually merge via GW emission. If \textit{all} DM is to be accounted for by such massive PBHs then the rate of expected GW events at LIGO and VIRGO would have been $10^4$ times higher than the measured rate. Conversely the observed rate of detected GW events can be due to mergers of ordinary astrophysical BH’s. Along with grav lensing data excluding the lower part  $M< M(Sun)$ of the high PBH's window, Ligo/Virgo almost closed this window for DM made of such PBH"s.\par
We have not mentioned the BH’s in the centers of galaxies of $10^6- 10^9$ solar masses. These modern incarnations of the "Quasars", the mysterious "Quasi-Stellar Objects" with huge, and in certain cases quickly varying, fluence can contribute at most $\sim 0.1\%$ to the critical density. Still the discovery of such BH’s, which earned Reinhardt Gentzel and Andrea Gez a recent Nobel prize, is a marvel of dedication, ingenuity and advanced experimental technology. By probing trajectories of stars close to the Schwarzschild radius the mass of the BH in our galaxy has been determined to an accuracy of one part in $10^4$ allowing to test GR modification of the trajectories due to the B.H rotation. A globe-wide net of radio dishes "imaged" a giant BH with secondary rings due to the part of the light arriving after encircling the BH. It has been noted that if our milky way BH was subsiding on its present accretion  “diet” inferred from its luminosity - it could not  grow on galactic time scales to its present mass. This suggests higher accretion rates in the past or that giant BH’s formed via coalescence of smaller Primordial BH’s. 
\par
Much more massive black holes are being discovered at higher redshifts and their early formation could indicate highly dissipative strongly interacting DM. Coalescence of supermassive BH's binaries produce GW's of long $~10^{8\pm 1}\ Sec$ periods. The integrated effect of many such binaries is the target of the ongoing "nanograv" project, monitoring over many years with high precision the periods of pulsars which should be affected by such GW’s. Most recently the Nanograv collaboration claimed to observe such GW’s with an overall rate and slope of number of events versus periods which may exceed values expected for astrophysical BH coalescences scenario. This  excited HE/particle physicists who suggested more exotic sources of these GW’s-some of which will be touched on below. See e.g. ref.  \cite{Barack:2018yly}.

\section{\hspace{\secspace}PBH’s DM- The LOW Mass Window}
\textbf{Where we discuss  the remaining window for DM made of PBH’s in the $10^{16}$-$10^{22}$ gr mass range and speculate on some lucky circumstances that may allow finding a PBH in this window.} 
\par
\vspace{0.2cm}
Our main focus in the following are special "Lucky" circumstances which may allow discovering PBH’s at the lower limit of the  window of $M(\text{PBH}) \sim 10^{16} - 10^{22} gr$ where the PBH’s could be DM or a non-negligible part thereof. Schwartzshield PBHs of lower mass Hawking radiate and disappear on Hubble time scale. The upper limit of the window avoids excessive micro-lensing and other dynamical effects \footnote{It has been suggested in \cite{graham} that an impact by BH’s of mass $10^{19} - 10^{20} gr$ would initiates a runaway fusion reaction in most white dwarf stars leading to a type $(1a)$ supernova explosion further narrowing the window for PBH DM. It is easy to verify that the  gravitational pull exerted by the PBH heats up nuclei starting fusion reactions in a cylindrical region around the straight line along which the PBH traverses the W.D.  For $10^{19} gr$ PBH the radius of this region is $R= ( \frac{c}{v_{\text{escape}}})^2 R_{\text{sw}} = 10^{-4}cm$. The harder to verify condition for runaway fusion explosion is that the rate of generating thermal energy by nuclear burning is faster than that of heat loss. It is believed that most standard astrophysical type 1-a supernovae are initiated by merger of binary W.D. induced via GW emission. The relatively slow rate of such events conforms to their contribution to the galactic Iron abundance and limits the new mechanism affecting \textit{all} W.D’s. Most works do not include this putative further limit on PBH masses.}.
Consider B.H's of masses $M(\text{B.H.}) \sim 3 . 10^{16} gr$ with H.R. (Hawking Radiation) temperatures $\sim 20$ MeV and lifetimes prescribed by the H.R of order t(Hubble). Typical astrophysical sources tend to generate monotonic $\gamma \hspace{0.2cm} \text{or} \hspace{0.2cm}{X \hspace{0.1cm}ray}$ spectra. A broad "thermal" peak seen in a particular direction may then suggest a black body as a likely source \footnote{BH’s are thermodynamically different from ordinary Black Bodies in having a negative specific heat with $T(\text{B.H})$ increasing rather than decreasing when losing energy via H.R. This traces to the ultra strong gravitational field. The ensuing strong redshifting of the frequency or energy of the emitted Hawking photons, as they recede from the BH horizon, yields an effective potential barrier which increases with increasing $E/T(BH)$ - an  important “gray body”’ correction to the simplistic Hawking radiation = BB radiation assumption. The modifications of the spectrum of the emitted radiation which depend on the angular momentum of the emitted particle and of the BH, are well known and accounted for in detailed calculation but do not change much our qualitative reasoning here.}. 
Avoiding/ Blocking known point sources, leaves a diffuse “background”  due to unresolved sources and Cosmic Rays (C.R.). These photons however cannot come from ordinary astrophysical sources which behave as B.B’s since a surface temperature as high as $T=20$ Mev will lead to an intrinsic brightness of $[T/{T \text{(Sun surface)}}]^4 = 10^{30}$ times higher than that of the sun. These objects must be BH’s and as any known stellar collapse requires a minimal mass of -$\sim 1/2$ solar mass, these BH’s should be PBH’s.\par
These light PBH's emit via H.R. approximate BB (Black Body) radiation\footnote{A convenient mnemonic is that as $M(\text{B.H})\rightarrow m(\text{Planck}) \sim 10^{19} \text{GeV}\sim  10^{-5} gr$ namely Planck mass black holes of radius $R=l_{\text{Planck}}\sim 2.10^{-33} cm$, temperature $T= m(Planck)$, the evaporation lifetime becomes $t(\text{Planck})= 6.10^{-44} sec$. The corresponding quantities for higher mass BH’s can be obtained by using  $R \sim M\sim 1/T$ and $t_{\text{evap}}\sim M/(\text{dW/dt})\sim M^3$.}  at rates:
\begin{equation}
    \label{15.4}
\frac{dW}{dt}\sim \frac{M}{t_{\text{evaporation}}}\sim 10^{19} \text{ergs/{sec}}
\end{equation} 

The lower mass PBH’s offer the best chance of seeing Hawking radiation: Both the \# density and temperature of heavier PBH’s which make up D.M are lower in proportion to $M^{-1}$ ,dramatically reducing the prospect of detecting their Hawking radiation. \par 
The local halo D.M. density requires an average number density $3.10^{-41} cm^{-3}$ of $3.10^{16}$ gr PBH’s or $\sim$ one such PBH at a distance d=3Au from their nearest neighbor PBH or from us. A detector of area $= \text{meter}^2$ collects from such a source $\sim 3. 10^{6} $ photons/year of $E=20 MeV$. \par
Halo PBH’s of the above masses and densities occupy a sphere of radius $R_{\text{eff}}(\text{Halo})\sim 20\ \text{KParsec}$ with average spacing of d= $3 \text{Au}$. Neglecting our $\sim 7$ Kiloparsec offset from the galactic center we find a halo radiation $R/d=7.10^9$ times stronger than that of a single source at distance d:
\begin{equation}
    \label{15.6}
        N(\gamma)_{halo} \sim 10^{15} \hspace{0.1cm}\textnormal{\text{photons of}} \hspace{0.1cm} E \sim 20 \text{MeV/Year}
 \end{equation}
This is reduced to 
$10^{15}.20^{-8/3} \sim 10^{11}$ photons of E $\sim$ 7 MeV if we assume PBH’s of mass $10^{17}$gr \footnote{That the nearest PBH has a negligible luminosity as compared with the collective effect of all the halo or cosmological PBH’s is related to Olber’s Paradox: that in a static, infinite, homogeneous universe the light from distant stars would make the whole night sky as bright as the sun. The paradox is evaded by the Hubble expansion which reduces the intensity of the radiation arriving from far (High z) sources by $1/(1+z)^3$ to a finite sum. However the main reason why the Olbers effect is absent in the original earth-sun context is that the Au distance to our sum is $10^5$ times smaller than the distance to the nearest star and the average distance between neighboring stars. This disparity need not happen in the case of the PBHs since the Earth and the Sun were not created jointly with the PBH’s.}.\par
The ability to resolve small angles $\sim \delta$ can in principle enhance the sensitivity to point-like sources by $En=4\pi/\delta^2$. However to benefit from this we need first to discover the specific source in searches which cover larger regions of the sky. Unfortunately the expected Cosmic ray background of such photons exceeds those expected from the Hawking radiation fluxes making the discovery of these PBH’s unlikely. Thus a "Lucky circumstance" of having a primordial B.H. near us offers the best chance of seeing its Hawking radiation. \par
Dynamical effects leading to capture of BH’s in the solar system (See.eg \cite{Khriplovich:2010hn}) can enhance the density of  near earth PBH;s with a much stronger flux of Hawking radiation far beyond the above estimate which used the average NFW CDM density in the halo near our location. \par
The solar system contains $\sim$ trillion asteroids of masses $\gtrsim 10^{17 }$gr residing in the Oort cloud. Early on there were many more asteroids and they also populated the inner solar system. Some of these were much larger than present day asteroids such as the object whose collision with earth generated our moon. When the sun ignited the radiation pressure ejected much of the gas and dust in the solar system  and most asteroids were eventually captured by the sun, Jupiter and the other planets. The PBHs however, freely traverse any  object that they collide with and are less susceptible to ejection due to near collisions with the much fewer similar mass asteroids.\par
It is conceivable that repeated traversals in earth's vicinity by PBH’s at that early period lead to the capture of some small fraction in gravitationally bound nearby orbits \footnote{The above is part of the general subject of  enhancement or other modification by the solar grav fields of the distribution of various types of DM. An early analytical discussion of massive wimps captured by the sun is presented in Thibault Damour and Lawrence M. Krauss \cite{Damour}.}. \par
A PBH bound in a sub lunar orbit of radius $\sim r R(E-\text{Moon}) = r\  360 .10^3 km$ rather than $\sim 10 Au$ as is the case for an average halo PBH, will yield a  $\frac{2.10^6}{r^2}$ larger H.R. flux and its direction should vary over a short period of $T=\text{month} \cdot {r^{3/2}}$. \par 
Precision data on orbits of satellites traversing the solar system limits the total DM (or other un-detected) mass in a sphere of 10 Au extending to Neptune's orbit to be less than M(earth) ~ 5.$ 10^{27} gr$ - a rather weak limit which is satisfied with a margin of $10 ^{9}$ for  the average local halo density of $~ 0.4 GeV/{cm^3}$ but excludes an enhanced population of PBHs of $10 ^{17} gr$ masses  with separation of $10 ^{-2}Au$  from existing all over the solar system. Indeed most of the early PBH’s may have been in dynamically unstable orbits and were kicked out by the gravitational fields of the various planets and their moons.
.\par
The above then translates into the question: "Can a BH in the inner solar system and in particular an "ace" PBH in a sublunar orbit survive until the present?" This may be the case if it is sufficiently close to earth at a fraction r of the Earth-Moon  distance $\neq$ but not near orbits with destabilizing resonant effects.
\par
Precise lunar observations were provided by the ingenious project of  retro-reflection of short laser pulses. It allowed measuring over a period of 40 years with few cm precision the distances to a small patch on the lunar surface where relatively large "retro-reflecting Corner Prisms" were planted in the 1969 lunar landing. (The weaker moon gravity g(moon) $\sim$ g(earth)/6 helped the astronauts carry the heavy prisms). This allowed precision tests of G.R., measuring continental drifts and moon-quakes
\footnote{$^*$ R. Dicke, an unsung hero of 20th century physics, suggested (with P.G. Roll) the retroreflection project. With D. Wilkinson, J. Peebles and P. G. Roll he initiated a search for the CMB \cite{Dicke1965}. The radio dishes used by them were smaller than those at IBM \cite{Penzias:1965wn}. However only thanks to Dicke and company the IBM duo realized what they discovered. To confront the Brans-Dicke tensor -scalar theory with Einstein's GR Dicke used many ingenious tests of the equivalence principle, searched for small changes in $G_N$ by using historical complete solar eclipses and did many other experiments. In the process he verified Einsteins GR to unprecedented levels ushering in precision GR along with I. I. Shapiro and others. He also suggested the supper-radiance encountered later in the review.}. As noted above the upper end of the allowed PBHs window is determined mainly by micro-
lensing. If by using the above or other techniques we will detect compact objects of masses in the $M\sim 10^{16}-10^{22}gr$ range, the latter need not be PBH’s and  could be baryonic or DM nugget types mentioned above. The careful visual monitoring of all near earth asteroids clearly excludes any object heavier than $10^{16} gr$ from being made of normal rock type material. Thus whether it is part of DM or not, discovering such a new compact object via its gravity will certainly indicate new BSM physics.\par
We next roughly estimate the effect of such objects within a spherical region of radius of up to $R=370$ $.10^3 Km$ (the Earth-Moon distance) and the likelihood of discovering a mystery object Z of mass M(Z) using the above lunar laser ranging via the retro-reflectors. Since earth is $\sim$ 100 times heavier than the moon we will first neglect the gravity of the moon and approximate Z to orbit Earth in a circle of radius R(Z) which stability considerations suggest lies in the plane of the Moon’s orbit. The fact that the searches need to focus only on this plane can be crucial in facilitating such a difficult project. Let the mass and distance ratios be $ f = M(Z)/{M(\text{earth})}$ and r=R(Z)/R . By Kepler’s third law the period T(Z)  of Z is $ T(Z)=Tr^{3/2}$ with T = month. The effect of Z on the moon's orbit is to increase the effective mass of the earth by a factor of 1+f with $f \ll 1$. The observable effects of interest are the periodic oscillations of the moon orbit  with period T(Z) generated by the extra radial acceleration due to Z  : $\delta(a)= G(N) M(Z) r/{R^2}$  beyond the constant acceleration due to earth: $a = \omega^2 R= G(N) M(E) /{R^2})$ with $\omega = 2\pi/{\text{Month}}$ the moon’s angular velocity. The extra acceleration $\delta(a)$ adds to the accelaration $a$ during the half cycle of Z when Z is closer to the moon and subtracts from it in the other half  when Z is further from the moon. The amplitude of the moon’s extra shift - collected during the first half of the cycle then is:
:
 $$
 \delta(R) =\frac{1}{2} f.a.r.\frac{2}{\pi} [\frac{1}{2} T(Z)]^2  =\frac{1}{2} r.f. \frac{2}{\pi} [r^{3/2} \frac{1}{2} T(Z) \omega ]^2. R = \pi r^4 f R  
$$
Using $f=2.10^{-10}$ (i.e $M(Z)\sim 3.10^{17} gr$), $r=1/2$ and $R\sim 3.7 \cdot 10^{10}cm$ we find $\delta(R)\sim 0.1 cm$. These tiny oscillations with a period $T(Z)\sim 10$ days period may have been missed but can be discovered by a repeated, careful analysis. Also an improved version of the retroreflection project has been suggested \footnote{Doug Currie from university of Maryland who along with the late Carrol Alley. and others  ran the retroreflecting project is currently pushing a new version with many more but smaller quarter prisms and shorter more intense Laser pulses which may improve the precision of distance measurements and allow for discovering Z like objects if they exist (Jordan Goodman P.C.)}. The above approximations apply when $r$ is small in which case the oscillations are almost purely sinusoidal and the problem reduces to that of a periodically forced harmonic oscillator. In the more likely event where $R(Z)$ is closer to $R$ we have a more complicated yet calculable trajectory of $Z$. The resulting modulation of $R$ will be larger and much easier to detect and follow up knowing the theoretically expected location of $Z$.\par

Unlike the questionable statement that “All humans are born equal but die different”, we can assert that “Black holes are born with different masses/sizes/temperatures but all die in the same way“. In the last phase of their life all BH’s share the same Hawking radiation that keeps getting hotter as $T\sim1/M =(1/\tau)^{1/3}$ where $\tau$ is the remaining lifetime, with increased intensity $dW/dt\sim 1/M^2\sim 1/\tau^{2/3}$ A further increase in the rate of the total mass loss $dM/dt$ is due to radiating more, heavier particles in the SM (and beyond!) once we cross the temperature = mass threshold for each of these particles. If this unique behavior can be observed it would most clearly prove the existence of evaporating primordial black holes.
The reason is that only BHs of mass $M(\text{PBH})\leq 3.10^{15} gr$ decay today efficiently via Hawking radiation and BHs of masses smaller than$\sim M(\text{Sun})$ cannot be generated via a collapse of any known astronomical object. Present bounds on the would-be Hawking radiation allow "light" PBHs of mass smaller than $2.10^{15}$ to make only a small fraction of at most $10^{-3}- 10^{-4}$ of DM,  implying a maximal density of such PBH’s :
\begin{equation}
    \label{15.7}
n(\text{PBH's}) \sim 10^{-3}\rho(\text{CDM})/M \sim 3. 10^{-43} cm^{-3}\sim 1/{ (10 \text{Au})^3}
\end{equation}

\par
This then raises the second independent question: "If BH’s of Hubble lifetime exist, can they be observed while in their last, dramatic "Rigor Morte" phase?"
The High Altitude Water Cherenkov(HAWC) device and its improved LAHASSA Chinese counterpart, have effective areas of $10^3 \text{Meter}^2- 10^4 \text{Meter}^2$ and can detect and locate within $\sim  deg^2\ \gamma$s of energies $\geq 100 GeV$ which can be separated from the background of hadronic CR showers. Since   $\tau\sim T^{-3}$ $\gamma$’s of 200 GeV  can be emitted from the above PBH's during the last year of their lifetime when they weigh
$\sim 10^{12} gr$. Since initially the lifetime of the PBHs of interest is $\tau \sim t(\text{Hubble})\sim 10^{10} \text{years}$ only a $10^{-10}$ fraction of them is presently in the last year phase. This decreases the number density of halo PBHs of interest from the expression in Eq.\ref{15.7} by $10^{-10}$ to:
\begin{equation}
    \label{15.8}
n(\text{PBH})_{M(\text{PBH})\sim 10^{12} gr} \sim 3.10^{-53} (cm)^{-3}\sim 1/(2.10^4 \text{Au})^3 
\end{equation}
and $d= 2.10^4 Au = 3.10^{17} cm$ is the average distance from earth and between nearest neighbor PBH’s of this type. A total of $N(\gamma)=   Mc^2/{E(\gamma)} = 10^{34}$ photons of energies $\sim 200 GeV$ will be  emitted in a year from one of these "light" PBH's. The flux from  a single PBH at the  minimal distance $d\sim 2. 10^4 Au$ of:
\begin{equation}
    \label{15.9}
\Phi(E(\gamma)\geq 100 GeV)_{\text{PBH at} d= 2.10^4 \text{Au}} = 10^{-2} cm^{-2} yr^{-1} 
\end{equation}
then yields in a large area $A=10^8 cm^2$ telescope up to $\sim 3.10^6$ events per year. In analogy with eq.(59) above we expect  that the integrated diffuse flux from the PBH’s in the halo is enhanced by $R(\text{halo})/d \sim 2.10^5$ to be $\sim 10^{12}$ events per year. As noted above, the GZK absorption of the HE $\gamma$s reduces the expected flux from the halo.\par
The final smoking gun evidence for the very light PBH’s will be provided by having -as the year progresses- the spectrum from that source hardening as  $T\sim E\sim [t-t(\text{final}) ]^{-1/3}$ where $t(\text{final})$ is the time when the Hawking radiation stops. Decays of SM particles such as the gluons (or rather the glue-balls), the bosons mediating weak interaction and all six quark flavors all of which are Hawking radiated at temperatures exceeding $\sim 1/2 TeV$ generate via the $\pi^0$ particles produced, many photons of energies considerably lower than $T(BH)$ These “secondary Hawking photons” suffer less absorption on the optical photons than the higher energy primary Hawking photons but have higher astro-physical backgrounds.

\section{\hspace{\secspace}PBHs "evaporation" via Hawking radiation}
\textbf{where we note that Hawking evaporation may offer a unique opportunity to test interesting BSM physics including a count of  all the light DoF  in the theory.} 
\par
\hspace{1cm}
Discovering “light” hot PBHs via Hawking radiation would be a monumental event though such PBHs can make up only a small fraction of DM. In particular its careful study may afford novel insights of BSM physics, which in some cases transcends any other approach.\par 
The evolution of the BH as the BH approaches the end of its life is the reverse of the evolution of field theories from the UV -high energy to the IR- low energy in the expanding universe after the big bang. While the effective number of DoF’s can change during cooling (though not increase according to the "A theorem" recently proved by Zohar Komargodski \& Adam Schwimmer \cite{Komargodski}), it need not do so. The rate of aging of the BH will then remain the same leaving no clue to possible compositeness of certain particles. An example is the SM evolution during the weak phase transition starting with massless $W^+, W^-$ and $Z^0$ and the complex Higgs doublet and winding up with massive $W^+$, $W^-$, $Z^0$ and the SM Higgs scalar particle having  altogether the same number of DoF.\par
The small black holes can also serve as the ultimate microscopes or  ‘vices” attempting to crack down and test the possible compositeness of various particles. Thus the Higgs particle could be composite at some distance scale $d$. Once the size of the Radiating BH  becomes smaller than $d$ we expect that the "big Gennie" - namely the  composite Higgs -will not be able to emerge from the tiny box and that the BH will radiate the constituents of the Higgs rather than the Higgs particle itself.\par
Light PBH’s at the lower end of the allowed window may afford in certain specific scenatios the \textit{only} way to find all the low energy DoF, namely all the zero or very low mass fields in the full theory-ordinary and dark sectors combined. There may be cases where the low 'exit temperatures' of certain particles from the primordial inflation and their extremely weak interactions with SM particles, make them unobservable. However, \textit{all} fields interact gravitationally and for a given mass and spin, all are equally Hawking radiated from BHs. To find these "Phantom" fields we can study the correlation between time $t-t(final)$ and temperature $T$ of the near-by "Dying" PBH. This correlation depends on the total number of DoF.namely  $N_{DoF} (T)$ and its increase in time $T(t)$, for which the SM contribution is known. Thus the fractional increase of  $N_{DoF} (t)$ will be \textit{slower} than the expectation from the standard model in the presence of extra, light, "Hidden" fields, a fascinating issue which we revisit towards the end of the review.\par
The "Vice" or the compression that gravity provides via the small black holes may break at some point. Thus if rather than remaining fundamental all the way to $l(Planck)$, gravity is “emergent” at some lower energy scale $\Lambda'$ or corresponding higher distance scale $d$, then  Hawking radiation from  PBH’s may stop as their size gets smaller than $d$ at a temperature $T'\sim \Lambda'\sim d^{-1} $.
\par
The interplay of PBH’s and other DM can be much richer than just via the overall mass-loss rate. When significantly lighter than the BH temperature, the emitted DM is relativistic and even comprising a small fraction of all the typically non-relativistic halo X particles with velocities $v_{Virial}\sim 10^{-3} c$, they  generate stronger, more readily detected recoils in the large underground set-ups. Also asymmetric halo DM X -particles can annihilate with the $\bar{X}$ part of $\bar{X}X$ pairs from the PBH. This and ALP-like DM which can decay to $2 \gamma$ can give rise to secondary, often distinguishable photons \cite{agashe}. \footnote{$^+$ An intriguing interplay occurs between BH’s and topological defects: Domain walls, cosmic strings (which could be part of DM) and monopoles. Thus various string networks  can at some late stage evolve very high density regions which collapse into black holes. Also pre-existing PBHs can keep swallowing the strings growing into the supermassive BH in galactic centers. Thus assume that a PBH off mass $M =\mu. 10^{15} gr$ attaches to a string with tension  $\lambda. [10^{12} GeV]^2$ where $\mu =1$ corresponds to a PBH with Hubble lifetime and $\lambda=1$ to the tension of an axionic string with $m(axion)\sim 10^{-5} eV$. The BH then proceeds to swallow the string with the speed of light increasing its mass at a rate of $\frac{dM(BH)}{dt} = c\lambda ^2.10^{24} GeV^2 cm =\lambda^2 10^{24} gr/sec$, which for $\lambda > 1$ much exceed the rate of decay via Hawking radiation. In particular over Hubble/galactic times the B.H. mass can reach $10^{41} gr \sim 10^8$ solar masses}.
Finally we have the possibility that small BH’s, if present at the time of some GUT or other symmetry breaking which generate via the Kibble mechanism magnetic monopoles, can assist the Kibble process by attracting to it regions of space where the higgs field direction varies so as to have monopoles centered on the BH. In turn the long range B field of the monopole can stabilize the B.H.

\section{\hspace{\secspace}Mechanisms for production of PBH’s}
We have not addressed the production of PBHs with specific mass distributions. Most scenarios for efficient PBH formations rely on fluctuation in the early inflationary universe. It can involve gravity only and may not require any additional fields. The computation of PBH formation (and possible GR wave emission associated with bubbles in first order early phase transitions) require detailed general relativistic calculations which are beyond the scope of this review . We still briefly recall here a relatively simple scenario of ref. \cite{marcos} utilizing a novel long range interaction between DM particles X of mass $M(X)$.\par
To avoid complete annihilation as the collapse proceeds we need asymmetric DM. Attractive rather than repulsive force between the same X particles then requires exchanging a \textit{scalar} particle $\phi$. Its light mass $m$ corresponds to a large range $R\sim 1/m$ of the resulting  Yukawa potential $V(r)=g^2 \exp^{-mr}/r$ with g the dimensionless $\phi \bar{X}X$ coupling. The collapse happens when the radiation temperature $T_c$ which we assume to be the same in the ordinary and dark sector, is below the “Freeze-out” temperature: $T=x M(X)$ with $x\ll 1$ so that we have non-relativistic X particles.\par 
Within a sphere of radius $R \approx \frac{1}{m}$, the attractive interaction due to the above scalar exchange is much larger than the gravitational attraction and concentrates the DM particles therein. However without the cooling effect due to the Bremsstrahlung of the light scalar particles the resulting enhanced velocities avoid  further collapse to a black hole. Careful considerations of these, along with limits on long range interactions, leave a range of $g, m$ and $M(X)$ values for which generation of a large component of DM in PBH’s in the above allowed window is feasible\footnote{$^+$It is interesting to speculate on the possibility that the light mediator couples to baryons/quarks and becomes at a later time and lower temperatures massive thanks to developing a VeV of $\phi$ or of some other field which couples to $\phi$. In this case the PBH’s can be generated without introducing the extra DM $X$ by the collapse of baryons in the early universe to PBH’s and with no dangerous long range "Fifth" interaction between the remaining baryons. The constraint
 $(g/m)\leq 10^{-13} cm$ avoids extra binding energy per nucleon in neutron stars with number density of $10^{39} cm^{-3}$ which exceed $\sim 100 MeV$ the typical energy of nucleons inside neutron star.}.
Unlike for Goldstone pseudo-scalar bosons we have no mechanism protecting the desired low mass of the $\phi$ field. (The derivatively coupled Goldstone boson generates spin dependent potentials that fall off as $1/{r^3}$ and cannot help the desired collapse.) In the case of scalar fields the Lagransian of $\phi$ includes the mass term $m^2\phi^2$ and the renormalizable quartic coupling $\lambda \phi^4$.   The $\phi$ field inside a sphere of the above radius $R$ has the value of $Ng/R\sim N gm$ with $N\sim M(PBH)/M(X) \sim 10^{42}$ of DM particles of mass $M(X)\sim GeV$ say, within the sphere, are required for making the $\sim 10^{18} gr$ PBH’s of interest. Thus even a tiny positive $\lambda$ can generate a large repulsive energy per particle; $U\sim \lambda$. $N^3 (gm)^4$ which may prevent the collapse.\par
Single loop diagrams with circulating $X$ particles and $n$ external $\phi$ legs generate local $\lambda(n)$ $\phi(x) ^n$ interactions. The smallest bubble loop yields a radiative mass $\delta(m^2)\sim [g M(X)]^2$ which is typically larger than the assumed $m^2$ requiring a finely tuned counterterm to cancel it.\par
An attempt to have "Technically" natural ultralight scalars - namely which once tiny masses are generated they can be naturally protected from getting large radiative corrections and where the scalar makes up DM was made by Abishek Banerjee, Csaba Csaki, Zamir Heller Algazi and Michael Geller. Their dark sector including the scalars of interest is "sequestered" namely connected to the SM only via gravity and they appeal also to Susy with a special breaking pattern. \par
In an inverse evolution scenario DM started as PBHs of masses $M(PBH)\leq 10^{9} gr$ and temperature of $T( PBH)\sim TeV$ so that they Hawking radiate all standard model and dark sector particles with masses of up to a TeV. The corresponding lifetime time obtained via scaling with $M(Pbh)^3$ from the $T(Hubble)$ decay time of $10^{15}gr$ and further shortening by due to the $\sim 20-200$ fold increase of the number of DoF, is :
$$
 \tau(\text{decay}) <  10^{-18}\frac{T(\text{hubble})}{20-200} \sim 10^{-3} - 10^{-2} sec 
$$
Such times may be sufficiently early so as not to adversely impact BBN which will occur at "Usual" expected time of $\sim 1 sec$ and temperature of $T\sim MeV$.
\par
Indeed a universe with initial matter domination for some period such as in the above scenario is assumed in many models allowing some early structure formation in addition to the conventional one at the present matter domination epoch.\par
The intriguing possibility that “light“ $M < 10^{15} gr$ extremal PBH’s with maximal angular momentum to mass ratio  $L/M =1$ which do not Hawking radiate constitute most DM has been considered in ref. \cite{taylor}. Such extremal BH’s have zero entropy and cannot be made in collisions of smaller BHs because the entropy is not being radiated away in the emitted classical GW’s. It seems therefore unlikely that extremal BH’s can be generated starting with BH’s with $L/M < 1$ in the process of Hawking radiation or otherwise. Also Primordial Black Holes with QCD Color Charge were suggested in ref. \cite{Alonso-Monsalve:2023brx}. 
 \footnote{$^{++}$ The group theoretic theorem of appendix G disallowing transitions between states whose masses are proportional to the square root of the quadratic Casimir operator in the corresponding representations may stabilize such PBH’s (and those of high angular momenta as well). The total Big Bang entropy $\sim$ number of photons in the observable universe and the total Bekenstein entropy in PBH’s of mass $\sim 10^{22} gr$ which make up all of DM are similar. This coincidence is however most likely fruitious.} 
 \footnote{In \cite{Loeb:2024tcc} Avi Loeb suggested that a recent analysis of the solar ephemeris (the collection of measured motions of solar system objects) implying  $[G_N M(sun)]^{-1} d/{dt}[G_N M(sun)] < 0.7 10^{-13} [Year]^{-1}$ can be used to exclude DM made exclusively of PBH’s of mass $10^{18} - 10^{20} gr$. If Indeed correct, this is most important.Lacking expertise and access to the details of the above analysis I cannot judge this. Note however that our main focus was on lighter PBHs of $m< 10^{17} gr$. To reach the above rate we need that $\sim 1000$ such PBHs cross the 50 Au sphere in a year. With roughly equal amounts of incoming and departing PBH’s this hardly builds up to anything significant. Still the very prospect of achieving such sensitivity boosts our hope that the conjectured discovery of a small PBH in a sublunar orbit (if it is indeed there…) is feasible. A more recent related paper \cite{Tran_2024} considers PBH's perturbing Mars’s orbit.}

\section{\hspace{\secspace}Readily detectable (if properly searched!) types of DM}
\textbf{Where we elaborate on very special "lamp-post" DM models where the DM cannot escape detection.}\par
In the preceding Sections we mentioned "Lucky circumstances" that made various scenarios with novel exciting physics observable. In this vein we note that there are DM types whose special construction renders readily discoverable. If they exist, then in the landscape of DM they provide "Lamp Posts" under which we are invited to search a ”key” physics element -the DM. The drunkard who keeps searching his lost keys under lamp posts may be laughable, yet ignoring such lamp-posts in the present context is both vain and foolish.\par
We next provide few examples starting with "Resonant DM". \par
The interaction between the DM and standard model particles is too weak to generate a resonance. In ref. \cite{Pospelov} it was noted however that $W$ mediated charge exchange in a WIMP-nucleus collision: $X^0 + (A,Z) \rightarrow X^- +(A,Z+1)^+ $- can lead to the formation of a "Coulombic" bound state of the final nucleus and $X^-$. The $X^-$ particle could be the negatively charged member of say, an $SU(2)_L$ triplet which includes the initial $X^0$.\par
This set-up was adopted in ref. \cite{Yang} to mitigate the then relevant DAMA/LiBRA anomaly. The mass difference $\Delta(M) \sim M(X^{-}) - M(X^{0})$, of order 10 MeV is tuned to allow \textit{one} intermediate “resonant” state -the lowest “Coulomb” bound state of $X^-$ and the $(A,Z+1)$ nucleus. Excited coulomb and/or nuclear states would manifest most dramatically by quickly de-exciting via energetic $\gamma$s and possibly also by eventual formation of new anomalously heavy, stable isotopes made of the lowest energy $X^- (A,Z+1)$ bound state. Experiments have not revealed either of these so we assume that the resonance decays back into the initial $(X^0 + (A,Z))$. By further tuning $\Delta(M)$, the authors of \cite{Yang} ensured that only DM particles with velocities $v=v(R)$ chosen to be close to the maximal-escape velocity, form the resonance. This happens mainly during the enhanced velocity summer phase of the annual modulation- dramatically enhancing the effect of modulations thereby helping understand the large annual modulations that DAMA reported.\par
Here we assume only that the collisions of the DM with nuclear targets proceed mainly via a resonance. Remarkably, as we show next, this often implies that all DM collisions deposit the \textit{same} energy in the detector dramatically enhancing detectability!.\par
In the resonance scenario most DM interactions occur for DM particles with velocities close to the resonant velocity $v= v(R)$. The starting total kinetic energy of the $X^0$ + target particle of mass $M(A,Z)= M(T)$ at rest in the Lab frame, is then fixed to be  $W= M(X^0)(v(R))^2/2$. 
After the collision this energy separates into that of  the center mass (CM) motion:
\begin{equation}
    \label{18.1}
    E(CM) =\frac{P^2}{(2 \hspace{0.15cm} \text{total mass})}=\frac {[M(X).v(R)]^2}{2[M(X)+M(T)]} 
\end{equation}
and the "internal" energy of relative motion inside the CM system which is:
\begin{equation}
    \label{18.2}
  E(int) =\frac{1}{2} k^2 \left(\frac{1}{M(X)} +\frac{1}{M(T)} \right)= \frac{k^2}{2 \mu}
\end{equation}
 $\mu= \frac{M(X).M(T)}{M(X)+M(T)}
$ is the reduced mass and $k$ is the momentum of $X^0$ or of the target nucleus in their CM Lorentz frame. The maximal cross-section at the resonance peak:
 \begin{equation}
    \label{18.3}
  \sigma(X^0 (A,Z))(peak) = \frac{4\pi}{k^2}(2J+1) \frac{\Gamma(el)}{\Gamma(tot)} \sim \frac{4\pi}{k^2} 
\end{equation} 
is very large. In the last part of Eq.\ref{18.3} used the lowest $J=0$ angular momentum for the elastic resonance with $\Gamma(el)=\Gamma(tot)=\Gamma$. For $m(T) = M(A,Z)\sim 100 GeV$ nuclear target say Xenon or Iodine in Zeplin X or DAMA we find using $k \sim Minimal [M(A,Z), M(X)]v(R)$ that  
\begin{equation}
    \label{18.4}
\sigma (peak)\sim 4.10^{-25} cm^2\ \ \ 
\mbox{if}\  \  M(X^0)\sim 100 GeV\sim M(A,Z)
\end{equation} 
and
\begin{equation}
    \label{18.5}
 \sigma(peak) \sim 4.10^{-21} cm^2\ \ \ \mbox{if}\ \  M(X^0)\sim  GeV \ll M(A,Z))
\end{equation} 
where we used for the initial resonance velocity the average $v\sim v_{Virial}\sim 10^{-3}c$.\par
The resulting peak cross section then is at least 13 orders of magnitude higher than the upper bound of $10 ^{-38} cm^2$ established over all the above mass range  by the low rate of events.[The lower values presented by the collaborations refer to the smaller scattering cross-section of $X^0$ off a single nucleon]. The low rate is reproduced by having only a small fraction of the incoming DM particles in a thin slice of size $\Gamma$ of the energy range:
\begin{equation}
    \label{18.6}
\Delta(E)\sim\langle E \rangle\sim v^2 \mu= 10^{-6} \mu = 5. 10^4 eV\ \ \  (or\ \Delta(E)\sim 5. 10^2 eV)
\end{equation}

share the peak resonant cross-section. The small observed rate then requires a very small width of the resonance
\begin{equation}
    \label{18.7}
\frac{\Gamma} {\Delta (E)}= \frac{\Gamma} { 5.10^4 eV} \sim  3.10^{-14} \Rightarrow \Gamma\sim  3.10^{-9} eV \hspace{0.2cm} (\text{or} \hspace{0.2cm} \Gamma\sim 3.10^{-11} eV \hspace{0.2cm} \hspace{0.2cm} \text{for} \hspace{0.2cm} M (X^0) \sim \text{GeV})
\end{equation}

The key point is that the ionization energy loss  of the  $[(X^- (A,Z+1)]$ system of the $X^-$ lodged inside a Xenon nucleus, brings it to a complete stop during the corresponding long life-time  $\tau(R) = 1/\Gamma\sim  0.2 -20\ \hspace{2mm}microseconds$. Indeed during this time the resonance could have traveled a distance of $10^3 cm$, far exceeding even for $X^0$ and  $X^-$ as heavy as 10 Tev, the rather short distance
\begin{equation}
    \label{18.8}
    l\sim 50 nm .\frac{M(X^0) + M(A,Z+1)}{ M(A,Z+1)}
\end{equation}
that needs to be traversed before stopping. This then guarantees that\textit{all} the translational kinetic energy of the motion of the center of mass is transferred to the $NaI$ crystal or Liquid Xenon and contributes to  the scintillation and  the ionization signals. The last piece of the argument is that when the resonance finally decays it is at \textit{rest} and the division of its (internal) energy between the escaping $X^0$ and the part given to the Nucleus (A,Z), which will also be deposited in the detector, is fixed by kinematics. This then completes the proof of our assertion \footnote{The bound system of target nucleus/atom/electron + $X^-$, interacts with the matter in the detector and deposits there the usually expected $dE/dx$ energy as when the DM recoils by itself.}.\par
We note that the very large $X^0$ - Xenon peak cross sections of Eqs.(\ref{18.4}, \ref{18.5}) do not bar the $X^0$ from entering 
the detector. This is due to the thermal Doppler broadening of the resonance by a large factor
\begin{equation}
    \label{18.9}
f_b =  \frac{v_{Thermal}}{v_{Vir}}.\frac{\Delta E}{\Gamma}.  
\end{equation}
which reduces the peak cross section by ($f_b)^{-1}$.
Here, $v_{Thermal}$ is the velocity of the thermal motion of the Xenon target atoms at a temperature of $T\sim 165 K =0.15 ~eV$. (At lower temperatures the Xenon freezes and experiments using it fail.)
\begin{equation}
    \label{18.10}
 v_{Th} = 3. 10^{-7} c \sim  10^{-4} v_{Vir} 
\end{equation}
Substituting $\Delta(E)$ from Eq.(\ref{18.4}) and the ratio of the velocity to the virial velocity $v/v_{Vir} = 10^{-4}$ from the above Eq. (\ref{18.10}) in Eq.(\ref{18.9}), we obtain
\begin{equation}
    \label{18.11}
     f(b) \sim 10^7 - 10^9
\end{equation}
restoring a m.f.p, mean free path, much bigger than the several meters size of the detectors. Note that this is the m.f.p. for the entering $X^0$ to interact with the Xenon and \textit{not} the much shorter stopping m.f.p of the $X^-$ Nucleus composite by ionization energy deposited along the path inside the detector  \footnote{$^+$ Crystalline detectors such as those used by Dama and many other groups can be cooled to much lower temperatures. However in this case the Zero Point Motion(ZPM) of the nuclei therein needs to be considered. In this context we recall that such (ZPM) of the emitting $^3H$ nucleus broadens the electron spectrum in decays of bound Tritium and smears out the endpoint. This irreducible quantum noise is smaller than the experimental resolution of the KATRIN experiment which uses $(^3H)^2$ Tritium molecules. It may however exceed the resolution of the planned "Neutrino physics with the PTOLEMY project" \cite{PTOLEMY:2019hkd} (originally suggested by Christopher Tuly) aiming to detect the Cosmic Neutrino Background (CNB) via the inverse $\beta$ process $\nu(e)$ + Tritium $ \rightarrow^3He^+  +e^-$ by utilizing Tritium adsorbed onto graphene. The more compact set-up as compared with the large  KATRIN $\beta$ spectrometer can achieve better resolution which might have allowed finding CNB neutrinos of  masses as low as $0.1eV$. Unfortunately the energy smearing induced by the dense excitations of the many-electron graphene generated by the "Sudden" disappearance of the $\beta$ electron may disable such a discovery when the neutrino masses are smaller than $0.12 eV$. see refs. \cite{Nussinov:2021zrj}, \cite{PhysRevC.91.035505} and \cite{cheipesh2021}).}. \par
In addition to nuclear recoils in DM interactions one might look for radiative de-excitation of nuclear levels generated in collisions with DM. The stability of the abundant isotopes which the DM encounters limits this to levels lying up to  $\mu v_{Vir}^2 /2 \sim 10-50 ~KeV$ above the ground state. Such low lying levels do exist in heavy nuclei such as lead but are absent in most common detector target nuclei \footnote{Bosonic CDM particles that couple to nucleons can be absorbed on nuclei if their/energy/rest mass exceeds the splitting between the ground and first excited nuclear state. The excited nuclei will then de-excite via emission of $\gamma$s which are much easier to detect than the tiny recoil in WIMP searches.}.\par
Analogs of the effect of F.N 69 on the Dark matter side are less restricted and the "Self Destructing" DM.(SDDM) of \cite{Grossman_2019} is a prime example. The idea underlying such models is that a fraction of dark matter consists of “molecular” $\bar{X} X$ or $X X$ states (for symmetric or asymmetric DM) in a high level so that annihilation or falling into the ground state releases very (moderately) high energy. Stabilizing this high state and preventing the coalescing of $\bar{X} X$ or $XX$ by tunneling to the ground state, is achieved by having levels with very large angular momentum $L$ with a large centrifugal barrier followed by alternating attractive -repulsive- attractive Yukawa potentials designed to generate the quasi-stable state. Collision with the A=16 Oxygen in the large water Cherenkov counters where searching for the SDDM is envisioned, can impart to the molecule energies $\Delta(E)$ of up to  $M(Oxygen).v(X^0)^2/2 \sim 20 KeV$. When $\Delta(E)$ exceeds the potential barrier which confines the molecule to its excited state an “explosion” is triggered as the system annihilates or jumps to its lower state. The resulting high enegy $\gamma$'s make this strikingly visible. \par
Even treating the couplings  $g_i$ and ranges $r_i$ of the Yukawa potentials as six free parameters, it is a non-trivial task to ensure that the potential barrier of height  $< 20 KeV$ will prevent tunneling and decay in Hubble time without increasing the size of the molecule and the rate of molecular disruption in collisions in the early universe with unbound $X(0)$’s.\par
Another rendering of the concept suggested in \cite{Geller2}, uses dark analoges of the rearrangement reaction $\mu^- +H \rightarrow[\mu^- +p]+e^-$  where the $[m(\mu)/{m(e)}]^2$ enhanced muon binding manifests as a sudden energy release. An interesting earlier work is presented in \cite{Giudice_2018}
\par
As noted in the previous section, quasi relativistic DM particles are emitted via Hawking radiation from PBH’s of the lowest allowed masses $10^{15} -10^{16} gr$. The collected galactic or cosmological supernovae provide a source of accelerated DM of masses $\lesssim 40 MeV$. Due to some Boltzmann suppression and the existence of 6 neutrino species, the DM accelerated flux is smaller than the integrated Neutrino flux from all supernovae calculated in \cite{Y_ksel_2007}.
Other mechanisms for accelerating DM on cosmological/Galactic solar and earth’s crust scales were suggested by Pospelov and collaborators. One \cite{PhysRevD.104.103026} uses light (sub MeV) DM collisions with electrons in the $O(KeV)$ hot solar core and the other \cite{pospelov2} 
actual acceleration of millicharged particles that concentrate at $O(km)$ depths underground. \par
While the fluxes of these accelerated particles are relatively small the prospects of their detection in large underground DM and neutrino detectors are greatly enhanced by the (much) higher resulting nuclear/electronic recoils\footnote{$^+$Mili-charged or other DM particles residing inside crystalline grains in earth’s crust can be accelerated  up to $10^{10} cm$ $sce^{-2}\sim 10^{7} g$ in ultra centrifuges. Such large accelerations and ensuing strains will not break the small crystals (or centrifuges), However heavy DM particles will be kicked out from the crystals- something that sensitive SQUID  loops can detect in the case of mili-charged DM and the small mass change due to escaping ultra heavy neutral particles may be be detected by ultra-sensitive weighing.}. Less exotic searches of light millicharged particles MCP’s (Milli Charge Particles) were suggested in \cite{Berlin:2022hmt} . A relevant source is Snowmass Whitepaper \cite{Krnjaic2022} \par
We close this section with two "Lamp-posts" of DM clustered in grains and in clouds.\par
If D.M. particles were completely clusttered inside equal grains, then the grain mass with the most dramatic signature is $\sim 3. 10^{- 5}  gr$. A single such grain passes through a $d= 2-3$ meter size large underground detector once per year carrying the same number of DM particles that would normally hit this detector in a year in the absence of  clustering (We assume that unlike in our sector, "Dark"  grains make up a sizable part of and potentially dominate DM).\par
If the fundamental DM- Nuclear cross section is close to, yet slightly higher, than the posted bounds, then we expect a few, say six DM interactions in a given detector during a year. In the grain dominated scenario these events will concentrate within $\delta(t)= d/{v(Vir)}\sim 10^{-5} sec$. They will be spatially aligned, temporally ordered with consecutive events separated by $\frac{d}{6}\sim 0.2-0.4$ meter. Finally, the direction and timing order will tend to conform to the WIMP wind. It is difficult to miss such events unless one keeps only spatially and temporally isolated nuclear interactions which are good candidates for the usual WIMPs collisions and excludes such exotic events (which if accounted for would lead to DM nuclear cross sections slightly exceeding the posted bounds)\footnote{Enroute to the underground detector the DM particles within the grain collide on average with  $10^9$ earth nuclei. With the average energy transfer in each collision being less than 20 KeV, the equivalent heat transferred  $Q=3.10^{-7} cal$ will not melt/vaporize the grain if its specific heat exceeds $0.01 cal/{(Deg..gr)}$ requiring a reasonable minimal binding of the DM particles in the grain.}. \par
DM grains/blobs and the interactions required for their formation at an appropriate past epoch have been addressed by many authors see e.g ref. \cite{Grabowska_2018} 
This is not the case for the DM clouds \cite{nussinov2020dark} which may well be the epitome of easy to discover DM. \par
To optimize detectability, the authors of \cite{nussinov2020dark} assume that DM forms earth’s size clouds $R(\text{Cloud)} = R(\text{Earth)} \sim 10^9cm$ of which one such a cloud passes once a year through any one of the large underground direct DM detectors and generates the same number of nuclear collisions as that expected in a year in the un-clustered case. To achieve this during the passage time $R(E)/{v(Vir)}\sim 30$ sec. $=10^{-6}$ years, $\rho(cloud)$, the density of DM particle inside the cloud, has to be $10^6$ times larger than the ``local” halo density yielding a cloud mass of $\sim 2.10^9 gr$. The earth-size cloud overlaps at the same time China, Italy and South Dakota where the three big Xenon experiments are being conducted. Consequently the handful of events in each experiment for which the energy estimates obtained by using the ionization and scintillation signals agree and which pass the other criteria for nuclear-DM interactions such as spatial isolation, will occur within the same $\sim 30$ second time period! The relative timing of the three events may point to the ``Wimp Wind'' direction, adding further indication for a DM source. Can such clouds be stable?\par
The surface gravity of the cloud $g_c=g .M(c)/{M(E)}=3.10^{-19}g =3.10^{-16} cm Sec^{-2}$ is extremely small and can be neglected when estimating the tidal deformation it suffers upon approaching earth. The tidal acceleration due to earth becomes when the cloud is at a distance of $R(cloud)\sim R(earth)$, very large $\sim g\sim 10^3 cm sec^{-2}$. Still since it traverses such a distance in less than $\delta(t)\sim 100 sec$ the resulting tidal distortion $g.t^2\sim 10^7 cm$ is hundred times less than R(cloud).\par
The sun tidal acceleration operates for a year time which is $10^6$  times longer than the 30 second earth traversal time above but being  $[R(C)/{Au}]^3 =10^{-13}$ weaker, the distortion $\sim a t^2$ is only one tenth as large.  
The fact that just like with earth, any given cloud “collides” with another cloud once a year is yet another potential source of disruption. Altogether in Hubble/galactic time we have $N_c\sim 10^{10}$ collisions. However the internal “heating ” up of the DM particles within the cloud is $\sim N(collisions) \Delta(v)^2 = N(collisions) [a\delta(t)]^2$. With a being $M(C)/M(E)\sim 3.10^{-19}$ times weaker than earth's $g= 10^3 cm (sec)^{-2}$ it is negligible.\par
Choosing the parameters of the cloud in order to maximize the effect of the DM clustering on the present direct DM searches is very artificial . The year’s time unit and the meter length have nothing to do with the fundamental parameters of the BSM theory that generates the clouds . Rather, a few years is the typical duration of most experiments and of the period required for obtaining a Phd in the context of such a project. It is also the duration of various grants allocated to experimental projects. Finally 1-3 meters is the linear size of the detectors containing several tonnes of liquid Xenon.
\par
The above scenario requires a mechanism for forming such DM clouds. In the standard $\Lambda -CDM$ cosmology structures form mainly after the universe becomes matter dominated. The earliest smallest, structures then are “Microhalos” more massive than earth that is $\sim 10^{17}$ heavier than the desired clouds. To achieve clustering on the desired smaller scales we need appropriate extra interactions in the dark sector. Finding such interaction, which are further limited by demanding that they will not cause excessive damage in mutual collisions of clouds (and not manifest experimentally otherwise) is a difficult though achievable task.\par
In certain models, axionic DM can form cloud-like structures, which for specific parameters indeed can reproduce also such clouds (Ref. \cite{Arvanitaki:2019rax}). Our purely statistical arguments apply to any kind of DM and to the widely separated sites where experiments searching Axions similar to say ADMX will be installed, in particular.\par
A spurt of events can be dismissed in each experiment separately as being due to some unexpected electronic or other noise. However, there are \textit{no} correlations between such noises in the three different continents and a coincidence within a time window of $\Delta(t)= R(cloud) / v\geq R(earth)/v\sim 30 sec$ is highly significant. Some minimal cooperation between the groups running the three big direct DM searches may thus be of crucial importance. Certain phenomena that jointly can be seen are likely to be missed by the individual collaborations.
The above cloud scenario is rather unlikely, however the extra little effort needed to test it is worth-while. Returning to the lamp-post parable it is unlikely that the “key” i,e  dark matter will be found under this glaring lamp-post. However if found there, then we will be able to figure out how it got there by finding the specific DM model where such dilute, yet stable, clouds can arise\footnote{Time coincidences between the widely separated detectors can, thanks to the excellent time resolutions, be searched for any type of events, not just those due to the rare DM candidates. Finding such time correlations would be of huge importance indicating a near-by supernova or other unique and unexpected BSM astrophysical effects. By now a carefully synchronized Multi-messenger network of different detector types has been set up in anticipation of supernovae, mergers of neutron star binaries and anything else unexpected. Hopefully the very competitive collaborations running the ultra-sensitive underground cryogenic detectors will follow this example.}

\section{\hspace{\secspace}DM and more generally BSM physics related to Neutrinos}
Such models have the advantage that neutrinos- the active left handed electron, muon and tau neutrinos and the hypothesized sterile right handed neutrinos are less exotic than most  DM types. Some BSM neutrino physics and related supernova physics are discussed next\footnote{So far only the neutrino sector firmly indicates specific physics beyond the truely minimal S.M. This may suggest that rather than pushing on all fronts of BSM models we should focus on this direction. A very different point of view was presented in the high intensity/high precision frontier conference in Rockville Md by Nima Arkany-Hamed telling an audience made up largely of neutrino physicists that the USA  long $\nu$ beam effort may be the least promising for exploring BSM  Physics.}.
In the early days of DM research when neutrino masses of $O(10-30)$ eV were allowed DM made of ordinary left handed neutrinos was considered. The neutrinos decouple from the rest of the radiation at temperature of order MeV when they are still relativistic and hence are HDM. For the mutual gravity between these light particles to overcome thermal fluctuations requires large minimal "Jeans" masses so that the structures they initially form are on supercluster scales. The observed structure is then built in an $up\rightarrow down$ pattern of fragmentation of the initial large structures (Termed Pancakes or "Blincy" in Russian by Y. Zeldovich). The tightening upper bounds on neutrino masses and the emergence of the $\Lambda CDM$ paradigm with $down\rightarrow up$ structure formation closed this possibility. Still the free streaming light neutrinos tend to destroy small scale structures seen today. The efficacy of this increases with the neutrino masses -leading to the remarkable bound of 0.12 eV on the sum of their masses.
\par
To qualitatively understand this, let's assume the standard cosmological CDM scenario where neutrinos rather than being massless, are endowed with some small mass $m(\nu)$. The Hubble expansion is controlled in the radiation dominated era by the  [$N_{DoF}(\gamma) +N_{DoF} (\nu)].T^4$ energy density. Once $T < m(\nu)$, the last expression becomes the larger $N_{DoF}(\gamma) \cdot T^4 + N_{DoF} (\nu) \cdot T^3 \cdot m(\nu)$ and the expansion rate increase. This delays the equality of the (mainly dark) matter and radiation. Hence by the time D.M. dominates larger perturbations entering the horizon tend to grow in conflict with detailed information from C.M.B and structure data. 


"What happens if neutrinos decay to lighter neutrinos and other massless particles around and after recombination so as to become radiation again?" 
\cite{FrancoAbellan:2021hdb} \cite{Serpico:2008zza}) The E.M. radiation decay of even one neutrino flavor would distort the perfect Planck's spectral distribution of the CMB. The radiative decays $\nu(3) \rightarrow \nu(2)$ (or $\nu(1)) + \gamma$ of width $\Gamma \sim (\mathcal{\mu}_{(2,3)})^2 \hspace{0.1cm} m_3^3 $  are slower than $t(Hubble)^{-1}$  due to the bounds on the mass $m(3) < 1 e.v$ and on the transition magnetic moment $\mu(3,2) <10^{-10} \mu(e) =10^{-10} e/{m(e)}$. A decay around the time of recombination into a different new light boson ( say a Majoron)  is however possible. Such decaying neutrino scenarios can relax the extremely strong constraint $\Sigma{(m(\nu_i)} \leqq 012 eV$ reviving hopes of detecting the neutrino masses with an inverted hierarchy and of finding NLDB decays. Yet the decay of the heavier neutrino(s) makes finding the CNB in Ptolomey-like experiments less likely. 
\par
A fourth sterile  neutrino, suggested by the LSND, experiment is much less motivated at present. A heavy fourth generation $t'$ quark may quadruple the Higgs production rate if the $t'$ triangle graph is added to that of $t$ in computing the glue-glue $\rightarrow$ Higgs process. Indeed a heavy Dirac fourth generation neutrino serving as DM was abandoned along with many scenarios (see Sec VII) where WIMP DM has ordinary weak interactions. Also neutrinos of mass smaller than 45 GeV are excluded by their (unobserved) contribution to the invisible width of the $Z^0$.  \par
The Higgs boson can couple massive right handed (R.H.) Majorana neutrinos $N^I_R$ with $\nu^i_l$ left handed neutrinos to generate their small majorana masses $m(i)_l = (y(i,I).v)^2/M(N(i
)_R)$. The original See-saw mechanism \cite{Bulbul_2014} explained the smallness of $m(i)$ without invoking extremely small dimensionless Yukawa couplings $y(i)$ by postulating large Right-handed neutrino masses $M(R)$. For $M(R) \sim 10^{9} GeV$ \text{and} $y(3) v\sim m(\tau) \sim Gev$  with $v$ the SM Higgs vev we obtain $m(3)\sim m(\nu(\tau))\sim eV$.\par
In general these right handed neutrinos decay quickly via the same interaction vertex $N({i})_R\rightarrow H+ \nu(i)$ at a rate $\Gamma\sim M( N(i)_R). y(i)^2= 10^{29} sec^{-1}$ where we used the numbers appropriate for the i=3 case above. This vastly exceeds the maximally allowed rate of decay if $N(i)_R$ made up the DM. \par
The RH Majorana mass term $M(R) N^i(R) N^i(R)$ violates, by two units, the conservation of lepton number $N_i$ and the double Higgs insertion transports this to the ordinary LH light neutrinos. This leads to the NeutrinoLess Double $\beta$ Decays
(NLDBD): $(A,Z+2)\rightarrow (A,Z) + e^+ e^+$  with a sharp energy deposition: $E(e_1) + E(e_2)= M(A,Z+2)- M(A,Z)$ by the two positrons. Only few nuclear isotopes which undergo neutrino-full double $\beta$ decay $(A,Z+2)\rightarrow (A,Z) + e^+ e^+ \hspace{0.1cm} \nu (e) + \nu (e)$ can also undergo NLDBD  at a rate  proportional to $m(\nu (e))^2$ - the induced majorana mass of the electron neutrino \footnote{$^*$This method of searching for NLDBD is due to Wendel Furry. An earlier, "Gedanken" version was suggested by Giulio Racah shortly after the original work of Etorre Majorana. Majorana, a true physics genius (of whom Fermi said that "There are good and very good physicists and there is Majorana”), disappeared under mysterious circumstances at the young age of 31 while traveling on a boat from Palermo to Napoli.}.
Given the mass (square) differences $m(2)^2-m(1)^2$ and $m(3)^2-m(2)^2$ measured in neutrino oscillations and the direct upper bound of $\sim eV$ on $m(1)$, NLDBD will be more readily observed if the two mass eigenstates mixing strongly with the electron neutrino which we refer to here as $\nu(1)$ and $\nu(2)$ lie above $\nu(3)$ -realizing the "Inverted Hierarchy". Various experiments suggest, albeit only weakly, a "normal" hierarchy of neutrino masses. \par
We can keep a specific linear superposition of RH neutrino states stable and (almost) massless by demanding the conservation of this linear combination of lepton numbers. This was used in connection with the 3.5 KeV X ray line observed by the Newton satellite and reported in ref. \cite{Bulbul_2014} which could arise \cite{King:2013iva} from the radiative decay    $\nu(R)\rightarrow \nu(L)+\gamma$ with $m(\nu(R)) = 2 E(\gamma)\sim 7 KeV$. Unfortunately the Hitomi X ray satellite, which could look for this line, self-destructed shortly after launch. The recent analysis of data from the NuSTAR \cite{Roach2022} and Xrizm satellite carried detectors did not confirm the $3.5 Kev$ X-ray line. We still find the RH sterile neutrino DM sufficiently attractive to discuss some of the issues involved.\par   
If such sterile weakly interacting RH neutrinos constitute most DM, then  they could not have been in thermal equilibrium in the early universe strongly violating the upper 0.12 eV bound on the neutrino masses. This cannot be avoided by making their couplings to SM ultra-weak so long as we entertain the possibility of their detection. Their production was explained in ref. \cite{Dodelson:1993je} by using a MSW resonance enhanced $\nu(e)\rightarrow N(e)_R$ transitions that occurred in the early universe at temperatures of a few tens of MeV. After a further cosmological expansion by $\sim 10^4$ the 7 KeV RH neutrino becomes non-relativistic and cools much faster -as  $\sim z^2$ rather than as $\sim z$, like the relativistic light neutrinos or photons. It then "freezes into" a sufficiently cold and "good" DM vis-a-vis its influence on the CMB and smaller structures. To achieve a sufficiently cold RH neutrino DM consistent with the ever tightening constraints we can have a degenerate Fermi Dirac distribution with a substantial excess of neutrinos over the thermally expected values or delay the MSW resonance by enhancing the usual potential term $G_F \hspace{0.1cm}n(e)$ (appearing in the $G_F \hspace{0.1cm} n(e)=  \hspace{0.1cm} \frac{m(N_{R})^2- m\nu^2}{2E}$ resonance condition) by a novel mutual neutrino interaction as in ref. \cite{DeGouvea:2019wpf} \cite{Abazajian2017}. To ensure that only the proper small fraction of $\nu(e)$ oscillate into the RH neutrino so as to generate the correct DM density requires some tuning of the initial inflation exit reheat temperature and of the mixing $U^2(4e)$ to be $\sim 10^{-7} - 10^{-5}$. Also tuning of the underlying model is needed to generate the magnetic moment $\mu(N(R) \rightarrow \nu(e))$ which will yield the correct radiative lifetime and resulting strength of the claimed 3.5 KeV line.\par
We next recall the-rather weak-upper bounds on  $U^2(4e)$ following from the effect of $\nu(R)$ emission on the neutrino pulses from SN 1987(a). Many similar themes originating in ref.\cite{PhysRevLett.60.1793} are taken up in the next section. While a complete understanding of SN explosions is not available it is believed that SN 1987(a) was a “standard” core collapse “Type 2'' supernova  where most of the gravitational collapse energy was emitted via neutrinos leaving a final neutron star.\par
The scattering cross section of the electron (anti) neutrino off nucleons and the hot trapped electrons, is  $G_F^2 E(\nu)^2 \sim 10^{-42} cm^2$ (and somewhat smaller for the other species which have only neutral current ($Z^0$ exchange) interactions). Along with the large density, this  makes the neutrino suffer up to  $N_c \sim$ million collisions spaced on average by their  mean free path $l_{m.f.p}\sim 10^3 cm$ on their way out from the forming neutron star. As neutrinos leave the hot plasma which is in thermal equilibrium more neutrinos are generated and diffuse out. Eventually after $O$(10 seconds) the core starts cooling and the neutrino emission peters off. If during the first 10 seconds, other, more weakly interacting particles which freely escape from the whole volume are being produced, then they form a competing cooling channel. These particles can - unless their production rate is too slow - "rob" most of the collapse energy from the neutrinos thereby weakening and shortening the observed signal. \par

In the case at hand the weaker interacting RH 7 KeV neutrino can be generated via oscillations of an initial LH neutrino. The probability of oscillating into the RH neutrino along the $l(m.f.p)\sim 10^3 cm$ traveled between collisions is given by the familiar expression:
$$
p (\nu \rightarrow \nu(R)) \sim \sin^2 (2\theta) \sin^2 [l(mfp) (\Delta(m^2)/{2E}] \sim \sin^2(2\theta)/2
$$
where $sin(\theta) = U(4, e)$, we used $E\sim 10 MeV$ for typical neutrino energies, $\Delta{m^2}\sim 50 KeV^2$ and the second factor of $\sin^2 [l(mfp) (\Delta(m^2)/{2E}]$ averages to a half. The conversion probability then is
$$
P (\nu\rightarrow \nu(R))= N_c.p (\nu\rightarrow \nu(R)) \sim (1/2 . 10^6) \sin^2(2\theta)
$$
with the total number of collisions $N_c \sim (r/l(m.f.p))^2= 10^6$ for $r=$ cores radius $\sim 10km \sim 10^6cm$. 
To avoid dramatic reduction of the neutrino signal we demand $\sin^2(2\theta)\leq 10^{-6}$. \par
The mixing generates also the $\nu(R)$ decay  $\nu(R) \rightarrow \nu(k) +\bar{\nu}(i) \nu(i)$ which would make DM decay in a Hubble time unless $\sin^2{2\theta}< 10^{-3}$. The $\nu(R)$ decay rate is obtained from that of the muon by scaling down with the $[m(\nu(R))/{m(\mu)}]^5\sim 10^{-21}$ kinematic factor and, being mediated by the Z exchange, also by $\sin^4(\theta(W))\sim 1/{20}$. \par   
We have focused on mixing with electron neutrinos for which the MSW effect is maximal. \par 
The branching fraction of $^3H\rightarrow$ $^3H^+e^-$$+ \nu(R)$ is suppresed by the phase space $[(Q-m(\nu(R))/Q]^5 \sim 0.1$ with $Q$ value of the beta decay being $M(^3H) - M(^{3} He)- m(e) =19$ Kev \text{and by} $\sin^2(2\theta)$   yielding $Br \sim 10^{-7}$. Searching for the RH neutrino via a kink in the spectrum of electrons is then rather difficult. Thus if the original $3.5 KeV$ line will be resurrected and have no atomic or nuclear explanation then it may be the only evidence for the dark matter.\par
Many neutrinos have been detected at Ice Cube and $\sim 200$ have UHE energies $E(\nu) \geq 200 TeV$. More than half of these UHE events which manifest in a shower-like pattern of Cherenkov light rings, are $\nu(e)$ and $\nu(\tau)$ events which cannot originate from decays of pions/kaons produced by cosmic ray interactions in the atmosphere. Also the upward going muons which point to the source of the neutrino are largely isotropically distributed rather than pointing to the galactic plane/center suggesting an extragalactic origin for the UHE neutrinos  \footnote{The first few PeV neutrinos detected in Ice Cube had similar energies suggesting that they originate from the decay $X^0\rightarrow \nu(i) +\bar{\nu(i)}$ of a heavy yet sufficiently long lived DM boson $X^0$ of mass $M(X)\sim PeV$. However, the $Z^0$ and the $W^+$ or $W^-$ from Sudakov Double log radiation, ref. \cite{Sudakov:1954sw} are emitted with probability $\sim 2 \left[ g((weak) \hspace{0.2cm} log \frac{M(X)}{M(W)} \right] ^2 \sim 0.3$, making this scenario untenable: the charged leptons that the neutrinos convert into upon emitting the W bosons and the leptons or hadrons that the W\& Z  decay into would have generated energetic e.m. showers which are much more readily detected than the neutrinos.}. This is not the case for less energetic neutrinos which are galactically oriented.
If the UHE neutrinos indeed come from cosmological distances of $\sim 10^{28} cm$ then a cross-section for (anti) neutrino- neutrino scattering
\begin{equation}
    \label{19.1}
    \sigma(\nu-\nu) \sim 10^{-30} cm^2  
\end{equation}
makes the UHE neutrinos scatter at least once off the CNB neutrinos where the number density of each of the six species $n\sim 100/cm^{-3}$ was used. The $\nu \nu$ CMS energy $W$ for 50 TeV incident neutrino is:
\begin{equation}
    \label{19.2}
       W \sim [100\ TeV. m(\nu(i))]^{1/2} \sim 1-100 MeV 
\end{equation}
where the lower/upper values obtain using a massless lowest $\nu(1)$ state of  momentum $p\sim kT \sim 10^{-4} eV$ for $T\sim 1.9 Kelvin$ and the highest mass $m(\nu(3))\sim 1 eV$ respectively.\par 
If the new interaction generating such cross sections stems from physics at a scale higher than $(W)$, then it can be written in a four Fermi form  $G' \psi^4(\nu)$. The mutual scattering cross-section $\sigma( \nu \nu) \sim G'^2 W^2\sim 10^{-30} cm^2$ at CM energy of $\sim 10 MeV$ is $10^{12}$ times larger than the weak scattering $\nu(e) e \rightarrow \nu(e) e$  cross section at this energy implying:
\begin{equation}
    \label{19.3}
    G'_F = 10^{6}\ G_F = g'^2 / {M'^2}= 10^{6} g_{weak} ^2 / {M(W)^2}
\end{equation}
 
where $M'$ is the mass of the mediator of the new interaction and $M(W) \sim 80 GeV$ is the mass of the weak interaction Boson.  For $M'> M(W) \hspace{0.1cm} \text{we need} \hspace{0.15cm} g'\hspace{0.1cm} ^2 \geqslant 10^6 \hspace{0.1cm} g^{2}_w $ which is way bigger than perturbative values. \par
The early universe and supernovae cores at temperatures of $O (3-10) MeV$ provide astrophysical settings with huge neutrino densities. Also $\nu \nu$ interactions were used to predict the non-linear oscillation of neutrino flavors due to the mutual interactions of the latter outside the S.N core studied by Raffelt and by Alexander Friedland (see e.g ref. \cite{Friedland:2010sc}). \par
Returning to the BSM large $\nu \nu$  cross sections of Eq.(\ref{19.1}) above the very short resulting m.f.p s of $10^{-4} -10^{-6} cm$ of neutrinos in the S.N. make the neutrino behave as a fluid rather than a gas. We will return in the next section to the question whether this does or does not delay the escape of neutrinos from the supernova and appreciably prolongs the expected neutrino signals from S.N. 1987(a) ).\par
Exchanging particles of mass $M'$ responsible for the above four Fermi interactions generates a Yukawa potential of range $1/{M'}$ and strength $g'^2$:
\begin{equation}
    \label{19.4}
         V(| r-r'|) \sim \pm  g'^2~ \frac{e^{-M'|r-r'|}} {|r-r'|}
\end{equation}

which for $\nu \nu$ interaction is attractive for a scalar but repulsive for a vector exchange. The sign of the interaction is important here as in many other cases. It played a crucial role in the MSW effect (Alexey Mikheyev \& Yuri Smirnov \cite{Smirnov_2005} and Lincoln Wolfenstein \cite{Wolfenstein:1979ni}). The effect was first discovered by Wolfenstein. Using the wrong (attractive rather than repulsive) sign for the electron- electron neutrino interaction, he did not realize its possible important effect on solar neutrinos. With the correct sign of the potential energy the electron neutrinos produced in the solar core travel towards the outer rarer layers where the reduced positive potential energy can make them degenerate with the slightly heavier muon neutrino thereby generating an "MSW resonance" and strong conversion of electron neutrinos into a superposition of muon and tau  neutrinos. The above, the beta function sign underlying asymptotic freedom and the repulsive sign of the magnetic Casimir force justify the saying by the late Joseph Sucher that "Understanding the sign is a sign of understanding".
\par
We next find the effective average potential U seen by a particle $P$ of momentum $p$ moving in a medium composed of "background particles" $P'$ of momenta p' and number density $n'$ each of which has the above interaction of eq.\ref{19.4} with $P-$ as e.g., for a WIMP inside a nucleus or the crystalline or liquid detecting medium. Approximating the Yukawa potential by a square well of size $R=1/M'$ and depth  $u=\pm g'^2 M'$  we have on average inside this well: \\
$N(P')= n . \frac{4\pi}{3} R^3$ particles each contributing interaction energy $u$. Particle $P$ then sees an averaged, spatially and temporally constant, effective potential of size:
\begin{equation}
    \label{19.5}
               U= N(P')u =  \frac{n(p')4\pi}{3} \frac{g'^2}{M'^2} = \pm 4 G'_F N(p')
\end{equation}
 When $P$ and $P'$ are different particles $g'^2\rightarrow gg'$, the product of the couplings of the exchanged particle with $P$ and $P'$. $G'_F$ refers to the "Fermi" coupling in the effective local four Fermi interaction. Note that the depth of the individual potential wells associated with each medium particle, $gg' M$ can be quite large as for ordinary weak interaction where $g_W^2\sim0.3$ and $M(W)\sim 80 GeV$ yields $u(W)\sim 30 GeV$. The "Weakness" of the weak interaction is due to the very short range of the interaction $d= 1/{\mu'}$. Vectorial interactions do not decrease as $P$ or $P'$ are boosted. The Lorentz contraction of the $\nu \nu$ interactions length is compensated by enhanced transverse $E$ field. For the scalar interaction the $m/E$ suppression remains. This is readily verified by computing the one vector or one scalar exchange diagrams in the Ultra relativistic limit $E\gg m$ \footnote{$^{++}$ A subtler feature is the vanishing of vectorial interactions between two mass-less, parallel moving fermions. In the Feynman gauge the diagram for the $PP' \rightarrow PP'$ scattering has a factor of $(2p$.$ 2p')$ which vanishes for forward scattering due to $p'^2 =p^2=pp'=0$. This may be related the fact that the two parallel, null fermionic vectors in the anomaly triangle graph behave as one massless pseudo-scalar a feature which survives all radiative corrections in nonabelian gauge theories as shown in \cite{Frishman:1980dq, Ando:2005ka, Adler:1969er}.}. \par 
The fact that the stellar core collapse requires that the bulk of the electron lepton number be emitted during a $\sim $ millisecond via a $N\sim 10^{57} \nu(e)$ pulse has dramatic consequences when the mass of the mediator of the $\upsilon \upsilon$ interactions tends to zero:
\begin{equation}
    \label{19.6}
 U(\nu)= g'^2 \frac {N^2}{R} \geq U( \text{Collapse}) = G_N \frac{M^2}{R} = \frac{m(N)^2}{M(\text{Planck})^2} \frac{N^2}{R}
 \end{equation}
\begin{equation}
    \label{19.7}
    \text{once} \hspace{0.1cm} \ g' > \frac{m(N)}{M(\text{Planck})} \sim 10^{-19}
    \end{equation}
This is the problem facing any long range interaction which is not gravity (or EM for which the net neutrality of matter is built in). To avoid the above difficulties and the precision tests of the Equivalence Principle (EP), extremely small coupling $g'^2$ are required -coupling which may be excluded by the "weak gravity" arguments \footnote{"Weaker than gravity" gauge interaction allows loading many particles charged under the new gauge interaction in BHs of mass (M= few m(Planck)) increasing the BH entropy beyond the Maximal Bekenstein value. A non perturbative "to gauge or not to gauge" dilemma arose in the CNN\cite{casher} collaboration that suggested the Flux-Tube model for multiple particle production - when the use of chromo-electric flux tubes for glue-ball production was debated. It was phrased by A. Casher as the immortal "Tube or no Tube"}. 
\par
After this extended detour let us go back to the (by now experimentally debunked) gap in the "ICE-CUBE PeV neutrinos". The local interaction yielding a linearly rising cross-sections cannot explain a putative gap in the spectrum, that initial data might have indicated (but never claimed by the Ice-Cube group!). The suggested explanations by Kfir Blum and John Beacomb invoked a new particle serving as a resonance $R$ in the $\nu_i \bar{\nu}_i$ channels with mass  $M_i(R)$ equal to the the cm energy $W$ of Eq.\ref{19.2} above, namely 1-100 MeV. The cross section at the peak of the Breit-Wigner distribution in the interval $M(R)$ -$\Gamma/2 \leq W \leq M(R) +\Gamma/2$ of width, $\Gamma = g^2 M(R)/{8\pi}$ is:
\begin{equation}
    \label{19.8}
  \sigma(max) \sim 4\pi M(R)^{-2} = 10^{-20} cm^2-10^{-24} cm^2\ \ 
  for\ M(R) = MeV - 100 MeV                                            
\end{equation}
These exceed the required cross-section of Eq.\ref{19.1} by six to ten orders of magnitude. Broadening the resonance beyond the above natural width by factors of $10^6-10^{10}$ is then required to reduce the cross sections to $10^{-30} cm^2$ allowing complete absorption of the extragalactic neutrinos in a broad energy region around the resonance. If the lightest neutrino  $\nu(1)$ is massless, then thermal CNB broadening by kT Induces the maximal spread of $\Delta(W)$ to cover the whole  $M(R)\sim 1 MeV$  interval which requires  $g^2= 10^{-10}$ with only a moderate effect of the t channel exchange of $R$ \footnote{The diffuse neutrino flux from all past supernovae considered in \cite{COLEMAN1982205} 
\textit{has} been discovered and could contribute to the "neutrino floor" in the large underground DM searches. We note that as indicated  by Eq.\ref{19.1} above the same cross section of $10^{-30} cm^2$ generated by the same mechanism between a typical SN neutrino of energy of 10 MeV and CNB neutrino of mass $\sim 0.1 eV$ mass would lead to several collisions with the CNB neutrinos. These in turn will reduce the energy of the diffuse SN flux to unobservable levels. Since the total number of CNB neutrinos exceeds that of the SN  neutrinos by a factor of more than $10^{11}$, the impact of the extra energy channeled into them is negligible.}. 

\section{\hspace{\secspace}Neutron stars and supernovae -the graveyards of (and hunting grounds for)- many DM types and BSM variants}
A dramatic title similar to the first half of the one above was used in a paper by Andrew Gould et-al \cite{20.gould} which excluded CHAMPS Charged Massive DM Particles, arguably the most daring type of DM proposed to date \cite{20.derujula}. \par
The high escape velocity  $v(escape) \sim 0.4 c$ and the huge number density of up to $10^{39} cm^{-3}$ causes particles with masses $M(X)$ as high as 20 TeV with $X$-nucleon cross-sections as small as $10^{-45} cm^2$ to have at least one collision upon traversing the neutron star. In such a collision they lose $\sim 10^{-4}$ fraction of their initial infall energy, they land in a bound orbit and after $\sim 10^4$ consecutive traversals fall into the star, collapse into a BH which gobbles up the rest of the NS into a black hole. As noted in \cite{20.goldman} this leads to occasional EM flare-ups and the absence of old cold neutron stars which are well known to exist. In adapting this method for excluding CHAMPs their efficient accretion on the progenitor star was used to enhance their concentration in the neutron star.\par
The high densities of many SM particles and BSM particles that can be generated at the collapse, the strong gravity manifesting in the high escape velocities, the high magnetic fields and the initial high temperature of the Super-nova (thanks to the initial 100 MeV Fermi energy of the electrons therein) make the title of this  section apply to a wide range of BSM scenarios and DM models. \par
For the magnetic moments of neutrinos and  the EM conversion of axions, the limits obtained from the many red giant and white dwarf stars respectively, are better than those from the SN 1987(a). 
We have already encountered S.N. limitations on the right handed ‘Sterile’ neutrinos. In this and in  the many other cases discussed below, one utilizes the neutrino pulses from SN 1987(a) at the LMC, the Large Magellanic Cloud neighbor galaxy. These pulses contain altogether $\sim 20$ neutrinos from the direction of the Sanduleak blue giant progenitor at a distance of $\sim 50 Kparsecs\sim 1.5\ 10^{23} cm$. The $\sim 10$ second duration and overall few $10^{53}$ ergs energetics of the neutrino pulses conform to the core collapse supernova theory. This  allowed deducing a wealth of constraints on new  particles and/or new BSM features of neutrinos that would, if realized, modify or even vitiate the observed neutrino signals\footnote{$^*$With much poetic license Churchill's famous saying on the British fliers who fought in the air battle over England in the second world war:  “never have so many owed so much to so few” may apply to the many who wrote so much about these $\sim 20$ neutrinos.}.
The rate of anti (electron) neutrino reaction with the protons in the water $\bar{\nu}(e) + p \rightarrow e^+ + n$ exceeds that of the neutral current interactions of the other five neutrino species and the strong nuclear binding  reduces the rate of charged current $\nu(e)$  interactions on Oxygen. The small proton recoil energy makes the final state positrons emerge with energies close to those of the incoming neutrino independently of the scattering angle.\par 
The $\sim 20$ SN1987a neutrinos display puzzling features such as a seven second gap between the arrival time in Kamiokande of the first eight neutrinos and the last three. This inspired the scenario of ref. \cite{Blum2016} with a remnant BH. In this Blum-Kushner model only the neutrinos emitted from SN1987(a) before the 7 second gap in the neutrino flow from Kamiokande conform to the pattern expected for the core collapse into a neutron star whereas to date the analysis of the neutrino pulses assume that \textit{all} the neutrinos are from such a source\footnote{The momenta of most positrons detected are pointing  in the forward (i.e SN to earth) direction contrary to the slightly backward peaked angular distribution predicted for the $\bar{\nu} (e) + p \rightarrow n + e^+ $reaction. It is generally believed however that all peculiarities are due to statistical fluctuations in the small sample and do not necessarily diminish the scope of possible deductions.}.\par
The most direct implication utilizing the \textit{extra} spread  $\Delta(t(i)= t(travel) \left[ m/2E(i) \right]^2$ of the time of arrival of  neutrinos of various energies $E(i)$ and mass $m$ to limit the neutrino masses was immediately noted by many authors. The lack of correlation between arrival times and neutrino energies implied a bound of roughly $m\leq 10 eV-$ far inferior to the present direct kinematic $\sim eV$ upper bounds but relevant at the time. Naively only the interacting  $\bar{\nu}_e$ is being limited. However neutrino mixing makes it apply to all three species. As emphasized by  Leo Stodolsky \cite{Stodolsky:1987vd} and by Michael J. Longo \cite{Longo:1987gc} this tests the equality of the velocities of and the applicability of equivalence principle to neutrinos and to photons. The time of arrival data test the correct relativistic dispersion $\omega^2= k^2+m^2$ for neutrinos which, in models with Lorentz Invariance Violations (LIV) may be modified into $\omega=[ k^2+k^4/M^2]^{1/2}$ where $M$ is the high scale of new physics that generated by the LIV \footnote{The existence of shorter, few millisecond time structures in Gamma ray bursts which typically originate at 1000 times larger distances than SN 1987 furnishes much stronger bounds on the scale $M$ \cite{Ellis2005sjy}.}. \par
The short spread of the neutrino arrival time does not furnish a direct test of Einstein's velocity addition rule. Neutrino waves - just like light waves - adopt the velocity in the new medium within a short "Ewald-Oseen Extinction Length" of $[n(i)-n(f)]\lambda(f)/c$. Thus their velocity becomes very close to that of light once they move from the core to the much rarer progenitor star and much more so when they leave to the ISM. In passing we note that \textit{flavor dependent} tiny phase shifts of \newline (n-1)l over a traveled distance l dramatically impact neutrino oscillations. This allows exquisitely sensitive tests of such hypothetical BSM interactions of the neutrinos with clouds of new light fields as in \cite{Brzeminski2022rkf}). Also Minakata pointed in neutrino 94 meeting the impact of tiny violation of the applicability of the EP to different neutrino flavors on solar neutrino oscillations.\par     
Recently it has been argued in \cite{Chang:2022aas} that frequent collisions between the neutrinos after emerging from the Proto- Neutron Star (PNS)  i.e. the Supernova core, into effectively empty space can dramatically prolong the arrival time.
We therefore consider the possible delaying effect of some new "secret" mutual $\nu \nu$ scattering. Following the last ref. we assume that the neutrinos leave in a single burst a sharply defined PNS of radius $R\sim 30 Km$ and focus on the \textit{extra} broadening/time delay incurred by  mutual collisions among the neutrinos. As they move from a radius $R$ to $R+ dR$, a tiny fraction $ dR/{l(mfp)}$ with $l(mfp)$ the mean free path for $\nu \nu$ collisions, collide with other neutrinos. The angle $\theta$ between a neutrino’s velocity and the radial direction is given in terms of the components of the momentum by: 
$$
\theta = \arctan  \frac{\sqrt{p(x)^2 +p(y)^2}}  {p(z)}
$$
where the local $+ z$ axis is taken parallel to the outward radial direction.
Neutrinos that move mainly radially out with small $\theta$, collide \textit{less} and therefore the average $\theta$ will decrease as we move outwards. This manifests the Liouville theorem that in a free dissipation -less and isolated Hamiltonian system the phase space volume of the $\mathscr{N}
$ ($\sim 10^{58}$ here!) particles is conserved. With the  $X$ and $Y$ the transverse coordinates growing with $R$ as $X\sim R$ and $Y\sim R$, $p(x)$ and $p(y)$ decrease as $1/R$. This is not the case for $p(z)$ because the sum over the colliding neutrinos of energies $[p(x)^2 + p(y)^2 +p(z)^2 ]^{1/2} = E$ is conserved in the elastic collisions. Thus $\theta$ decreases as $1/R$. Since we can take the mfp to be arbitrarily small, the only relevant distance in the system is $r(PNS)$, the initial radius of the proto-neutron star, so that $\theta \sim r(PNS)/R$. Such radial focusing occurs in free propagation as can be verified by simple geometry. All this implies minimal extra time broadening $\delta(t)$ over and above $\Delta(t)$ due to the in PNS delay. Since it is difficult to precisely predict the latter, “peeling off” the extra $\delta(t)$ from the observed total of 5-10 seconds is challenging\footnote{A more detailed analysis by Damiano  Fiorillo, Georg Raffelt and Eduard Vitaglio reached similar conclusion.}. 
A beautiful application by Bernardo Barbiellini and Giuseppe Cocconi \cite{Barbiellini:1987zz} uses the magnetic fields of our and the LMC galaxies as giant spectrometers to bound the charge of the electron neutrino by  $q( \bar{\nu}(e))\leq 10^{-17} q(e)$. Otherwise the higher energy more ‘Rigid” neutrinos, whose trajectories curve less, would arrive before the lower energy neutrinos, a trend that the data fails to show. This \textit{direct} bound is however weaker than the indirect bound of $\sim 10^{-21}$ deduced from experiments testing the neutrality of (regular and heavy) water and Charge Conservation  in the neutron $\beta$ decay.\par
Most of the SN1987(a) bounds invoke the possible deterioration of the neutrino pulse due to the competitive emission of some other particles which often are DM candidates. These particles couple more weakly to nucleons/electrons than the neutrinos and therefore escape faster -but not too weakly coupled so that they are abundantly produced in the hot core. This approach was already used above in connection with a sterile RH DM neutrino, and was applied also to  put  lower bounds on the PQ symmetry breaking scale which is inversely proportional to the axion -photon g(a, $\gamma\gamma$) coupling, to limit mixing of our and a dark photon and or to production of new BSM particles of masses up to 50 MeV.\par
Limits were derived also for the charge radius and magnetic moments of neutrinos. If neutrinos are majorana (self charge conjugate) particles, then they cannot have any em properties -though transition magnetic moments (MM’s) connecting neutrinos of different flavors are allowed. Joint flavor and spin oscillations of solar neutrinos induced by an $\nu_e \hspace{0.1cm} \nu_\mu $ transition MM, were considered when data from the Homestake mine (Davis) experiments hinted at a possible correlation between sun spots/ magnetic activity and paucity of detected electron solar neutrinos, an anomaly which was not reproduced by later higher statistics experiments. \footnote{$^*$Ref.\cite{Barbieri:1988av} and Arnon Dar (1986 unpublished) suggested that for an appreciable magnetic moment (M.M) the initial left handed $\nu_{(l)}$ in the forming S.N. core can oscillate into right hand neutrinos $\nu^e_{R}$ which readily escape early on. Roughly half of these $\nu_{e}(R)$ oscillate back to $\nu_e (L)$ en route to earth. The energy of these neutrinos generated in the early stage of neutronization via: $e^-+ p \rightarrow  \nu_e +n $ is that of the high "Fermi" energy $E \nu_l \approx  E_e \approx E_f \approx 100 MeV$. This early pulse of anomalously high energy $\nu$ 's would manifest by the large cross-sections $\sigma{(\nu - A,Z)} \approx E_\nu^2$ on nuclei and on oxygen in particular. To avoid this unobserved signal, strong limits on the $\nu _l$ magnetic moment have to be imposed. A different approach was used in ref. \cite{Goldman:1987fg} to suggest even stronger limits. In ref. \cite{Hidaka_2006} it was pointed out that re-conversions of $\nu_R \rightarrow \nu_l$ in outer layer of the progenitor can help the S.N. explosion. A similar motif appeared in a recent work in ref. \cite{Fiorillo:2022cdq} where the authors note that two $\nu$'s can combine into a weakly interaction Majoron. The latter escapes the S.N. core and decay into two energetic neutrinos outside. }
The limits deduced from the physics of the  horizontal branch red Giants -which maintain constant luminosity while shrinking and becoming hotter noted by G.Raffelt, are competitive with Supernova bounds, see ref. 
 \cite{Barbieri:1988av} and \cite{Goldman:1987fg} and superior to the direct terrestrial laboratory upper bound. The latter is deduced from low energy neutrino scattering on nuclei where the constant cross section due to the charge -magnetic moment  interaction: $\sigma(MM) \sim \alpha  \mu(\nu)^2$  exceeds the  \textit{weak} cross-section $\sim G_{Fermi}^2 E(\nu)^2$ \quad  if  $\mu(\nu) > 10^{-10} (e/m(e))$.\par
That trace amounts of captured DM X particles can affect properties of stellar objects was mentioned towards the end of Sec III and is particularly clear for the accretion of very heavy D.M. particles inside neutron stars \footnote{Yueh Zhang suggested that trace amounts $\Delta(M)\sim  10^{-12} M(NS)$ of the mass of  DM  $X^+ (and\ X^-)$ charged under the dark photon can, for a large scattering cross-section with the dark photon, stop the escape of the latter from SN 1967(a) and evade the resulting very stringent upper bounds on $\epsilon -$ the kinetic mixing with the ordinary photon. As the dark photons may carry $\sim 1/4$  of the total gravitational collapse energy which is $\sim 0.1\ M(NS)c^2$ the outward radial pressure exerted by the $\gamma$’s which are trying to escape may kick out the much fewer $X'^+$ or $X'^- $ of much smaller total mass and gravitational binding and evade any trapping of the dark photons by these  “guards”.}.
\par 
Even if a small fraction of the energy of the escaping species converts into photons, then the much easier detection of the e.m. signal allows enhancing bounds on the BSM  species - axions, dark photons, etc beyond the "standard" cooling limits. The reason is the following: At the time $t=t(0)$ of the gravitational collapse and neutrino emission from the SN1987(a) a Solar Maximum Measurements (SMM) X-ray satellite had the Supernova in its field of view. Still, no enhanced photon flux from this direction above background was measured during the first hour after $t(0)$ (Chuppe et - al \cite{PhysRevLett.62.505}). (Later the SMM measured the $\gamma$ decay line of Cobalt empowering the month -long light curve).
Here we focus on the consequences of \textit{not} seeing early em signals expected in certain BSM/DM scenarios. In ref. \cite{Dar:1987nq} the strong limit on radiative neutrino decays $\nu(i)\rightarrow \nu(j) +\gamma$ was used to exclude a broad range of neutrino masses and transition magnetic moments which at the time were allowed and would enable such decays enroute from SN 1987(a). Kinematics and simple geometry imply that irrespective of the angle $\theta$ relative to the line of sight with which the initial $\nu(i)$ was emitted at  $t=t(0)$, the arrival of the final photon of average energy $E(\nu(i))/2$ is delayed by the extra, small, length of the  path traversed by the neutrino $\Delta L \sim L\theta^2 /2 \sim L [m(i)/{E(i)}]^2$ by:
\begin{equation}
    \label{20.1}
    \Delta(t) = L/c .[m(\nu(i)/E(\nu(i)]^2 \leq  O(10) sec
\end{equation}
The shortness of $\Delta(t)$ and of the duration of the neutrino emission then “squeezes” the expected decay photon signal to be within $\sim 20 sec$ interval starting at $t=t(0)$, the time when the first neutrino was detected - greatly enhancing the deduced bounds\footnote{The bounds deteriorate (in proportion to $m(\nu(i))^2$) as the mass of the decaying neutrino increases . The radiative decay rate scales as $m^3$, and the lorentzian $m/E$ factor is due the prolonging of lifetime for the decay in the lab yielding together $m^4$. Finally the shortening (in proportion to $m^2$) of the spread of arrival times of the expected photons make the net time spread $\sim m^2$.}. \par
Similar reasonings was used in \cite{Goldman:1987fg} to improve the SN 1987a bound on $\epsilon$, the mixing of the photon with a dark photon of mass $50 MeV> m' >2 m(e)$ and a range of mixing $\epsilon$ values. Such $\gamma$’s can (a) be amply produced in the supernova and (b) freely propagate through the progenitor star of SN 1987a of radius $R(PS)=3\cdot10^{12} cm$. Moving typically with speeds $v>1/2 c$ they leave the progenitor in less than 3 minutes. Finally $(c1)$ a sizeable fraction of the Dark photons decay outside the progenitor star within a shell of outer radius $\sim 4 R(PS)$ radius into $e^+ e^-$ pairs in which case what happens next is $d(1)$ the dense mix of electrons and positrons annihilates and generates a gamma ray burst (GRB). Alternatively $(c_2)$ a sizeable fraction of the dark photons decay inside the outer shell $R(PS) >= r >=0.8 R(PS)$ of the progenitor star where only a small fraction of it’s mass resides, in which case ($d_2$), the radial pressure exerted by  the $e^+$ and $e^-$ decay products of the escaping dark photon unbinds the outer shell and kicks it out. In both cases a remarkable ‘“Firework” of $\gamma$, X rays and optical photons should erupt within minutes after the neutrino emission from SN1987a. The lack of any indications for this allows excluding a significant region in the $log(m(\gamma')) - log (\epsilon)$ plane beyond $\epsilon\leq 10^{-12}$ already excluded by the standard cooling argument\footnote{For $\epsilon$ values smaller than $10^{-14}$ most decays occur further out and the expected GRB is much weaker as most positrons will escape the dilute $e^+ e^-$ system. If $m(\gamma ')$ is comparable or a bit larger than the average energy (ot temperature), then the mildly relativistic $\gamma$’s arrive over a longer time window decreasing the expected luminosity. 
If $m(\gamma' )\leq 2 m(e)$ ,then $\gamma'\rightarrow 3\gamma$ proceeds at a much slower rate $\Gamma \sim \epsilon^2 [m(\gamma')]^9/{(m(e)^8 4096\pi^2)}$ and only the accumulated effect of all past supernovae is measurable. \par Very similar arguments limit BSM (pseudo)scalar particles of masses $< 50 MeV$ which can be produced inside the collapsing core and decay towards the outer edge and or outside the progenitor via $X^0\rightarrow 2 \gamma$. The case of very small masses and $\epsilon $ coupling will be revisited again in the next section. The last comment applies, in particular to Axions and Axion-like particles (ALPs) referring to pseudo scalars heavier/lighter than the original “ QCD axion” of mass $m(a) \sim f(\pi)^2)/ F(PQ)$. Thanks to their  
$\frac{1}{M} \cdot F \Tilde{F} = \frac{1}{M} \vec{E}. \vec{B}$ couplings with $M=8\pi F(PQ)/{\alpha}$, 
the light Axions convert into photons in $\vec{B}$ fields. at a rate:
\[
    \Gamma[(a +\vec{B})\rightarrow \gamma]\sim  M^{-2} (B.L)^2 .m(a)\   for\ L \leq L(Coh) 
\]
\begin{equation}
    \label{20.2}
    \Gamma[(a +\vec{B})\rightarrow \gamma]\sim M^{-2} (B.L(Coh)^2 .m(a)L/L(Coh)\ for\ L\geq L(Coh)
\end{equation}

with the coherent oscillation length
\begin{equation}
    \label{20.3}
    L(Coh) \sim 2E(\text{axion})/(m(a)^2-m(\text{plasma})^2)
\end{equation}
m(plasma) is the plasma frequency in the medium in question and $E(a) = E(\gamma)$. These issues were discussed early on in \cite{Grifols_1996}.} \par
Strong magnetic $B\sim 10^{12} gauss$ fields in the collapsing core of size  $2 R(NS)\sim 20 Km$ are generated by compressing the original fields in the progenitor star so that B grows as  $\sim 1/R^2$. However the resulting $axion \rightarrow \gamma$ conversion photons are absorbed in the progenitor and escape only if generated beyond the surface of the progenitor star where no enhanced B fields exist. This suggests that isolated pulsars, i.e. neutron stars where all matter was ejected or fell back onto the compact N.S, and which retained the large magnetic fields, are ideal hunting grounds for axions. The B field near the Neutron star converts up to $50\%$ of the  axions into photons with the same energy of the parent axions. The latter axions are generated via the Primakoff process in the hot ($T\sim 2-10 KeV$) core of the old NS and the conversion photons hopefully may be separated from the photons emitted directly from the  pulsar. There were  some indications for this, however, only in two of the seven nearby old pulsars studied in ref. \cite{Dessert:2019dos}. With later searches yielding no further positive results this line of research has been abandoned.\par
A more sophisticated approach to the interplay of axion and N.S./ pulsars is illustrated by the most recent work \cite{Caputo:2023cpv}.
It uses the ever improving understanding of the E.M. fields and plasma around the N.S.to find features of the well measured pulses which can be affected in axion scenarios.
Of particular interest are the polar cap regions from where the pulsar jets emanate and which have very strong, \textit{parallel} $ \vec{B} \hspace{0.2cm} \text{and} \hspace{0.2cm} (\vec{E}) $  fields. The back reaction of the oscillating axion field generated therein may imprint these oscillations on the emitted radio pulses. At the present time, the analysis of many white dwarfs which have weaker but more extended  and better studied magnetic fields than in supernovas, yields the best bounds on photon-axion mixing. \par  
In several other BSM/DM scenarios, the accretion of DM onto the N.S. is far more efficient when the DM is self interacting and tends to heat up old pulsars- providing that the energy gained  per accreted particle $\sim 1/2 M(X^0) v(escape)^2 \sim 1/8 M(X^0) c^2$ is not emitted via dark rather than ordinary, photons.\par  
The upper bound on photonic emission from old neutron stars was recently used by D. McKeen, M. Pospelov, and N. Raj, \cite{McKeen_2021} in an attempt to limit putative mixing of neutrons and mirror neutrons. In devising tests of mirror models  R. Bondi  M. Mannarelli, Z. Berezhany, and F. Tonelli, Proceedings of the Nordita ESS Workshop (2018) noted that while loss of nuclear binding forbids $n\rightarrow n'$ transitions in nuclei, the predominance of gravitational interactions and corresponding high density and high Fermi energy, favors such $n \rightarrow n'$ transitions inside neutron stars. Thus for $\epsilon(n\rightarrow n')$ mixings still allowed by laboratory experiments ,neutron stars can convert over relatively short time spans of 10-100 million years into more compact, lighter stars with an equal mix of ordinary and mirror neutrons.\par
Some 50 cold pulsars with few KeV core temperatures do not cool by neutrino pair emission which scales as $T^8$ \cite{Yakovlev_2004}. The existence of cold pulsars with minimal BB like em radiation, excludes scenarios with efficient novel sources of energy inside the NS. - pulsa. It was argued that the existence of a particularly cold pulsar implied that the above neutron $\rightarrow$ mixed star transition must be extremely slow. This then would suggest an upper bound on $\epsilon(n,n')$ mixing smaller than what planned terrestrial experiments can achieve. However in \cite{Goldman:2022rth} it was noted that in the context of almost exact mirror models (which are required for the planned terrestrial experiments in the first place), the subsequent decay of the mirror neutrons $(n's)$ yields $p'$ and $e'$ and the resulting cooling of the star via $\gamma'$ emission can evade the claimed  bounds even for dark photon mixing $\epsilon$ as small as $10^{-12}$.

\section{\hspace{\secspace}Axions, ALPs and Dark photons}

\textbf{In which we sketch the motivation for axions and axion DM , describe some of the methods for generating and discovering axions and various   approaches to detection of  axions or other forms of light DM}
\subsection{\hspace{\secspace}Introduction}
Axions, introduced into high energy physics by R. D. Peccei and H. R. Quinn \cite{Peccei:1977ur}, followed by .S. Weinberg "A New Light Boson?" \cite{PhysRevLett.40.223} and F. Wilczek, "Problem of Strong P and T Invariance in the Presence of Instantons" \cite{PhysRevLett.40.279}, to resolve the "Strong CP problem" are most elegant and subtle constructs.
\footnote{This CP violation is distinct from the well measured and understood CP violation in weak decays which traces to the irreducible complex part of the 3x3 CKM Quark mixing matrix.}
Axions tie with many aspects of gauge theories such as the axial $U(1)$ anomaly, nontrivial topological configurations of instantons and corresponding periodic classical vacua with integer values of the integrated topological charge. They are connected with the $E \leftrightarrow B$ electric magnetic duality and with the "Chern Simon" term.
 The "Strong CP problem" is that a CP violating term: \hspace{0.1cm} $\theta \hspace{0.1cm} \tilde{G}G  \hspace{0.15cm} \text{with} \hspace{0.15cm} \tilde{G}^{\mu\nu} = \epsilon^{\mu\nu\rho\sigma} G_{\rho,\sigma}$ - the $E\leftrightarrow B$ dual field strength, can be added to the QCD lagrangian. Apriori $\theta$ can be of order 1. The large CP violation is translated into the quark Lagrangian and generates a neutron electric dipole moment (EDM). An upper bound of $\theta < 10^{-11}$ needs then to be imposed in order to respect the experimental upper bound on the neutron's EDM (Electric Dipole Moment)
\footnote{$^+$ To estimate $d(n)$ we view the neutron magnetic moment $\mu(n)\sim 2 e/{m(n)}$ as due to two hypothetical opposite magnetic charges $\pm e$ separated by $r(n) \sim 1/2$ Fermi. Witten showed in ref. \cite{Witten:1979ey} that a $\theta$ E.B term induces an electric charge $\theta$ $e$ for monopoles of magnetic charge $1/{\alpha}$ times bigger. This suggests an induced electric dipole moment $d(n) \sim \theta \alpha(em) \mu(n)$ - similar to the correct theoretical value despite the fallacy of the above reasoning: Magnetic moments of SM particles are generated by electric currents with continuous closed B field lines - not by monopoles. The presence of the $\theta$ term in the Lagrangian rather than Hamiltonian makes the magnetic monopole pick up a small electric charge but \textit{not} vice-versa maintaining the $e(1)g(2) -e(2)g(1)= n \bar{h}$ Dirac  quantization rule for dyons with both electric \& magnetic charges. Most recent, extremely precise measurements of the electrons electric dipole moment, provide exquisite tests of the S.M.}.    
A Cosmological Constant ($10^{60}$ times too large) CC $\sim M(Planck)^2$ is naturally generated by the vacuum fluctuation of the various fields, dwarfing the strong CP problem. (the above huge factor is reduced to $10^{30}$ if SUSY is broken at TeV energies). Also, a small $\theta$ does not have the huge "Anthropic" value of a small CC. That the axion which in some models is the Holomorphic Twin of the dilaton, can "relax" from its initial $O(1)$ value to the present tiny value, suggested similar ideas for explaining the CC. see e.g ref. \cite{Graham:2019bfu}. \par
A full account of how the $\theta$ term generates observable CP violation, how the axion field mitigates such effects, the axion mass and potential required to produce a good axion CDM, and the detailed generation of such CDM are beyond the scope of this review. The following few brief comments can be readily skipped by those who are familiar with axions or wish to focus on their possible role as DM and on experimental methods devised for their detection.
\begin{enumerate}
\item[a)] The infinitesimal version $G^{\mu,\nu} \rightarrow G^{\mu,\nu}+\epsilon \tilde{G}^{\mu,\nu}~$ of the $E\leftrightarrow B$ duality transformation generates a common chiral rotation of the three light quarks: q(j) = u,d,s:
$q(j) \rightarrow [1+\epsilon \gamma(5)]q(j)$ .The finite version of this induces a CP violating phase in the determinant of the mass matrix of the u,d,s quarks which in diagonal form, is given by the product  $m(u) m(d) m(s) \exp {3 i\epsilon}$. While $m(u) < m(d)$, the apparent nonvanishing of $m(u)$ excludes the possibility of canceling $\theta$ by a chiral rotation when $m(u)=0$, arguably the simplest solution of the strong CP problem.

\item[b)] The $G\tilde{G}$ term in QCD is a derivative 
$ \partial _\mu J_\mu$ of the Chern Simon current:
$$
J_{\mu}=\epsilon(\mu , \nu , \alpha , \beta)[A_{\nu}.G_{\alpha,\beta}-\frac{g}{3} f. A_\nu A_\alpha A_\beta]
$$
with $f=f(a,b,c)$ the structure constant of the gauge group with color indices suppressed. If the $A_{\mu}$ fields vanish at infinity then the $d^4(x)$ integration of $G\tilde{G}$ makes a vanishing contribution to the action. However the instantons in QCD (or other non-abelian gauge theories) have non-vanishing gauge potentials at infinity and contribute an integer value of topological charge which is  $\pm 2\pi N \int G\tilde{G}$ to the action, defining an infinite periodic series of degenerate QCD vacuums .After summing over this series only the Mode $2\pi$ part of the coupling in front of $\widetilde{G}G$ survives making it an angle $\theta$.
\item[c)] The constant $\theta$ was promoted into a dynamical D.o.F, by R. D. Peccei and H. R. Quinn - the field $a(x,t)=\exp{i\theta}.\Lambda(P,Q)$. This axion is the Nambu Goldstone massless particle associated with breaking of an axial $U(1) \hspace{0.1cm} P.Q$ symmetry at some high scale $\Lambda(P.Q)$. It is the zero mode of the imaginary - pseudo scalar part of the axion field along the bottom rim of the corresponding "Mexican Hat” potential. It gets a mass thanks to the Adler and Bell Jackiw axial $U(1)$ anomaly in QCD just like  the ‘Ninth goldston pseudo-scalar $\eta'$ gets a mass of order $\Lambda(QCD)$. This mass is transmitted from the high $\Lambda(P.Q)$ sector. The resulting axion mass is suppressed to a see-saw like value: $m(a) \backsimeq \Lambda(QCD) ^2/ {\Lambda(P.Q)}$.
\item[d)] Initially (before, during or shortly after inflation) $\theta$ could be appreciable. The axion field would then slide on its (Periodic) potential of the now slightly tilted "mexican hat" towards the minimum at $a = 0$ where the $m^2a^2$ mass term makes it oscillate with frequency $m$. This  relaxes $\theta$ towards zero and generates a mass for cold DM axions. Choosing $\Lambda(PQ) \sim10^{12} GeV$ or, a QCD axion $m(a) \sim10^{-5} eV$, yields a sufficiently small $\theta$ and a correct DM density.  
\item[e)] The original QCD axion model satisfying the above see-saw relation with an electroweak $O(TeV) PQ$ scale, was ruled out by direct searches and other DSFZ: \cite{Dine:1981rt} and \cite{Zhitnitsky:1980tq} 
and KSVZ  \cite{Shifman:1979if} variants were introduced with  much higher $\Lambda(PQ)$ scale and lower $m(a)$. Also different - Axion Like Particles $(ALPS)$ which deviate from the see-saw relation above are being studied. Most axion phenomenology traces to the $(\frac{1}{M}) aF \tilde{F}=(\frac{1}{M})a E.B$ interaction term with $M\sim \Lambda.{\alpha(em)^{-1}}$ - the electromagnetic analog of the $G\tilde{G}$ axion interaction. It leads in particular to photon - axion mixing in the presence of strong magnetic or electric fields: \cite{Raffelt:1987im}.
The $U(1)$ anomalies in both QCD and QED  arise via triangle loops and depend on the color/ em charges of the fermions circulating therein which are different in  the KSVZ and DFSZ models.    
\item[f)] A key point is that the optimal (and allowed!) axion DM is not hot DM even when its mass m(a) is lower than $T(CMB)(z=1)\sim 5.10^{-4} eV$ - the present CMB temperature. The axion cosmic density is generated via a "remaining mis-alignment" i.e by having a very shallow tilt in the potential of the axion field so that the initial $\theta$ does not fully relax to zero but maintains some finite value. Non-perturbative large instanton effects, the axion mass and ensuing tilting of the Mexican hat potential, all happen at the QCD phase transition. The $m (a)^2 a^2$ mass term fixes the curvature of the axion potential at its minimum leading to the requirement that $m(a)$ be small enough to keep the residual misalignment which in turn fixes the energy density of the axion DM. For a misalignment energy density $V (a)\sim m^2a^2$ the residual a is given by $a=V^{1/2}/m$ and the effective "number" density  of axions is $n \sim V/ m \sim m a^2$. The Euler Lagrange equation leads to the time dependence $a(t) \sim exp{-iwt}$ with $w \sim m$ so that the general expression for the number density of a scalar Klein Gordon field $n\sim a d/{dt}$ a indeed yields $n\sim ma^2$.
Even when treated as a field rather than a particle the axionic DM has an  active (and equal inertial) energy density of N.R particles! Hence at present, axions in our halo (or halos of other galaxies) have virial $rms$ velocity of $v\sim 300 Km/{Sec}$ as any CDM.
\end{enumerate}

The above  introduction omitted many relevant features. An early PQ symmetry breaking can yield a cosmologically uniform $\theta$ or different $\theta(i)$ in different causally disconnected patches. This and incomplete breaking of the  symmetry into Z(N) subgroups lead to a bewildering multitude of  scenarios. Strings formed by a two dimensional analog of the Kibble monopole generation in three dimensions, can make complex cosmic string networks contributing to the cosmic energy density\footnote{Super horizon strings  stretch upon Hubble expansion and  their contribution to the cosmological energy density scale as $R^{-2} \sim T^2$ rather than $R^{-3}$ as for CDM.}. The cutting and reconnecting of strings keeps shortening them and radiating axions. Large misalignment and attendant large CP violations may arise after the QCD phase transition;. For an extended review of Axion physics see ref. \cite{DiLuzio:2020wdo}

\subsection{\hspace{\secspace}Generation/ detection strategies for axions and other light particles}

Axion and dark photon searches divide into a) searches where we try to generate and detect axions in the lab b) where we detect in the lab axions generated astrophysically or cosmologically and c) where the inference is indirect via the impact of axions on various stellar objects and/or on the CMB ..Detection is optimized when the axions act coherently as a classical field in their production or detection and preferentially in both. In the following we present a potpourri of detection approaches which seem elegant.

\subsubsection*{\hspace{\secspace}THE SHINING THROUGH WALL (STW) APPROACH }
In its axionic version it is shining an intense, pulsed laser beam on a region with strong transverse magnetic field B and detecting some radiation in a resonant Fabry-Perot cavity beyond a thick wall impassable for photons but not for axions. (See e.g. "Axions and other similar particles" by A. Ringwald, L.J. Rosenberg \& G. Rybka \cite{P.PDBook}.
 If light axions exist, a tiny fraction $\sim [eBL/F_{PQ}] ^2$ of the coherent laser photons convert into axions in the first magnetic field and then convert back  with the same probability behind the wall into photons of the original frequency. The rate is suppressed by $F(PQ)^{-4}  \sim m(a) ^4$  yet the large magnetic field, the large coherent initial laser field and the high occupation of the desired mode in the detecting cavity conspire to make it quite sensitive. The suggestion of producing (via the Schwinger process) milli charged particles or using "Millicharged relics to reveal massless dark photons" \cite{Berlin:2022hmt} which are STW to a detecting cavity and generally of using collisions of SM particles within a collider or near an accelerator to produce BSM DM weakly interacting particles which penetrate the beam dump "Wall", are closely related.
 Many other detection methods of DM are variations on the STW theme. The "wall" being a stellar object with the energy produced in its interior escaping via the particles which most quickly shine through it - such as axions rather than photons or neutrinos. This can happen from the solar core or from old or just forming neutron stars. The weakly interacting penetrating particles then convert back into photons in the magnetic field inside the bore of an LHC discarded magnet in the SOLAX experiment or in the magnetic fields surrounding the neutron and other stars. Also the more energetic R.H sterile neutrinos rather than the left handed neutrinos from the supernova core escape and then  convert back later to the more strongly interacting left handed neutrinos outside are examples of STW. Finally a putative cosmological STW could manifest via PeV photons from a gamma-ray burst associated with supernovae in high redshift galaxies. The wall in this case is the CMB in the Giga-parsec size intervening space. The initial photons convert to dark photons, or aided by the galactic B fields, to axions. The latter freely traverse the large distance and convert back in our galactic B fields to the observed photons.\par
\begin{figure}[h]
\begin{center}
 \includegraphics[width=0.4\textwidth]{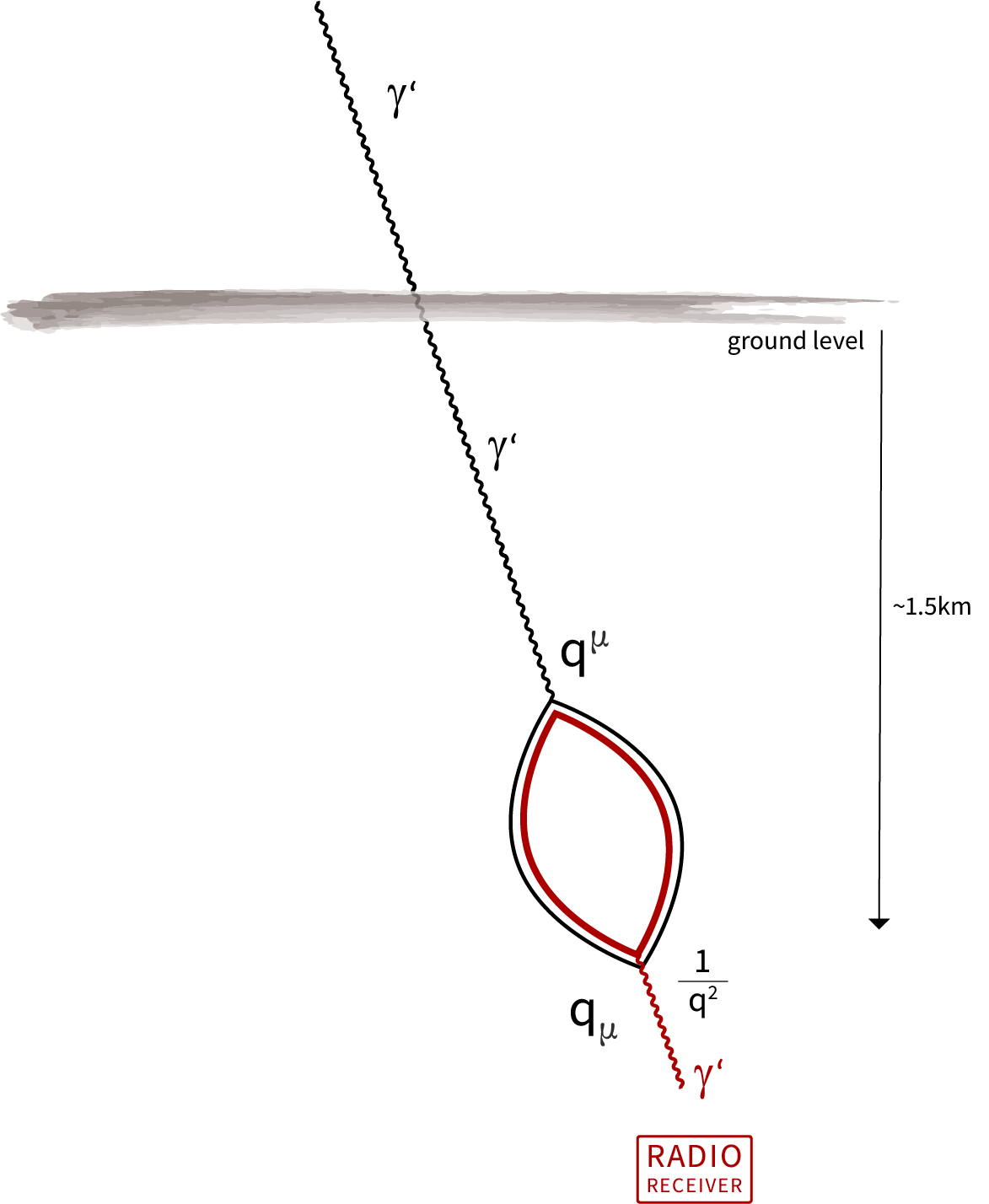}
    \caption{Detecting dark photons via underground segregated sensitive radio receivers}
\label{fig:011 ground line}
\end{center}
\end{figure}

 
The "Nasduk" collaboration in ref. \cite{Bloch:2022kjm} and previous works cited therein suggested extending initial efforts to detect axions to also exclude Dark photon DM in a wide domain in the $\epsilon-m(\gamma')$ plane. The dark $E'$ fields of frequency $\omega =m'$ or wavelength $\lambda ' =c/{m'}$ due to the DM $\gamma$’s flux, penetrate the experimental set-up of size of order $ 100 m^3$. This region is enclosed by a metallic shielding which excludes ordinary electric fields and also varying  magnetic field allowing only the dark $E'$ field to penetrate. After $E'\rightarrow E$ conversion the radio receiver  -a  RLC  circuit enclosed therein is excited. The latter becomes ultra sensitive when tuned to the correct frequency which corresponds to the mass of the dark photon or its energy. For $m' \neq 0$  the latter varies only over a small interval $[ m' + \frac{m'\beta^2}{2} ]$  with $\beta \sim 10^{-3}$ allowing circuits with quality factor up to $Q \sim 10^{6}$ to be used with advantage. The response of the device to dark $E'$ field is identical to that involving ordinary $E$ field at the same time frequency apart from an overall $\epsilon$ factor. This factor manifests in the Feynman diagram fig.\ref{fig:011 ground line} where the photon propagator of $\frac{1}{\omega^2}$ cancels the two  $\omega$ factors from the epsilon $F_{\mu \nu} F'^{\mu \nu}$ mixing term.

To have a sense of the projected future sensitivity of such designs we compare it with that of radio emission and reception in a space mission planned to probe the earth like planet in the exo-solar planetary system around our nearest neighbor M star Proxima Centauri four light years= $4.10^{18}$ cm away. Radio waves broadcasted from there by putative small radio devices with power of $\sim K.watt \sim 10^{10} ergs/{Sec}$ will hopefully be detected on earth where the corresponding flux, ignoring possible directionality of the radio signal, is: 
\begin{equation}
    \label{21_B_1.1}
    \Phi \sim  \frac {10^{9} ergs} {(3.10^{18})^2\ cm^2. sec}= 10^{-28}  ergs.cm^{-2}.sec^{-1}
\end{equation}
We assume that the flux of dark photons or dark radio energy hitting our secluded chamber to be the same as that of the CMB: 
\begin{equation}
    \label{21_B_1.2}
\Phi' \approx 10^{-2}\hspace{0.15cm} \frac{erg s} {cm^2sec} 
\end{equation}

Thus the effective flux impinging on the device $=\epsilon^2 E'^2 \sim \epsilon^2 \Phi'$ with $\Phi' \sim 10^{26} \Phi$. 
Assuming that the sensitivity needed for the astronomical communication can be achieved and with all other factors being similar but with the $\frac{E'^2}{E^2} = \epsilon ^2$ suppression this would then allow excluding $\epsilon$ values as small as $10^{-13}$. (This sensitivity may be  exaggerated as we ignored possible directionality of artificial signals). The above rather qualitative discussion hardly does justice to the many theoretical and experimental elaboration in the above ref. 

\subsubsection*{\hspace{\secspace}AXION detection methods utilizing photon polarization}
The following utilizes the different polarization of photons induced via the $\frac{1}{M} \cdot \overrightarrow{E} \cdot \overrightarrow{B} $ interactions with the ambient galactic Axion field which interfere with the original large photon amplitude. The suppression of the sought signal in laboratory and astronomical detections is then only $\sim 1/M^2$ rather than $1/M^4$ in the above STW approach.
To introduce the general idea in its simplest settings, consider the case of a very strong $\overrightarrow{B}$ field and focus on the amplitude of the photon polarization (in the direction of its $\overrightarrow{E}$ field) parallel to $\overrightarrow{B}$. It can be readily shown that the Raffelt Stodolsky axion photon effective $2\times 2$ hamiltonian yields $\ket{\gamma_\parallel} \pm \ket{a}$ eigenstates split by energy 
\begin{equation}
    \label{21_B_2.1}
    \delta(E) \sim \frac{B}{M}. 
\end{equation}                 
This holds when this B field induced splitting is larger than the difference $[m(a)^2- \omega(plasma)^2] /\omega$ (with $\omega$ the photons energy) - between the diagonal elements of the above $2\times 2$ evolution matrix. This may not be the case in many laboratory applications but the modifications involved do not vitiate the following general arguments.\par
\vspace{0.2cm}
\textbf{Measurements utilizing changes of vacuum birefringence}\par
\vspace{0.2cm}
That a magnetic field $\overrightarrow{B}$ rotates the polarization of light propagating in matter in the direction of $\overrightarrow{B}$ is the famous Faraday effect. Virtual $e^+ e^-$ pairs cause analog "Vacuum Birefringence" \footnote{A photon propagating in vacuum can encounter a transient virtual $e^+ e^-$ pair which forms a closed loop in a pependicular plane. In the absence of the external $B$ field there is no net clock or anti-clock wise circulating current in this loop. An external field $B$ induces an A.B. (Aharonov Bohm) phase and some net circulation thereby rotating slightly the polarization of the incident photon. This is the vacuum birefringence effect computeable via the light by light scattering diagram.}. 
The key point is that in the presence of $\gamma_{\parallel} \rightarrow a$ mixing, the vacuum birefringence is changed by the depletion of the polarization parallel to the external B field. Experiments attempt to detect this via an induced elliptical polarization oscillating with frequency $\omega$ of the external magnetic field. With the limited laboratory $B$ fields and distances of propagation therein a rather small effect was predicted for $M \sim 10^{10} GeV$. An early PVLAS experiment found a much larger effect, eventually attributed to instrumental effects. A variant utilizing faster $B$ field rotations to lower the noise was suggested in \cite{Creswick:2020amx}. Skipping a related "Vacuum Dichroism" based class of experiments we move to. \par
\vspace{0.2cm}
\textbf{A Stern Gerlach analoge for astrophysical axion fields}\par
\vspace{0.2cm}
While the practical applications of the following idea is limited by poorly known Magnetar parameters, it is truly beautiful. The expression in Eq.\ref{21_B_2.1} above for the energy splitting of  $\ket{\gamma_\parallel} \pm \ket{a}$ in a strong $B$ field is analogous to that of the two spin states of a spin $1/2$ particle with magnetic moments $\mu =\pm 1/M$ in the direction of the $B$ field. The idea then is to use the strong $B$ fields with strong gradients around magnetars, in order to spatially separate the two beams \cite{Chelouche:2008ax}
\footnote{Magnetars are pulsars with huge magnetic fields of order $10^{15} Gauss$ extending over $R \sim 10 Km$ radius.}.
It was estimated that the splitting angle and the time delay between the resulting split radio pulses can be resolved even for $1/M \sim 10^{-13}$, far smaller than the bound established by SOLAX and many other astronomical measurements. However the region near the stellar surface may be dominated by plasma rather than magnetic fields and the above scenario will become relevant only at say $R \geq 100 Km$ with much reduced ($B \sim 1/R^3$) fields. If Magnetars source FRB'st (Fast Radio Bursts), then optimizing parameters to achieve this may impact their ability to aid this axion search. The fast (O millisecond) radio bursts have a carrier frequency in the expected range of D.M axions which led to efforts in the H.E. community to relate FRB to axion physics. 

\subsubsection*{\hspace{\secspace}Axion induced $\frac{1}{r}$ potentials in dense nuclear environment} 
As in some of the previous example the following non-trivial encounter of the ace (hypothetical!) particle -the axion- and the ace neutron star, can lead to most interesting results. \par
The exchange of some massless particles such as the photon or graviton yields long range $1/r$ potentials. This is not so for massless Nambu-Goldstone bosons $\phi(i)$ whose Lagrangians include only derivative terms and, in particular, no mass $m^2\phi(i)^2$ terms. Thus both the QCD pion and the axion pseudo-scalars generate nucleon-nucleon spin dependent potentials between non-relativistic nucleons which for distances $r<1/m$ with m the mass of the pseudoscalar are:
$$
V=\sigma (1). \overrightarrow \nabla (1/r) \sigma (2). \overleftarrow \nabla=r^{-3}[\sigma(1).\sigma(2)- 3 \sigma(1).\textbf{n}\sigma(2).\textbf{n}] 
$$
with $\sigma(i)$ the spins of the two nucleons and $\overrightarrow r = r. \overrightarrow n$ is their relative separation. Detection of such interactions between systems of aligned nuclear spins have been discussed by Wilczek and Moody.
V dramatically changes when we consider nucleons in the very dense interior of neutron stars \cite{21B3_hook}. Very light axions of mass $m(a) < 10^{-11} eV$ corresponding to almost planckian $F_{(P,Q})\sim 10^{18} GeV$ can generate therein $1/r$ inter-nucleon potentials which can compete with gravity over distances of $\sim$ 20 Km the size of neutron stars and/or terminal separation in merger events of two N.S. While we do not reproduce the detailed arguments of the paper the following is a ( very)  heuristic short-cut. The new element is the conversion of the $\bar{\psi}\lambda^5\psi$ pseudoscalar axion quark vertex into $ \bar{\psi} \psi$ so that the long range force generated becomes an attractive $V\sim 1/r$ potential. (A similar conversion of the potential due to the exchange of a singlet Majoron  was suggested in \cite{Chang:1985mu}). In dense nuclear matter it is energetically favorable to flip the tilted "Mexican Hat" from $\theta=0$, to the maximal parity violating opposite tilt with minimum at $\theta=\pi$. The axion then couples to the "sigma term"
$$
\sigma=\bra{N}\sum_j {m(j)} \bar\psi(j)\psi(j)\ket{N} \sim 50 MeV\sim m(N)/20  
$$
where we sum over the light $u,d$ quark. The ratio of the nucleon couplings to this “scalar" axion and to the graviton then turns out to be roughly $$r\sim[\Lambda(PQ) / M(\text{Planck})] ^2 \sim [\frac{1}{20}]^2$$ and a more careful evaluation yields $r\sim 0.1$\footnote{$^+$The authors of ref \cite{21B3_hook} applied this to binary neutron star merger. Such mergers can be studied via multi-messengers” in many- gravity waves, optical and radio channels but are much rarer than ordinary type 2 Supernovae.  It may be worthwhile to study possible reduction of the maximal mass of neutron stars that are stable against a collapse into a BH,  due to the extra attractive axionic force. Hopefully future more precise E.o,S of dense nuclear matter will allow doing this in a reliable manner. Finally the inverse effect of the nuclear density on the axions mass persists also for smaller densities such as that of earth and effectively does not allow $m(a)$ much smaller than $10^{-13}$ eV}.  

\subsubsection*{\hspace{\secspace}Using resonant cavities  and precise atomic clocks to detect axions ALPs, light dark photons, and/or dilaton clouds}
After the above purely astrophysical cases we return to more” down to earth” existing and suggested Laboratory searches of light axion, dillaton and or dark photon DM. Resonant cavities tuned to wavelengths of $10^{1\pm 1} cm$ present the oldest most mature approach where the proximity of the axion mass and the resonant frequency enhances the $a\rightarrow\gamma$ transitions in a strong external magnetic field. Axion masses in the range $\sim 10^{-6}-10^{-5} eV$ bordering the QCD seesaw line of the KSVZ and DFSZ models for an “Invisible axion” may hopefully be soon probed by such or similar experiments. In ADMX, changing the cavity resonant frequency is acheived by moving an internal conducting rod. In other cases such as in solax it can be done by inserting a gas so that $\omega(Plasma) = m(a)$\footnote{This was suggested by Von Bibber for the Solax experiment. Amusingly the pressure gradients due to the tilting of the 10 meter long LHC magnet used by the Solax collaboration to align it in morning and evening with the direction of the sun, change the density of the $NH(3)$ gas inside the bore. The attendant plasma frequency then changes beyond the very narrow resonance band helping scan over a range of axion masses (ref. \cite{Creswick:2008as})}.
\par
In the next class of measurements one tries to detect  small periodic changes in the value of atomic/ nuclear parameters such as magnetic moments or level splitting and clock levels in  particular,  induced by the oscillating axionic or other field coupled to quarks or electrons. 
To illustrate this we consider large coherent axionic clouds extending over $\sim 1000 Km$ (corresponding to $m(a) \sim10 ^{-10} eV$ and $k=\beta \omega = \beta m = 10^{-13} eV$) affecting the nuclear mass and charge radius and thereby the clock levels of the atom \footnote{Atomic clocks utilize the classical beating of $N$ atoms in a superposition  $\ket{g}+\ket{e}$ of a ground and a long lived, excited level. The beating frequency $= E(e) - E(g)$ and it’s phase $\phi$ with $\delta(\phi)\sim 1/N$ are extremely well defined allowing to mark time with a precision of nanosecond per year! Nuclear isomer levels with transitions slowed by high angular momentum barriers  and $\sim10^5$ times higher frequency may soon allow corresponding improved precision..}. Using the gluonic coupling of the axion which are less model dependent than the photonic couplings, the pions mass shift computed in ref. \cite{kim2024oscillationsatomicenergylevels} by XPT, Chiral Perturbation Theory is \footnote{XPT is an effective low energy theory where the pseudo NG pseudoscalar pions are the active light DoF with nucleons treated as almost static sources, see e.g. \cite{Gasser:1983yg} The relevant QCD physics is subsumed into $8$ terms in the lagrangian all of which contain at least one derivative as required by the Goldstone theorem. It provides a systematic expansion in the momenta- or more precisely in $\frac{p}{ 4\pi f(\pi)}$. It is a Chiral effective lagrangian of the type first introduced by S. Weinberg \cite{Weinberg:1978kz} to account for all low energy soft pion / current algebra theorems. With the coefficients of the above operators fixed by fitting 8 observables in the pion -nucleon low energy regime the system is “trained” to treat other low energy issues such as the axionic impact here.}:

\begin{equation}
    \label{21_B_4.1}
    \frac{\delta(m(\pi))}{m(\pi)} \sim   \theta  ^2 \frac{m(u)m(d)}{m(u)+m(d)} 
\end{equation}
Nuclear physics is then used to estimate the resulting changes in the mass of the $(A,Z)$ nucleus and in its charge radius. The latter are then translated into changes of the clock levels by careful atomic physics calculations as in \cite{banerjee2023oscillating}.

\par
As a last example of using precision AMO/CM/Nuclear physics to search for "Dilaton like", potentially D.M. fields I briefly mention the following new work "Probing (Ultra-) Light Dark Matter Using Synchrotron Based Mössbauer Spectroscopy" \cite{Banerjee:2024bkp}.\par
A measure of the sensitivity of any device is given by its $Q$ (= Quality) value. The Mossbauers effect manifests the unusually high $Q \sim \omega/{\Gamma} \sim 10^{12}$ of the Iron $14.4 KeV$ nuclear level when the $\gamma$/hard X ray photons are emitted from non-recoiling atoms in a lattice. At resonance a matching Fe nucleus in its ground state has the huge:
$$
\sigma(res) \sim (2J+1)  4\pi/{(\omega)^2} 
$$
absorption cross section making  for almost $100\%$ absorption within $10^{3}$ atomic layers.\par
Having the Emitter $E$ and Absorber $A$ in even slightly different milieu can throw the system off resonance due to slight effects of surrounding atoms, magnetic field etc. Even moving $E$ in the direction of $A$ at speeds of $0.03 cm/sec \sim 10^{-12} c$ pushes the $\gamma$ off resonance  by the tiny doppler effect.\par
The new idea is that for ideal exactly matched $E$ and $A$ but at a separation $L$ between them, a BSM scalar dilaton like $\phi$ field manifests in slight changes of the absorption coefficient.
The coupling $\frac{\phi}{f} G G$ shifts nucleon masses by $\delta M/M \sim \frac{\phi}{f}$.In turn this translates into a shift  of the resonant transition by

\begin{equation}
    \label{21_B_4.2}
\delta E(R) =  M^{*}\frac{\phi}{f}.
\end{equation}

 $\phi$ is \textit{not} constant but oscillates with frequency $\sim m$, the $\phi$ field mass, so that $\delta \phi \sim \phi \omega t = \phi m L/c$ is the change of $\phi$. The photon arriving at A  after a time $t = \frac{L}{c}$ will meet an absorber of frequency shifted by 
\begin{equation}
    \label{21_B_4.3}
\delta(E(R)) = \frac{M^*}{f} \delta \phi = \frac{M^*}{f} (\phi. m) \frac{L}{c}
\end{equation}

If we have $N= 10^{20}$ such photons at our disposal (say a flux of $10^{13}$ photons/{s}   for $10^7 s$) then even 
\begin{equation}
    \label{21_B_4.4}
\delta(E(R)) \sim \frac{\Gamma}{(N^{\frac{1}{2}})} = {10^{-10}}{\Gamma} \sim 10^{-19} eV
\end{equation}

will yield a noticeable shift of the resonance. To maintain coherence over such lengths we need small $\phi$ masses $m(\phi) < 10^{-10} eV $. \par
If the $\phi$ field  makes up all of DM with a local halo density of
$$
\rho= m^2 \phi(0)^2\sim  10^9 eV .cm^{-3} \sim10^{-5 } eV^4 
$$
then we find that 
\begin{equation}
    \label{21_B_4.5}
m\phi(0)\sim 3.10^{-3} eV^2.
\end{equation}
substituting this and $L=10^4$cm in Eq(\ref{21_B_4.4}), we find that f values as large as $10^{25} M^*$ may be detectable.
 
The technology of manipulating X rays has greatly advanced lately. We do not have the Fabry-Perot resonators and lasers/masers of optical and longer waves but can use crystal lattices with angstrom spacings as "gratings" to direct and monochromatize the X rays.\par

If the correct $M^*$ in Eq(\ref{21_B_4.3}) relating the change of the resonant level difference to the $\phi$ field is 10-100 MeV, then the  condition $f > 10^{25} M^*$ dramatically improves bounds on light dilaton derived from tests of the equivalence principle. Such $M^*$ values follow if the anomalously small Mossbauer level splitting is due to a cancellation of large Nuclear (QCD) and EM contributions. If however the latter small splitting is due to peculiarities of the nuclear levels then $M^*$ can be much smaller. (Neal Sobotka P.C.)
 \par
We omit discussion of "Millicharged Dark Matter Detection with Ion Traps" \cite{21b4_PRX} and many other related approaches some connecting also to GW's detection as in "Searches for New Particles, Dark Matter, and Gravitational Waves with SRF Cavities" \cite{21b4_SRF}.\par
We note that most recently there have been remarkable experimental breakthroughs in nuclear clocks which may open new vistas for searching for ultra light DM as in "On the sensitivity of nuclear clocks to new physics" \cite{caputo2024sensitivitynuclearclocksnew}

\subsubsection*{\hspace{\secspace}Using CMB observations to measure the electro-magnetic induced axial anomaly when we have axion strings}
The next exotica involving axionic strings and CMB polarization is not the optimal method for discovering putative axion BSM physics yet its beauty warrants the following mention.The axionic cosmic string is a 1-dim topological defect (aligned with, say, the local $z$ direction), where the PQ condensate vanishes. Its phase  $a/\Lambda(P.Q)$ varies between the different radial ($\rho$) directions emanating from the string with $\phi$ -the spatial azimuthal angle. The interaction term  CaE.B induced by the E.M. $U(1)_A$ anomaly yields Equations of Motion (EoM) which prescribe different polarizations of CMB photons passing on the right or the left of the axion string. The relative rotation $\delta(\phi)$  between the two polarizations is rather small but has the unique feature of being proportional to the coefficient C above of the $U(1)_A$ electromagnetic axial anomaly. As noted in \cite{Agrawal2019lkr} the value of C depends on the charges of the extra, heavy particles circulating in the loop which vary between the various models, hence the "Milikan" in the title of \cite{Agrawal2019lkr}.

\subsubsection*{\hspace{\secspace}Superradiance and interactions of very light bosonic DM with a Kerr BH}
The ingredients of the following fascinating scenario accumulated over 50 years. A key element is Roger Penrose’s observation that one can extract some of the rotational energy when a particle falls onto and reflects from a rotating Kerr B.H. The Kerr BH has effectively “two horizons” in the form of two axially symmetric roughly concentric ellipsoids. The switched signature of the metric in the "Ergozone" between these surfaces allows the initial timelike four vector of the infalling particle to switch to space like so that upon reflection, the transverse momentum imparted to it become added energy. A suggestive analog is the blue shift of light reflected from a rotating cylindrical mirror.  
The second ingredient is quantum superradiance: coherence in spontaneous radiation processes \cite{Dicke:1954zz} manifesting when atoms in an excited state jump to a highly populated lower levels. It applies not only to photons but to any boson. If furthermore that boson has finite mass, then any number of such bosons can reside in the various gravitational “atom”  bound states $[nl(z)]$ with a binding of the order of the rest mass. To ensure coherence over all the ergosphere which is also required for the de-excitation of these states the boson must have a very small mass. 
Rotating BHs found in LIGO-VIRGO or elsewhere exclude such light bosons since otherwise \textit{any} rotating BH would lose all its angular momentum in the following steps:
\begin{itemize}
\item[I] some axions fall on the BH or are spontaneously generated and  gravitationally bind to the B.H.
\item[II] once $N$ such axions populate the state, further creation and addition of another axion is enhanced by  the "Bose factor" of $N$ leading to a run-away scenario where.
\item[III]the angular momentum providing a-la-Penrose the energy required for the particle production is stored in  the atomic $[n,l(z)]$ states populated. Due to the cylindrical symmetry only the z component of the angular momentum along the $z$ axis of rotation is a good quantum number. 
\item[IV]the angular momentum is radiated away via a gravitational wave which also is a coherent transition of bosonic gravitons!
\end{itemize}
Detection of these gravity waves at specific frequencies fixed - in analogy with  the Hydrogen atom- by the splitting between various $E(n,l_z)$ states, may be feasible. A century after the splitting of Hydrogen  levels ushered in the Bohr atom and quantum mechanics such splittings may be at play again - this time for the analog gravitational radiation from gravitational atoms!\par
The main conclusion then is that rotating BH’s exclude stable (or long lived) particles of mass lighter than $10^{-11} eV$ which in particular covers most of the range of Feeble DM.\footnote{Asimina Arvanitaki, Savas Dimopoulos,Ken  VanTilburg, Masha Bakhtiar, Asher Berlin and others are connected with the fascinating modern astrophysical version of supper radiance.}

\subsubsection*{\hspace{\secspace}Superradiance of Dark photons and  emission of cosmological dark magnetic vortices from Kerr BH’s}
In the following we consider the above super-radiance scenario for a non-zero mass dark photon when $m(\gamma')=e'\langle \phi \rangle= e' v$ is obtained via the original Goldstone mechanism: the spontaneous breaking of the dark $U'(1)$ symmetry by the condensate a.k.a. the VeV $\langle \phi \rangle=v$ of a light boson field $\phi$ that carries the $U'(1)$ charge. In this field theory analog of ordinary superconductivity, supercritical $B'$ fields penetrate into the superconductor via vortices. The analog vortices are termed here “Cosmic strings” with the “superconductor” being all of empty space. The radius of the magnetic vortices/cosmic strings- the analog of the London penetration length is  $R'=m(\gamma')^{-1}$. The minimal quantized flux is $\Phi'= B'\pi R'^2  =2\pi/ {e'}$ so that: 
\begin{equation}
    \label{21_B_7.1}
    B'=\frac{\Phi'}{\pi R'^2}= (2 /{e'})  m(\gamma')^2= B'(crit)
\end{equation}

and the string tension- the energy per unit length is: 
\begin{equation}
    \label{21_B_7.2}
\sigma'=\frac{B'^2}{8\pi}\times \pi R'^2= \frac{1}{2} \frac{m'^2}{e'^2} = \frac{1}{2} v^2.   
\end{equation}
If the magnetic field build-up via the above superradiance reaches a sufficient level, then Kerr black holes can spawn and emit closed magnetic’ loops. After being emitted from a Galactic BH such loops can be detected by a terrestrial magnetometer or other sensitive device\footnote{This short summary and discussion below of detailed calculation in ref. \cite{East:2022rsi} and \cite{Brzeminski:2024drp} hardly touch the many subtle issues involved. }.
The  thickness $R'= m'^{-1}$ of the vortices can be much larger than R(SW) the Schwartzschild radius $\sim 10 Km$ for say three solar mass black holes. This is reminiscent of soap bubbles emitted from a thin pipe of small radius mimicking R(SW), and then grow outside to a size $R'$ -where the internal pressure fixed here by $B'^2$ is matched by the surface tension - the analog of the VeV $v$ here.\par
Following \cite{Brzeminski:2024drp},  we take the new $U'(1)$ to be the SM anomaly free B-L gauge group and further assume that $R'  \geq R(Earth) \sim 10^9 cm$. The precision tests of the EP equivalence principle on such scales by the Microscope experiment imply that $e'^2 \leq G_{New} m(N)^2 10^{-13} =10^{-51}$.\par  
This coupling strength is (much!) weaker than that of gravity but for now we ignore the weak gravity conjecture. Terrestrial detection of such vortices would be most dramatic and offer a more stringent limit on Dark Photons than that obtained by requiring the absence of rotating BH’s. 
As argued in ref. \cite{Brzeminski:2024drp} this may be possible despite the smallness of $e'^2$.
\par 
When a B’ vortex hits one of Ligo’s mirrors it exerts on it a Lorentz force F:
$$
F= Q'B'\left(\frac{v'}{c} \right) \sin{\theta'} \sim \mathscr{N}e' B'v'/c \sim \mathscr{N}e'm'^2/{e'} \left(\frac{v'}{c} \right) = M/{m(\mathscr{N})} m'^2 v'/c 
$$
with $v'\sim c$ the velocity of the vortex, $\theta'$ a relevant angle, $\mathscr{N}(M)= M/m(Nucleon)$ is the number of nucleons in the mirror and $\mathscr{N}(M).e' \sim Q'$ is the mirror’s total B-L charge. 
The resulting displacement $\delta(L) =1/2 a t^2$ with $a = F/M$ the acceleration of the mirror and $t= R'/{v'} = 1/({m'v'})$ the time during which the force is exerted on the mirror, then is:
$$
\delta(L) \sim 1/{m(Nucleon) v'/c}\sim 1/m(Nucleon) \sim 10^{-14} cm. 
$$
A remarkably simple expression which is independent g everything - the B-L couplings $e'$ and mass $m'$ of the $U(B-L)$ photon \hspace{0.15cm} \textit{and} the mass of the mirror! Also with $1/m'= R'\sim R(\text{earth})\sim 10^9 cm$, the above pulse duration $t \sim 3 .10^{-2} sec$ translates to an effective frequency of $\sim 100 \text{Hertz}$ where Ligos sensitivity has its maximal value of $\delta(L) \sim 10^{-17}cm$.
The estimated $\delta(L)$ above exceeds Ligo’s limit by $10^{3}$. Such pulses, simultaneous within 10 milliseconds or less in Ligo and virgo and in the two Ligo branches respectively are very unlikely to be accidental. 

A necessary condition for a detectable effect follows from energetics. Let us assume that $\sim 1\%$ of the matter in the galaxy is in $\sim 3 M(Sun)$ B.H. s  and that $10\%$ of these are Kerr BHs  with $\sim 20\%$ of the mass of each of the latter in the rotational energy. By assumption all this rotational energy has to be emitted from the Kerr BHs on the galactic lifetime $\sim 5 \hspace{0.05cm} Billion \hspace{0.05cm} Yr$. Most optimistically it will be exclusively via the magnetic’ loops and we need to verify that an earth -magnetic vortex encounter happens at a sufficient rate.  
There is a different “Stueckelberg Mechanism” for generating the  Dark photon mass for which the above scenario does not apply. We will not elaborate it here.

\subsubsection*{\hspace{\secspace}Accumulation of axions (or other particles) in the solar gravitational basin}
 The following ingenious suggestion that Newtonian gravity enhances concentration and D.M. detection prospects is due to Ken Van Tilburg  see ref. \cite{VanTilburg:2020jvl}. It applies to any stable weakly interacting particle of mass of few (KeV) or less that is produced in the solar core irrespective of being part of DM or not. We assume that the particle in question has a radiative decay channel. For concreteness let the particle X be an ALP of mass $m$ with a $(1/M) \phi \bar{E}. \bar{B}$  interaction. The latter generates $\phi\rightarrow \gamma + \gamma$ decays and $\phi$ production by the scattering of a $\gamma$ on a charged nucleus in the solar core. With no energy transfer by the static Coulomb field the required energy of the core $\gamma$ is:
\begin{equation}
    \label{21_B_8.1}
    E[\gamma(c)] = [m^2+ p^2]^{1/2} \sim m + KE(p).
\end{equation} 
Assume that $dN/{dt}$ \hspace{0.1cm} X particles are emitted from the sun per second. $dN/{dt}$ depends on the solar core temperature and its composition which did not change much over the last Billion years. For $M(X) \leq T(core)=T(c) \sim 1.5 KeV$ most of the radiated X particles have relativistic velocities close to $c$ escaping along straight lines radially away from the sun yielding a number density of X particles  
\begin{equation}
    \label{21_B_8.2}
n(X)=  \frac{dN}{(dt \cdot c)} \hspace{0.1cm} \frac{1}{r^2} \mbox{\hspace{0.3cm} \text{with}}\ r= |\vec{r}(\mbox{\text{sun}})- \vec{r}(X)  \hspace{0.2cm}| \cong A(u) = 1.5 \hspace{0.15cm} 10^{13}cm
\end{equation}
A terrestrial observer attempting to detect the X-ray photons from radiative decays of the ALPs outside the sun will then look towards the sun (Blocking its direct radiation). The volume of X particles whose decay will contribute to this signal is a sphere of radius Au $\sim 1.5 \hspace{0.15cm} 10^{13}cm=$ Earth-sun distance. 
\begin{figure}[h]
\begin{center}
 \includegraphics[width=0.70 \textwidth]{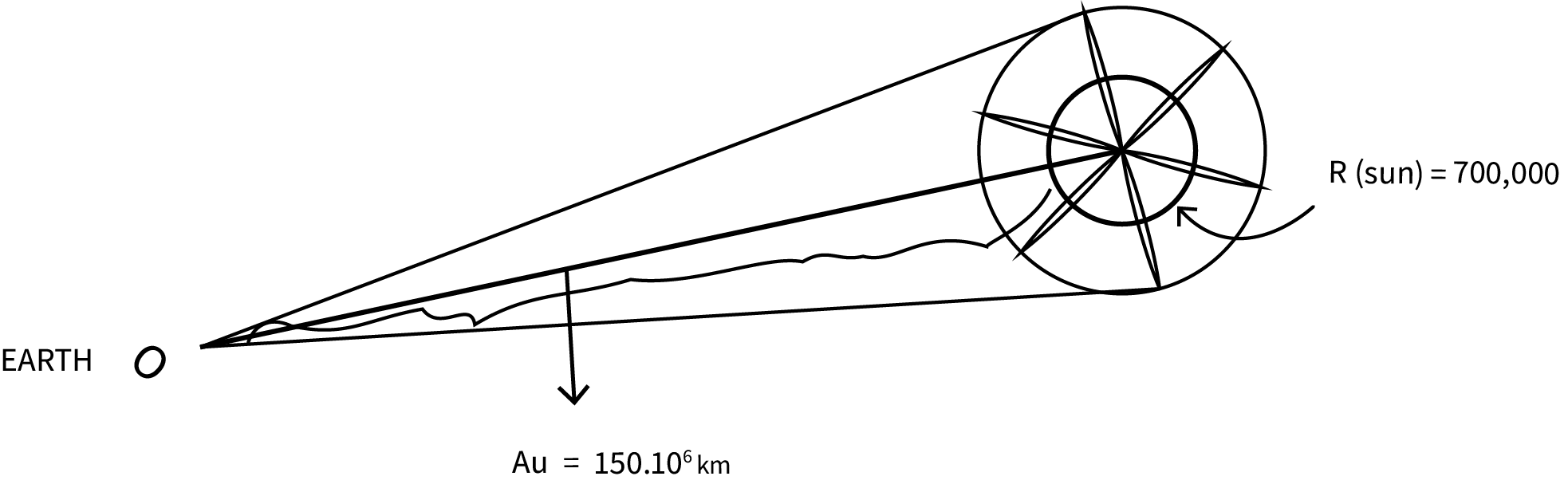}
    \caption{The geometrically narrow photon beam from decays of ALPs moving in tightly bound trajectories near the sun}
\label{fig:012}
\hfill
\end{center}
\end{figure}
The total rate of decays therein will then be:  
\begin{equation}
    \label{21_B_8.3}
\sim dN(\mbox{\text{decaying X}})/{d(t)} \sim 
\Big( dN(X)/d(t) \Big) \cdot (1000~\mbox{sec}/\tau)
\end{equation}

where 1000 Sec is the time for crossing the sphere and $\tau^{-1}$ is the {partial} radiative decay rate of X. The corresponding expected flux is approximately: 
\begin{equation}
    \label{21_B_8.4}
 \Phi(\mbox{X ray}) \mbox{ $ \text{from solar emitted particles}$} = dN(X)/{d(t)} .1000 ~\mbox{sec}/({\tau} \cdot {Au^2}).
\end{equation}
The surprising observation is that trapping in the gravitational solar basin of the small fraction of particles emitted with velocities smaller than the escape velocity $\sim 2-3 .10^{-3} c$  dramatically helps their detection. We next indicate how this comes about:
Eq.(\ref{21_B_8.1}) along with the Planck distribution of photons in the solar core of temperature $ T(c)$ maximizes the number of ALPs captured in the solar basin if  $m \sim T(c)$ which we will assume first. The photons generating the slow ALPs with $v\sim (1-3) 10^{-3}c$ \hspace{0.15cm} which will be captured gravitationally then originate from a region around the peak of the Planckian distribution at $E(\gamma) = T(c)$. For the non-relativistic ALPs emitted  the width of this region is  $\Delta( E) \sim m v^2/2 \sim 10^{-6} m$. The probability $P$ of binding the emitted ALP in the basin is 
\begin{equation}
    \label{21_B_8.5}
    P [for\ m\sim T(c) ]= \Delta(E)/{ T(c)} \sim \Delta(E)/m \sim (v/c)^2 \sim 2 . 10^{-6}. 
\end{equation}
In the more general case of $m < T(c)$ we need to use lower energy photons on the Raleigh- Jeans side of the distribution and the corresponding smaller  value of P is:
\begin{equation}
    \label{21_B_8.6}
  P(m< T(c) ) \sim 10^{-6} .[m/{T(c)}]^3.  
\end{equation}
Most of the fall-off of the solar gravitational potential occurs at distances  $r= 3R/2- 7R/2$ and the probability that the aphelion $\sim 2a$ with a the major axis of the ALps orbit will be in this interval is:
\begin{equation}
    \label{21_B_8.7}
P  ( 7R/2  >a > 3R/2  )  \sim 1/3. 10^{-6} [m/{T(c)}]^3. 
\end{equation}  
With $R = 7.10^{10}$cm the radius of the sun. These small orbits, dominated by the sun's gravity, are stable against disruptions. This allows -according to detailed simulation- for the population of such ALPs to keep building up for $\sim 1-2$ Billion years. Since this time is $3. 10^{13}$ times longer than the $10^3$ seconds appearing in eq.\ref{21_B_8.3} above the total relevant population will be for $m \approx T_c$ : $3.10^{13} \cdot (1/3). 10^{-6} \sim 10^7$ times larger. Another helpful factor comes from considerations of Doppler broadening. The energy of the putative decay photons from the non captured, essentially relativistic ALPs peaks at the midpoint $E(\gamma = \frac{E(\text{alp)}}{2}$, yet  the distribution is very broad with $\Delta [E(\gamma)]\sim E(alp)$. However for the very slow $v< 10^-3 c$ ALPs considered here the doppler broadening $\delta(E) \sim 10^{-3} E$ is miniscule and X ray detectors of high energy resolution would therefore enhance the S/B ratio by another factor of $10^3$.
Finally the gammas in question originate from  $r= 3R/2- 7 R/2$ as indicated in fig. \ref{fig:012} The flux coming in from the specific direction implied is therefore enhanced by an angular resolution of $\sim \frac{1}{5}$ degree which allows excluding X-rays of direct solar origin. This will further enhance the signal by $10^3$. The overall improvement by $10^7 \times 10^3\times 10^3= 10^{13}$ applies for $m\sim few\ T(c)$ with $T(c) = 1.5 KeV$ and as indicated by Eq.\ref{21_B_8.7} for smaller m’s it is reduced by $[m/{T(c)}]^3$. A huge swath in the plane of $m$ and coupling (or of the radiative decay lifetime) is excluded by applying this method. As it happens similar bounds can be obtained by considerations of ALP radiative decays in the early universe (see ref. \cite{Langhoff2022bij}).  
\par
Creation of particles somewhat heavier than T(core) using  the higher end of the Planck energy distribution where $f(E) \sim E^3 \exp(-E/T) \sim m^3 \exp{-m/T(c)}$ is possible (even for $m=7 T(c)$ we have $f \sim 0.2$). The Boltzman suppression becomes prohibitive for mass higher than 15 KeV. The number density $n(X)$ of light fermions that the solar basin attraction can lead to is Pauli exclusion limited by $n(X) \sim k^3 \sim(m(X) v(escape))^3$ which is $10^{-9} m^3$ for the case of the sun.
\par
In principle non-solar gravitational attraction basins can be used. The terrestrial basin has the advantage of being closer and also the pendulum like earth bound orbits may be even more stable. Unfortunately the very small escape velocity ($\sim 11 Km/sec$) dramatically decreases the efficiency of this basin.  
It has been noted by Yuval Grossman, Itay Bloch. Margarita Gavrilova, Mitrajyoti Ghosh and Jeffrey Vincent Backus that for inverted hierarchy glactic neutrinos can be captured in earth’s gravity via the reaction $\bar{\nu} (2) + e^- \rightarrow  \bar{\nu} (1) + e^-$ - thanks to the approximate equality of the (1,2) mass difference and the kinetic energy $T(\nu)$ of the neutrino  $\delta(m(1,2)) = \delta(m^2) (1,2) / {2(m(2))}$  $\sim 3 \text{Temperature} (\nu) \sim 8 .10^{-4} eV$. However the small $v(escape)_{earth}$ and the smallness ($\sim0.1 eV$) of the mass considered do not allow helpful grav basin enhancement\footnote{In view of the difficulty of discovering the CNB via the two tiny peaks in the Tritium $\beta$ decay spectrum any local (galactic, solar system or terrestrial) enhancement could be extremely useful. A careful evaluation by Pascoli-et-al suggested such enhancement but only by $\sim 30\%$. While the discovery of the CNB will be of great importance we note that unlike for the massless CMB photons, the CNB spectrum and angular distribution can be modified by "local" gravitational fields detracting from its value for cosmology (Robert Shrock P.C.).}.

\subsubsection*{\hspace{\secspace}horizontal Deflection of falling atoms by DM kicks}
We have mentioned above sophisticated DM searches utilizing tiny delays or interference of atoms falling in high vacuum, low temperature, 100 meter long vertical pipe (Magis) \cite{MAGIS100:2021}. Here we consider the far simpler idea of using the transverse deflections of such atoms by DM kicks. We can readily show that a nuclear collision with a DM particle of mass M(X) and halo velocity of $v(halo) \sim 3.10^7 cm /{sec}=10^{-3}c$ with an atom of mass $Am(N)\sim 100 GeV$ deflects the latter during $t(f) = 5$sec fall by 
$\frac{10^8 \cdot M(X)}{A GeV} cm$. Thus a measurable deflection of $\sim 0.1 cm$ obtains already for $M(X)$ as small as $=100 eV$. The tiny product of the nuclear radius and momentum transfer $\sim M(X)v(Halo) \sim 10^{-11} GeV$ for the above mass makes the assumed, spin independent scattering, isotropic and coherent: $\sigma(XA) \sim A^2 \sigma(XN)$. Can this scattering yield significantly more deflected atoms than the expected background? 
The probability $p$ that while falling during time t(f) a nucleus is impacted by a DM particle X then is: 
$$
p=\Phi(X)\sigma(XN)A^2  t(f) \sim n(X)v(X)10^4  \sigma. t(f) = t(f) .10^{14} \sigma(XN) 
$$
There are three different possible backgrounds:
\begin{itemize}
\item[a] Neutrinos. The Solar neutrino flux at the typical MeV energies replacing the $10^{14}$ in Eq.\ref{21_B_8.3} is $\sim 10^{10}$ in CGS units and the $\nu-$ nucleon cross-section $\sigma(\nu-N)$ at $E(\nu) \sim $, $G_F^2 E(\nu)^2/{\pi} \hspace{0.2 cm} \text{is} \hspace{0.1 cm}  \sim  10^{-44}cm^2$ times smaller. Also for a CNB neutrino $E_{\nu} \sim  10^{-4} \hspace{0.1 cm} e.V$ and the cross section is $10^{-20}$ times smaller.
\item[b] Photons. At temperature of $T \sim 4 Kelvin$ the kinetic energy an average photon imparts upon an isotropic collision to a $100 Gev$ atom  is $\sim10^{-19} eV$. To achieve the minimal velocity of $v= 0.02 cm/{sec}$ or energy of $10 ^{-13} eV$ required for an observable deflection the falling atom should experience $10^6$ collisions, namely
$$ \Phi(t) \hspace{0.1cm} t(f) \sigma \Big( \gamma (A,Z) \Big) \sim 10^6 \text{where} \hspace{0.2cm} \Phi = n(\gamma).c \sim 10^{13} cm^{-2}$$ 

with a density of $\sim 400 photons/{cm^3}$ and flux of $\Phi=10^{13} /({cm^2 .sec}).$ 
The CMB photon energy is far lower than any excitation energy $E^*$. Thus scattering occurs in second order via excited intermediate states leading  via an effective Euler Heisenberg lagrangian
$\sim \frac{1}{(M^3)}  F^2 \hspace{0.1cm} \overline{(A,Z)} \hspace{0.1cm} (A,Z)$
 to an elastic cross section of $\sim 10^{-38} cm^2$. With $\gamma$ flux $\sim 10^3$ times higher than that of the MeV DM and the need to have $10^6$ collisions this becomes  a relevant background only if $\sigma(XN)<  10^{-41} cm^2$.
\item[c]The most serious background is generated by scattering with the gas  atoms in the “vacuum” pipe. With atomic cross-sections $\sim 10^{-15} cm^2$ even a very high vacuum of $10^3 atoms/ {cm^3}$ yields a probability $p = 10^{-8}$ for colliding while falling $10^4 cm$. However only in a tiny fraction $[mm/{(100 meter)}]^2 \sim 10^{-10}$ of the deflections the isotropic scattering leads to the tiny scattering angles mimicking the collisions with the light DM looked for. This leaves some room for restricting $\sigma(XN)$ below the rather weak bounds existing for such light DM but only if we manage to track many ($\sim 10^{12}$) falling atoms. 
\end{itemize}

\section{\hspace{\secspace}Indirect detection via DM annihillation into S.M. particles}
\textbf{In which we mention the attempts to detect DM via $\bar{X} -X$ annihilation in the galactic core or other locations producing GeV  gammas, antiprotons and antideuterons.}\par
\vspace{3mm}
Indirect detection via annihilation into anti-protons, positrons and/or photons of present day DM  is usually studied in the context of symmetric DM  which was in thermal equilibrium. The rate of annihilation which ensures the correct “ Freeze out” relic density is:
\begin{equation}
    \label{22.1}
    Rate = v.\sigma(XX\rightarrow SM+ SM ) \sim 3.10^{-27} cm^3 {sec}^{-1}
\end{equation}
Recalling that for the exothermic annihilation reaction the product $v\sigma $ is constant and using $v= 3. 10^ 7 cm [sec^{-1}]$ we find that $\sigma(XX\rightarrow SM+ SM) (now) \sim 10^{-33} cm^2$. To optimize detection, the DM annihilation signal should maximally differ from ordinary astrophysics backgrounds - be it in the location of the annihilation and/or in the type and spectra of the annihilation products. This motivated in particular the Fermi-LAT searches for enhanced $\sim GeV \gamma$’s fluxes from dwarf spheroids/ Milky way satellite galaxies and other regions which are DM rich and baryon poor\footnote{Other appealing search directions (Peter W. Graham, Surjeet Rajendran, Ken Van Tilburg, Timothy D. Wiser Phys.Rev.D 91 (2015) 103524 are towards galaxies that "recently" collided such as in the bullet and other clusters. Specifically we should target the outer regions where gravitational lensing indicates the presence of DM but there are no X rays from gas which has largely segregated in the central region between the galaxies. For $\gamma$ rays it is hard to attain the very  high angular resolution required at these distances to subtract the signals from stars which also are collisionless and populate part of the regions of interest.}.
Evidence for an excess of $\gamma$s from the G.C. (Galactic Center) was discussed in ref. \cite{Daylan_2016}. 
A contrasting point of view was presented in  \cite{Blum:2017qnn} \par
The rate of annihilation is $\sim n(X)^2$, and spiked NFW like DM profiles help generate this excess. Also it is conceivable that after a sufficiently long time SIDM can undergo a Gravo-thermal collapse leading to such enhanced central concentration.\par
An excess of photons and of slow anti-particles relative to what is expected from Cosmic Rays $(CR)$ interacting with the ISM was suggested some time ago by the Pamella and Attic collaborations and presumably also by the satellite-carried AMS magnet spectrometer. The argument that such excesses need not be of DM annihilation origin but rather can be generated by young very active pulsars has been  contested by Dan Hooper and others as such pulsars also emit TeV $\gamma$ and no G.C,excess of these was found. \par
MultiI TeV, monochromatic $\gamma 's$, emerging from annihilation of heavy DM particles would be the most dramatic signatures of such DM. In particular it has been argued that if the DM is made of WIMPs in high $SU(2)_W$ representations both present and early universe annihilations can be enhanced to conform to the relic density and to lead to a signal observable in future telescope arrays \footnote{Recent works aiming to go above the G.K. bound, suggested that in DM models with multi Tev DM particles, annihilating via "Wimponium" states (Eric Braaten, Evan Johnson and Hong Zhang JHEP 05 (2018) 062), bound by $Z^0$ exchange, the mono-chromatic UHE $\gamma$ line is partially revived. For these higher energies, large area Cherenkov counters such as HAWC and LASSO can be used along with telescope arrays.   
While such "Wimponium" bound states dramatically enhance the rate of DM annihilation \textit{now} they are unlikely to strongly affect the early universe annihilation rate and the resulting freezeout density and ensuing GK bound on $M(X)$ since the Wimponium binding $\alpha(\text{Weak})^2 \frac{M(X)}{4}$ is less than the freezeout temperature $T(f.o) \sim \frac{M(X)}{20}$. and they readily break. The $XX \rightarrow \bar{b} b$ model may be related to the enhanced $B \rightarrow D^{(*)} + \tau + \nu$ decays found in TBabar. Belle \& LHC B.- the only surviving anomaly therein.}. \par
The $\gamma$ and antiproton/positron are very different messengers: inside the ISM the $\gamma$’s travel to us largely unscathed from, and point to their source, whereas the antiprotons (and positrons) follow magnetic field lines. Unlike the single passage of the photons, the population of the antiprotons builds up for several million years before they diffuse out of the galaxy or annihilate. \par
It was suggested that \textit{slow} anti-deuterons may be the best indicators of annihilating DM. We consider this in the following using simple arguments, in the framework of a  particular model of DM of mass  $m(X) \sim 50 GeV$ annihilating mainly via $XX \rightarrow \hspace{0.1cm}b\bar{b}$, which was extensively studied. 
The $\gamma 's$ from decays of $\pi^0$  measured in (and calculated for) decays of $Z^0 \rightarrow \bar{b} + b $ with a  mass $m(Z) = 90 GeV \sim 2 m(X)$ at LEP, have an energy spectrum similar to that of the GC $\gamma $ "excess".\par 
During 3 Myr of galactic residence the antiprotons travel $\sim 10^{25} cm$ which for the average $\sigma( ann) \sim 10 mb$ and an average ISM density $n(ISM) \sim 1 cm^{-3}$ amounts to an optical depth of only 0.1 (the total cross-section is larger but in most collisions the anti-proton survives albeit with somewhat reduced energy). In contrast the slightest encounter breaks the fragile anti-deuterons with only 2.2 MeV binding energy. This implies that such $\bar{D}$’s - \textit {if} they reach the detecting satellite, have their original kinetic energies. To obtain distinctive slow $\bar{D}$’s we need to produce them with low kinetic energy which will not change until detection and also will slow down their diffusion out of the galaxy and prolong the residence time during which their population can build up. In the following we argue that this is indeed the case when $X^0X^0 \rightarrow b\bar b $ with $m(X^0) =.50GeV$ \par
The direct detection bounds on X-Nucleon elastic scattering from recent direct searches are particularly severe at $m(X) \sim 50 GeV$: $\sigma(XN)\leq 10^{-48} cm^2$. How is this consistent with the cross-section $\sigma [(XX) \rightarrow \Bar{b}b] \sim 3\cdot10^{-34} cm^2$ required here? We assume that the vector or scalar mediator in the $XX\rightarrow \bar{b} b$  annihilation has a mass $M$ and couplings of g’ and g(b) to a (fermionic) X particle and b quark respectively. The annihilation cross-section corresponding to the Feynman diagram with M exchange in the s channel is:
\begin{equation}
    \label{22.2}
    \sigma( ann) \sim g(b) ^2 g'^2 / {M^2}  =3.10^{-34} cm^2 
\end{equation}
The X- Nucleon elastic scattering cross-section $\sigma(el)$ due to the t channel exchange of the same mediator is for $M \sim M(X) \sim 50 GeV$:
\begin{equation}
    \label{22.3}
\frac{g(N)^2 g'^2}{M^4} \hspace{0.15cm}. [ M(N)\beta] ^2 \sim \sigma(ann) [g(N)/g(b)]^2.10^{-10} 
\end{equation}
where g(N) is the coupling of the mediator to the nucleon and the $10^{-10}$ is the familiar momentum transfer square factor appearing in weak scattering cross-section$\sim Q^2 .G(Fermi)^2$ where $Q \sim m(N) \beta \sim10^{-3} GeV$\par
The remaining $\sim3 10^{-5}$ gap required to make $\sigma(el)\leq 10^{-48} cm^2$ is readily explained by the suppressed coupling of the mediator to the nucleon via the tiny admixture of the b quark pairs in the nucleon\footnote{The t channel propagator in Eq.\ref{22.3} for the elastic scattering is $\sim 1/{M^2}$. It can be much bigger $\sim1/[(2m(X))^2-M ^2]$ in the annihilation(s) channel. A mild tuning of $M$ making $M- 2M(X)< 0.1M$ enhances the ratio of $\sigma(ann)/\sigma(el)$ at the present time by another factor of a 100.}.\par
A putative discovery of antideuterons can suggest a DM annihilation source only if its rate of production via the DM annihilation in the specific kinematic region of slow $\bar{D}$’s much exceeds the competing SM production by collisions of HE Cosmic Rays (CR) and ISM in our galaxy. The productions in pp collisions of photons, antiprotons \& antideuterons of various energies were studied over the last 60 years at many accelerators in fixed target experiments. Folding in the estimated gas densities, cosmic ray fluxes at various locations in our disc, the propagation in the galactic fields to earth and even in the magnet and detectors in the planned satelite, the estimated count rates were calculated along with the anti deuteron yield in the DM annihilation scenario.\par 
"The discoveries of yesterday are the backgrounds of today". Physicists in the DM field often view the Pythia code for multiple particle production in high energy collisions and the one for coalescing $\bar{p} \bar{n}$ into an antideuteron as "Black Boxes" containing the dirty hadronic/ nuclear physics details. Indeed, just like the code propagating cosmic rays in our galaxy- such codes are needed to break into where the hoped for true gold of the new BSM DM physics is hidden. In the spirit of this review we try to present more physical \& intuitive understanding of some results of such codes and Pythia in particular. 

In a precursor to the Lund model of multiple particle production and ensuing Pythia code, CNN \cite{casher} suggested that the confining chromo-electric flux tube that stretches between the separating  quark and antiquark generated in $e^+ e^-$ collision is the natural arena for producing additional $\bar q q$ pairs which keep breaking the initial tube. A proper extension of Schwinger’s vacuum tunneling mechanism fixes, with no fitted parameters, the transverse momentum distribution, the ratio of strange to non strange quark production and predicts a uniform distribution in rapidity of the pairs and eventually of the mesons. Most importantly unlike in the abelian case it allows generating a quark rather than an antiquark next to an existing quark forming a diquark which is a $\bar{3}$ of QCD color leading next  to a $qqq$ baryon and an antibaryon on the other side. The effective chromo electric field in the first stage is only half that for the usual $\bar{q}$ production suppressing baryonic production by roughly a factor of $\sim10$.\par
An alternative approach to multiparticle production focuses on the "evolution" (in rapidity) of the leading initial quark and antiquark by brehmstralling gluons which in turn further fragmented into softer gluons and $\bar{q} q$ pairs. The fact that bremsstrahlung of gluons is  reduced for the heavy $\bar b\&b$ quarks makes the CNN approach better suited to this case.\par
Over the years, the fragmenting flux tubes along with jets, partonic evolution, fragmentation functions, and more were incorporated into the Pythia code for multiple particle production making it a most powerful and reliable tool which confirmed the above $\sim1/{10}$ suppression of events with baryon \& anti-baryon produced. Here we estimate the fraction of $\bar{b} b$ states at 100 GeV which evolve into states with slow $\bar{p}$ and $\bar{n}$ that in turn can combine to slow anti-deuterons by using existing data and simple arguments.\par
Prior to that we note that the raw total rates of CR -ISM collisions and $X  \overset{\textbf{\fontsize{3pt}{3pt}\selectfont(---)}}{X} $ annihilations happen to be similar. The rate of XX annihilations with an average X number density of 
$\rho(X)/ m(X) \sim0.5 GeV cm^{-3} / 50 GeV = \frac{1}{100}cm^{(-3)}$, and  $\sigma(ann)v$ having its "Wimp miracle value" is:   
\begin{equation}
    \label{22.4}
n(X)^2 \sigma . v \sim 1/{(100)^2} \sigma.v = 10^{-30} 
\hspace{4mm} cm^{-3} \hspace{0.2mm} sec^{-1}
\end{equation}
The  rate of collisions of CR protons of energy $E (lab) > 50 GeV$ (corresponding to $W (cms) \sim 10 GeV$ needed for reasonable $N\bar{N } -N \bar{N}$ pair production) with ISM in a generic $cm^3$ of the galactic disc is:     
\begin{equation}
    \label{22.5}
n(ISM)\times \sigma(pp)\times \Phi{CR}(E\geq 50 GeV)\sim 1\times 4.10^{-26} \times 5.10^{-4} = 2. 10^{-29} \hspace{0.2cm} cm^{-3} sec^{-1}
\end{equation}  
Recalling that the DM extends much beyond the baryonic gas into a spherical halo, the rates of the DM annihilation and SM processes are roughly the same.\par
The key difference is that $X \bar{X} \rightarrow \bar{b} b$ is \textit{much} more likely to produce \textit{slow} antideuterons, than collisions of a CR- proton with a proton from the ISM. This is so despite the CM energy in $X \bar{X}=100 GeV$ being \textit{higher} than that in the colliding pp system. The point is that the $b\rightarrow B$ or $b\rightarrow\Lambda_b$ fragmentation is ”hard” and the stable b hadrons of masses $\sim 5.5 GeV$ take most of the$\sim50 GeV$ forward/backward momenta in the $b/\bar{b}$ jets. This is even more so for the various excited $B^*, B^{**}$ , etc all the way to  $B$ or $\Lambda_b$ excitations of $\Delta\leq 4m(N)$. Neither the hadronic decays of these states to the ground state nor the subsequent weak $b\rightarrow c$ decays can produce the required \textit{two} pairs of $\bar{N}N$ $(N= \hspace{0.1cm} n \hspace{0.1cm} or \hspace{0.1cm} p) $. Also pairing into an antideuteron of a left moving $\bar{p}$ from the b jet with the spatially separated $\bar{n}$ from the opposite moving $\bar{b}$ jet is extremely unlikely. In the main $B\rightarrow D^{(*)}$ or $\Lambda_b\rightarrow \Lambda_c$ decays there is not enough energy to make \textit{two} $\bar{N} N$ pairs of minimal $4 m(N)=3.5 GeV$ mass. To allow this the weak decays should involve a $\bar{b}\rightarrow\bar{u}$ transition -suppressed by the CKM matrix elements ratio $|V(ub)/V(cb)|^2 \sim 10^{-2}$ relative to the main decays. Furthermore an anti-deuteron produced in the $\bar{B}$  or$\bar{\Lambda _b}$ decay will on average carry a fraction of $2m(N)/{m(B)} \sim 0.35$ of the very high ($\sim 40 GeV$) momentum of the B, namely 14 GeV, hardly qualifying as slow!\par

\begin{figure}[h]
\begin{center}
 \includegraphics[width=0.5\textwidth]{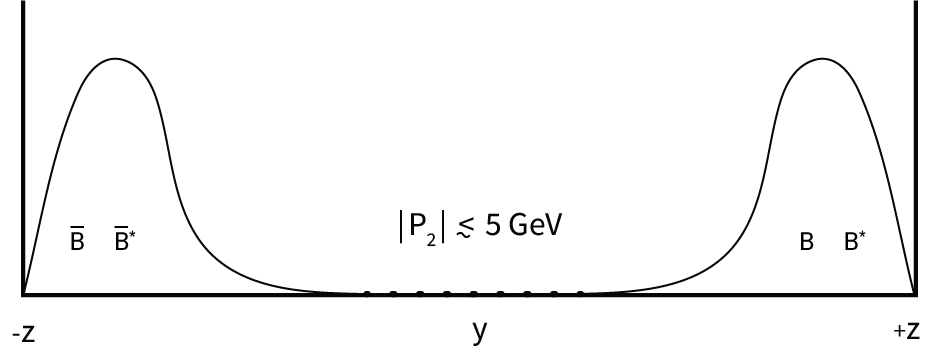}
    \caption{The small central rapidity region in the final states generated by \\ $e^+ e^- \rightarrow b \bar{b}$ remaining after the hard fragments in the b forward and $\bar b$ backward jets are subtracted.}
\label{fig:013}
\end{center}
\end{figure}

\par 
As shown in fig.\ref{fig:013}, what remains then are the few (up to$\sim 10$) GeV of the "unused" part of the central rapidity section. Both $\bar{N} N$ pairs then come from the central region which readily produces the \textit{slow} anti-nucleons required. A dedicated analysis of the many  $Z^0$ -decays of mass $M(Z)\sim 91 GeV\sim 2m(X)$- at the Lep  $e^+ e^-$, or a future $Z^0$ collider where a $B$ and $\bar {B}$ mesons are produced, searching for additional two $N \bar{N}$ pairs and extra pions could provide a complete answer. Absent this we note that after subtracting the forward and backward stable b hadrons and their corresponding entourage of excited states, the remaining energy is $W \sim6- 12 GeV$. The probability of creating two nucleons and two antinucleons is dramatically falling at the low $W'$ - cms energy values, but the probability of coalescing the produced $\bar{p} \bar{n}$ into a slow antideuteron is much higher.\newline 
$e^+ e^-$ colliders collected much data at $W=s^{1/2} \sim m(\Upsilon (1,2,3,4)-S \sim 9.5- 10.5 GeV$ and measured the inclusive antideuteron production which indeed is very large!. (See results from B. Hamilton PHD thesis published as a Babar collaboration paper in Phys.Rev.  D 89 111102). Further the distribution tends to peak at low x low momenta (anti) particles which are needed here. This then suggests that as claimed, the DM scenario with $XX-> \bar{b} {b}$ can be sensitively searched via  the detection of slow antideuterons.\newline
Most recently it was noted  that the annihilations of symmetric D.M inside early structures attracting baryons may affect the ensuing formation of the earliest stars. Too intensive annihilation may impede such formation; there are however circumstances where the annihilation may help explain the observations of early stars/ galaxies formation at high $z$ see ref. \cite{Qin2023kkk}

\section{\hspace{\secspace}Searching DM/BSM Particles Produced by Accelerators}
Searching new particles in accelerators in fixed target or collider experiments dominated HE physics for more than 70 years. All entries in the Particle Data Group (PDG) apart from the nucleons, Hyperons, $\pi$, K mesons and the electron and muon were found in accelerators and some of these findings, such as of the weak interaction bosons, charmed particles, third family fermions and the Higgs Boson helped usher in the present S.M. of particle physics\par
Most previous and present accelerators have been and are being used in the search for DM and BSM physics. We will largely skip over the most extensive search for SUSY particles and Susy LSP in ATLAS and CMS detectors at the giant accelerator LHC, which along with space missions and the human Genome project is a crowning achievements of humanity.\par
The search for DM candidates at the LHC focused on events with jets carrying unbalanced high transverse momenta indicating a "Hard" collison where new physics is more likely to manifest. These may generate escaping particles of lifetime:
\begin{equation}
    \label{23.1}
    \tau \ge L(\mbox{apparatus})/ {c \gamma}\sim 1-10\ \mbox{nanoseconds}.
\end{equation}
Without knowing the energies and momenta of all the particles produced in the collision including the fairly numerous forward ones with soft transverse momenta, we cannot have a precise estimate of the associated missing mass: \newline $MM^2 = \Delta(E)^2 -\Delta(P)^2$. Pairs of massless neutrinos from prompt kaon decays can conspire to have large MM form backgrounds. For this and other reasons we will not discuss the ingenious ideas and as yet unsuccessful efforts to ``fish” out non neutrino escaping particles from the many ($\gtrsim 10^{16}$) pp collisions at the LHC.\par
Detecting Long Lived Particles (LOLIPS) produced at CMS at the diametrically opposite ATLAS  (or vice versa) is impractical. The $\sim 8km$ distance between the two detectors and the corresponding lifetime of the particles involved relatively long but mismatch of the difference between bunch crossings makes the travel time and the tiny solid angles subtended by ATLAS at the location of CMS (or vice versa) make this impractical \par 
The large and tall hangar above the CMS detector covering $\sim 5 \%$ of the $4\pi$ solid angle, inspired the following simple suggestion in \cite{Chou2016lxi}. It was to look for new upward moving particles coming from CMS, in sync with bunch intersections, which penetrated $\sim 100 meter$ earth/rock and decayed into SM particles and possibly additional D.M. particle between the hangers floor and ceiling with both surfaces covered by scintillating sheets.\footnote{$^+$ METUSALA- the acronym of this project, is the name of the biblical forefather with the longest lifespan. The biological “metusala” effect involves people who lived to an advanced age thanks to strong immune systems and therefore are likely to live longer yet. "Color transparency" in high energy QCD physics is rather similar to this as illustrated by the following Gedanken experiment. Let a beam of pions hit a wall of nuclear matter of density $n(N)\sim(Fermi)^{-3}$. Usually the fraction $f(d)$ of the pions traversing the wall unscatted falls exponentially with the wall’s thickness d: $f(d) \sim \exp{-d/l(m.f.p.)}$. All relevant hadronic distances are $\sim$ Fermi which is  the pion’s size and the mean free path in nuclear matter for pion nucleon collisions  $l(m.f.p)= [ n(N) \sigma(\pi \hspace{0.4mm}N)]^{-1} \sim R(pion) =R$. 
Surprisingly, as the energy of the pions increases and $\gamma=E(\pi)/{m(\pi)} > d/l(m.f.p.)$ this fails and $f(d)$ decreases only as a power of d! The point is that mesons which consist of a quark- a 3 of color and an antiquark  a $\bar{3}$ of color-interact as color dipoles of size g.r. The separation $r$ between the quarks is a dynamical variable and the probability of a particular $r$ is given by a wave- function $\Psi_{\pi}(r)^2$. If R is the average $\bar{q} -q$ separation in the pion we have a probability $\sim (r'/R)^3$ that at the time it entered the wall the pion was small and  $r<r'$. The pion cross section for this particular configuration is $\sigma' \lesssim r'^2 \ll R^2 \sim \sigma$ as long as the $\bar{q} \hspace{1mm} and \hspace{1mm} q$ remain close to each other within $r'$. The mean free path $l'(m.f.p.)$ for these special pionic configuration is larger than the average $l(m.f.p)$ above, by $R^2/{r'^2}$ allowing the small “Metusalah “pion to penetrate the wall. The probability for this being $r'^3/{R^3} \sim [l(m.f.p)/d]^{3/2}$ is then power suppressed only as advertised. It takes a time of $R/c$ for the $\bar{q}$ and $q$ in the pion to travel from $|r| \sim0$ to $|r|=R$. A time dilation factor $\gamma= E(\pi)/{m(\pi)}=[R/{r'}]^2  =d/R$ will ensure that if the pion was small at the starting point, its color dipole moment stays tiny throughout the wall traversal. A facility to Search for Hidden Particles at the CERN SPS described in" "The SHiP physics case"  \cite{Alekhin2015by} with a similar goal, rather than Metusalah, is presently being funded.}.  
The search for LOLIPS and their possible longer lived decay products would much improve if a longer decay path was available. Short of digging a cavern the general approach is to use some sort of a beam dump. An early experiment of this type was carried out at SLAC by Mel Shwartz who shared the Nobel prize for discovering $\nu(\mu)$. Since Schwartz means black in German it was jokingly referred to as "Schwart’s black hole"\footnote{This theme was pursued by the late James Bjorken, a true Giant of HE Physics \cite{Bjorken:2009mm}. The Philip Schuster- Natalia Toro duo who, along with Reuven Essig, contributed much to fixed target/beam dump searches, are presently at Stanford where an ambitious dark photon search, parasitically using the existing SLAC beam to deliver $10 ^{16}$ tagged, time resolved, photons, is being contemplated.}.
The common idea underlying most of the above is to have a decay vacuum pipe aligned with the fixed target beam or - as in the faser concept due to Jonathan Feng using one of the LHC beams at the CMS intersection and the pipe is followed by a mound of earth shieldings. (Like for the CEBAF fixed target used for dark photon searches, the shielding transmits only very weakly interacting long-lived particles which are hopefully new BSM entities). The very forward - relative to the primary beam particles have a very high energy (of $\sim TeV$ in the faser case)and make a uniquely efficient set-up. The huge $\gamma$ factors of up to $\sim 10^4$ for pions with this energy makes their average decay path $\sim 5000$ meter long. The resulting  reduction of neutrinos emerging from pions which decay before being absorbed is largely compensated by the linear increase of the neutrino nuclear cross section in proportion to the same factor of energy. Pulsed primary beams and the intersection of short packets at LHC reduce backgrounds. Finally the very forward motion requires instrumenting a limited area in this direction. \footnote{In many cases the power of a high energy collider beam is used to investigate BSM physics much below it’s maximum reach of $W \sim 2E$ . Thus ISR or FSR (Initial or final) state radiation were used in $e^+- e^-$ colliders to explore light new states recoiling against a H.E monochromatic $\gamma$. Recently the LHCB collaboration. investigated unusual nuclear states by having a gas jet cross one of the LHCb beams getting an effective fixed target set-up of very high luminosity.}
In this context we note that large natural underground caverns or part of unused mines with instrumented floor, ceiling and walls with X ray and $\gamma$ ray detectors can be used in the DM indirect search. This will be much cheaper and provide far larger detector than the various satellites and will have almost zero stellar and or cosmic ray background \footnote{Bhaskar Dutta Doojin Kim \& Hyunyong Kim 2305.16383 [hep-ph] noted that the hadron and EM Calorimeters nearer to the intersection point can serve as “Dumps” for the many particles produced in the main LHC collisions. Such particles can then decay along the $\sim 3$ meter path and be detected in the muon chambers. This in particular improved limits on the dark photon - photon  mixing parameter.}.\par

\section{\hspace{\secspace}Some comments on E.T.s In which we defend SETI type projects}
The following discussion of ET’s is motivated by a possible BSM Quirk model which may facilitate communications over galactic distances and may also provide a light $(m(X) \sim O(10-100)eV)$ DM which is extremely strongly self-interacting.\par
The Billions of years long lifetimes of stars and planets and the short$\sim 200$ year period required for the rise of our technological civilization, strongly suggest that \textit {if} other technological civilizations exist, they are likely to be older and far more advanced than us. It is then much more probable that they will discover us before we discover them and in turn suggests focusing on SETI-rather than  METI-like projects, that is Searching for ExtraTerrestrial Intelligence by listening rather than declaring our existence by messaging them\footnote{The reciprocity between the source and sink of EM radiation i.e. the fact that the probability of a pair-wise communication depends on the product of the Q values and directionality of emitters and listeners suggests that METI and SETI are equally hard. This holds when both parties know the direction to and the frequency used by their partner which for us is not the case here. The ET’s had a much longer time to discover us or more generally our solar system and their emitted radio waves are likely to be much more intense than what we can generate.}.\par
Arguments 1)-5) listed below are often made suggesting that SETI like projects are doomed to fail. This makes SETI less likely to be funded than "active" searches of exso-solar planets and studies using new telescopes to investigate their atmospheres.\footnote{The Breakthrough Starshot initiative tries proving the feasibility of traveling to the nearest star alpha-centauri exo-solar system. The new concept uses a larger more durable solar sail. If this works, it could allow reaching Alpha Centauri in as little as 20 years by accelerating the probes sent to $20\%$ of the speed of light. To find if the planet in the habitable zone of Alpha Centauri at a distance  of 0.1 Au from its dim star can  support intelligent life the probes sent have to radio back to earth enough relevant information which has to be collected in a few minutes of close by passage  posing a significant challenge.}.
\begin{itemize} 
\item[1)] While prokaryotic primitive forms of life may have evolved elsewhere, the requirements for intelligent \textit{technological} society are manyfold. It is then possible that the latter formed only on our "rare earth" which lies in the habitable zone around its relatively quiescent parent star, has reasonable surface gravity and a molten core allowing for magnetic fields shielding dangerous cosmic rays and solar wind. Our large moon stabilizes earth's rotation axis, slows down its rotation and enhances tides, facilitating the transition from aquatic life to land where large radio transmitters and detectors can be built. With $15\%$ more water there would be no land, without the large asteroid hitting earth and causing the “great extinction” $\sim60$ million years ago earth could be ruled by dinosaurs and without Jupiter deflecting dangerous asteroids, early hominid civilizations may have been destroyed. Finally, even if in the short span of human history,  east would have dominated  west we may have not  arrived at our present technological society.   
\item[2)] Projecting our expansionist behavior on the much older and far more advanced galactic civilizations led Fermi to ask: "Where are they?" or "Why the extraterrestrials who by now must have colonized most of our Galaxy did not arrive here?" the single most potent argument against E.T's. 
\item[3)] Even if technological societies which sent strong enough signals that we could detect existed somewhere for some time in our galaxy, the chance that we will detect these signals \textit{now} are small. It is possible that the E.T’s have been sending us messages for a very extended period - say $\sim$ one million years and finally gave up. Also if they are similar to us they may self-destruct in short order via a nuclear holocaust or ecological disasters. The 30 years duration of the SETI project is miniscule on cosmic time scales. Evidence for our own terrestrial technology exists only for $\sim$ 50 years when short radio waves which can penetrate the Ionosphere were emitted in TV broadcasts , strongly restricting the distance to planetary systems which could detect this signal. 
\item[4)] The sophistication of the E.T. s may be so high that we may be unable to decipher the content of their communication. We can even miss the very existence of such attempts in the unlikely case that they bothered to send us signals in the first place  
\item [5)] Such advanced civilizations may not use just radio waves - that we technological "babies" use, but rather some other known interaction, or altogether new methods relying on novel BSM physics unknown to us\footnote{This last widely mentioned possibility, was brought to my attention by Jonathan Devor - then a student of Tsvi Mazeh - who is one of the earliest hunters for extrasolar planetary systems}. In a closely related great allegory due to Carl Sagan members of wild new guinea tribes (representing us) are communicating by drums and fires on mountain-tops completely oblivious to the overflying cell phone wi-fi T.V and other radio communication of advanced humans (representing the numerous E.T.s civilizations). I am indebted to Hagai Netzer for this example.
\end{itemize}. \par
 Properly addressing all these issues requires much more than my amateurish exposure. Still my deep conviction that SETI type projects should have the highest priority among human endeavors, motivated my following attempts to counter the above arguments: 
\begin{itemize}
\item[1] The uniqueness of our existence as an intelligent technological civilization would be  vastly accentuated if we adopt a version of the Anthropic principle with a huge “multiverse” consisting of $O(10^{500})$ universes. Only one or very few may have parameters which allow the formation of galaxies, long lived shining stars, light and heavy chemical elements-all of which are prerequisites for life and intelligent life in particular. It would be a monumental waste if in such a very special universe and possibly also in a special, favorable galaxy, only one planet among the $\sim10^{12}$ planetary systems in the milky way harbors intelligent life. Applying most questionable statistics we argue that what happened once in our galaxy is likely to happen again. Inter-galactic distances are almost $10^3$ times larger than interstellar distances in the galaxy, requiring a million times stronger broadcasting power so we consider only communications from E.T.s in our galaxy. 
\item[2] To address Fermi's question we note that the extraterrestrials may be very different from us. With mainly inward oriented advances they may have improved consciousness and brain activity by genetic modifications, interfacing with artificial intelligence and/or other means and they do not share our expansionist instincts. Darwinian natural selection, operating over our hunter gatherer phase (and in more recent times), has favored aggressive individuals with such tendencies and conditioned many of the rest of us to obey these Atilas, Ghengis Khans, Napoleons and Hitlers. An opposite selection of long living, non self destructing civilizations, operating much more slowly but for vastly longer periods, may have selected societies with spiritual and peaceful tendencies which lack the urge to colonize the galaxy, and also to self-destruct but still are curious to find if other intelligent beings exist.\par 
A different partial explanation was offered in \cite{nussinov2009}. The E.T.'s may use far more advanced  methods than those which we presently have to investigate planetary systems in the galaxy. These include direct imaging of planetary systems with the light of the “parent star” completely blocked. They will not refrain however from using the partial eclipse of the parent star by a transiting planet whenever it is possible. Sighting of the transit removes the $1/{sin(i)}$ (with i the inclination angle) ambiguity and maximizes the radial doppler shift. Furthermore, the information gained from such transit events on the planet’s size and atmosphere is invaluable for assessing the likelihood that it hosts (intelligent) life. The large angle of $60^o$ between our ecliptic and galactic planes and the relatively large aspect ratio of the galaxy reduces the number of stars in our galaxy from which such events in our solar system can be seen as compared with much higher visibility of eclipsing events in the many planetary systems with smaller inclinations. 

\begin{figure}[h]
  \centering
  \subfloat[]{\includegraphics[width=0.5\textwidth]{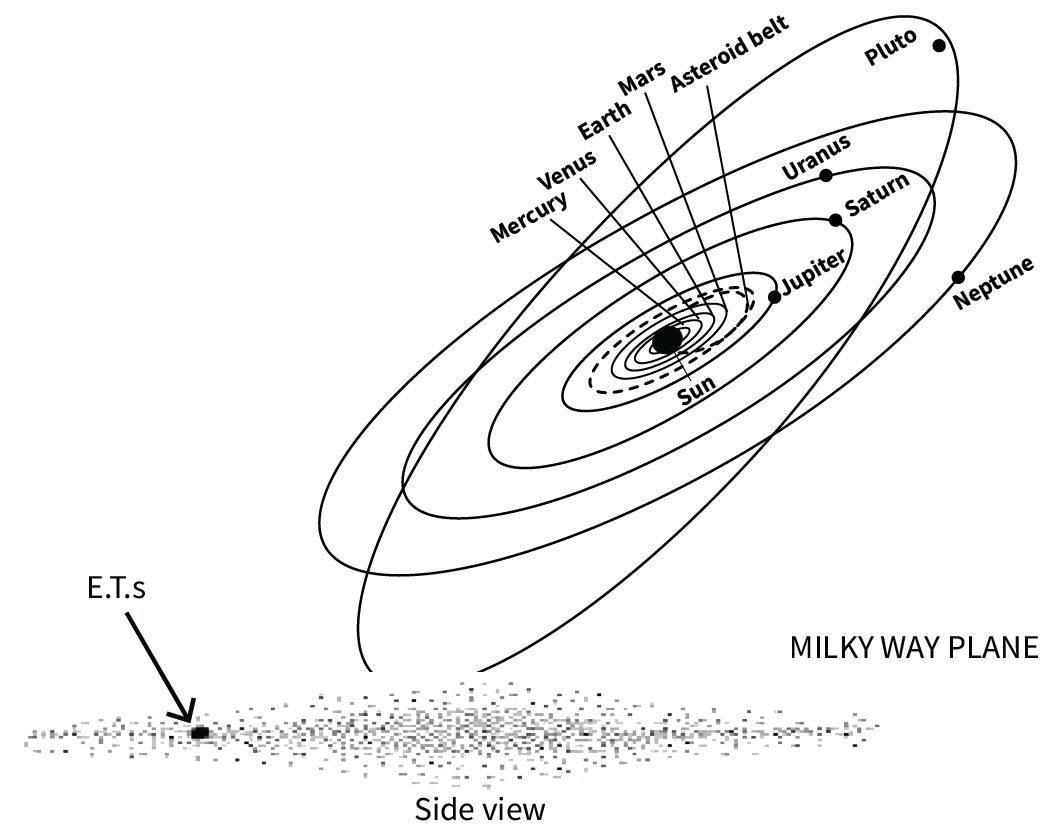}\label{}}
  \hfill
  \subfloat[]{\includegraphics[width=0.4\textwidth]{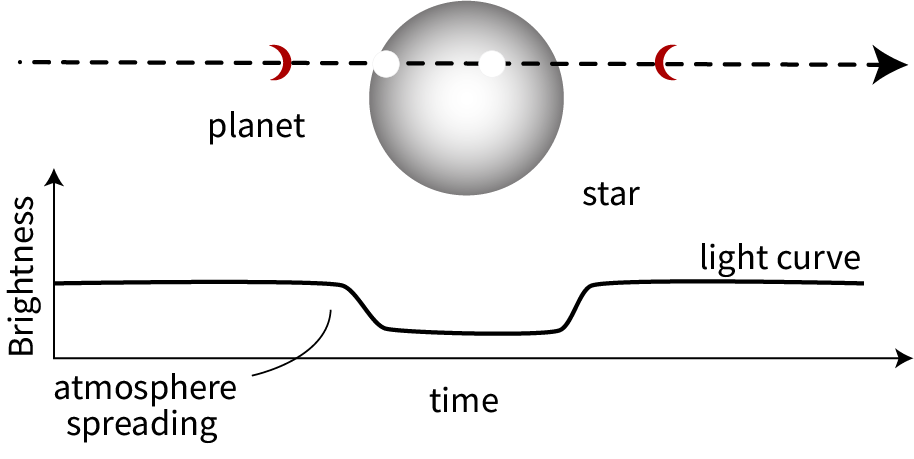}\label{fig:014ab}}
  \caption{The Ecliptic vs. Galactic planes reducing the probability of detecting transit events fig. (14) and  showing the signal of a transit event}
\end{figure}

Fig. 14 illustrates this and shows the typical transit signal. This makes these planetary systems more likely targets than us for further investigations, being broadcast to and/or for being visited by the E.T 's. On the flip side, the enhanced likelihood that ET's in the relevant limited zone enabling the finding of the earth/solar system via the transit method- will discover us and broadcast to us, suggests that our SETI-like searches focus more on signals from these directions. \par
The popular answer to Fermi’s question that the E.T's have been here, left various relics and are encountered in sightings of UFOs (recently addressed by the U.S. congress) will not be elaborated here, nor the more serious Trans-Spermia concept of Hoyl that life originated elsewhere, traveled to, and  “infected” our planet (a feat which most likely is feasible only for very primitive forms of life). 
\item[3] E.T.  societies which can communicate with us, if they exist, are likely to be far more advanced. Most likely they managed not only to discover the solar system and our planet earth and verify that we live in the habitable zone of our sun. By carefully following the evolution of our atmosphere and earth's coloring they could have inferred changes in earth's foliage and concluded that life on earth is likely. If there are very few other such planets in our galaxy  this makes the E.T’s  distance from us larger, say of order $\sim$ few K-Parsecs, and  communications more difficult. It implies however that failing to discover many candidates for intelligent societies, our E.T neighbors will focus on messaging us\footnote{Radio communication over a distance of 3 K-parsecs $\sim 10^{22} cm$ is challenging. For reception dishes at “room” temperature, the minimal energy required for transferring a bit of  information is $T\sim 300^o \sim 0.025 eV\sim 4.10^{-14} erg$.   The required energy for milliradian angular divergence and a $km^2$ reception area then is:  $4.10^{-14} 10^{-5} [10^{22}] ^2 \sim 10^{26} erg$ roughly the total solar energy hitting earth in a second !}.
To answer the question  "Why would they bother to communicate?" we offer the following hypothetical scenario: Observing their “sun” and other stars, they found indications that destruction due to a giant flare-up of their sun is iminent. They want their scientific/other achievements to survive. To this end they broadcast these, including the sequences of the DNA of their better specimens, towards candidates for other intelligent societies that they discovered, hopefully including us. Being aware of  the small probability that their messaging will coincide (modulo travel time) with our listening period, they will  spread out the arrival time of this information. To this end they can send many small explorers towards our solar system with different velocities that will periodically broadcast the same messages when close to earth. This along with laser signaling or the sending of small physical probes is a simpler method of civilization transfer then by actual rocket space trips.
\item [4] While humility becomes us as most likely the E.T.s are far more advanced, we should not overdo it. Some $10^{31}$ communications between neurons occurred in our and in our predecessors' Homo-Sapiens and Primate brains, in millions of years and natural selection honed the pattern of neuronal connections to a high degree of perfection. While only a tiny fraction of these neural communications were involved in creative, scientific or artistic achievements, the latter are remarkable. The E.T.s may not be impressed by the music of Bach or drawings of Escher. Their different atmosphere and color of their sun can make them deaf/ blind to our music/paintings, a remote possibility as their technological advances will allow them to access all regions of the audio/optical spectrum. Most likely they will appreciate the fact that humans proved Godel’s incompleteness theorem-namely that an infinite class of mathematical conjectures cannot be proved or falsified\footnote{"Agadmator", a popular chess commentator presented a century old chess problem of great beauty and difficulty. He then suggested that "when the aliens arrive, we present it to them as proof that we did not waste our time". Clearly different earthlings can have very different views of the intrinsic value of various intellectual achievements.}.
Most likely the messages of the E.T.s will use the universal language of mathematics and we will be able to decipher them.\par
The gap between us and the ET’s may not be just technological or quantitative manifesting in the strength of signals generated and more importantly in rates of information processing. At some stage these may result in their higher levels of cognition, logic and reasoning. Can this prevent our understanding of their message? While this cannot be ruled out, we note that the ETs would know better than to address us “babies” in the manner that they address each other. Indeed should we run into a primitive hominid from million years ago we would not address him (or her) in english or chineese. Still we could communicate by sounding two notes on a drum and later repeat them in quick succession thus stating that 1 +1 =2 which is precisely the way via which the E.T.s could radio this message to us.\par
To broaden the time window for other civilizations to "hear"  their messages the E.T.s may re-broadcast them after given time intervals and hopefully do so without waiting for a response from us. Once we find the direction from which the message is coming and the carrier frequency we can vastly improve our listening abilities allowing us to decipher the long report on all their achievements. This may include most useful information on whether (and how!) controlled fusion is achievable, how genetic improvement of mental powers in synergy with artificial intelligence can be safely practiced, and how we can live peacefully with each other and with our planet. They may also tell us whether Reimann's hypothesis on the zeros of the $\zeta$ function and other long standing math conjectures are correct and how to prove them, of evidence for physics beyond our SM, what is DM and how to discover it and what is the best TOE candidate. Even failing this, conclusive evidence for E.T.s may unite earth's inhabitants and stop all wars.  
       
\item [5] The above philosophical and subjective (counter) arguments were, and will be made by many individuals. However, the suggestion of non-E.M galactic communications can be addressed by known physics showing that Gravity waves or neutrinos cannot enable it. (Recalling Sagan's alegory we can "tell" the tribesmen that we communicate with better more advanced methods but should not suggest that they can communicate better by developing familiar techniques such as blowing horns or shooting arrows). Gravitational waves originating at up to 150 Megaparsec distances were detected at Ligo/ Virgo. However these waves originated from merger events of 15-50 solar masses BH’s. The $\sim 10^4$ times smaller distances in the galaxy and the \textit{linear} fall-off of the GW amplitude with distances d: $h\sim R (SW)/d $, allow $10^ 4$ times lighter, smaller objects to serve as a source.( $R(SW)\sim \frac{G M}{{c^2}} $). Another factor of 1000, say, can be gained by having a sharp (non-chirping) driving oscillation frequency and directed emission. To optimize detection one needs to oscillate these $\sim$ terrestrial mass objects with $\sim$ 50 Hertz frequencies. To achieve the required intensity of the varying Gravitational fields these oscillations have to be at velocities close to the speed of light. The required energy $\sim M(earth)c^2\sim10^{48} ergs$ exceeds the solar output in $10^6$ Yr! With the rate of information transfer $\lesssim$ 50 bits/sec communication via G.W is impractical.\par
In appendix I we argue that the inevitable spreading of neutrino beams prevents using neutrinos for continuous communication on galactic scale. \par
An interesting approach to non-radio galactic communication involves catapulting small objects into space like someone cast on an island floating a message in a bottle that may be picked up eventually. \par 
The possibility that the E.T.s employ a novel communication method using technology based on BSM physics which we do not know of, is the main motivations for the "bit more" part of this mini-review exploring such a possibility. Most of the material in the following 8 pages and attendant appendices was addressed in \cite{Nussinov2014xua}. The present shorter version is much clearer.
\end{itemize}

\section{\hspace{\secspace}THE BSM QUIRK Models}
\textbf{Where the BSM Quirk model of $\sim TeV$ Quarks with an additional confining low scale gauge interactions is introduced.}\par \vspace{4mm}

 While BSM new physics is likely to exist, the possibility that it will be technologically useful seems remote. All particles discovered after the neutron live at most a few microseconds.
\footnote{$^*$ In particular this debunked muon-catalyzed fusion as energy source. Such events occur when slow muons are captured in Deuterium and form molecules by replacing one electron. Since $m(\mu) \sim 200 m(e)$, the Bohr radius of muonic Hydrogen is $200$ times smaller and a similar decrease occurs also in the size of the $D \mu D$ molecule. Bringing the deuterium nuclei that much closer enhances the rate of fusion $D \mu D \rightarrow \alpha + \gamma +\mu$ to the point of  making commercial fusion feasible -had it not been for the relatively short, 2 microseconds, lifetime of the muon.}   
Also new particles of mass $M \geq TeV$ can be produced (if they carry color) only in a tiny $\sim 10^{-11}$ fraction of the pp collisions at the LHC or other future similar hadron colliders.\par
However, in a specific BSM scenario a handful of such massive, stable, particles can be extracted and enable a new communication method! These are Luty’s Quirks \cite{25.Luty'sQuirksi} - new fermions in the fundamental representations of our $SU(3)_c$ color group and another confining $SU(N')'$ gauge group. Interesting phenomena arise when $\Lambda'$, the scale of $SU(N')$ is much smaller than the $O(TeV)$ Quirk masses $M_i(Q')$
\begin{equation}
    \label{25.1}
    \Lambda'=  [10^{-3}-10^{-11}] M(Q')
\end{equation}
As emphasized before there is no direct connection between $\Lambda'$ and Quirk masses  so that such ratios are "$natural'$".\par
The WIMP miracle, where annihilation of $TeV$ WIMP via SM weak interactions left the correct relic density, implies that stronger annihilation rates into two gluons prevent Quirk DM. The remaining Quirks do form $Q'\bar{q} = M'$, fractionally charged strongly interacting mesons (and their $\bar{M'}$ anti-particles). Evading strong bounds on such particles requires a further dramatic reduction of their abundance. As elaborated in \cite{25.Junhai} and in \cite{Jacoby:2007nw} this is achieved by the late color' confinement occuring when the temperature in the new sector $T'$  falls below  $\Lambda'$. Along with the attractive SM color interactions, it ensures a sufficiently high rate of annihilation. In particular it also avoids conflict with the  upper bounds on heavy $Q'+ Hydrogen$ and $Q'+ Oxygen$ isotopes with $Q'$ s produced over $\sim$ billion years by UHE cosmic rays interacting in ocean water.\par
Taking $N' \le 3$ avoids potential difficulties due to the $N'^2-1$ light D.o.F. present at BBN when $T'\sim \Lambda' < MeV$ (we implicitly assumed that the reheat temperature in the $Q'$ sector is similar to or lower than in ours: $T'\lesssim T)$. For $N'=2$, color singlet scalar di-Quirks and the conjugate di-anti-Quirks readily form. However, pairs made of such a baryonic boson and its conjugate keep rearranging into pairs of Quirk -anti-Quark mesons and the  $Q'_i- \bar{Q}'_i$ quickly annihilate making this a viable scenario. An added advantage is having $N'^2-1 =3$ versus $N'^2-1 \cdot =8$ for $\cdot N'=3 \cdot$ gluonic light D.o.F contributing to $\Delta(Neff)$. \par \vspace{0.3cm}
The Quirk model was not suggested to resolve puzzles in the SM or cosmology but rather by the following: If Quirks exist with mass $M(Q')\gtrsim TeV$ and  $\Lambda' \leq MeV$ , then the unique signatures associated with a quirk pair production in LHC could be missed. Present LHC searches are largely focused on "standard" extensions of the SM such as SUSY, a right handed or Mirror/Twin Higgs sector, KK modes, etc. The unexpected Yo-Yo motion of the Quirk -anti-Quirk pair which we will elaborate soon, tends to exclude their discovery in the present set-up of the LHC detectors. Over the last 80 years almost every new particle or high energy phenomenon was anticipated. It is high time for new surprising experimental findings that are \textit{not} suggested by mainstream theory! \par  
As will be shown our suggested application further requires very small  $\Lambda'\leq 200 eV$, corresponding to string’ tension:
\begin{equation}
    \label{25.2}
    \sigma'\sim \Lambda'^2\sim 4.10^4 eV^2 = 20\  eV / {\text{Angstrom}}
\end{equation}
After the $Q'\bar{Q'}$ produced at the LHC separate  by $\sim$ 1 Fermi a $\bar{q}q$ pair of ordinary QCD light q= u,d, or s quarks is generated between them making $M' \text{and} \bar{M'}$, very heavy analogs of the SM charmed D or bottom B mesons. Color and color‘ conservation renders $Q'$ and the lightest such M’ meson completely \textit{stable}. The unique novel feature is that a color‘  string still stretches between the two Meson’s. This color’ flux can be broken only via a $\bar{Q'}Q'$ pair creation. However the rate of $\bar{Q'}Q'$ production by the Schwinger's mechanism \cite{schwinger1951gauge} for $M(Q') > 10^3 \Lambda'$, 
\begin{equation}
    \label{25.3}
        e^- \Bigg\{\frac{M(Q')^2} {\Lambda'^2}\Bigg\}< e^{-10^{6}},
\end{equation}
is negligible. Thus string's are "Forever", disappearing only when the $\bar{Q'}$ and ${Q'}$ at their ends mutually annihilate. The following describes how this happens for  $\Lambda’^2$ values exceeding $(200eV)^2$. In the absence of external magnetic fields or mediums the occurrence of events is an invariant concept and so is the number of close $Q' \bar{Q'}$ encounters preceding their annihilation and we will first present the discussion in the $Q'\bar{Q'} \sim M' \bar{M'}$ center mass Lorentz frame.
\par
The heavy mesons $M'$ and $\bar{M'}$  produced at the beam intersections in LHC will eventually stop and turn around due to the constant pull or string tension $\sigma'\sim\Lambda'^2$, when separated in their center mass Lorentz frame by a distance of:
\begin{equation}
    \label{25.4}
\Delta(L)=\frac{E'}{\sigma'} =\frac{E'}{\Lambda'^2} = \frac{M(Q')\beta^2/2}{\Lambda'^2}
\end{equation}
E' is the maximal kinetic energy of each of the Quirks. For $M(Q') \sim TeV$ and $\beta'\sim$0.14 or 0.5 respectively we find that for
$$
\Lambda' =(10 eV,100 ev,1 KeV,10 KeV,100 KeV )
$$ 
we have
\begin{equation}
    \label{25.5}
 \Delta(L)=(10^{3}, 10;0.1;0.001;10^{-5}cm) \hspace{0.3cm} \text{or} \hspace{0.3cm} \Delta(L)=(10^{4};100;1;0.01,10^{-4} cm) \hspace{0.1cm} \textnormal{\text{respectively}} \footnote{For $Q'$ kinetic energies $E'=M(Q')\beta'^2$ much smaller than $W\sim 13 TeV$, the total energy of the colliding protons at the LHC, the $Q'-\bar{Q'}$ production rate is proportional to the $Q'$ and $\bar{Q'}$ two body phase space factor$\sim \beta'$ -the relative velocity of the Quirks in their center mass Lorentz frame. This ensures that in $\sim$ 1/2 or 1/7 of the cases $\beta' \leq 1/2$ (or 1/7) the $\beta$ values used above.}   
\end{equation} \par
\begin{figure}[h]
\begin{center}
 \includegraphics[width=0.65\textwidth]{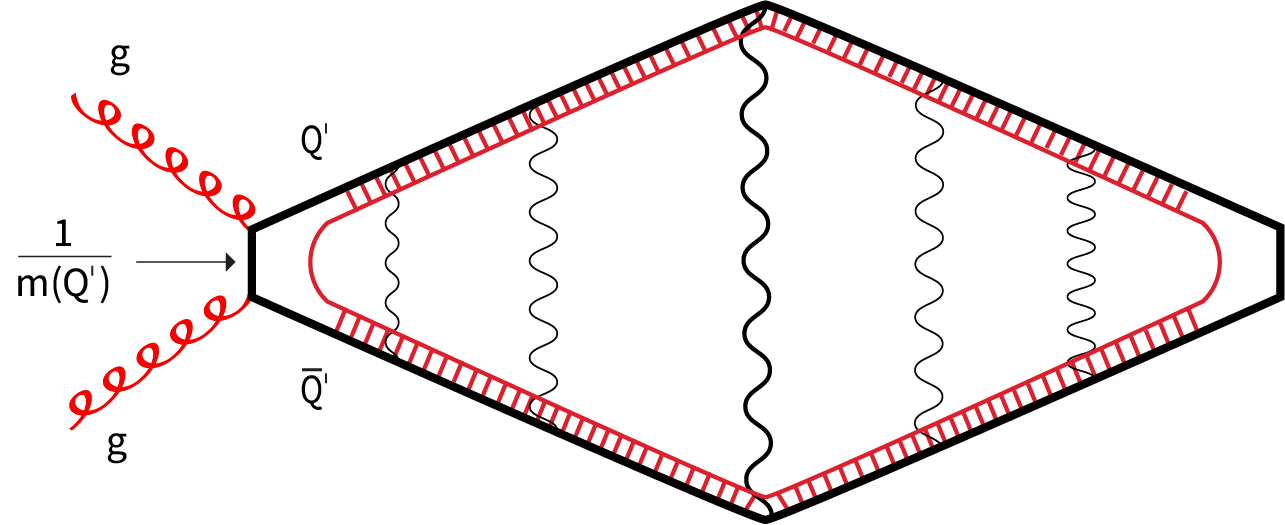}
    \caption{ A schematic configuration space Feynman diagram illustrating the production of the $\bar{Q'}Q$ pairs in gluon $gg$ fusion and yo-yo like motion of the $M'= Q'\bar{q}$ and $\bar{M'}= \bar{Q'}q$ "nesons" all the way to the first close traversal on the right. The $SU'(N')$ weak string is indicated by the long black lines.}
\label{fig:015}
\end{center}
\end{figure}

 \vspace{0.3cm}
The above $\Delta(L)$ is the part of the separation projected on the plane transverse to the beam (z) direction. It is roughly the same in the lab frame. The initial invariant mass of the $Q' \bar{Q'}$ and of the $M' \bar{M'}$ systems is \\ $W=2(M(Q')+E')= \bigl( [x(1) x (2)s]^{\frac{1}{2}} \hspace{0.2cm} \text{with} \hspace{0.2cm}  s^{\frac{1}{2}} = 14 TeV \hspace{0.2cm} \text{and} \hspace{0.2cm} x(1), x(2) \bigr) $ the fractions of the momenta of the colliding proton(1) and proton(2) carried by the two fusing gluons. The different distance/momenta involved are depicted in fig. \ref{fig:015}

For the symmetric case $x(1) =x(2) \sim W/{14 TeV} \sim 1/7$. If, furthermore the $Q'$ and $\bar{Q'}$ are emitted at $90^\circ$ to the collision axis then $\beta = (x(1) -x(2))/2$ becomes their (vanishing) relative velocity. In general $[x(1)-x(2)]/2$ is the longitudinal drift velocity of the $M' \bar{M'}$ system. The relatively low longitudinal (and transverse) boosts make for small (and negligible in most of the following) time dilation and Lorentz contraction effects  when we go from the $M’\bar{M'}$ CMS system to the lab system - the rest frame of the colliding $pp$ beams.\par
The $M'\bar{M'}$ mesons keep passing by each other at time intervals of:
\begin{equation}
    \label{25.6}
\Delta(t)=\frac{2\Delta(L)}{\beta c} = \frac{M(Q')\beta^2}{\Lambda'^2\beta c}= \frac{M(Q')\beta}{\Lambda'^2c}
\end{equation}
for the $\Delta L$ values listed above and $\beta=0.3$ we find that $\Delta (t)$ varies between $10 ^{-9} - 10^{-14}sec$.
The average energy $\Delta(E)$ lost in any such encounter fixes the total time t during which $N(tr) \sim E(initial)/ \Delta(E)$ traversals are completed. During these $N(tr)$ most initial energy is dissipated, the $Q'$ \& $\bar{Q'}$ bind and very quickly, cascade down via a series of gluon and/or gluon’ emission to the Quirkonium 1S ground state and annihilate after time.
\begin{equation}
    \label{25.7}
        t= N(\mbox{traversals}) \cdot \Delta(t) = \frac{E(initial) \cdot \Delta(t)}  {{\Delta(E)}}
\end{equation}
We first assume that there is no magnetic field at and near the beam intersection.  
The 10-100 GeV internal kinetic energy of the initial $\bar{Q'} Q'$ system allows $\pi^0$ or several pion emission in each $M'-\bar{M'}$ collision. This energy tansfer from the heavy $\bar{Q'}Q'$ system to the light quarks and emitted pions happen as follows:
As the distance $r(Q' \bar{Q}')$ during the close approach becomes shorter than the size of the $M'$ mesons $r(M') \sim 0.5 Fermi$ the $\bar{Q'} Q'$ start screening each others color field so that the light $\bar{q}$ and $q$ in the respective mesons become effectively free. This holds so long as $r( Q' \bar{Q'}) \le 2r(M') \sim Fermi$. Accounting for the relative $Q'\bar{Q'}$ velocity of $\beta \sim  0.3$, the resulting time when the screening lasts is close to that typical of the light quark, suggesting that this emission is not  adiabatically suppressed.\par
Viewed in configuration (r) space the Feinmann diagram in fig.\ref{fig:015} for the perturbative production of $Q'$ and $\bar{Q'}$ from fusion of two gluons suggests a tiny initial separation:
\begin{equation}
    \label{25.8}
    \Delta(r) \sim 1/{ M(Q')} \sim 2. 10^{-4} Fermi.
\end{equation}
The initial Quirk momentum of $p=\beta \cdot M(Q')=0.3 M(Q') \sim 300 GeV$ corresponds to an initial  kinetic energy of:
 \begin{equation}
    \label{25.9}
   E(\text{initial}) \sim \frac{p^2}{2m(Q')} = 50 ~Gev.  
\end{equation}
and an initial orbital angular momentum of the $ \bar{Q'} \bar{Q'}$ (or $M' \bar{M'}$) systems of:
\begin{equation}
    \label{25.10}
    L \sim  p . \Delta r \sim 300. 2. 10^{-4} GeV. \text{Fermi} = 0.06 \times 5 = 0.3
\end{equation}
The small fractional value implies that the initial state has mainly L=0 with a small admixture of higher angular momenta. If an average traversal leads to energy loss via pion(s) emission $\Delta(E) \sim 300 MeV$, then the total number of traversals required for dissipating the energy is:
\begin{equation}
    \label{25.11}
                 N= E(initial) /\Delta(E) \sim 200.
\end{equation}
Each emission also entails an average angular momentum change of $l_i = 1$ of the system. The resulting "random walk" then yields:
\begin{equation}
    \label{25.12}
\vec{L } =\sum_{i=1}^{N} \vec { l _i}\sim N^{1/ 2}\sim 14.
\end{equation}
Such an L still allows a closest approach of  $b= L/p\sim 0.02 \text{Fermi}$ so that the above scenario where color screening leads to the release of the light quarks still applies.\par

The total distance traveled from production at the beam intersection vertex to the point where the annihilation happens, varies between $\sim$ 5 meters and 5 microns as $\Lambda'$ varies from 100 eV to 100 KeV. For most of this range the annihilation occurs inside the CMS or ATLAS detectors. The release of $2 M(Q') \sim 2 TeV$ of energy in the form of QCD gluon jets and the many charged pion prongs along the line connecting the beam intersection and $Q'\bar{Q'} $ annihilation point provide a truly striking signature. This then may allow detection despite the small cross-section for $\bar{Q'}Q'$ production. This cross-section of $Q' \bar{Q'}$ production at the LHC is extrapolated using perturbative QCD, from the measured rate of $\bar{t} t$ pair production yielding an inclusive $\sigma [pp\rightarrow Q' \bar{Q'} + X]  = 3 \hspace{0.2cm} \text{pico-barn}$. \par
The dramatic reduction of the expected signals by the effect of magnetic fields near the primary interaction is elaborated in Appendix J. More  features such as the planarity of events even for $B \neq 0$ were noted by Michele Papucci and others. A related work is in ref. \cite{Harnik_2011} \par
For $\Lambda' > MeV$ the annihilation vertex cannot be separated from the production vertex.
For very low $\Lambda'$ which is our main focus and the arena for the new applications, $\Delta(L)$ exceeds the size of the detectors and, as we describe in detail in the next section the $Q' \bar{Q'}$ stop in the rocks outside. The possibility of tracking such Quirks using the Faser setup was recently noted in \cite{25.feng} \par 
Dangerous cosmological manifestations of Quirk models can occur if the reheat temperature  of  the new sector after inflation - $T'_{reheat}$ is similar to that of  SM reheat temperature $T$. The initial gluon’s then are as copious as all other standard model particles and after confinement at $T'\sim \Lambda'$ form various glueballs.
The key relevant points are:
\begin{itemize}
    \item [a] The $0^{++}$ lightest scalar glue' -ball S' of mass $4-7\hspace{0.15cm} \Lambda'$ found in quenched (i.e. quarkless) lattice calculations - which are  completely justified here for the heavy quirks - is stable over cosmological times, and
    \item [b] The comoving number density of the S' glue’-balls after the confinement phase transition is similar to that of its constituent gluons just a bit before the PT namely $n(S') \sim T'^3 \sim \Lambda'^3$. $n(S')$ slightly decreases further in later evolution (see ref. 295). This yields a relic density which may exceed the required  $\Omega(DM) h^2 \sim 0.12$.
    \item[c] The cross-section $\sigma(S'S')\sim \Lambda'^{-2}\sim 10^{12} Barns$ expected for collisions of slow DM particles, exceeds the upper bound $\frac{\sigma (XX)}{M(X)} \leqq \frac{Barn}{GeV} $  by $10^{20}$, resulting in a tiny (m.f.p.)$\sim 10^3 cm$ for S-S elastic collisions. This can make S' an unsuitable D.M. candidate.
\end{itemize}
In \textbf{Appendix J \& K} we discuss possible ways of alleviating these difficulties.

\section{\hspace{\secspace}Quirks and a new communication method over galactic distances}
\textbf{where we note that novel communication systems more efficient than radio waves, exists in extreme variants of Quirk models.}\par \vspace{4mm}
 
Next we address the unique aspect of some Quirk models which motivated this  and the two preceding sections. It is that for a certain range of $M(Q')$ and $\Lambda'$ values, the model may afford a new method of communications over galactic distances. For simplicity, we consider the "minimal" version where the Quirks carry just the new color and our QCD color. Most arguments apply also to charged Quirks. \par
The expressions in Eq.\ref{25.4} of the previous section for maximal distance $\Delta(L)$ by which the $\bar{Q}'Q'$ or corresponding $Q'\bar{q}$ and $\bar{Q'} q$ mesons separate before turning around for a relative velocity $\beta'$, apply when these particles move in vacuum.
After travelling a total distance of  $N(\text{encounters}) \Delta(L)$ of several meters the heavy, fractionally charged Quirky mesons encounter the shieldings and the surrounding rocks before the  $Q'\bar{Q'}$ could annihilate. Energy losses due to strong and EM interactions slow them down and they bind to some of the nuclei encountered.
The $\beta$ decays converting the $d$ or $\bar{d}$ in the mesons to $u$ or $\bar{u}$ respectively, are very slow and unlikely to happen before the mesons join the ambient nuclei where the d quark is stabilized by the binding. This is important as the stable $\bar{Q'}u$ is repelled from nuclei by the Coulomb interactions.
However the many collisions suffered cause $50\%$ of the $\bar{Q'} u$ to convert into $\bar{Q'} d$, which, thanks to the $-1/3$ charge of the $d$ quark will bind to $ A= 20$, $Z=10$ nuclei with bindings $\sim 10 MeV$. The $Q'\bar{u}$ at the other end of the string has 4 times stronger coulombic binding \textit{and} also hadronic bindings - the analog of $\bar{K}^0$ binding to nuclei.\par
The envisioned communication device consists of say $Q'\bar{u}$ and $\bar{Q}' d$, each bound to a nucleus (A,Z)  where the two nuclei are several meters apart. To use these and the connecting string as a communication device, we need to limit $\Lambda'$. Specifically, we have to ensure that the string' tension - $\sigma' = \Lambda'^2$  will \textit{not} pull the $M'\bar{M}'$ mesons along with the nuclei and atoms/heavy ions in which they are embedded out from the material grains where they reside. Using the lattice binding of the Ion of U $\sim$ 50 eV  to compute the force U/a retaining the ion in its location, we then find the condition:  
\begin{equation}
    \label{26.1}
   \Lambda'^2  < \frac{U}{a} \sim 50 ~eV/ {Angstrom}=  5.10^{4}~ eV^2 \ \mbox{or} \  \Lambda'<  200 ~eV. 
\end{equation} 
Amusingly, avoiding excessive relic densities of DM made of glue'-balls and ensuring confinement of Quirks inside material grains requires roughly similar low $\Lambda'$ values. \par
\begin{figure}[h]
\begin{center}
 \includegraphics[width=0.7
\textwidth]{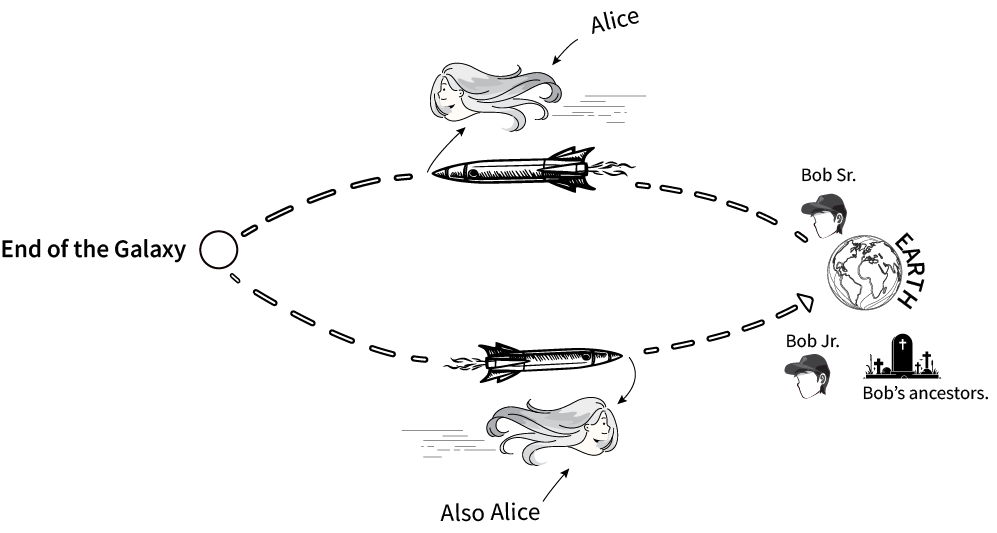}
    \caption{Alice travelling with velocity $\sim c$ to the end of the galaxy keeping in touch with many generations of Bob descendants before returning to earth}
\label{fig:16}
\end{center}
\end{figure}

The basic observation is that with the Quirk (and anti-Quirk) confined in separate small chunks of matter or "grains", we and the ETs, can put one end on a spaceship, travel thousands of light years and keep communicating with the “person” at the other end of the string as in fig.  \ref{fig:16}.
\par
\begin{figure}[h]
\begin{center}
 \includegraphics[width=0.5\textwidth]{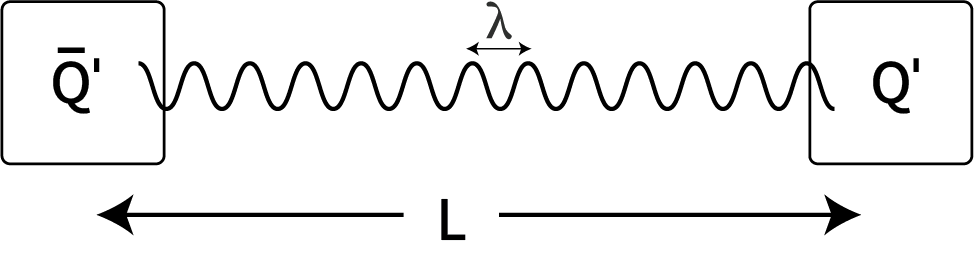}
    \caption{String’ communication by shaking the Quirk and anti-Quirk inside their respective grains at the end of the string and sending transverse phonons}
\label{fig:17}
\end{center}
\end{figure}

The communications are done by mechanically shaking the grains in which the Quirks are embedded, thereby sending transverse phonons along the string (fig. \ref{fig:17})

The string is pulled tight by its tension and lies along the geodesics connecting the two ends. With equal tension and energy density, the phonons propagate with speeds extremely close to the speed of light c. The wave lengths $\lambda \sim c/f \sim3.10^{7\pm 2} cm$ corresponding to carrier frequency f $\sim10-10^5$ Hertz, are much larger than the transverse size of the string: 
\begin{equation}
    \label{26.2}
  d \sim \frac{1}{\Lambda'} \sim  10^{-7} cm  [ 200 ~eV/{\Lambda'}].
\end{equation}
As a result, modulations of the transverse amplitude propagate with minimal dispersion. The finite thickness d of the string causes small wavelength dependent variations $\delta(c)/c$ in  the speed of propagation along the string:
$$\delta(c)/c\sim[d/{\lambda}]^2=[10^{-7} \times (200 eV/\Lambda') /(3 \times 10^7 cm)]^2 =10^{-29} \times [(200 eV/\Lambda')]^2$$
 Even for  $L=3 \hspace{0.2cm} \text{KiloParsecs} \hspace{0.2cm}  = 10^{22} cm$ communication string the arrival of simultaneously sent waves of wavelengths $\lambda$ and $2 \lambda\sim$  will be separated by a tiny distance of:
$\delta(L)=L \delta(c)/c \ll \lambda = 3.10^7 cm $ 
which does not destroy the signal embedded in carriers of such wavelengths. \par 
A unique aspect of such communication is that unlike E.M. waves whose intensity falls as the square of the distance traveled, the transverse phonons are restricted to the string' and their intensity does \textit{not} fall at all! Another advantage is that both the transmitting and receiving units are identical.  Hence the receiving "Microphone" is automatically tuned to the correct frequency even when operated by one who was not informed of the frequency of the carrier waves. Furthermore, the reception is not reduced as in the case of radio waves when the receiving antenna is not directed towards the transmitting source. Ignorance of carrier frequency and of the sky location of the putative broadcasters greatly hinders the SETI project and its future versions and may be the reason why SETI did not find evidence for E.T's even if they exist and messaged us at the appropriate times in the past. \par
To allow useful communication devices various potential hurdles should not kill this "Project", hopefully adopted by our galactic neighbor E.T.s (If Quirks do exist, $\Lambda'$ is sufficiently small, and $M(Q')$ is not too large - so as to allow production of $Q' \bar{Q'}$-  by LHC type colliders or other devices that the E.T.s possess). In Appendix $M$ we show that some apparent hurdles to the proposed communication are in fact harmless.\par
To use Quirks and strings attached for galactic communications, we (or our E.T. friends) have  to find some of the hand-full of Quirk-anti-Quirk paired with the connecting string produced in the many ( $O(10^{16})$ for LHC) pp collisions which occurred during its running lifetime after the LHC (or the E.T.s equivalent accelerator) shuts down. The $Q'$s and $\bar{Q'}$'s  could reside in the rocks or in the water surrounding the detectors of the LHC (or the hadronic accelerator on the planet of the ET's).\par
Appendix $N$ offers a Qualitative picture of a setup which may overcome the above tremendous challenge. It appeals to the unique feature of the constant pull by the string' connecting the Quirk and anti-Quirk and the fact that ordinary matter is completely transparent to these strings but \textit{not} to the Quirks and grains at their ends. (which is why the Quirky transmitter and microphone readily provide the means of communications between terrestrial antipodal points mentioned in appendix I). The remaining task of delivering the grain in which the Quirk is embedded along with a small broadcasting device directing us to its location seems achievable. Indeed, as noted above, accelerating small probes using lasers or other-wise to some fraction of the speed of light and sending them towards our nearest neighbor star is presently being contemplated.
Even the remote possibility of unexpected new technologies which only super accelerators such as the LHC and later higher energy successors can uncover, justifies, along with many other considerations, the continuation of LHC and related efforts in the "High energy frontier".

\section{\hspace{\secspace}The Multiverse, special D.M , and physics analogs of Godel’s Theorem}
At the present time we cannot answer basic physics questions such as: "why do we have the $SU(3)_c \times SU(2)_L \times U(1)$ gauge groups and what extensions -via RH gauge interactions ,SUSY, GUT, or Quirks- are there?". 
This frustrates many in the theoretical community who long for a unified, ultimate Theory of Everything (ToE). The (to be argued for) impossibility of verifying or excluding extensions of the SM such as variants of the Quirk model, reminds of Godel's theorem\footnote{$^*$ When at the IAS in  Princeton Goedel went with friends including Einstein and his secretary Helen Dukas, in the late 1930’s to Trenton to get US citizeship. The officer interviewing him was impressed by Godels’ knowledge of the constitution and bragged that  it protects the US from having a Hitler type tyrant take over as happened in Germany. To this Goedel responded saying that  he found some inconsistencies that may allow this to happen…. To prevent a confrontation  his friends quieted him and  we do not know what he had in mind .(Helen Dukas  PC Circa 1976.)}. In mathematics, a related statement is that we can have consistent theories where a certain conjecture holds or is false. This conjecture can then become an additional "axiom" just as the axiom of parallels can be added so as to define Euclidean geometry. Unlike the clear-cut cases in mathematics we cannot "prove" such physics undecidability.\par   
A potential example is provided by the multiverse. The multiverse can include many ,$10^{500}$ universes sharing gravity, QM and other consistency principles, but having different sets of fields, couplings and VeV.s. We live in one particular universe where these parameters, including the enigmatic Cosmological Constant (CC), have values which jointly allow intelligent life. Short of finding convincing theoretical models which allow predicting most of these essential parameters we cannot disprove the multiverse assumption nor has a completely credible approach to proving it been proposed to date\footnote{Two great theoretical physicists, Leonard Susskind and Paul Steinhardt, have very different  attitudes towards the Anthropic principle, eternal inflation and the multiverse. When I told  Paul that I "converted" to the Anthropic principle he asked me if I ceased to be Jewish… hinting that many believe the framework Susskind helped build (and until recently) strongly advertised, without appreciating the issues involved. As an alternative he helped suggest a periodic universe as in ref. \cite{Khoury2001bz}.}. Countering physics analogs of the Godel indeterminism one may argue that physics ,unlike mathematics, is an experimental science and that "A theory which is not experimentally refutable is not physical". I suggest, however, some theoretically acceptable models can be neither proved nor refuted by any experiment.
An example may be provided by a "Quirk-less" variant of the above Quirk models. Reheat (inflation exit) temperatures $T'$ in the dark sector above an  $SU(N')$  $\Lambda'$ scale  allow for DM consisting of the stable, lightest $0^{++}$ scalar glueball $S'$ in an $SU(N')$ pure Yang-Mills theory. With \textit{no} Quirks, we will not be able to discover this sector via the production of Quirks and Low tension $SU(N')$ strings attached in the LHC or any future collider. Also the vertex $gb' \rightarrow 2 SU(3)_{color}$ gluons does not exist in the absence of the box diagram with circulating Quirks making the gb’s absolutely stable. Conversely they cannot be produced in any accelerator. The extremely strongly interacting  SIDM made of such gb’s can be - as noted in Sec III - inconsistent. It is however consistent as a small, sub-leading fraction of DM. This can be ensured by tuning their reheat temperature after inflation $T'$ to be significantly lower than the corresponding reheat temperature in the SM sector T or by having the inflaton field largely decouple from the SU'(N) gluons. Having no Quirks in the theory ensures that they will not be produced via coupling to our $SU(3)_{color}$ gluons even when the SM reheat the temprature is high.\par
The  extension of the SM to include an $SU(3)'$ and corresponding gb’s may then furnish an example of physics analog of the Godel’s theorem: the new predicted particles cannot be detected as they interact only gravitationally with standard model particles and cannot be ruled out by unacceptable terrestrial/astrophysical predictions\footnote{In the special case where “light” $O(10^{15}- 10^{16})$ gr PBH’s form (a fraction of) DM - the Hawking emission of the eight almost massless gluons of the new $SU(3)'$ would double the number of DOF emitted as compared with those in the SM. The resulting shortening of the lifetime of the PBH’s by a factor of two  may allow us to experimentally infer the existence of the $SU’(N)$ even when we can detect only the ordinary photons emitted.}. \par
An earlier suggestion by Yakir Aharonov, Aharon Casher \& myself \cite{27.Aharonov} of DM made of "Planckons" -particles of Planck mass $M(X)= M(Planck)\sim 10^{19} GeV$ and Planck size $l_P= 10^{-33} cm$ may furnish another Godelian example. The small number density of such DM particles (one per 10.000 $Km^3$) and the smallness of their purely gravitational scattering off any target ensure that such DM will never be directly detected \footnote{The Planckons carry no charges and have no gauge interactions. To see why, let's assume that they carry a charge $Q'$ of some gauge interaction mediated by vector bossons of mass $m(V')\leq M(Planck)$. This generates in the $r<m(V')^{-1}$ neighborhood of the Planckon a $\frac{q}{r^2}$ Coulombic field which extends inwards to $r=l_{Planck}$ and produces, via the Schwinger mechanism a $q'\bar{q'}$ pair. The $q'$ and $\bar{q'}$ then separate and the infalling $\bar{q}'$ neutralizes the Planckon. Most recently it has been argued by Alejandro Perez, Carlo Rovelli and Marios Christodoulou that quantum entanglement and the fact that coherent superposition of mesoscopic particles leads to a superposition of their gravitational fields (i.e,  superposition of geometries central to various emerging gravity theories) may allow detecting Planckons. To achieve this they suggest using large arrays of Josephson junctions. The idea of utilizing the coherence of the bosonic Cooper pairs for sensitive measurements (Advocated by Armen Gulian and collaborators over the last decade and a half) is most appealing. However the required packing of $10^{17}$ Josephson junctions within a cubic meter of detector while maintaining coherence seems impossible.}. 
The Planckons were motivated by the "information paradox" arising from the complete Hawking evaporation of black holes. This problem has been with us for some 40 years, and inspired a lot of new physics. By now it is widely believed that the information can subtly leak out via the Hawking radiation. This largely undermines our suggestion that once a BH shrinks to a Planck mass \& size, it becomes stable where the stability follows from the need to retain the information in the many
$$
 N'= \frac{Area(BH)}{l_P^2}\sim \frac{[M_0(BH)]^2}{[m_{Pl}]^2}
$$
quanta that fell onto and made up the initial "Parent" BH of mass $M_0(BH)$. The phase space for emitting from the Planckon $N'$ quanta each with momentum $p\sim M(Pl)/{N'}$ is tiny:
$(l_{Pl} \hspace{0.1cm} p)^{3 N'}$. The lifetime of such Planckons is then not the naively expected
$$ 
 t_{Pl}= \frac{l_{Pl}}{c}  \sim 10^{-44} sec
$$
but $N'^{3N'}$ times larger. This time exceeds $t_{Hubble}$ already for $N'\sim 36$, which, using the expression for the black hole area entropy, corresponds to $M_0(BH) \sim 3 M(Planck)$. Similar arguments were used earlier by Jacob Bekenstein \cite{27.bekenstein} to prove that small objects can carry only limited information, a proof that may not apply to the singular case encountered here. \par
To use this as an example supporting our suggested physics analogs of the Goedel related conjecture, one needs a reliable scenario where the Planckons form and provide the right DM density. I and Aharon Casher, inspired by Pawel Mazur, argued in \cite{Casher1997rr} that "elementary" particles cannot be accelerated to Planck energies and no new B.H.s of few Planck masses, the potential "parents" of our stable remnant Planckons, can be created. An exception to the above is the "Quantum Gravity era" in our early universe, an era during which R. Brout -et-al  \cite{Brout} suggested a gravity based inflation. 
Assume that the evaporation of the initial BHs happens at some equilibrium temperature $T_0$. We have to ensure that at any later time the correct red shifted temperature, expected in the standard big bang scenario, is retrieved. At the present the total energy density in the CMB radiation is $\sim 4.10^{-4}$ that of the DM which is $\sim1/5 \rho(critical) \sim 1/2 KeV cm^{-3}$.  For this to happen despite the huge red-shift by $M(planck)/T(now) \sim 10^{25}$ only a tiny portion of the initial BH’s should have the mass of 3 M(Planck) required to yield stable Planckons. The accretion of radiation onto the BHs and the Hawking radiation from the BHs could keep for a while these B.Hs in equilibrium at the initial temperature $T_0$. If $T_0$ is some fraction of T(Planck); $T_0= T(Planck)/ n$, then the density of the BHs whose evaporation will lead to the stable Planckons is suppressed by a Bolztaman factor and the correct DM density/radiation density arises when $n \sim 16$.\par
None of the above physics analogues of Godel's theorem: the multiverse, the Quirkless light glue’-ball’s, and DM made of stable Planckons is convincing. Still they suggest possible limitations and a more humble approach of not attempting to explain everything but to separate the explainable from that which is not (at least within the present framework of physics). \footnote{Closer analoges of the true Godel incompleteness may be provided by theoretical physics models that are too hard to allow deciding if specific solutions are indeed correct.}\par

\section {\hspace{\secspace}Furthur discussion of the Anthropic principle}
The AP (Anthropic Principle) started from the observation that many physical constants seem to be tuned to allow our existence. More recently it was promoted to a "principle" by the emergence of the multiverse. The later - a multitude of largely disconnected universes with different gauge interactions, couplings and masses, was inspired by the many possible different degenerate ground states (vacuums) of string theory. Furthermore these universes can be continuously created by “eternal inflation”. Also, despite decades of effort, no universally accepted mechanism protecting the small cosmological constant from huge quantum corrections was found. As emphasized by S. Weinberg such a protection is of utmost “Anthropic value” ,as otherwise a universe with large C.C. would have been extremely short lived hosting no stars and life. In addition, we need to avoid the “Hierarchy problem” and maintain the “smallness” of the scale of weak interactions -by preventing v=the v.e.v. of the Higgs particle and attendant Higgs and W and Z masses as high as $m(Planck) \sim 10^{19} GeV$ and very high masses for the electron and nucleon. \par 
This general approach was challenged by Rony Harnik, Graham D Kribs and Gilad Perez in an outdated yet interesting paper \cite{N.Harnik}. The puzzle of life, with so many ‘moving parts', is taken apart and after some parts are modified (or omitted), they are recombined in a different way to what is claimed to be another life supporting set-up. This most difficult problem is further exacerbated by the fact that some features that may be crucial for forming intelligent life were not yet realized as such and, conversely, some which are believed to be essential are not. In the above analogy this means that some of the parts of the puzzle may be missing and others are redundant. \par
HKP pointed out that if an unsuppressed Higgs vev which sends the masses of $W^+$ $W^-$ and $Z^0$ to very high values rendering our universe "Weak-less", is accompanied by a correspondingly large, scaling down of the Higgs Yukawa couplings to $u$ $d$ and $s$ quarks and to the electron, then the latter can retain their masses leaving Atomic, molecular physics and chemistry unchanged. In general to leave our paradise world which is optimized for our existence and move in the complex landscape of the multiverse to another “Oasis” allowing the same, we need to simultaneously change many fundamental physic parameters and astrophysical initial conditions. If this allows such a radical change as having only $SU(3)_c \times U(1)_{E.M}$ gauge groups, then we cannot infer even the basic symmetry group and interactions from the A.P.\par
Physicists have widely different attitudes to the A.P. While some view it as a cop-out, others think that it helps explain parameters or other aspects that we should not try to explain ab-initio. Thus we should not try and predict from basic principles (as Kepler famously tried) what seemed once to be of fundamental importance, namely the distances between the sun and the various planets. A modern analog is that in some models with extra dimensions the pattern of quarks and leptons, masses and mixings is mapped into the locations in a compact internal space of the various left and right handed fermions. This does not exclude the possibility that some future deeper physics will explain the fermionic masses and mixings. Still this tends to demote the latter from being “fundamental” and discourages efforts to explain them.\par
The HKP scenario was challanged by many authors and is indeed very unlikely. Most criticism centered around the modified BBN. Thus Luigi Clavelli and R.E. White\cite{N.clavelli} claimed that the modified BBN would suffocate life by depriving us of the required amount of Oxygen. Our own mild criticism here is that in the HKP scheme the charged pions cannot decay weakly and are stable. They completely disappear early on via $\pi^- + p \rightarrow n \pi^0$ ( or $\pi^+ n \rightarrow p \pi^0$ ) followed by $\pi^0 \rightarrow 2\gamma$. However high energy charged pions produced by cosmic rays in our atmosphere will induce these most dangerous reactions in our body. The C.R pions cannot be stopped by a thicker atmosphere or magnetic fields as the former can block sun light and both decrease the flux of cosmic rays which enhance rainfall and genetic mutation responsible for evolution.\par 
A different class of challenges to the anthropic- multiverse package is illustrated in \cite{N.Fedrow}. It addresses the "Cosmic coincidence" - the rough present equality between the constant C.C. contribution to the total critical energy density of "Dark energy" and the strongly varying contribution of matter density. The authors acknowledge the importance of this coincidence to formation of life. However they suggest that rather than having many universes with different evolutions spanning all possible ratios of dark energy and mass density at the present time, the observed value can be obtained by appropriate collection of inflaton-like scalar fields instigating inflationary spurts at different times. \par

\section {\hspace{\secspace}MOND Modified Newtonian Dynamics}
We cannot conclude this review without commenting on Mordechai Milgrom's suggested MOND. It avoids DM by modifying Newtonian Dynamics at very small accelerations $a(0) \sim c/t(Hubble) \sim 10^{-8} cm/sec^2$  \cite{Milgrom:1983ca}  \hspace{0.5mm}\footnote{The acceleration considered is \textit{not} a relative acceleration - such as between two astronauts floating in space at 10 meter separation. Rather it is the acceleration  relative to an "absolute" cosmological frame like that in which the CMB is isotropic. Thus in the above example the acceleration of each body is dominated by the much larger gravitational pulls towards the sun  or earth.
In principle MOND could be tested in  widely separated binaries. They should be near us to allow to closely monitor their orbits. They then participate in the rotation of the disc with $v(rot) \sim 220 Km/sec$ and acceleration $\sim a(0)$. If the relative acceleration is not much  larger, then some deviations from Newtonian motion are expected in MOND. To avoid further perturbation the region between and around the binary members should contain no stars leaving very few candidates and  statistically inconclusive results.}. \par
An alternative conformally invariant GR action quadratic in the Weyl Tensor, was suggested in \cite{Mannheim:1992tr}. The resulting attraction is enhanced at large distances providing some of the DM effects. \par 
In the particle physics community, the Bullet and other clusters with colliding galaxies and discrepant results from X ray emitted by the heated gas and from Gravitational lensing which is sensitive to all masses, dark and baryonic alike, is viewed as a proof of DM. The MOND aficionados have different interpretations consistent with MOND. The original N.R  MOND did not aspire to, and does not explain gravitational lensing which DM models readily do. Attempts to realize MOND in a relativistic framework with extra scalar and vector fields(TEVES) \cite{Bekenstein:2004ne}, failed. For electro-magnetic waves i.e on shell massless photons - the trace of the energy-momentum tensor = $E^2 -B^2$  vanishes and so do Feynman diagrams where free photons couple to a scalar field . Also charge conjugation forbids their coupling to vectorial mediators. This suggests that only the tensor part of gravity bends light causing gravitational lensing. Scalar “gravitons”, unlike ordinary GW, would be copiously emitted in spherically symmetric core collapse supernovae.\par
I wrote  this review on DM because I believe that DM exists. However in physics history, there have been cases where both MOND-like and DM-like approaches were correct. Neptune was predicted by insufficient gravity (as DM), and the anomalous precession of mercury was explained by a MOND (which GR is!) \cite{27.Milgrom}. Conceivably, a similar combination will explain anomalies at different scales \cite{27.Edmonds} with MOND applied to the "Low" galactic scales.
However ref. \cite{27.lisanti} used the radial (halo induced) and the transverse (disc induced) acceleration of stars oscillating around the milky way disc, to test MOND. Both of the above accelerations are small enough so that the initial suggestions of Milgrom should apply to both. The fact that there is no good fit with a \textit{single} value of the critical acceleration for both motions does not bode well for MOND even on its original ‘home turf' of single galaxies. 

\section{\hspace{\secspace}Summary and conclusions}
In this review I tried to present many types of DM and the inter-relations between DM models, various BSM extensions and experiments/astronomical observations which may detect them. Any corner of this vast landscape can be investigated in more depth as indicated by the discussion of the unitarity bound on annihilation cross-sections of massive elementary symmetric DM\par
I mentioned a selection of beautiful methods using temporal/directional variations and stellar observations to enhance DM searches and suggestions for detecting dark photons,milli- charged particles, axions and ALPs.\par
To keep this work relatively short I some-times mention subjects in “standard” HE particle physics and cosmology without explaining them (as I eventually attempt to do in footnotes). I still hope that it will be accessible to a wider audience than just DM aficionados and that some readers will find (some of) it amusing and interesting.\par
Regardless of how topical a particular suggestion is by now, I may present it so long as it has some beauty or interesting physics. The many physics issues touched may be helpful when facing further novel BSM/ DM models. In choosing subjects to present I was guided by the beauty and elegance of the particular item and by my ability to present it in a simple, intuitive way. This included mainly topics that I am familiar with and like, resulting in a strong bias towards works of my own or of close collaborators. \par
I tried to justify discussing, SETI type projects by the Quirk BSM scenario which may allow a novel, extremely efficient, communication method on galactic scales.\par
When presenting simple points/elaborations I often refrain from taking or giving credit as most likely these have been suggested by some-one else before the person I know of.\par
Some material presented here which appears to be new is not entirely so. Jonathan Rosner told me that analogs of Quirky communication strings were suggested by Lev Okun and named by him “Thetons” as, like Witten’s cosmic strings, they are associated with the breaking of a $U(1)$ symmetry with one $\theta$ parameter. Okun suggested that they connect foreheads of couples in love, an idea beautifully (and independently!) rendered in the “Avatar” movies.
From Lenny Susskind I learned that focusing in the SETI project on directions from which partial eclipsing of our sun by earth or other solar planets can be seen, was  suggested five years before me, by Seth Shostak and collaborators from the SETI project.   David Latham told me that this idea featured in an undergraduate summer physics project he mentored ten years before Shostak’s paper. Five years after my posting I was told by Subo Dong that he also had  the same idea. All this indicates that the idea may be trivial but sound.\par
The "Six domain model" for DM + matter made from quarks of the three families was influenced by comments made by Yoshio Nambu in an early “Rochester” conference\footnote{$^*$ Yoshio Nambu - Famous for the Nambu-Goldstone bosons- also introduced precursors of QCD Color, suggested the string action, soft pion physics and hadronic vector mesons. As a referee he noted that the SSB of weak interactions requires a NG boson, which turned out to be the celebrated Higgs boson. The fact that he got his Nobel prize so late and that Jeffery Goldstone and Nicola Cabibbo were left out is most puzzling. Also the 65 year old Aharonov Bohm effect manifesting in flux quantization, the Quantum hall effect and which features in Q.M. textbooks, has not yet received a Nobel recognition.}. \par
In conclusion I express my hope that unlike the extinct Aether, a central subject in late 19th- early 20th century physics, DM involving exciting BSM physics will be discovered and also that improved SETI projects will receive useful messages from friendly E.T’s.

\section{\hspace{\secspace}Acknowledgements}
I will be able to mention only a small subset of the many physicists who contributed in some way to this review. This includes long-time collaborators Itzhak Goldman, Rabi Mohapatra  Robert Shrock, Arnon Dar, Graciela Gelmini,Andrzej Drukier (the late) Roberto Peccei, Frank Avignone and R, Creswick. Others that my joint work with influenced me are A. Casher, Y, Aharonov, Herbert Neuberger, Howard Georgi \& S.L.Glashow, Armen Gulian, Ram Cowsik, Pham Quang Hung, Marco Roncadelli, Konstantin Zioutas, Demos Kazanas, Tomer Volansky, Yongchao Zhang and Lenny susskind.  Many others helped and inspired me including Maxim Pospelov, Kathy Zurek, Kathy Freese, Asimina Arvanitaki, Surjeet Rajendran, Amarjit Soni, Marc Kamionokowsky, Ely Kovetz, Viktor Teplitz, Nissan Itzhaki, Paweł Mazur, Nima Arkani Hamed, John Bahcall, Savas Dimopoulos, David beacomb, David Curtin, Yuval Grossman Keith Dienes Anson Hook, Kfir Blum, Zackaria Chacko, Kaustubh Agashe, Tom Cohen,  Raman Sundrum, Yossi Nir, Gilad Perez, Marek Karliner, Karl Rosenfeld, Sekhar Chivukula, Jordan Goodman, Greg Sullivan, Cobi Sonnenschein Shimon Yankielowicz,  Oren Slone, Nadav Outmezguine, Itay Bloch, Ushin Tsai, William Buckley, Francesc Ferrer, Bhupal DeV, Martin Israel, Robert Binns, Marc Alford, James Bjorken, Steve Adler, Scott Tremaine, Glenn Starkman, Richard Mushotsky, Glennys Farrar, Joshua Ruderman, Philip Manheim, Waly Greenberg, Alfred Goldhaber and Sandip Pakvasa. 
I am  grateful to S.L. Glashow for deciding at the last moment not to give the summary talk of the Neutrino 94 conference in Eilat and to Arnon Dar, the organizer, for asking me to do it at short notice. Having done it then in $\sim$ 3 days encouraged me to write the present review over the last three years\footnote{$^*$ Comparing this review and my neutrino 94 summary \cite{NUSSINOV1995497}  I find that much has changed and  much has stayed the same. Gravitational lensing just came online and while the atmospheric neutrino anomaly was gaining ground it was not universally accepted that there is no astrophysical solution to the solar neutrino problem. Young Saul Perlmutter described a future project using  1-a supernovae as standard candles to calibrate distances and $O (GeV)$ D.M -a bound state of a gluon and a light gluino was suggested by G. Farrar.}. Special thanks are due to Robert Shrock, Itzhak Goldman, Zohar Nussinov and Pawel Mazur for many Emails and discussions which greatly helped me write this review.\par
Finally I wish to thank Moty Milgrom, Ken Brecher, Sidney Coleman, Sekhar Chivukula \& Andrew Cohen \& Martin savage \& Terry Walker, Mat Strassler and Alonso Botero who pointed out mistakes in early versions of works I coauthored which are relevant to the review.

\begin{appendices}
\section*{Appendix A - Some comments on mirror like SIDM models}
\addcontentsline{toc}{section}{\protect\numberline{}Appendix A - Some comments on mirror like SIDM models}
 We assume that $m(n') = 5m(n)$ is achieved by having  $\Lambda'$ - the scale of the mirror QCD'  $\sim 5 \Lambda(QCD)$. As repeatedly noted the scale of a non-abelian gauge theory can be much larger than that of our QCD -such as for Techni-color- or vastly smaller, as in extreme versions of Quirk models. Thus $\Lambda' \sim  5 \Lambda$  and hence $m(n') \sim  \Lambda' \sim  5 m(n)$ is readily achieved. We Keep the mirror Weak vev and couplings and the Yukawa couplings of the Higgs' and most fermion masses in the mirror sector the same as in the SM Sector. The only small change made is to switch the masses of the up and down quarks in the mirror sector: $m(u')= m(d) ; m(d') = m(u)$  resulting in $m( p') - m(n') \sim  7 MeV$.\par   
The masses of most mirror meson's - the color' singlet $q'\bar{q'}$ bound states, scale linearly with the same factor of $r=\Lambda'/{\Lambda}\sim 5$  as the baryon masses.   
 However the masses of the light, pseudo Nambu-Goldstone pion's  scale as 
\[
m(\pi'^0)\sim [\Lambda'. (m_0(d')+m_0(u'))]^{1/2}\sim  \Lambda'^{1/2}
\]
with the bare u and d masses are switched in the mirror section so that:
\[
m_0(d')+m_0(u')= m_0(d)+ m_0(u)  
\]
$m(\pi')$ is only $r^{1/2}= 2.25$ \hspace{1mm}rather than five times heavier than the mass of our pion\footnote{The 0 suffix indicates Lagrangian or "Current” quark masses which for our lighter uds are much smaller than the constituent quark masses. The latter masses subsume much of the relevant  non-perturbative QCD effect and are the key to the  successful NQM ( Naive Quark Model) which uses them.}. It is well known that the small nuclear bindings of deutron like states in our sector result from cancellations between the strongly repulsive short range $\omega$ exchange force (which embodies Pauli repulsion between quarks of equal flavor and spin), and the attractive pion and the two pion- broad $0^{++}$ “ sigma” state exchanges at roughly twice the range of the repulsive potential. The doubling of the range of the attractive part of the potential relative to that of the repulsive part due to the light mirror pions is likely to tip the balance in favor of binding. Stable $n'^2$ bound states of spin paired mirror neutrons - and more generally $n'^{2k}$ states will  then form. The Fermi energy is lowered by having two types of fermionic constituents yielding a critical $k^*$ which depends on $m(p')- m(n')$ beyond which the $n'^{2k^*-1}p'$ "Hydrogen" nucleus with unit dark charge is stable.
In our SM sector growing Coulomb barriers and the finite lifetime of the neutron are important in BBN and later formation of heavier elements in stars. In the specific Mirror model considered here with initially no Coulomb barriers and stable mirror neutrons, nuclei with growing $A'$ (and occasionally also  growing $Z'$) keep forming in collisions with ambient $n'$s and with other nuclei. This process is similar to of the growth of bubbles in a liquid as they migrate towards the upper surface or of water droplets falling in air \footnote{$^{++}$ Stoke's expression for the drag force $F=6\pi \eta v R$ suffered by a spherical ball of radius R falling with velocity v in a liquid with viscosity $\eta$ when compared with the gravitational minus the  buoyant force implies a terminal velocity:
\[
v_{final} \sim \frac{4\pi}{3} R^2 g [(\rho(ball) -\rho(liquid) ]/{6\pi \eta}
\]
which grows with the area of the falling ball. An analog expression (with an intriguing extra factor of 1/2) holds in the “inverted” set-up where a bubble of gas with density lower than that of the liquid is accelerated upward. A larger, faster, bubble then overtakes smaller ones and in close encounters the two  coalesce into a larger spherical bubble minimizing the total area and surface tension energy.
Neglecting the reverse process of bubble bursts and for a container which is infinite in the vertical direction, the bubble coalescence becomes an avalanche terminating when the number density of the final large bubbles becomes small enough avoiding further bubble collisions. The avalanche may then become critical with universal features (S.Nussinov \& Z. Nussinov 2003 unpublished). This may serve as an analog for a particular SIDM. For mirror nuclear dark matter we have to include the extra dilution due to the cosmological expansion. Similar ideas have been elaborated for dark matter by Hai-Bo Yu}.\par
A new "Bubbly" version of excited Dark matter may then arise as follows. If the dark photon has a mass of few Mev, then adding say one extra $n'$ from our halo to a pre-existing large dark nucleus can yield an excited “phononic” nuclear state which cannot dexcite via dark photon emission. Upon encountering an ordinary (A,Z) nucleus in some underground device a constituent $n'$ does not behave as a free $n'$ with tiny average kinetic energy of $m(n') \hspace{1mm}\beta'^2/2 \sim  KeV$. Rather with its nuclear' excited energy $\sim$ few MeV it will lead to much larger recoil energy of our target nucleus or even generate an excited $(A,Z) ^*$ nuclear state with a dramatic emission of Mev $\gamma$ ray upon de-excitation. This DM is similar to the other DM species designed to maximize the chances of their detection. To claim a new type of DM we need (at the very least!) to a) verify that in the early universe in the epoch analog of our BBN various mirror nuclei  form, and b) make sure that the required couplings of the few MeV dark photon are consistent with various bounds on such Photons.\par 
Another interesting "branching" of the present model is related to the "Nucleon Portal" involving a new interaction with $\Delta(B') = -\Delta B$ or $n\leftrightarrow n'$ mixing. Such transitions have been discussed and looked for experimentally in the case where $n$ and $n'$ masses are sufficiently close to allow coherent build-up of the $n'$ component in neutron beams. In the present case and more generally for $m(n') \gg m(n)$ the same $1/{\Lambda}'^{5} q'^3 q^3$ effective interaction induces the decay $n'\rightarrow udd$ yielding a decay rate which scales with $[m(n')]^{11} \sim [\Lambda(QCD')]^{11}$. This process can be treated  in perturbative QCD and the extensive information on the latter allows predicting the resulting $\pi$'s and $\gamma$ spectrum expected. \par
For a discussion of the dark sector and cosmological impacts in the Tween Higgs model which shares some aspects with mirror models see ref. \cite{Chacko2018}

\section*{Appendix B- Neutron mixings and mirror models}
\addcontentsline{toc}{section}{\protect\numberline{}Appendix B - Neutron mixings and mirror models}
The "Neutron lifetime anomaly" is the $\sim  0.6 \%$ smaller $\tau(n)$ measured for Ultra Cold Neutrons (UCN) in a bottle relative to the value found by monitoring the actual decays in a neutron beam, This anomaly could  be related to  $n\leftrightarrow n'$ mixing and presently is the only data which may suggest mirror models. Energy conservation forbids  $n\leftrightarrow n'$ transitions in nuclei as the n's are devoid of the attractive nuclear interactions. Such transitions can happen in neutron stars converting them to stars made of equal amounts of ordinary and mirror neurons. It has been noted that the energy gained is emitted radiatively. Along with very low temperature neutron stars observed, this suggested strong upper bounds on the microscopic $\epsilon(n'.n)$ (\cite{McKeen_2021}). Quite remarkably this may \textit{not} prevent measuring such small mixings in planned terrestrial experiments Goldman et-al \cite{Goldman:2022rth}.\par
The need for$\sim$ 5 times as much DM as baryonic matter rather than equal amounts and for a lower temperature $T'$ in the mirror sector required in order to avoid too fast expansion at the time of BBN -Big Bang Nucleosynthesis strongly suggest that exact or almost exact mirror symmetry is untenable. Indeed in Halo DM made of mirror Hydrogen and Helium atoms we would have large $H'-H'$ -or $He'-He'$ cross-sections of order $10^{-16} cm^2$, vastly exceeding the bound of $\sigma =10^{-24} cm^2 . {M(X)}/ GeV$ and very dissipative interactions. Most workers in the field therefore use fairly strongly broken mirror or "Twin Higgs" models. Two physicists - Z. Berehaiani and R. Foot and respective coworkers are a notable exception and their heroic efforts to make exact mirror symmetry consistent with the ever mounting relevant data are quite fascinating. See e.g. ref. \cite{BEREZHIANI_2004} and ref. \cite{Foot_2014}.

\section*{Appendix C- S matrix motivated  bounds on annihilation amplitudes threshold}
\addcontentsline{toc}{section}{\protect\numberline{}Appendix C - S matrix motivated  bounds on annihilation amplitudes threshold}
The $s,t$ and $u$ variables for the $A(X+\bar{X}\rightarrow X+\bar{X})$ amplitude describing scattering of DM particles are the Lorentz invariant squares of the sums of four-momenta in the different channels:  
\[
s=(p_i(X) + p_i(\bar{X})^2=W^2= 4(k^2+M(X)^2)
\]
\[
t = (p_i(X) - p_f(X))^2=- 2 k^2(1-cos(\theta))
\]
\begin{equation}
    \label{C.1}
u=(p_i(X)-p_f (\bar{X}))^2 = -2 k^2(1+cos(\theta))
\end{equation}
with $k$ the $X$ (or $\bar{X}$) initial three momentum in their CMS Lorentz frame defining the $s$ channel. The $t$ and $u$  channels have the final $X$  (final $\bar{X}$) as incoming particles.\par \vspace{0.2cm}
If the mass of the lightest particle or system of two or more particles, that can be exchanged in the $t$ channel of the $X \bar{X}\rightarrow X \bar{X}$ scattering is $\mu$, then the corresponding amplitude $A(k^2, cos \theta)$ is analytic in the complex $z= cos(\theta)$ plane inside the "Lehman-Martin ellipse" shown in fig.19: 
\begin{figure}[h]
\begin{center}
 \includegraphics[width=0.5\textwidth]{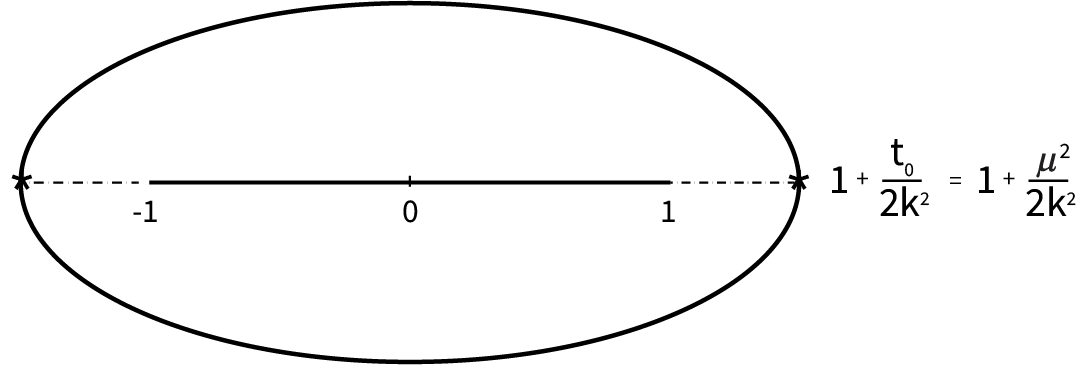}
    \caption{The Martin-Lehman ellipse in the complex $z= cos(\theta)$ plane}
\label{fig:019}
\end{center}
\end{figure}

The ellipse has foci at $+1, -1 $ and a semi-major axis of $1+{\mu}^2/{2k ^2}$.
Indeed a pole at $t_0={\mu}^2$ due to exchange of a particle of mass $\mu$:
\begin{equation}
    \label{C.2}
    A= \frac{g^2}{t-\mu^2}
\end{equation}
or a cut due to the exchange of  two(or more) particles of minimal, total mass mass $\mu$: 
\begin{equation}
    \label{C.3}
    A=\int_{\mu^2}^{\infty} \frac{\rho(t')}{t'-t}d(t')
\end{equation}
comprise a singularity at $t= \mu^2$ or at $cos(\theta) = 1+ \frac{\mu^2}{2k^2} $ on the rightmost point of the ellipse. Two mathematical results used here are :
\begin{itemize}
\item [a] Ellipses with foci at +1 and -1 and varying semi-major axes are the "natural" domains of convergence of series of Legendre polynomials: $\Sigma _l{ a(l) (2l+1) P_l(z)}$ -just like circles centered at the origin are the natural convergence domains of ordinary power series. In both cases there is a singularity on the convergence ellipse/circle .
\item[b] Any function can be expanded in terms of the Legendre orthogonal polynomials. For physical scattering amplitudes this is the partial wave expansion Eq (11) Sec IV.
\end{itemize}
Using $cos(\theta) =x$ for real physical $-1\leq cos(\theta)\leqq 1$, the pole contribution of Eq.\ref{C.2}: 
\begin{equation}
    \label{C.4}
A=  \frac  {g^2} { 2k^2(1-x) +\mu^2 }
\end{equation}
can be expanded for $k\ll \mu$ in a geometric series in $1-x$:
\begin{equation}
    \label{C.5}      
A=\frac{g^2} {\mu^2} + \frac{g^2} {\mu^2} \hspace{0.15cm} \sum_{n>0}(1- x)^n (\frac{2k^2}{\mu^2})^n
\end{equation}
where we exhibited the x independent $n=0$ term. 
\begin{equation}
    \label{C.6}
a(0) \sim  \frac{g^2}{\mu^2} \hspace{0.2cm}\cdot
\end{equation}
The "partial wave" coefficients $a_l(k)$ in the expansion in terms of Legendre polynomias are:
\begin{equation}
    \label{C.7}
    a_l(k)=\int_{-1}^{1} A(k,x)  P_l(x) dx
\end{equation}
The completeness and orthogonality of the Legendre polynomials imply that the $n=l$ term in the  geometric series of Eq.\ref{C.5} is the first to contribute to the integral, the amount
$ \left(\frac{2k^2}{(2 \mu^2)} \right) ^l \hspace{0.1cm} I (l)$  with $I(l) = 2 \int_{0}^{1} P_l (x) x^l$. 
Using the Rodrigues formula for $P_l(x)$ and integrating $l$ times by part yields 
$I(l)\leqq 2^{-l}$. 
Finally, we have 
$a_l(k) \leq (\frac{k^2}{(\mu^2)}^l$ 
and the corresponding partial wave cross-section  satisfies $\sigma_l(k) \sim \left(\frac{k^2}{\mu^2} \right)^{l}  $.
For $k^2 <<\mu^2$ the $l=0$ partial wave  dominates the contributions of the higher $l$   
\begin{equation}
    \label{C.8}
    \frac{\sigma_l(k)}{\sigma_0(k)} \leq \left(\frac{k^2}{4\mu^2} \right)^{2l} \cdot
\end{equation}
Thus in the threshold region where $k<<\mu$ we can keep only the $l=0$ term. By expanding in a geometric series the integrand in the case of a cut in the t plane we can repeat the argument and find the same suppression factor of Eq.\ref{C.8}.
As noted in Sec IV such suppression reflects the centrifugal barrier in potential scattering. A pole or cut terms translate to a Yukawa potential or a superposition of Yukawa potentials:
\begin{equation}
    \label{C.9}
 V(r)= g^2 \frac{\exp(-\mu.r)}{r} \hspace{0.4cm} or \hspace{0.4cm}  V(r)= \int_{\mu^2}^{\infty} dt'  \rho(t') \frac{\exp(-\sqrt{t'}\ r)}{r} 
\end{equation} 
The above refers to the elastic $A_{ X \bar{X} \rightarrow X \bar{X}}$ ($s$, $z=cos{\theta}$) scattering amplitude. The imaginary part of the forward ($\theta =0$ or $x=z=1$) amplitude is related by the optical theorem to the total $X \bar{X}$ cross section and both are infinite for Coulomb scattering due to the exchange of massless photons.\par
We are interested however only in the part contributed by annihilations which for elementary X particles are usually dominated by the two body processes    
\begin{equation}
    \label{C.10}
    A{(X \bar{X} \rightarrow \bar{x}x)}(s,t)
   \hspace{0.2cm} \text{or}  \hspace{0.2cm} A{(X \bar{X} \rightarrow \bar{x'}x')}
\end{equation}
 where $x$ or $x'$ is a SM light fermion/gauge boson or a light dark sector particle with $m(x) \hspace{0.1cm} \text{and} \hspace{0.1cm}m(x') \ll M(X)$.  In the final state made of a pair of light particles energy conservation leads to a large center mass momentum $k' \sim  M(X)$. For the initial CM momentum  $k\sim 0$ the momentum transfer is $t=-t(0) + 2kk'(1-cos(\theta)$ with  $t(0)=-M(X)^2$ rather than the $t(0) =0$ for elastic forward scattering. In the main text we show that the minimal mass $M(Y)$ exchanged in the $t$ channel in annihilation, exceeds M(X) so that:  
\[
A _{ \textnormal{annihilation}} = \frac{g^2}{M(Y)^2 - t(0) + 2kk'(1-x)} < \frac{g^2}{2 M(X)^2 + 2kM(X)(1-x)} 
\]
The arguments above can be repeated to bound the ratio of  $a_{l,ann}(k)$ partial wave to the x independent S wave by :
\[
\frac{a_{l,ann}(k)}{a_{l=0}(ann)(k) } \leq  \left(\frac{k}{M(X)}\right)^{2l}          
\]

so that the corresponding partial cross-section satisfies:      
\[
\frac{\sigma_(l,ann) (k)}{\sigma_(l=0,ann) (k)} \leq \left(\frac{k^2}{M(X)^2}\right)^{2l}  
\]
Using $k^2 \sim  M(X)T$ at temperature $T= T_{f.o}=M(X)/f$ the suppression factor for the early universe annihilation becomes $[\frac{1}{f^2}]^l$. Recalling that $f \sim  25 \pm 5$ this geometric suppression by powers of$\sim 600^{-l}$ justifies the restriction to the $l=0$ wave underlying the G.K bound. 

\begin{figure}[h]
  \centering
  \subfloat[]{\includegraphics[width=0.35\textwidth]{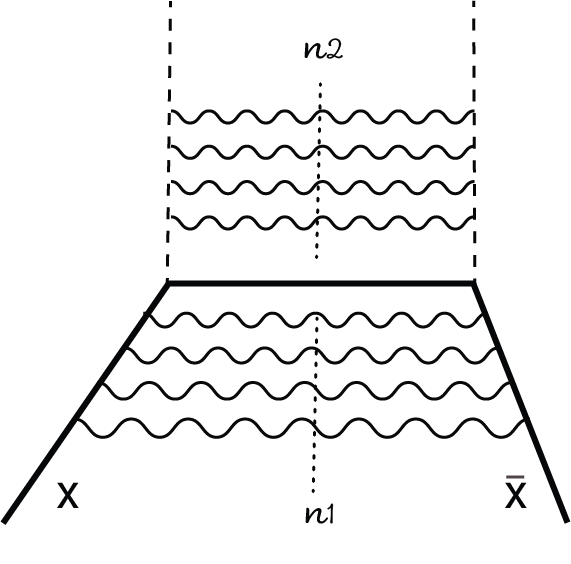}\label{fig:20a}}
  \hspace{0.15\textwidth}
  \subfloat[]{\includegraphics[width=0.35\textwidth]{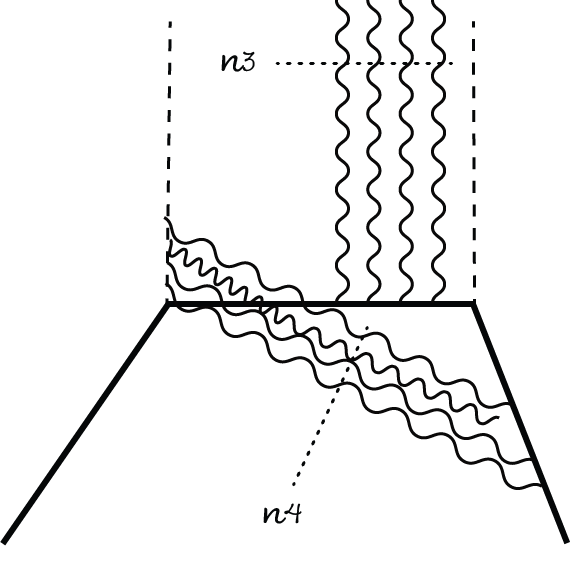}\label{fig:20b}}
  \caption{The annihilation diagrams dressed by further t channel exchanges of n(1) and n(2) light particles between the initial $X$ and $\bar{X}$ and he final $x$ and $(\bar{x})$ [or $x’$ and $\bar{x’}$]  And  n(3) from and n(4) light particles across the exchanged heavy particle in the t channel}
\end{figure}

The use of the lowest order annihilation diagrams does \textit{not} imply that our arguments are perturbative. As shown in fig. \ref{fig:20a}, \ref{fig:20b} we can "Dress up" the diagram by adding any number $n(1)$ of exchanges of some particles of mass $m_1$ between the $X$ and $\bar {X}$ initial DM particles, $n(2)$ particles of mass $m(2)$ exchanged between the $x$ and $\bar{x}$ or $x'$ and $\bar{x'}$ light final particles.

and by $n(3)$ particles of mass $m(3)$ and $n4$ particles of mass $m(4)$ exchanged across or emitted from the heavy exchanged line respectively. The net effect of the extra  exchange in fig.\ref{fig:20a} is to increase the total "mass" exchanged in the t channel to $M(Y) +n(1)$ $m(1) + n(2)$ m(2) and further soften the original Pole term to a high $n(1) +n(2)$ order cut. (Still when $m(1)$ is a massless or very light vector particle the sum over such diagrams leads to  Somerfeld enhanced annihilation rate discussed in the text).
The $n(4)$ particles in the u channel will change the factor  $k' \sim  M(X)$ to $k' =   M(X)+ n(3) m(3)$ which only decreases the amplitude. Further these particles should couple appreciably both to the initial Dark sector particles X and to the final pairs. The optimal possibility is exchanging in the s channel a single such particle to mediate the annihilation, an alternative which we discussed towards the end of Sec VII in the text .

\section*{Appendix D - The S.E. - Sommerfeld enhancement and related processes at different epochs}
\addcontentsline{toc}{section}{\protect\numberline{}Appendix D - The S.E. - Sommerfeld enhancement and related processes at different epochs}
At later epochs just before the "Cosmic Dawn", that is the formation of the first stars and galaxies at temperature $T_{z=25} \approx  70 ^o\ Kelvin \sim  0.006 eV$,  the SE $\sim (M(X)/T)^{-1} \sim 10^{16}$ for $M(X)\sim  10 TeV$ is quite large. Still the annihilation rate is proportional to
 $n(X) \sim  T^3 \sim  z^{-3}$ and  falls faster than  the Hubble expansion rate which is proportional to $\rho(T) ^{1/2} \sim  T^2\sim  z^{-2}$ in the radiation dominated era. At present gravity imparts to halo dark matter kinetic energy K.E$\sim  M(X)\beta^2\sim 10 TeV \hspace{1mm} 10^{-6} \sim 10 MeV$. 
The enhancement by $\frac{\alpha'^2}{\beta^2} \sim \alpha'^2.10^{6}$ is relevant for "indirect"  detection via $\bar{X}X$ annihilation in overdense regions in the halo only if the range of $\gamma'$ exchange = $m_{\gamma'}^{-1}$ exceeds $r_0 (S.E)^{\frac{1}{2}}$.\par \vspace{0.3cm}
In principle other mechanisms beside S.E can enhance the annihilation rate. The radiative capture 
\[
    X+\bar{X} \rightarrow (X\bar{X})_{n,l} + \gamma'
\]
where $n,l$ indicate any one of the Hydrogen-like bound states, can form once the temperature falls below the binding $\frac{M(X)\alpha'^2}{4 n^2}$ \hspace{0.15cm}. This is then followed by a fast cascade to the ground $(n=1,l=0)$ state and $\bar{X} \hspace{-0.15cm}- \hspace{-0.15cm} X$ annihilation therein. On dimensional grounds the cascade down the ladder of excited $(n,l)$ states to the ground state is accelerated in proportion $M(X)$ and so would the capture above from the low lying continuum to the nearby $(n,l)$ states. A formal derivation of the S.E appears in ref. \cite{Blum_2016}

\section*{Appendix E- Why $\bar{p}- p$ annihilation is really a quark rearrangement -explaining its large value}
\addcontentsline{toc}{section}{\protect\numberline{}Appendix E - Why $\bar{p}-p$ annihilation is really a quark rearrangement -explaining its large value}
The $\bar p p$ annihilation into pions is \textit{very} different from $e^+ \bar e \rightarrow 2 (or 3)$ photons or $Q\bar Q \rightarrow 2(or)3$ gluons where the electron or Quark propagates in the t channel and both electron + positron or $Q +\bar{Q}$ disappear. However in most  $\bar pp$ annihilation events the initial $q^3 + \bar{q}^3$  state rearranges into a $(q \bar{q})^3$ state of three mesons, see ref. \cite{Rubinstein1966zza}
No exchanges of the heavy nucleons in the t channel are involved and the only suppression of this process comes from the finite size of the proton and antiproton of$\sim 1/{\Lambda(QCD)^2}$ which is much larger than the $[1/{2m(N)}]^2$ expected in the original Yukawa theory where both nucleon and pion are elementary. Factoring in also the S.E due to the Coulomb attraction of the slow $p$ and $\bar{p}$ and the fact that the (real part of) the hadronic potential between the $p$ and $\bar{p}$ is strongly attractive, complete annihilation of the $\bar{p}$ minority is expected. The absence of relic anti-protons along with the fact that anti-protons produced by collisions of cosmic ray protons with ambient interstellar Hydrogen tend to be energetic, suggested using the slow anti-protons in the AMS (Alpha Magnetic Spectrometer on satellite) project as possible indicators for annihilating/decaying DM particles.

\section*{Appendix F- Why $O(0.1-1 mb) X-N$ scattering cross sections may impede the detection of X}
\addcontentsline{toc}{section}{\protect\numberline{}Appendix F - Why $O(0.1-1 mb) X-N$ scattering cross sections may impede the detection of X}
The nuclear recoil energy in a collision of a DM of $O(GeV)$ mass with the $A>20$ nuclei in the atmosphere or in the underground detectors:  $\delta(E) \sim \frac{KeV}{A} \sim 50 eV$ or $\sim 10 eV$ in the large liquid Xenon detectors - is way smaller than the detector's threshold of few KeV. Still, if some fraction of DM makes it to the large underground detectors, then it can generate a significant unobserved "noise". Also newer detectors are sensitive to lower recoil energies. However the relatively strong nuclear interactions damp the DM flux even at relatively shallow detectors. Detectors on Balloon and rocket missions suggested by Paul Steinhardt, are immune to this and may have already provided negative albeit inconclusive evidence. \par
The small momenta $\beta GeV \sim  MeV$ of the halo's GeV mass particles allow only S waves to contribute to the scattering on the $A \sim  16- 30$ nuclei in earth's crust. With the CMS and Lab system almost coinciding for light ( $O(GeV)$) DM particles, such as some $n'$ or the sexaquark, almost half of them are scattered backwards. After  $M(A,Z)/M(X) \sim A$ scatterings the vast majority of the infalling DM particles are reflected or stopped. The energies of DM particles that still reach the big underground detector will be largely depleted making direct detection via nuclear recoil impractical.\par
More subtle effects such as the accumulation of the WIMPs in earth's crust (or lunar surface) heating up cryogenic devices and spreading the LHC beams have also been dealt with in the paper by David A. Neufeld, Glennys R. Farrar \& Christopher F. Mckeen. 
\footnote{$^+$ The following idealized experiment could look for a new weakly decaying or stable S state (\textit{if}
we have a very long running time in a high precision experiment of pp collisions in a Hydrogen bubble chamber or of an intense beam colliding with a transverse Hydrogen Jet). We look for final states of $K^+ K^+ $ + Missing $X^0$ where the mass of the missing system satisfies: $2m(\Lambda) > m(X^0) \geq m(S)$. The initial pp CMS (Center of Mass System) energy W is tuned to be in the interval $2m(\Lambda) + 2 m(K^+) \geq W \geq 2m(n) + m(S)$. It is conceivable that the very different character of Sexa-quarks will suppress the rate of this reaction. However with complete identification of the charged Kaons there is no background in the specific kinematic region of the final state region whose size increases directly with the boldness of the assumed
$m(S) < 2m (\Lambda)$
}

\section*{Appendix G$^+$$^*$ - A Group theoretic theorem prevents decays of a class of  Kaluza Klein (KK) excitations}
\addcontentsline{toc}{section}{\protect\numberline{}Appendix G$^+$$^*$ - A Group theoretic theorem prevents decays in a class of  Kaluza  Klein excitations}
"K.K" excitations first appeared in a model with one compact circular dimension 
suggested about a century ago by Theodor Kaluza and by Oscar Klein in an attempt to unify gravity and EM. In this model the angular momentum excitations in the extra circular dimension correspond to the conserved and quantized electric charge. Other compact, internal manifolds appear in many extra-dimensional models where the "fundamental particles" belong in irreducible representations of the symmetry group of that manifold. The transition $A+B\leftrightarrow X$  is allowed by this symmetry only if the representation of $X$ is included in the C.G. (Clebsch Gordan) decomposition of the direct product $A\times B$ of the representations of particle A and B. The masses of the KK recurrences in the original KK model are linear in the internal angular momentum l(z). More generally the masses are given by the square root of the quadratic Casimir operator in the representation where that KK particle in question belongs. 
Thus consider the $3+1+D$ dimensional generalization of the relativistic wave operator: $\frac{\partial^2}{(\partial ^{2}(t))} + \nabla^2$.  The $D+3$ dim "spatial" part of the new Laplacian separates into two parts: the ordinary 3 dim  $\nabla^2= {\left(  \frac{\partial}{\partial x} \right )^2} + {\left(  \frac{\partial}{\partial y} \right )^2} + {\left(  \frac{\partial}{\partial z} \right )^2}  $ Laplacian and the Beltrami- Laplace operator acting on the wave functions on the internal D dimensional manifold in the representation of interest..By definition this yields the value of the quadratic Casimir operator in this representation $C_{2}(A)$. The wave equation in the  3+1 dimensional case has plane wave eigenstates  ($\vec{k} = k_x,k_y,k_z$) and $E^2 = |k|^2$. where $E$ is the energy and $\vec{k}$ the three dim momentum or wave number vector. With the extra $D$ dimensional part the last relation is modified to:$E^2=|k|^2 + C_{2}(A)/{a^2}$ where a is the scale of the compact $D$ dimensions and $C_{2}(A)$ the quadratic Casimir operator in the representation A . $E^2=|k|^2 + M(A)^2$ then implies that the mass of the particles in this representation is:
\begin{equation}
    \label{G.1}
    M(A) = C(A) . \frac{1}{a} ; \hspace{0.15cm} \text{with} \hspace{0.15cm} C(A) = \bigg(C_2(A)\bigg)^{\frac{1}{2}}
\end{equation}  
The decay $X\rightarrow A + B$ is kinematically allowed only if $M(X)\geq M(A)+M(B)$. This however cannot be true. The reason is a theorem stating: \par
"If a representation D(X) appears in the CG series of the direct product $D(A)\times D(B)$ of two other irreducible representation of the same symmetry group -which we assume to be one of the Lie groups-, then the inequality :
\begin{equation}
    \label{G.2}
    [C_{2}(A)]^{1/2} + [C_{2}(B)]^{1/2} \geq [C_{2}(X)]^{1/2}
\end{equation}
holds between the square roots of the corresponding quadratic Casimir operators".\par
It is easy to verify it in the special case of SU(2). The maximal Casimir operator among the representations in the direct product  $D(j_1)XD(j_2)$ is $D(J=j_1+j_2)$. It corresponds to the case  when we add the maximal m values $m_1=j_1$ and $m_2=j_2$ to obtain the representation with the maximal $M =m_1+m_2$ and with $C_{2}(j) =j(j+1)$.
A proof of the theorem was provided by Joseph Bernstein. The following is a sketch of the proof: The representations of a rank r group are described by the Cartan weight diagram in r dimensions. The representation $D(X)$ in the $D(A)\times D(B)$ product of maximal Casimir $C^2(X)$ obtains when we add vectorially the maximal weights of $D(A)$ and of $D(B)$. The computation of $C^2(X)$ for any representation (See e.g. \cite{Slansky1981yr})  formally amounts to squaring a $d(X)$ dimensional vector $\vec{R}(X)$ where $d(X)$ is the dimensionality of the $D(X)$ representation, using a metric that depends on the group only  and not on the representation. The desired result then follows as the triangular inequality $| R(X)|< |R(A)| + |R(B)|$ when ${\vec{R}(X)}= {\vec{R}(A)} + {\vec{R}(B)}$. \par
The above does not apply for more general internal spaces which include Orbifolds, support fermionic representations  and allow for non-positive metric with vanishing and or negative squares.  The mass inequality is muted when the masses of all excitations are very high (corresponding to a very small compactification scale) and all accessible particles are zero modes. The issue came up in the ADD universal large extra dimensions scheme  \cite{Mohapatra2003ah}. This ADD scheme also predicted modified submillimeter gravity $V\sim r^{-(D)}$ which so far experimental searches did not find.\par
 The RS model has only one extra dimension but the extended GR  “Warps” it in non-trivial and physically meaningful ways. In particular the above KK non-decay problem is evaded by having non -periodic boundary conditions.

\section*{Appendix H- Intra-generational mass relations}
\addcontentsline{toc}{section}{\protect\numberline{}Appendix H - Intra-generational mass relations}
We briefly mention the other-intra-generational relations which may reflect radiative effects. Thus quarks tend to be heavier than leptons in the same generation  
\begin{equation}
    \label{I.1}
    m(b)\geq 3m(\tau) \hspace{0.3cm}  m(s)  \sim  m(\mu) \hspace{0.3cm} m(d)  \geq 6  m(e)
\end{equation}
Gauge interactions respect the Xiral invariance and cannot induce the $f(Left)\leftrightarrow f(Right)$ transitions required by a mass term. This is rectified in the standard model by  the Higgs Weak doublet with $\langle H\rangle = 250\hspace{1mm}GeV = v$ (where \textit{v} is the VeV vacuum expectation value), which spontaneously breaks the $SU(2)_L$ symmetry. The Higgs Yukawa couplings to quarks and leptons initiate the mass generation. The "renormalization group running" of the Yukawa couplings explains the ratio of $m(b)/m(\tau)\sim  3$ if the Yukawa couplings of the Higgs to the  $I(Weak)=-1/2$ $b$ and $\tau$ members of their weak iso-spin multiplets, have the same value in a GUT framework at a high $\sim 10^{15}GeV$ scale\footnote{$^*$ The "Renormalization group" in its fundamental form, developed for statistical mechanics and quantum field theory by Leon Kadanoff, Michael Fisher and Kenneth Wilson, expresses the change in the Hamiltonian describing the system as a function of the distance (or momentum) scale. In renormalizable theories this can amount to the change of the effective coupling constant with the (log of the) momenta of the particles entering the interaction vertex as found in ref. \cite{Gell-Mann:1954yli} and ref. \cite{StueckelbergdeBreidenbach1952pwl}. 
Since the rate of this logarithmic change is fixed by the corrections to the vertex and vacuum polarization diagrams which determine the lowest terms in the relevant $\beta$ function -we can view this as "radiative corrections".}. 
Another intra-generational mass pattern is that the "up" members of the quark weak isospin doublets with charge $q= 2/3$ are heavier than the lower members with charge $q=-1/3$:
\[
m(top) > 40 m(bottom) ; \      m(charm) > 20 m(strange)
\]
In SUSY separate Higgs fields $H_u$ and $H_d$ couple to the upper and lower members of the weak Iso-spin doublets with a ratio of vev’s:
\begin{equation}
    \label{I.2}
    tag{\beta} = \langle H_u\rangle/ \langle H_d\rangle  \sim  5
\end{equation}
helping  explain the above $m(t)/ m(b)$ and $m(c)/m(s)$ ratios.\par
Our first, light generation provides a counterexample to the intra-generational quark mass hierarchy by having  $m(u) < m(d)$. This yields a positive $m(Neutron) -m(Proton) \sim  1.2 MeV$. If the Fermions mass/mixing pattern will eventually be  accessible to exhaustive theoretical analysis this "Deviant" fine detail may be the last feature explained. We recall however its immense bearing  on the "Anthropic principle". The resulting stability of the proton against $\beta$ decay or $K$ capture: $e + p \rightarrow n +\nu(e)$ is essential for atoms ,chemistry and life. Also the $m(n)-m(p)$ mass difference yields the $\sim$ 12 Minutes lifetime of the neutron - a key in the  “correct” BBN  generation of light nuclei which serve as stepping stones for synthesizing inside stars the heavier elements all of which are crucial to our life.\par
Also the second generation analog of  $m(b) > m(\tau)$, namely $m(s)\geq m(\mu)$ is at best, marginal. Thus the clear patterns in the heaviest family get blurred as we go down to the lighter families and in this last case with no obvious Anthropic advantage!\footnote{$^*$$^+$ The fermion masses seem to be correlated with the number of \textit{massless} gauge interactions the fermion has. Thus we have (almost) massless neutrinos with no EM and $SU(3)_c$ gauge interactions, moderately massive EM charged leptons and higher masses of quarks which have both EM and $SU(3)_c$ gauge interactions. Lacking any explanation for this we view it as an amusing curiosity.}.

\section*{Appendix I- Neutrino beams and galactic communications}
\addcontentsline{toc}{section}{\protect\numberline{}Appendix I - Neutrino beams and galactic communications}
Our goal is to prove the impracticality of neutrino communications on galactic  Kilo-parsec scales. We preface this by the simpler question of whether we can use neutrino beams to transmit information between antipodal points A and B on earth (Fig. 21) 

\begin{figure}[h]
\begin{center}
 \includegraphics[width=0.3
\textwidth]{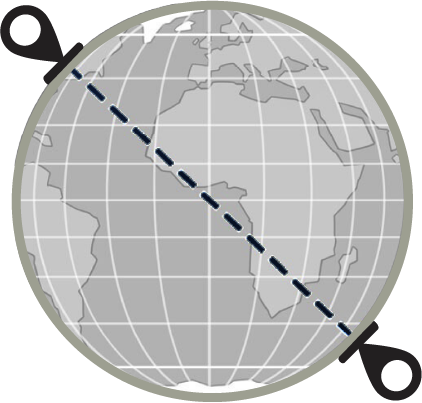}
    \caption{Antipodal points on earth}
\label{fig:21}
\end{center}
\end{figure}

Such a transmission will be faster than by light/radio signals encircling the earth by $\delta(t) = (\pi-2) R(\text{earth}) /c \sim  27 \hspace{1mm} \text{milliseconds}$ and the head-start on, say, stock market information, could be invaluable.\footnote{$^*$ Some French journalists claimed that in 1814 the London Mayer Rothschild got first word of the victory in Waterloo via postal pigeons, bought much of the lacking british stock market to become richer yet. An EM signal propagating radially between diametrically opposite points arrives faster than the proton beam traveling with almost the speed of light along the perimeter of  the circular ring in the Cern $p-p$ collider, was used to cool the circulating proton beams, helping the discovery of the W and Z bosons and Van-Der Meer and Rubbia to get a Nobel Prize.}
Neutrinos are produced in decays $\pi\rightarrow \mu +\nu(\mu)$ of the many $\pi^+$, $\pi^-$ mesons produced by high energy proton beams hitting a fixed target at A . H.E neutrinos minimize the spreading of the neutrino beam when arriving at  B and enhance the cross sections which scale as $E(\nu)$. The $p\rightarrow \pi\rightarrow \mu + \nu (\mu )$ chain requires protons of $\sim  10$ times higher energy than that of the final neutrinos. To date the  $E(p) \sim  7 TeV$  energy proton beams at the LHC collider is highest. To separate genuine difficulties from “technical” issues that sufficient expenditure and effort can resolve, we assume that  a Multi-billion LHC-like set-up is used in a fixed target mode to communicate between two specific antipodal points. To proceed we make the (Optimistic!) assumptions that :1)in $f(1) =10\%$ of the collisions a "leading" $\pi^+$ with $E(\pi)\sim 0.2 E(p) = 1.4 TeV$ is produced. 2)$f(2)\sim 10 \%$ of these pions are focused and directed downward exactly in the direction of the detector at B  3) With a Lorentz factor of $\gamma(\pi) \sim 10^4$,  a Km vertical vacuum pipe allows $f(3)=1\%$ of these pions to decay.\par
The key observation is that the $\pi$ decays yield neutrinos of average energy:
\begin{equation}
    \label{I.11}
    E(\nu) \sim  E(\pi)/3\sim 0.5  TeV
\end{equation}
with a spread of transverse momenta of 
\begin{equation}
    \label{I.12}
    \Delta [p(T)]  \sim \bigg( m(\pi)^2-m(\mu)^2 \bigg) /{2 m(\pi)}\sim 35 MeV
\end{equation}
 The ensuing angular spread 
\begin{equation}
    \label{I.3}
        \Delta(\theta)= \Delta[p(T)] /{E(\nu)}\sim   10^{-4} 
\end{equation}
\textit{cannot} be corrected since the motion of the neutrino cannot be controlled. Traveling a distance $L$ to its target the neutrino beam widens to a circle of radius $r= L.Delta(\theta)$. For antipodal points $L \sim  2 R(Earth) \sim  10^{4} Km $ and $r \sim  2Km$.\par
The neutrino nucleon cross- section at $0.5 TeV$ is $\sim   10^{-36} cm^2$. Thus  after traversing a Km long target detector of average $1 gr/{cm^3}$ density or $\sim  6.10^{23} nucleons/{cm^3}$ a fraction $f(4) = 10^{-7}$ of the neutrinos will interact and be detected.\par
Even with the huge detector at B and accelerator at A, transferring just 100  bits of information in the relevant few millisecond period i.e at a rate of $f(4)= 10^5 Hertz$ is challenging. The energy of the proton-beam cannot be ramped up to $7$ GeV in a few milliseconds so that we continuously operate the system modulating the flux to transmit information, with any single bit encoded in extra $\Delta(N) \sim 5$ \textit{detected} neutrinos over a background of $N\sim 10$ in the relevant time of a millisecond. Combining all the above factors we find that we need  $N \cdot [f(4) f(1)\cdot f(2)\cdot f(3)] ^{-1}.\sim 5.10^{15} $ seven TeV protons interact in the primary target each second, amounting to a $\sim$ 4 GigaWatt power. Such a beam will melt most fixed targets. \par
These difficulties become insurmountable for galactic communications. With distances of $L \sim  3 \hspace{0.1cm} \text{kilo-parsec} \sim  10^{22} cm \sim  10^{13} R(Earth)$ the spreading of the beam reduces the flux at B by a further factor $10^{-26}$(!) and no enlarged detector can compensate for this. Also unlike when both A and B were fixed on earth, here we need to rotate the Km long decay pipe to correct for the daily rotation of the earth to keep pointing in the same direction.


\section*{Appendix J- L.H.C.Magnetic fields affect Quirks for $\Lambda’ < KeV$}
\addcontentsline{toc}{section}{\protect\numberline{}Appendix J - L.H.C.Magnetic fields affect Quirks for $\Lambda’ > KeV$}

How is the evolution of the $\bar{Q}' Q'$ (or $\bar{M}' M'$) systems modified by magnetic fields?  The $M'\bar{M'}$ carry charge of $\pm (2/3 e)$ if they contain $u \bar{u}$ quarks and $\pm 1/3$ for $\bar{d} d$ or strange quarks. To simplify we use 1/2 e in all cases.
Here we discuss "Large"  $\Lambda'> KeV$ in which case the $Q'\bar{Q'}$ annihilate inside the LHC detectors. This can dramatically change due to the $B= B_z = 4 Tesla$ field in the CMS detector, which is parallel to the beam(s) axis. Small $\Lambda' < 100ev$ which are the main focus in this work thanks to the possible applications, imply excursion of the Quirks out of the vacuum region into the shielding and rocks. The physics there is very different and the present discussion does not apply.\par
The $\bar{Q'} Q'$ (and initial $M' \bar{M'}$) Lorentz frames are boosted along the z axis by: 
\begin{equation}
    \label{I1.1}
    \beta(L)= [x(1) -x(2)] 7 TeV/ {2m(Q')}\sim 7/2 [x(1)-x(2)] 
\end{equation}
where the Feynman parameters $x(1)$and $x(2)$ are the fractions of the longitudinal momenta of  the two colliding protons carried by the two gluons fusing to make the $\bar{Q'} Q'$ pair of invariant mass $W \sim  2m (Q')= 2M' \sim  2 TeV$. (The low velocities of the heavy Q's (or M's) allow the neglect of relativistic corrections and using Galilean velocity addition. The total (rest frame) energy of the heavy particle system is: $W= 2M'+ p^2/{2M'})\sim 2M'$).\par
The key observation is that only \textit{transverse} boosts of the $M'\bar{M'}$ system can modify the previous analysis of the nearby $Q'\bar{Q'}$ traversals in sec XXIII above. Thus If  $x(1) =x(2)$ and there is no transverse boost then any magnetic deflection of $M'$ during the first quarter of the oscillation period, while it is moving outward from the beam intersection -initial production point (the origin) is canceled during the second quarter when it retraces its trajectory backward to the Origin - and the same holds for $\bar{M}'$. Also any further velocity imparted by the $\beta(L)$ boost along the z (beam) axis is not affected by the $B_z$ field parallel to it.\par
In reality we do have a transverse boost due to unequal transverse momenta $P(T(1))$ and $P(T(2))$ imparted to the $\bar{Q'} Q'$ system by the two fusing gluons. The transverse momentum keeps building up during the $N(tr)$ nearby $Q'-\bar{Q'}$ traversals by the transverse momenta $\vec{P}_T(i)$ of the pions emitted so that eventually it reaches:
\begin{equation}
    \label{I1.2}
\beta(T) \sim \frac{\left| \vec{P}_T(1)- \vec{P}_T(2) + \sum_{i= 1}^{N(tr)} { \vec{P}_T(i)} \right|}{2M}
\end{equation}
With each of the $N(tr)$ momenta $P( T_i)$ being of order $\sim  0.2 GeV$ the random diffusive sum yields for: $N(tr) \sim 100$
\begin{equation}
    \label{I1.3}
\beta(T) \gtrsim  [N(tr)]^{1/2}\frac{0.2\ GeV}{2\ TeV} \sim  10^{-3}
\end{equation}
For transversal boost the magnetic force moving the $M' \bar{M'}$ along tangent but oppositely curved circles of radius $R$ is most effective in separating them.
If the Quirk string tensile attraction:
\begin{equation}
    \label{I1.4}
 F(\text{attractive})= \sigma'=  \Lambda'^2 = \frac{\Lambda'^2}{(Kev)^2} \hspace{0.1cm}.\hspace{0.1cm} 0.05 \hspace{0.1cm} \text{dyne} 
\end{equation}
exceeds the repulsion due to the magnetic field B (in cgs/ gauss units): 
\begin{equation}
    \label{I1.5}
F(\text{mag-rep}) \sim eBv=eB.\beta(T) ,c \sim 1/2 (4.8 \hspace{0.1cm} . \hspace{0.1cm} 10^{-10}\hspace{0.1cm} . \hspace{0.1cm} 4.10^4 \hspace{0.1cm} . \hspace{0.1cm} 3.10^7) cm/sec=300 \text{dyne} 
\end{equation} 
then the magnetic separation is ineffective and the described option in Sec XXVI and XXVII still holds. However this requires that  $\Lambda'> 200 KeV$. The discussion in section XXVI-XXVII leaves then a very limited range where the annihilation is not averted by the magnetic field \textit {and} the annihilation vertex is sufficiently far from the production point to enable experimentalists to distinguish between the two. For a few months CMS had no magnetic field. These $B=0$ periods may reoccure in future upgrades and/or malfunction offering better Quirk  detection opportunities.

\section*{Appendix K- The Relic stable g'g'=glue(ball)'=S' remnants}
\begin{figure}[h]
  \centering
  \subfloat[]
  {\includegraphics[width=0.25\textwidth]{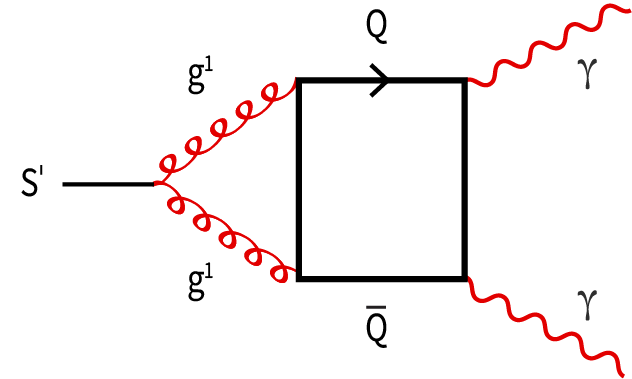}\label{fig:22a}}
  \hspace{0.15\textwidth}
\subfloat[]
  {\includegraphics[width=0.4\textwidth]{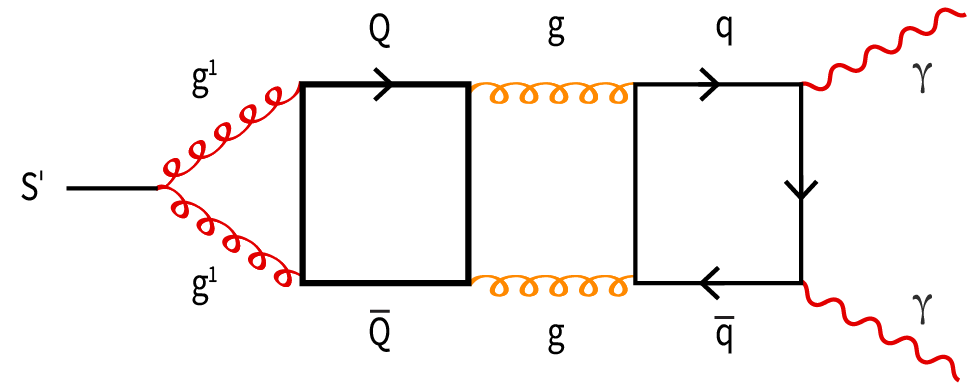}\label{fig:22b}}
  
  \caption{The decay of the $S'$ Glueball via a box diagram with circulating Quirks indicated by the heavy lines and the decay of a $S'$ Glueball via two box diagrams in tandem when we have only neutral Quirks}
\end{figure}

\addcontentsline{toc}{section}{\protect\numberline{}Appendix K - The Relic stable g’g’ = glue(ball) ‘ = S remnants}
Let's first assume that Quirks are electrically charged which minimizes the lifetime of the $S$ glueballs. In this case the $S' \rightarrow 2  g'g' \rightarrow 2 \gamma$ proceeds via the $Q'$ box diagram of fig.\ref{fig:22a}

This analog of the electron box diagram in light by light scattering, then yields the effective Euler- Heisenberg Lagrangian describing the decay:
\begin{equation}
    \label{J.1}
    \mathcal{L}  =  K M(Q')^{-4}  \alpha \alpha'.[F(em) _{\mu,\nu}]^2 [G'^{\mu\nu)}]^2  
\end{equation}
with $F(em)$ and $G'$ the em and color'  fields and K a numerical factor of order $10^{-4}$. We then find that for $M(Q')\geq Tev$ and $\Lambda' < MeV$ that the $S'$ lifetime which is proportional to  $K^{-2} M(Q')^8/{\Lambda'^9}$, exceeds Hubble time by many orders of magnitude. In the minimal version with Quirks carrying the $SU(3)_c$ color and no electric charges the $S'\rightarrow 2 \gamma$ decay proceeds via \textit{two} box diagrams in tandem as in fig.\ref{fig:22b}. One loop is for $g' g' \rightarrow g_{color} g_{color}$, and the second is for $g_{color} g_{color}  \rightarrow\gamma \gamma$ yielding far longer lifetime yet. \par
We next estimate the residual relic density of the stable $S'$ glueball. At the confinement phase transition (P.T.) of $SU(N')$ at temperature $T'\sim \Lambda'$, the dark gluons disappear, forming color singlet glueball's. All the higher gb's decay or transform into the lightest one via  $gb'^* + gb'^*\rightarrow S'S'$ reactions where $gb'^*$ is any excited gb'. 
Thus essentially all the "latent heat" energy density $\sim  T'^4 \sim m(s') \hspace{0.1cm}  T(s') ^3$ is channeled into the S' particles of mass $(4-7) \Lambda'$. Shortly after the PT is completed the temperature of the glueballs is slightly lowered according to $T'^4 \sim  m(S')T(S') ^3$. Even then with $N^{'2}-1 \sim 8-3$ we have as many glueballs as original gluons and this form of dark matter will lead to ecessive $\Omega(D.M)>0.25$ if $\Lambda'> 30 eV$ and $m(S') > 100 eV$.\par   
Since the S' bosons carry no quantum numbers they can "self-cannibalize", their comoving number density decreasing  via S'+S'+S' $\rightarrow $ S'+ S' , 3 $\rightarrow$ 2 collisions while the inverse $2 \rightarrow 3$ process is suppressed by the red shifting of the S' kinetic energy. This was carefully studied by E.D. Carlson, M. E. machacek \& J. L. Hall \cite{Machacek:1992yf} who found only a mild suppression by a factor of  $\log \frac{{\Lambda'}}{(T'_{\text{now}})}$. Thus for $\Lambda'= 30 eV$ and $T'_{\text{now}}\sim T_{\text{now}} \sim 2.4 \hspace{1mm}10^{-4} eV$ we have a relic density of $m_{eff} T^3$ with an effective mass (for $N'=2$ and  $m(S) \sim  4 \Lambda'$) $m_{\text{eff}} = \Lambda'. 4/{14} = 30 .4/ {14}\sim 10 eV$.\par
Since $\sim  1/2$ of all the entropy in the $\sim  60$ DOF at high temperatures $T$ is eventually channeled into the neutrino sector and the latter can have masses of $\sim  8 eV$  $m_{eff}$, the above $m_{eff}$ of $8 eV$ and the corresponding $m(gb')=30 eV$ can be scaled up to $m_{\text{eff}} \sim  240 eV$ and $\Lambda'$ to $60 eV$  without having the $S'$ dark matter exceed the maximal C.D.M energy density  $\sim h^2\Omega(DM)\sim 0.12$. These or smaller relic S' energy density may still conflict with observations due to their large mutual scattering.\par
The "reheat" temperature in the SM sector should exceed $T=MeV$ if we wish to explain the abundance of Helium and light nuclei via BBN. (TeV)- temperatures of the Weak interaction phase transition are required for several scenarios of baryogenesis. The abundant presence of $Q'$ ensures that thermal equilibrium is achieved between the SM "radiation" and the gluons' of $SU(N')$ as $Q'$ couples to both, making $T'= T$. Conversely if $T_{\text{reheat}} < M(Q')$ \textit{and} the reheat temperature in the $SU(N')$ sector, $T'_{\text{reheat}}$ vanishes, as would be the case if the inflaton couples to SM fields only, then thermal equilibrium between the S.M. and the Quirk sectors is never established and no restrictions on $\Lambda'$ arise.

\section*{Appendix L- Dark matter made of S- nlueball. may be unacceptable}
\begin{figure}[h]
  \centering
  \subfloat[]{\includegraphics[width=0.39\textwidth]{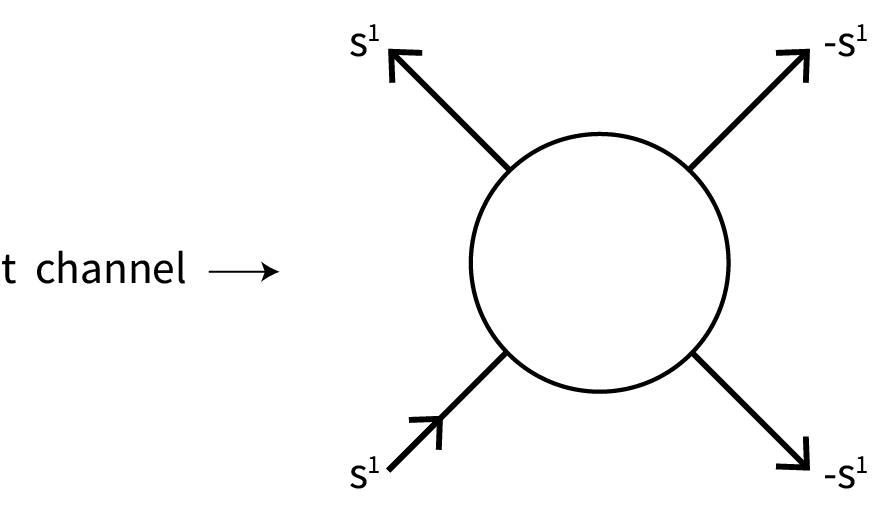} \label{fig:23a}}
 \hspace{1cm}
  \subfloat[]{\includegraphics[width=0.20\textwidth]{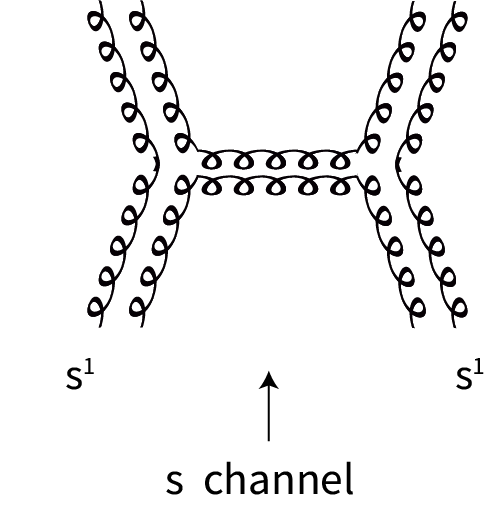}\label{fig:23b}}
  \caption{The $S’ (=g’g') - S’(=g’g')$ scattering amplitude as generated by two gluons corresponding to the exchange in the t channel of $S’$ and the tower of Even $JC \hspace{0.1cm} \text{and} \hspace{0.1cm} P \hspace{0.1cm} gb’$ states”}
\end{figure}

\addcontentsline{toc}{section}{\protect\numberline{}Appendix L - Dark matter made of S- nlueball. may be unacceptable}
While a formal proof that at the P.T. Yang Mills theories transmute into gapped, confining theories is still pending, most members of the HE community believe that this is true.
Exchanges of $S'$ or any one of the higher allowed even spin, parity and charge conjugation glue’-balls, generate attractive potentials and this is likely to hold in the full fledged theory, see fig. \ref{fig:23a} \ref{fig:23b}

The resulting $S'S'$ cross section would then have a large value of  $\sigma\sim  1/{m(S)^2}$ due to strongly coupled $S'$ exchange, vastly exceeding the $\sigma/m \leq \frac{Barn}{GeV}$ upper bound from the "bullet" and other galaxie clusters.\par 
The lowest state of a many-body N.R. system of bosons with attractive interaction obtains when they are in the same quantum ground state. If the attraction clusters the S particles and  each cluster contains a large number of DM particles $N_c$ in a not too large volume $V_c \sim  4 R_c^3$, then the mfp for cluster cluster collisions may become large enough so as to be consistent with the upper limits on $\sigma/M$. \par
For elementary bosons such as axions, the attractive interaction can operate down to very short distances. In particular the cluster radius $R_c$ can get small enough, so that the clusters collapse into B.H. s. Here there is a $\Lambda'^{-1}$ or $m(S')^{-1}$ lower bound on the inter-particle distance within the cluster and for lower distances/higher densities the system reverts to a gluon plasma. For appropriate $\Lambda'$ value, these objects behave as huge collisionless DM "clouds" of size $R \sim 3.10^9 cm \sim  R(sun)/20$ but extremely dilute of mass $\sim  10^{17}$ gr and density of $\sim  10^{-12} gr/{cm^3}$.

\section*{Appendix M- Some potential hurdles of the quirky communication }
\addcontentsline{toc}{section}{\protect\numberline{}Appendix M - Some potential hurdles of the quirky communication}
The analog of the radio background noise for the proposed quirky string communication is the noise generated by the $S'$ glueball's if it makes up the putative cosmic "gb' background" hitting the $SU(N')$ string. 
The huge mutual $S’ S’$ cross sections may require that such a background be largely absent. This can be guaranteed by assuming a reheat temperature $T < M(Q’)$ so that the $g’$ sector cannot be activated via Quirks in the early universe, and $T’$ the corresponding reheat temperature in the $g’$ sector much smaller than $m(gb’)$ as in the case when the $g’$ do not couple to the inflanton. Still to be conservative and because the arguments presented are relatively simple and instructive we show next that even in the case when the $S’$ glueballs constitute DM  and we assume a typical carrier frequency of $f = kiloHertz$ and $\Lambda’ \geq 10 eV$ this noise does not present any serious problems. \par
\begin{figure}[h]
\begin{center}
 \includegraphics[width=0.5
\textwidth]{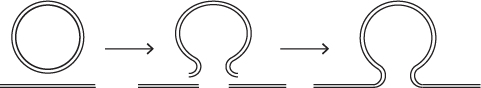}
    \caption{The stages involved in incorporating a glue-ball shown as a closed color loop into the communication string}
\label{fig:24}
\end{center}
\end{figure}

At first sight, this noise looks problematic. 
Using $m(S') \sim 7 \Lambda' \sim 70 eV$ and $\rho(DM) \sim 0.35 GeV cm^{-3}$ for the local DM density, the local glueball number density is 
\[
n(S') = \frac{\rho(DM)}{m(S')} = 5.10^7 cm^{-3}
\]
and its flux is
\[
\Phi(S') \sim v_{vir}.n(S') = 1.5 \hspace{1mm} 10^{15} cm^{-2} sec^{-1}
\]
The string bit carrying one bit of information has a minimal length of $l= 2\pi/f .c \sim 10^{8} cm$. For $\Lambda’\sim  10 eV$ the diameter of the string is $ d\sim 2/[\Lambda'] \cong 10^{-6} cm$ and the corresponding area is:
\[
 A= l.d = 100 cm^2 
\]
The time the signal propagates to a potential receiver at a distance of $\sim 3 kparsecs$ is:
\[
 t_{prop}=10^4\ years \sim 3.10^{11} Sec 
\]
During this time the string bit of interest is exposed to glue-balls’ flux $\Phi(S’)$ and suffers 
\[
N= A . \Phi. t_{prop} \cong  10^{30} 
\]
hits. Upon each hit the glueball may be incorporated into the long communication string and in the process transfer to it the full rest-mass energy $\Delta(E) \sim m(S) \sim 15 eV -300 eV$ Alternatively the glue-ball S may elastically scatter off and reflect from the communication string. In this case the maximal energy delivered is the far lower kinetic energy $\delta(E) =\Delta(E) \beta(Vir) ^2 \sim 0.3 mev$, per hit (see fig.\ref{fig:24}):\par

The first process is extremely rare and quite negligible. If the glue ball is viewed see fig.\ref{fig:24} as a closed string, its incorporation in the communication string requires a temporary "opening - up" of both strings. This generates a "topological potential barrier" of $\Delta(E) \sim m(gb)$. The "thermal" kinetic energy available is only $T\sim m(gb)\beta^2/2\sim 5 10^{-7} m(gb)$ leading to a Boltzman suppression by $ e{-\Delta(E)/T} \sim e^{-[2.10^6]}$. A careful evaluation (Nissan Itzhaki P.C.) yields a suppression $\sim e \big(- t(in)\Delta(E) \big)$ where $t(in)$, the time of incorporation is the relevant distance $d\sim 1/{Lambda'}$ divided by the velocity $\beta = \big(T/m(gb) \big)^{1/2}$. The suppression then is "only" by $\sim e^{(-10^3)}$ which still allows neglecting this branch.\par
The many elastic collisions/ reflections of gb' s off the string segment of length $\sim 10^7 cm$ carrying one bit of information, can eventually- in the worst case scenario - lead to  thermal equilibrium of the glueball gas and the phononic modes of the string. Specifically, the whole communication string will "heat up" to the temperature of the dark glue balls $T(eff) \sim m(gb) \beta^2/2\sim 10^{-4} eV\sim 1 Kelvin$. Treating the communication string as a "box" confining transverse phonons, the latter then have a one dimensional Planckian black-body spectrum. The frequency of interest $\omega = 10^3 Hertz$ is much smaller than $T/\bar{h} \sim 10^{12} Hertz$ and is deep in the Rayleigh -Jeans part of the spectrum where each mode has energy of $kT$. The total thermal power in the low frequencies of interest is then suppressed by $(\omega/ T) ^2 = 10^{-18}$ as compared with the total thermal energy in the string-bit rendering it harmless, fig. \ref{fig:026}.
\begin{figure}[h]
\begin{center}
 \includegraphics[width=0.5\textwidth]{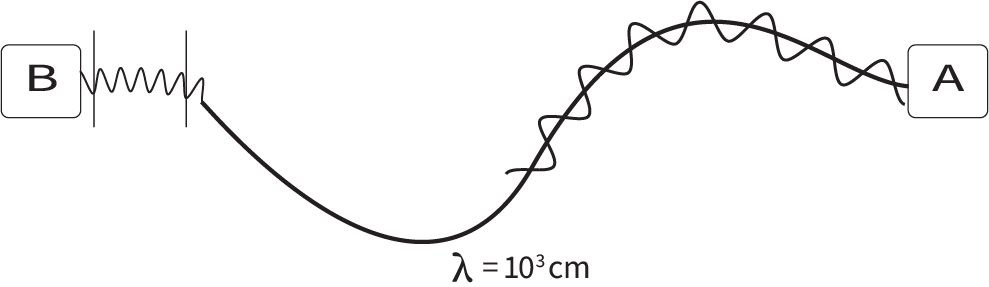}
    \caption{the signal and carrier waves dressed up by many shorter wave fluctuations which make a treatable background by appropriate sieving of the desired band width}
\label{fig:025}
\end{center}
\end{figure}
Illustrates how the main carrier wave can still be recognizable even when dressed up by many fast oscillating shorter waves which extend all the way to $\Lambda^{-1}$ the natural cutoff of the string thickness. \par
To supply the equivalent of the thermal energy kT to the string we have to shake up and down (along the ,say y direction) the $Q'$ of mass $m(Q')\sim TeV \sim 10^{-21} gr$ at the string's end with frequency $w$ and amplitude $\Delta(y)$ of  $\sim Angstrom$. We note that:
\begin{itemize}
\item [a] The force exerted on the $Q' ; \hspace{2mm} F=T/{\Delta(y)} \sim \hspace{0.1cm} 10^{-4} eV/$ Angstrom where $\Delta(y) \sim \Lambda^{-1} \sim $ Angstrom, is $\sim$ million time smaller than the string tension which was bound by $50 eV/ {Angstrom}$ and cannot kick the $Q'$ out of its hosting grain.
\item [b] If the grain weighs a $ \text{Nanogram}\sim 10^ {12} M(Q')$, the actual power required for generating the carrier waves is $\sim 10^3 \hspace{0.1cm}\text{Hertz}. T = 10 \hspace{0.1cm} \text{Nanowatt}$ and may be attained in state of the art mesoscopic cantilevers.
\end{itemize}
\par 
Our communication string freely traverses the sun, earth etc... The only way the string can be cut is via interchange of color fluxes upon encountering another $SU'(N')$ string. Such new string's stretch between a Quirk and anti-Quirk , which can be pair produced in collisions between UHE cosmic ray protons and ambient ISM Hydrogen. 

\begin{figure}[h]
\begin{center}
 \includegraphics[width=0.5\textwidth]{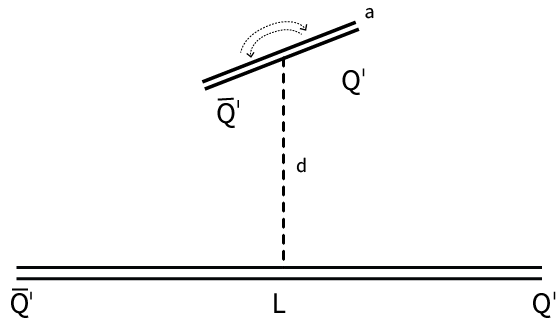}
 \caption{The oscillating "rouge" string produced by a UHE CR - ISM proton collision on his way to cut the long communication string}
\label{fig:026}
\end{center}
\end{figure}

The cross section for the string string encounter is the huge product of $l(c)\sim O(Kparsecs)$ the length of the communication string and $\Delta(L)$ - Yo-Yo ing amplitude of the newly produced string which for $50 eV \leq \Lambda'\leq 4 eV$ is $3-200 meters$. 
Can these "Rouge" strings cut our communication string and hamper the communication project? 
The efficacy of this mechanism in cutting the communication string is reduced by the following factors:
\begin{itemize}
\item[1] The small number density of n $\sim 1 /{cm^3}$ of target interstellar hydrogens 
\item[2] The $\sim 10^{-14}$ fraction of incident protons at the tail of the cosmic ray spectrum of energy $\geq 10^3 TeV$ required to produce $Q'\bar{Q'}$ for $M(Q')\sim TeV$.
\item[3] The tiny $\sim 10^{-11}$ fraction of these UHE C,R.- Hydrogen collisions where the Quirks are produced. 
\end{itemize}
Finally the Yo-Yo-like relative motion of $ Q' \bar{Q'}$ in vacuum is damped by the $\sim 0.1 Gev$ average energy lost in each near -by traversal. This shortens the string connecting the $Q'$ and $\bar{Q'}$ to zero length in $\sim 100$ traversals with each traversal having (for string tension $\Lambda'=50 eV$) a length of $\sim 20 meters$ between subsequent collisions. The total distance traveled before stopping after the 100 traversals is 600 meters.\par
The velocity of the Yo-Yo motion $\beta'=(E'/{M(Q')})^{1/2}$ of the $Q'$ is about $\sim 1/{10}$ of the speed of light: the velocity with which the whole rouge string is traveling relative to its length and yo-yo motion. Thus the distance from the communication string where the small new strings can be produced via $Q'-\bar{Q'}$ pair production and still reach the long communication string and cut it before it itself disappears is:
\[
 l_{tr} = 600 meters .5 \times 10\times10^3 =3000 Km
\]
where the last $10^3$ factor represents the Lorentz time dilation factor $\gamma = 10^{3}$ due to the motion of the string frame in the galaxy.\par
Simple considerations show that the expected number $\mathcal{N}_{cut}$ of cutting events during the time $t_d\sim 3 Kparsec \sim  10^4 Years= 3.10^{11} seconds$ of the string deployment:
\[
 \mathcal{N}_{cut} =\Phi_{CR}(E \geq 10^{15} eV) . 10^{-11} .l_c. \Delta L .l_{tr} n .\sigma (pp).t_d 
\]
is only $\sim 10^{-5}$ !\par
It is crucial to verify that the Lorentz force due to the galactic field will not overcome the $\Lambda'^2$ attraction. Failing that, the $Q'-\bar{Q'}$ produced by UHE C.R's collisions with ISM protons tend to miss each other in subsequent near encounters, retain their kinetic energy and avoid annihilation. This in turn may allow the rogue strings between them to survive and cut our communication string as in (fig.\ref{fig:026}). Fortunately the B field of interest $\sim 2 10^{-6} \hspace{0.1cm} Gauss$ is $\sim 510^{-10}$ times weaker than the 4 tesla $B_z$ considered in Appendix I and the corresponding $F( Lorentz) \sim 1/2 e. Bc$ is $5.10^{-10}$ weaker. Thus even for the $10^{-6}-10^{-7}$ weaker attraction due to the $10^{-3}$ smaller $\Lambda's\sim 100- 200 ev$
 considered the magnetic effect is negligible and the above analysis is unchanged.

\section*{Appendix N- A  scheme for finding and segregating $Q' \bar{Q'}$ pairs connected by strings}
\addcontentsline{toc}{section}{\protect\numberline{}Appendix N - Propagation of quirks in matter and how we can collect some of them}
The Quirky mesons produced at the LHC or a future and better (alien?) accelarator $Q'\bar{q}=M(1)$ and $q\bar{Q'}= M(2)$ lose energy while traversing matter and slow down much faster than when performing the Yo-Yo motion in vacuum. The amplitudes of these Yo-Yo motion $D(1)$ and $D(2)$ of the mesons in vacuum are
\begin{equation}
\label{M.1}
D_i = E'_i /{\Lambda'^2}
\end{equation}
The total lengths traveled in the transverse directions (relative to the longitudinal beam axis)  by each meson' inside matter
\begin{equation}
\label{M.2}
L'_I = E'_i /({dE/{dX}})
\end{equation}
depends on their common CMS energy $E(cms) = M(Q')\beta^2/2 \leq 5 GeV$ for $M(Q') \sim TeV$ and $\beta \leq 0.1$ which holds in $\sim 10\%$ of the $Q'\bar{Q '}$ pair production events.\par 
At the low $\beta\gamma \sim 0.1$ values of interest the em energy loss allows stopping in rocks after traveling on average a total distance of 
\begin{equation}
\label{M.3}
L\sim  10 \text{meter}
\end{equation}
The lab energies $E '_i$ depend on the angle between the direction of the relative motion in the cms frame and the beam direction chosen to be the z axis and also on the rapidity y of the $Q' \bar{Q'}$ CMS frame in the lab. Due to the different $E 'i$ and different em (ionization) energy losses due to different (2/3 e or 1/3 e) charges, the two Quirky mesons will stop after traveling different total path lengths $L '_i$ at different distances $D '(1)$ and $D '(2)$ from the beam intersection and in different locations even when traveling in the same material.\par
Approximating the overall cms motion to be along the z (beams) axis, the $Q '_i$ are eventually lodged in the rock at points roughly uniformly distributed inside a cylinder aligned with the z direction of length $L \sim 10 meter$ and radius smaller than the transverse extent of the vacuum motion:
\begin{equation}
\label{M.4}
         D(trans) \sim E '(trans)/{\Lambda '^2} <1.5 meter 
\end{equation}
for $E '(trans)< E(cms) \sim 5 Gev \hspace{1mm} and \hspace{1mm} \Lambda' \geq 30 eV$. At for $\beta\gamma \sim 1$ the energy lose per unit length $d(E)/{dx}$ is given essentially by Bethes formula in the section on "Passage of Particles Through Matter" in the PDG ref. 233 above. (We used an earlier version of the PDG to pay homage to the many individuals who devoted much of their time to support this critical project to establish the S.M. and beyond). It implies stopping on average in rocks after traveling transversely inside the rock a distance of $D(transverse) \sim 1.5 meter$ in the transverse direction (and $\sim$ 3 times more total distance!). Thus the Quirks stop within a volume $Vol= \pi D(trans)^2.L \sim 45 meter ^3$. This volume contains some $10^{31}$ atoms and the task of searching therein for the $O(1000)$ Quirks produced over years of running the LHC is daunting. However, as we argue next it may be achievable.\par
First we can grind these $100$ Tons of material into small grains of size $l^3$. The size $l$ is limited by demanding- for reasons that will soon become clear -that the $SU'(3)$ string tension : 
\begin{equation}
\label{M.5}
     T\sim \Lambda '^2=2.10^{-4}dyne [\Lambda '/{30 eV}]^2 
\end{equation}
exceeds the weight of the grain:
\begin{equation}
\label{M.6}      
     T\geq F_g=3.l^3 gr 
\end{equation}
where the 3 in front is the average rock density in $gr.(cm)^{-3}$. For $\Lambda '= 30 eV$ the weight of each grain should then be smaller than $2. 10^{-7} gr$ and $l \leq 6.10^{-3} cm$.\par
Grinding to grains of this size reduces the number of objects that need to be individually manipulated from $10^{31}$ atoms to "only" $\sim 10^{18}$ grains - which still is impractical. We can overcome this difficulty as follows. \par
Consider a double conical water container made of stainless steel or another material with a very smooth internal surface. 
\begin{figure}[h]
\begin{center}
 \includegraphics[width=0.5\textwidth]{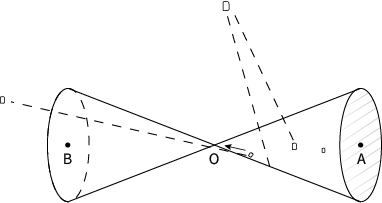}
    \caption{A schematic description of the double conical - initially open and then closed container designed to let the string between a grain containing a and one containing an anti-Quirk pull the grain in the container or both if they are in opposite sides and make it slide to the tiny partition at the conical apex}
\label{fig:27}
\end{center}
\end{figure}
Optimally OA (or OB) the height of each cone is equal to the expected separation between the points where the $Q '$ and $\bar{Q '}$ stopped and lodged into the grains in which they reside. As indicated above this a-priory unknown distance could be of order of a  few meters. Let the container be filled with a say $10\%$ : $90 \%$ mixture (by volume) of grains and water and assume it contains n Grains which are loaded with $Q '$ or $\bar{Q '}$, each. Consider one particular such grain located in the right cone which the string attached to it points in the solid angle subtended by the left cone. As indicated in fig.\ref{fig:27} the pull of this string will make it slide on the smooth internal surface of the right cone until it will reach the tiny region near the apex O where it will be stopped. The same happens for a grain in the left cone pulled to the right by a string which will then slide also towards the apex region but from the opposite side. Note that most of the grains which have no Quirks embedded are free to fall under gravity and slide to the opposite sides of the conical bases.
Next consider all of the grains paired by strings. Where both members of the pair reside in the same (left or right) half of the container the pull of the string will make them coalesce and finally fall to and be distributed over the lower internal surface. We will be interested however in those pairs the members of which lie in the two different conical halves. Assume next that the double cone is laid horizontally .The unbreakable strings connecting such grains in the two cones will keep pulling them towards each other. Thanks to Eq.\ref{M.5}, they overcome their gravity making them slide along the smooth internal conical surfaces and concentrate in the tiny volume at both sides of and near to the conical apex. By assumption the tension force $\Lambda '^2$ is unable to pull Quirks out of the grain - and the grains cannot tunnel from one cone to the opposite cone. Thus all we need is to carefully collect the pairs in this small vertex region.
\par \vspace{0.2cm}
We note that all the steps required to achive communications via Quirk strings are helped if $\Lambda '$ is larger.: the thinner strings reduce the finite width distortion of the signal and the noise due to the impinging cosmic glueballs whereas the allowed rate of bit transfer is increased. The string tension behaving as $\Lambda '^2$ is enhanced allowing to manipulate bigger and fewer grains. It also shortens the distance traveled by the Quirk after production at the LHC or by UHE cosmic rays in proportion to $1/{\Lambda^2}$ decreasing the probability of stray strings cutting the communication string. Finally it reduces the volume where the LHC produced $Q '$ and $\bar{Q '}$ get lodged and need to be searched. This helps facilitate the last stages of the search for and separation of pairs of grains containing a Quirk ( anti-Quirk) respectively, which, as described above, utilizes the continuous mutual pull of the Quirky string connecting them. \par
We have no control on the Quirky string tension $\sigma '$ (If Quirks indeed exist...). We can however, look for optimal grains for which the Quirks will not be ripped out of the grains .Such grains should be made of high Z elements suppressing the tunneling of a $Q'$ (or $\bar{Q'})$ residing at some vacancy to neighboring interstitial location. Optimaly the small grains should be perfect mini-crystals without planes or lines of dis-located atoms along which the Atom /Ion hosting the (anti-) Quirk can more readily be pulled out.\par
Since it is unlikely that the grinding of the rocks will produce such ideal grains one may want to put pre-fabricated grains made of optimal rigid crystalline material in the above cylindrical volume where the Quirks are likely to be trapped prior to the operation of a future LHC like accelerator.\par


\end{appendices}

\bibliographystyle{unsrt}
\bibliography{bib_arxiv}

\begin{thebibliography}{100}

\bibitem{Rotverschiebung}
F.~Zwicky.
\newblock {Die Rotverschiebung von extragalaktischen Nebeln}.
\newblock {\em Helv. Phys. Acta}, 6:110--127, 1933.

\bibitem{rubinVC}
Vera~C. Rubin, W.~Kent Ford, and Norbert Thonnard.
\newblock Rotational properties of 21 sc galaxies with a large range of luminosities and radii, from ngc 4605 /r = 4kpc/ to ugc 2885 /r = 122 kpc/.
\newblock {\em The Astrophysical Journal}, 238:471--487, 1980.

\bibitem{Zeldovich:1968ehl}
Andrzej Krasinski and Ya.~B. Zeldovich.
\newblock {The Cosmological constant and the theory of elementary particles}.
\newblock {\em Sov. Phys. Usp.}, 11:381--393, 1968.

\bibitem{DMC}
Mark~W. Goodman and Edward Witten.
\newblock Detectability of certain dark-matter candidates.
\newblock {\em Phys. Rev. D}, 31:3059--3063, Jun 1985.

\bibitem{Drukier:1984}
A.~Drukier and Leo Stodolsky.
\newblock {Principles and Applications of a Neutral Current Detector for Neutrino Physics and Astronomy}.
\newblock {\em Phys. Rev. D}, 30:2295, 1984.

\bibitem{CPviolation}
R.~D. Peccei and Helen~R. Quinn.
\newblock $\mathrm{CP}$ conservation in the presence of pseudoparticles.
\newblock {\em Phys. Rev. Lett.}, 38:1440--1443, Jun 1977.

\bibitem{Sikivie1983}
P.~Sikivie.
\newblock {Experimental Tests of the Invisible Axion}.
\newblock {\em Phys. Rev. Lett.}, 51:1415--1417, 1983.
\newblock [Erratum: Phys.Rev.Lett. 52, 695 (1984)].

\bibitem{Shrock1980}
R.~E. Shrock.
\newblock {New Tests For, and Bounds On, Neutrino Masses and Lepton Mixing}.
\newblock {\em Phys. Lett. B}, 96:159--164, 1980.

\bibitem{ZurekStrassler}
Matthew~J. Strassler and Kathryn~M. Zurek.
\newblock Echoes of a hidden valley at hadron colliders.
\newblock {\em Physics Letters B}, 651(5–6):374–379, August 2007.

\bibitem{AlexanderVilenkin}
Alexander Vilenkin.
\newblock {Cosmic Strings and Domain Walls}.
\newblock {\em Phys. Rept.}, 121:263--315, 1985.

\bibitem{WITTENEDUARD}
Edward Witten.
\newblock Superconducting strings.
\newblock {\em Nuclear Physics B}, 249(4):557--592, 1985.

\bibitem{SidneyQ}
Sidney Coleman.
\newblock Q-balls.
\newblock {\em Nuclear Physics B}, 262(2):263--283, 1985.

\bibitem{neutrino}
F.~Reines and C.~L. Cowan.
\newblock Detection of the free neutrino.
\newblock {\em Phys. Rev.}, 92:830--831, Nov 1953.

\bibitem{Fischbach1985}
Ephraim Fischbach, Daniel Sudarsky, Aaron Szafer, Carrick Talmadge, and S.~H. Aronson.
\newblock {Reanalysis of the Eotvos Experiment}.
\newblock {\em Phys. Rev. Lett.}, 56:3, 1986.
\newblock [Erratum: Phys.Rev.Lett. 56, 1427 (1986)].

\bibitem{greisen}
Kenneth Greisen.
\newblock End to the cosmic-ray spectrum?
\newblock {\em Phys. Rev. Lett.}, 16:748--750, Apr 1966.

\bibitem{ZK}
G.~T. Zatsepin and V.~A. Kuzmin.
\newblock {Upper limit of the spectrum of cosmic rays}.
\newblock {\em JETP Lett.}, 4:78--80, 1966.

\bibitem{snowmass}
Elcio Abdalla et~al.
\newblock {Cosmology intertwined: A review of the particle physics, astrophysics, and cosmology associated with the cosmological tensions and anomalies}.
\newblock {\em JHEAp}, 34:49--211, 2022.

\bibitem{Freedman_2021}
Wendy~L. Freedman.
\newblock Measurements of the hubble constant: Tensions in perspective*.
\newblock {\em The Astrophysical Journal}, 919(1):16, September 2021.

\bibitem{Freedman2024}
Wendy~L. Freedman, Barry~F. Madore, In~Sung Jang, Taylor~J. Hoyt, Abigail~J. Lee, and Kayla~A. Owens.
\newblock {Status Report on the Chicago-Carnegie Hubble Program (CCHP): Three Independent Astrophysical Determinations of the Hubble Constant Using the James Webb Space Telescope}.
\newblock 8 2024.

\bibitem{superluminal}
Andrew~G. Cohen and Sheldon~L. Glashow.
\newblock Pair creation constrains superluminal neutrino propagation.
\newblock {\em Phys. Rev. Lett.}, 107:181803, Oct 2011.

\bibitem{PattyAbdus}
Jogesh~C. Pati and Abdus Salam.
\newblock {Lepton Number as the Fourth Color}.
\newblock {\em Phys. Rev. D}, 10:275--289, 1974.
\newblock [Erratum: Phys.Rev.D 11, 703--703 (1975)].

\bibitem{PhysRevLett32}
Howard Georgi and S.~L. Glashow.
\newblock Unity of all elementary-particle forces.
\newblock {\em Phys. Rev. Lett.}, 32:438--441, Feb 1974.

\bibitem{UGT}
H.~Georgi, H.~R. Quinn, and S.~Weinberg.
\newblock Hierarchy of interactions in unified gauge theories.
\newblock {\em Phys. Rev. Lett.}, 33:451--454, Aug 1974.

\bibitem{Simpson:1985xc}
J.~J. Simpson.
\newblock {Evidence of Heavy Neutrino Emission in beta Decay}.
\newblock {\em Phys. Rev. Lett.}, 54:1891--1893, 1985.

\bibitem{dienes2022differentnonminimaldarksectors}
Keith~R. Dienes and Brooks Thomas.
\newblock More is different: Non-minimal dark sectors and their implications for particle physics, astrophysics, and cosmology -- 13 take-away lessons for snowmass 2021, 2022.

\bibitem{Creque_Sarbinowski2018}
Cyril Creque-Sarbinowski and Marc Kamionkowski.
\newblock Searching for decaying and annihilating dark matter with line intensity mapping.
\newblock {\em Physical Review D}, 98(6), September 2018.

\bibitem{Fundamental}
Georg~G. Raffelt:.
\newblock "stars as laboratories for fundamental physics. the astrophysics of neutrinos, axions, and other weakly interacting particles". university of chicago press, chicago, 1996. theoretical astrophysics. isbn 0-226- 70271-5.
\newblock {\em Science}, 274(5288):733--734, 1996.

\bibitem{Feng2017}
Jonathan~L. Feng, Iftah Galon, Felix Kling, and Sebastian Trojanowski.
\newblock {ForwArd Search ExpeRiment at the LHC}.
\newblock {\em Phys. Rev. D}, 97(3):035001, 2018.

\bibitem{Feng_2023}
Jonathan~L Feng, Felix Kling, Mary~Hall Reno, and et~al.
\newblock The forward physics facility at the high-luminosity lhc.
\newblock {\em Journal of Physics G: Nuclear and Particle Physics}, 50(3):030501, January 2023.

\bibitem{arkani}
Nima Arkani-Hamed and Juan Maldacena.
\newblock Cosmological collider physics, 2015.

\bibitem{zeldovitch}
Ya.~B. {Zeldovich}.
\newblock {The Universe as a Hot Laboratory for the Nuclear and Particle Physicist}.
\newblock {\em Comments on Astrophysics and Space Physics}, 2:12, January 1970.

\bibitem{Sunyaev:1970er}
R.~A. Sunyaev and Ya.~B. Zeldovich.
\newblock {The Interaction of matter and radiation in the hot model of the universe}.
\newblock {\em Astrophys. Space Sci.}, 7:20--30, 1970.

\bibitem{Amelino-Camelia1997}
G.~Amelino-Camelia, John~R. Ellis, N.~E. Mavromatos, Dimitri~V. Nanopoulos, and Subir Sarkar.
\newblock {Tests of quantum gravity from observations of gamma-ray bursts}.
\newblock {\em Nature}, 393:763--765, 1998.

\bibitem{ADMX_Collaboration}
S.~J. Asztalos et~al.
\newblock {A SQUID-based microwave cavity search for dark-matter axions}.
\newblock {\em Phys. Rev. Lett.}, 104:041301, 2010.

\bibitem{marion2003}
H.~Marion, F.~Pereira~dos Santos, Michel Abgrall, S.~Zhang, Yvan R.~P. Sortais, S.~Bize, I.~Maksimovic, D.~Calonico, J.~Gruenert, C.~Mandache, P.~Lemonde, G.~Santarelli, Ph. Laurent, A.~Clairon, and C.~Salomon.
\newblock {A Search for Variations of Fundamental Constants using Atomic Fountain Clocks}.
\newblock {\em {Physical Review Letters}}, 90:150801, 2003.

\bibitem{carney}
David~Carney et~al.
\newblock Mechanical quantum sensing in the search for dark matter.
\newblock {\em Quantum Science and Technology}, 6(2):024002, January 2021.

\bibitem{Mohapatra:1998rq}
R.~N. Mohapatra and P.~B. Pal.
\newblock {\em {Massive neutrinos in physics and astrophysics. Second edition}}, volume~60.
\newblock 1998.

\bibitem{science1996review}
John~N. Bahcall.
\newblock Review of theoretical astrophysics.
\newblock {\em Science}, 274(5288):733--734, 1996.

\bibitem{cirelli2024darkmatter}
Marco Cirelli, Alessandro Strumia, and Jure Zupan.
\newblock Dark matter, 2024.

\bibitem{epr}
A.~{Einstein}, B.~{Podolsky}, and N.~{Rosen}.
\newblock {Can Quantum-Mechanical Description of Physical Reality Be Considered Complete?}
\newblock {\em Physical Review}, 47(10):777--780, May 1935.

\bibitem{SCHRAMM1979}
David Schramm and Gary Steigman.
\newblock Lepton degeneracy and cosmological constraints on the number of neutrinos.
\newblock {\em Physics Letters B}, 87(1):141--143, 1979.

\bibitem{Brout}
R~Brout, F~Englert, J~M Frère, E~Gunzig, P~Nardone, C~Truffin, and P~Spindel.
\newblock {Cosmogenesis and the origin of the fundamental length scale}.
\newblock {\em Nucl. Phys. B}, 170:228--264, 1980.

\bibitem{kasanas}
D.~{Kazanas}.
\newblock {Dynamics of the universe and spontaneous symmetry breaking}.
\newblock {\em apjl}, 241:L59--L63, Oct 1980.

\bibitem{guth}
Alan~H. Guth.
\newblock Inflationary universe: A possible solution to the horizon and flatness problems.
\newblock {\em Phys. Rev. D}, 23:347--356, Jan 1981.

\bibitem{guth_tye}
Alan~H. Guth and S.~H.~H. Tye.
\newblock Phase transitions and magnetic monopole production in the very early universe.
\newblock {\em Phys. Rev. Lett.}, 44:631--635, Mar 1980.

\bibitem{Preskill}
John~P. Preskill.
\newblock Cosmological production of superheavy magnetic monopoles.
\newblock {\em Phys. Rev. Lett.}, 43:1365--1368, Nov 1979.

\bibitem{drukier}
A.~K. Drukier and S.~Nussinov.
\newblock Monopole pair creation in energetic collisions: Is it possible?
\newblock {\em Phys. Rev. Lett.}, 49:102--105, Jul 1982.

\bibitem{kibble}
T~W~B Kibble.
\newblock Topology of cosmic domains and strings.
\newblock {\em Journal of Physics A: Mathematical and General}, 9(8):1387, aug 1976.

\bibitem{Dashen}
Roger Dashen and Herbert Neuberger.
\newblock How to get an upper bound on the higgs mass.
\newblock {\em Phys. Rev. Lett.}, 50:1897--1900, Jun 1983.

\bibitem{albrecht}
Andreas Albrecht and Paul~J. Steinhardt.
\newblock Cosmology for grand unified theories with radiatively induced symmetry breaking.
\newblock {\em Phys. Rev. Lett.}, 48:1220--1223, Apr 1982.

\bibitem{linde}
A.D. Linde.
\newblock A new inflationary universe scenario: A possible solution of the horizon, flatness, homogeneity, isotropy and primordial monopole problems.
\newblock {\em Physics Letters B}, 108(6):389--393, 1982.

\bibitem{Frieman_2008}
Joshua~A. Frieman, Michael~S. Turner, and Dragan Huterer.
\newblock Dark energy and the accelerating universe.
\newblock {\em Annual Review of Astronomy and Astrophysics}, 46(1):385–432, September 2008.

\bibitem{DesyCola2024}
M.~Maus, Y.~Lai, H.~E. Noriega, and S.~Ramirez-Solano et~al.
\newblock A comparison of effective field theory models of redshift space galaxy power spectra for desi 2024 and future surveys, 2024.

\bibitem{Mace}
Charlie Mace, Zhichao~Carton Zeng, Annika H.~G. Peter, Xiaolong Du, Shengqi Yang, Andrew Benson, and Mark Vogelsberger.
\newblock {Convergence Tests of Self-Interacting Dark Matter Simulations}.
\newblock 2 2024.

\bibitem{palubski2024}
Igor Palubski, Oren Slone, Manoj Kaplinghat, Mariangela Lisanti, and Fangzhou Jiang.
\newblock Numerical challenges in modeling gravothermal collapse in self-interacting dark matter halos, 2024.

\bibitem{Lynden-Bell1966}
Donald Lynden-Bell.
\newblock {Statistical mechanics of violent relaxation in stellar systems}.
\newblock {\em Mon. Not. Roy. Astron. Soc.}, 136:101--121, 1967.

\bibitem{Spergel_2000}
David~N. Spergel and Paul~J. Steinhardt.
\newblock Observational evidence for self-interacting cold dark matter.
\newblock {\em Physical Review Letters}, 84(17):3760–3763, April 2000.

\bibitem{Navarro_1997}
Julio~F. Navarro, Carlos~S. Frenk, and Simon D.~M. White.
\newblock A universal density profile from hierarchical clustering.
\newblock {\em The Astrophysical Journal}, 490(2):493–508, December 1997.

\bibitem{Butler:2023glv}
Joel~N. Butler et~al.
\newblock {Report of the 2021 U.S. Community Study on the Future of Particle Physics (Snowmass 2021)}.
\newblock 1 2023.

\bibitem{pomeran}
I.~Yu. Kobzarev, L.~B. Okun, and I.~Ya. Pomeranchuk.
\newblock {On the possibility of experimental observation of mirror particles}.
\newblock {\em Sov. J. Nucl. Phys.}, 3(6):837--841, 1966.

\bibitem{arnik}
Z.~Chacko, Hock-Seng Goh, and Roni Harnik.
\newblock Natural electroweak breaking from a mirror symmetry.
\newblock {\em Phys. Rev. Lett.}, 96:231802, Jun 2006.

\bibitem{senjanovic}
Rabindra~N. Mohapatra and Goran Senjanovi\ifmmode~\acute{c}\else \'{c}\fi{}.
\newblock Neutrino mass and spontaneous parity nonconservation.
\newblock {\em Phys. Rev. Lett.}, 44:912--915, Apr 1980.

\bibitem{WayneHu}
Wayne Hu, Rennan Barkana, and Andrei Gruzinov.
\newblock Fuzzy cold dark matter: The wave properties of ultralight particles.
\newblock {\em Phys. Rev. Lett.}, 85:1158--1161, Aug 2000.

\bibitem{LamHui}
Lam Hui, Jeremiah~P. Ostriker, Scott Tremaine, and Edward Witten.
\newblock Ultralight scalars as cosmological dark matter.
\newblock {\em Phys. Rev. D}, 95:043541, Feb 2017.

\bibitem{blum2024axionh0graphyhuntingultralightdark}
Kfir Blum and Luca Teodori.
\newblock Axionh0graphy: hunting for ultralight dark matter with cosmographic h$_0$ bias, 2024.

\bibitem{Yonit}
Yonit Hochberg, Eric Kuflik, Tomer Volansky, and Jay~G. Wacker.
\newblock Mechanism for thermal relic dark matter of strongly interacting massive particles.
\newblock {\em Phys. Rev. Lett.}, 113:171301, Oct 2014.

\bibitem{yonit2}
Yonit Hochberg, Eric Kuflik, Hitoshi Murayama, Tomer Volansky, and Jay~G. Wacker.
\newblock Model for thermal relic dark matter of strongly interacting massive particles.
\newblock {\em Phys. Rev. Lett.}, 115:021301, Jul 2015.

\bibitem{David}
David Tucker-Smith and Neal Weiner.
\newblock Status of inelastic dark matter.
\newblock {\em Physical Review D}, 72(6), September 2005.

\bibitem{NimaArka}
Nima Arkani-Hamed, Douglas~P. Finkbeiner, Tracy~R. Slatyer, and Neal Weiner.
\newblock A theory of dark matter.
\newblock {\em Phys. Rev. D}, 79:015014, Jan 2009.

\bibitem{Fan2013}
JiJi Fan, Andrey Katz, Lisa Randall, and Matthew Reece.
\newblock {Double-Disk Dark Matter}.
\newblock {\em Phys. Dark Univ.}, 2:139--156, 2013.

\bibitem{geller}
Michael Geller and Zamir Heller-Algazi.
\newblock Boosting asymmetric charged dm via thermalization.
\newblock {\em Journal of High Energy Physics}, 2023, 03 2023.

\bibitem{walker2019}
Andre Walker-Loud.
\newblock On the cottingham formula and the electromagnetic contribution to the proton-neutron mass splitting, 2019.

\bibitem{Cottingham1963}
W.~N. Cottingham.
\newblock {The neutron proton mass difference and electron scattering experiments}.
\newblock {\em Annals Phys.}, 25:424--432, 1963.

\bibitem{Press1985CaptureBT}
William~H. Press and David~N. Spergel.
\newblock Capture by the sun of a galactic population of weakly interacting massive particles.
\newblock {\em The Astrophysical Journal}, 296:679--684, 1985.

\bibitem{osti_5939847}
J~Faulkner and R~L Gilliland.
\newblock Weakly interacting, massive particles and the solar neutrino flux.
\newblock {\em Astrophys. J.; (United States)}, 299:2, 12 1985.

\bibitem{Lee:1977ua}
Benjamin~W. Lee and Steven Weinberg.
\newblock {Cosmological Lower Bound on Heavy Neutrino Masses}.
\newblock {\em Phys. Rev. Lett.}, 39:165--168, 1977.

\bibitem{Griest1989}
Kim Griest and Marc Kamionkowski.
\newblock {Unitarity Limits on the Mass and Radius of Dark Matter Particles}.
\newblock {\em Phys. Rev. Lett.}, 64:615, 1990.

\bibitem{hall}
Lawrence~J. Hall, Karsten Jedamzik, John March-Russell, and Stephen~M. West.
\newblock Freeze-in production of fimp dark matter.
\newblock {\em Journal of High Energy Physics}, 2010(3), March 2010.

\bibitem{kramer}
Eric~David Kramer, Eric Kuflik, Noam Levi, Nadav~Joseph Outmezguine, and Joshua~T. Ruderman.
\newblock Heavy thermal dark matter from a new collision mechanism.
\newblock {\em Phys. Rev. Lett.}, 126:081802, Feb 2021.

\bibitem{Smirnov2019}
Juri Smirnov and John~F. Beacom.
\newblock {TeV-Scale Thermal WIMPs: Unitarity and its Consequences}.
\newblock {\em Phys. Rev. D}, 100(4):043029, 2019.

\bibitem{Acharyya2023}
A.~Acharyya et~al.
\newblock {Search for Ultraheavy Dark Matter from Observations of Dwarf Spheroidal Galaxies with VERITAS}.
\newblock {\em Astrophys. J.}, 945(2):101, 2023.

\bibitem{derujula}
A.~De~R{\'u}jula and S.~L. Glashow.
\newblock Nuclearites---a novel form of cosmic radiation.
\newblock {\em Nature}, 312(5996):734--737, Dec 1984.

\bibitem{herrin}
E.~T. Herrin and V.~L. Teplitz.
\newblock {Seismic search for strange quark matter}.
\newblock In {\em {Orbis Scientiae 1997: Celebration of 25 Coral Gables Conferences and Their Impact on High-Energy Physics and Cosmology}}, pages 145--151, 1 1997.

\bibitem{Starkman2022}
Nathaniel Starkman, Glenn~D. Starkman, Harrison Winch, and Jagjit~Singh Sidhu.
\newblock {A Straight Lightning Bolt?! Observation of a Predicted Macro Dark Matter Signature}.
\newblock 2 2022.

\bibitem{Kusenko1997}
Alexander Kusenko, Vadim Kuzmin, Mikhail~E. Shaposhnikov, and P.~G. Tinyakov.
\newblock {Experimental signatures of supersymmetric dark matter Q balls}.
\newblock {\em Phys. Rev. Lett.}, 80:3185--3188, 1998.

\bibitem{Carr:1974nx}
Bernard~J. Carr and S.~W. Hawking.
\newblock {Black holes in the early Universe}.
\newblock {\em Mon. Not. Roy. Astron. Soc.}, 168:399--415, 1974.

\bibitem{Flores_2021}
Marcos~M. Flores and Alexander Kusenko.
\newblock Primordial black holes from long-range scalar forces and scalar radiative cooling.
\newblock {\em Physical Review Letters}, 126(4), January 2021.

\bibitem{ilie2023}
Cosmin Ilie, Katherine Freese, Andreea Petric, and Jillian Paulin.
\newblock Uhz1 and the other three most distant quasars observed: possible evidence for supermassive dark stars, 2023.

\bibitem{Gemmell2023}
Caleb Gemmell, Sandip Roy, Xuejian Shen, David Curtin, Mariangela Lisanti, Norman Murray, and Philip~F. Hopkins.
\newblock {Dissipative Dark Substructure: The Consequences of Atomic Dark Matter on Milky Way Analog Subhalos}.
\newblock {\em Astrophys. J.}, 967(1):21, 2024.

\bibitem{Dhakal:2022rwn}
Pawan Dhakal, Steven Prohira, Christopher~V. Cappiello, John~F. Beacom, Scott Palo, and John Marino.
\newblock {New constraints on macroscopic dark matter using radar meteor detectors}.
\newblock {\em Phys. Rev. D}, 107(4):043026, 2023.

\bibitem{Yennie:1987sa}
D.~R. Yennie.
\newblock {Integral quantum Hall effect for nonspecialists}.
\newblock {\em Rev. Mod. Phys.}, 59:781--824, 1987.

\bibitem{Baumgart2023}
Matthew Baumgart, Nicholas~L. Rodd, Tracy~R. Slatyer, and Varun Vaidya.
\newblock {The quintuplet annihilation spectrum}.
\newblock {\em JHEP}, 01:158, 2024.

\bibitem{Cirelli:2005uq}
Marco Cirelli, Nicolao Fornengo, and Alessandro Strumia.
\newblock {Minimal dark matter}.
\newblock {\em Nucl. Phys. B}, 753:178--194, 2006.

\bibitem{nussinov}
Shmuel Nussinov and Melissa~A. Lampert.
\newblock Qcd inequalities.
\newblock {\em Physics Reports}, 362(4):193–301, May 2002.

\bibitem{De_Luca_2018}
Valerio De~Luca, Andrea Mitridate, Michele Redi, Juri Smirnov, and Alessandro Strumia.
\newblock Colored dark matter.
\newblock {\em Physical Review D}, 97(11), June 2018.

\bibitem{Geller2018}
M.~Geller, S.~Iwamoto, G.~Lee, Y.~Shadmi, and O.~Telem.
\newblock Dark quarkonium formation in the early universe.
\newblock {\em Journal of High Energy Physics}, 2018(6), June 2018.

\bibitem{arkani2005supersymmetric}
Nima Arkani-Hamed and Savas Dimopoulos.
\newblock Supersymmetric unification without low energy supersymmetry and signatures for fine-tuning at the lhc.
\newblock {\em Journal of High Energy Physics}, 2005(06):073, 2005.

\bibitem{Sakharov:1967dj}
A.~D. Sakharov.
\newblock {Violation of CP Invariance, C asymmetry, and baryon asymmetry of the universe}.
\newblock {\em Pisma Zh. Eksp. Teor. Fiz.}, 5:32--35, 1967.

\bibitem{Zeldovich:1976vw}
Ya.~B. Zeldovich.
\newblock {Charge Asymmetry of the Universe Due to Black Hole Evaporation and Weak Interaction Asymmetry}.
\newblock {\em Pisma Zh. Eksp. Teor. Fiz.}, 24:29--32, 1976.

\bibitem{cui}
Yanou Cui and Raman Sundrum.
\newblock Baryogenesis for weakly interacting massive particles.
\newblock {\em Phys. Rev. D}, 87:116013, Jun 2013.

\bibitem{nussinov1985}
S.~Nussinov.
\newblock Technocosmology — could a technibaryon excess provide a “natural” missing mass candidate?
\newblock {\em Physics Letters B}, 165(1):55--58, 1985.

\bibitem{Susskind:1978ms}
Leonard Susskind.
\newblock {Dynamics of Spontaneous Symmetry Breaking in the Weinberg-Salam Theory}.
\newblock {\em Phys. Rev. D}, 20:2619--2625, 1979.

\bibitem{Weinberg:1977ma}
Steven Weinberg.
\newblock {A New Light Boson?}
\newblock {\em Phys. Rev. Lett.}, 40:223--226, 1978.

\bibitem{Zurek2014}
Kathryn~M. Zurek.
\newblock Asymmetric dark matter: Theories, signatures, and constraints.
\newblock {\em Physics Reports}, 537(3):91–121, April 2014.

\bibitem{murgui}
Clara Murgui and Kathryn~M. Zurek.
\newblock Dark unification: A uv-complete theory of asymmetric dark matter.
\newblock {\em Phys. Rev. D}, 105:095002, May 2022.

\bibitem{Brandt:1969gh}
Stephen Adler.
\newblock {Axial-vector current in spinor electrodynamics}.
\newblock {\em Phys. Rev.}, 180:1490--1502, 1969.

\bibitem{Bell:1969ts}
J.~S. Bell and R.~Jackiw.
\newblock {A PCAC puzzle: $\pi^0 \to \gamma \gamma$ in the $\sigma$ model}.
\newblock {\em Nuovo Cim. A}, 60:47--61, 1969.

\bibitem{Politzer1973fx}
H.~David Politzer.
\newblock {Reliable Perturbative Results for Strong Interactions?}
\newblock {\em Phys. Rev. Lett.}, 30:1346--1349, 1973.

\bibitem{Gross1973id}
David~J. Gross and Frank Wilczek.
\newblock {Ultraviolet Behavior of Nonabelian Gauge Theories}.
\newblock {\em Phys. Rev. Lett.}, 30:1343--1346, 1973.

\bibitem{randall}
Lisa Randall and Raman Sundrum.
\newblock Large mass hierarchy from a small extra dimension.
\newblock {\em Phys. Rev. Lett.}, 83:3370--3373, Oct 1999.

\bibitem{arkani2}
Nima Arkani–Hamed, Savas Dimopoulos, and Gia Dvali.
\newblock The hierarchy problem and new dimensions at a millimeter.
\newblock {\em Physics Letters B}, 429(3–4):263–272, June 1998.

\bibitem{Arkani-Hamed2000}
Nima Arkani-Hamed and Martin Schmaltz.
\newblock Hierarchies without symmetries from extra dimensions.
\newblock {\em Phys. Rev. D}, 61:033005, Jan 2000.

\bibitem{KALUZA_2018}
T.~Kaluza.
\newblock On the unification problem in physics.
\newblock {\em International Journal of Modern Physics D}, 27(14):1870001, October 2018.

\bibitem{Klein:1926tv}
Oskar Klein.
\newblock {Quantum Theory and Five-Dimensional Theory of Relativity. (In German and English)}.
\newblock {\em Z. Phys.}, 37:895--906, 1926.

\bibitem{Hannestad2001}
Steen Hannestad and Georg~G. Raffelt.
\newblock New supernova limit on large extra dimensions: Bounds on kaluza-klein graviton production.
\newblock {\em Physical Review Letters}, 87(5), July 2001.

\bibitem{bodas}
Arushi Bodas, Manuel~A. Buen-Abad, Anson Hook, and Raman Sundrum.
\newblock A closer look in the mirror: Reflections on the matter/dark matter coincidence, 2024.

\bibitem{An2010}
Haipeng An, Shao-Long Chen, Rabindra~N. Mohapatra, and Yue Zhang.
\newblock Leptogenesis as a common origin for matter and dark matter.
\newblock {\em Journal of High Energy Physics}, 2010(3), March 2010.

\bibitem{witten}
Edward Witten.
\newblock Cosmic separation of phases.
\newblock {\em Phys. Rev. D}, 30:272--285, Jul 1984.

\bibitem{Farhi1984}
Edward Farhi and R.~L. Jaffe.
\newblock {Strange Matter}.
\newblock {\em Phys. Rev. D}, 30:2379, 1984.

\bibitem{zhitnitsky}
Ariel~R Zhitnitsky.
\newblock `nonbaryonic’ dark matter as baryonic colour superconductor.
\newblock {\em Journal of Cosmology and Astroparticle Physics}, 2003(10):010–010, October 2003.

\bibitem{Kharzeev:2007tn}
D.~Kharzeev and A.~Zhitnitsky.
\newblock {Charge separation induced by P-odd bubbles in QCD matter}.
\newblock {\em Nucl. Phys. A}, 797:67--79, 2007.

\bibitem{farrar2018stablesexaquark}
Glennys~R. Farrar.
\newblock Stable sexaquark, 2018.

\bibitem{kolb}
Edward~W. Kolb and Michael~S. Turner.
\newblock Dibaryons cannot be the dark matter.
\newblock {\em Phys. Rev. D}, 99:063519, Mar 2019.

\bibitem{Moore2024}
Marianne Moore and Tracy~R. Slatyer.
\newblock {On the cosmology and terrestrial signals of sexaquark dark matter}.
\newblock 3 2024.

\bibitem{fritzsch}
Harald Fritzsch and Peter Minkowski.
\newblock {Unified Interactions of Leptons and Hadrons}.
\newblock {\em Annals Phys.}, 93:193--266, 1975.

\bibitem{Glashow:1970gm}
S.~L. Glashow, J.~Iliopoulos, and L.~Maiani.
\newblock {Weak Interactions with Lepton-Hadron Symmetry}.
\newblock {\em Phys. Rev. D}, 2:1285--1292, 1970.

\bibitem{Kobayashi:1973fv}
Makoto Kobayashi and Toshihide Maskawa.
\newblock {CP Violation in the Renormalizable Theory of Weak Interaction}.
\newblock {\em Prog. Theor. Phys.}, 49:652--657, 1973.

\bibitem{Struppa2013}
Daniele~C. Struppa and Jeffrey~M. Tollaksen, editors.
\newblock {\em Quantum Theory: A Two-Time Success Story: Yakir Aharonov Festschrift}.
\newblock Springer, Milano, 2013.

\bibitem{avignone}
F.~T. Avignone and R.~L. Brodzinski.
\newblock {A Review of Recent Developments in Double Beta Decay}.
\newblock {\em Prog. Part. Nucl. Phys.}, 21:99--181, 1988.

\bibitem{Fiorini:1983yj}
E.~Fiorini and T.~O. Niinikoski.
\newblock {Low Temperature Calorimetry for Rare Decays}.
\newblock {\em Nucl. Instrum. Meth. A}, 224:83, 1984.

\bibitem{Chiles2022}
Jeff Chiles, Ilya Charaev, Robert Lasenby, Masha Baryakhtar, Junwu Huang, Alexana Roshko, George Burton, Marco Colangelo, Ken Van~Tilburg, Asimina Arvanitaki, Sae~Woo Nam, and Karl~K. Berggren.
\newblock New constraints on dark photon dark matter with superconducting nanowire detectors in an optical haloscope.
\newblock {\em Physical Review Letters}, 128(23), June 2022.

\bibitem{Drukier:2014rea}
A.~K. Drukier, Ch. Cantor, M.~Chonofsky, G.~M. Church, R.~L. Fagaly, K.~Freese, A.~Lopez, T.~Sano, C.~Savage, and W.~P. Wong.
\newblock {New class of biological detectors for WIMPs}.
\newblock {\em Int. J. Mod. Phys. A}, 29:1443007, 2014.

\bibitem{silk}
Joseph Silk, Keith Olive, and Mark Srednicki.
\newblock The photino, the sun, and high-energy neutrinos.
\newblock {\em Phys. Rev. Lett.}, 55:257--259, Jul 1985.

\bibitem{jungman}
Gerard Jungman, Marc Kamionkowski, and Kim Griest.
\newblock Supersymmetric dark matter.
\newblock {\em Physics Reports}, 267(5–6):195–373, March 1996.

\bibitem{feng}
Jonathan~L. Feng and Jason Kumar.
\newblock {The WIMPless Miracle: Dark-Matter Particles without Weak-Scale Masses or Weak Interactions}.
\newblock {\em Phys. Rev. Lett.}, 101:231301, 2008.

\bibitem{Bjorken:2009mm}
James~D. Bjorken, Rouven Essig, Philip Schuster, and Natalia Toro.
\newblock {New Fixed-Target Experiments to Search for Dark Gauge Forces}.
\newblock {\em Phys. Rev. D}, 80:075018, 2009.

\bibitem{Essig2012}
Rouven Essig, Aaron Manalaysay, Jeremy Mardon, Peter Sorensen, and Tomer Volansky.
\newblock First direct detection limits on sub-gev dark matter from xenon10.
\newblock {\em Physical Review Letters}, 109(2), July 2012.

\bibitem{maity2023}
Tarak~Nath Maity, Akash~Kumar Saha, Sagnik Mondal, and Ranjan Laha.
\newblock Neutrinos from the sun can discover dark matter-electron scattering, 2023.

\bibitem{sensei2023}
SENSEI~Collaboration Prakruth Adari~et al.
\newblock Sensei: First direct-detection results on sub-gev dark matter from sensei at snolab, 2023.

\bibitem{bottaro2023}
Salvatore Bottaro and Diego Redigolo.
\newblock The dark matter unitarity bound at nlo, 2023.

\bibitem{Bottaro_2022}
Salvatore Bottaro, Dario Buttazzo, Marco Costa, Roberto Franceschini, Paolo Panci, Diego Redigolo, and Ludovico Vittorio.
\newblock Closing the window on wimp dark matter.
\newblock {\em The European Physical Journal C}, 82(1), January 2022.

\bibitem{Drukier1986tm}
A.~K. Drukier, Katherine Freese, and D.~N. Spergel.
\newblock {Detecting Cold Dark Matter Candidates}.
\newblock {\em Phys. Rev. D}, 33:3495--3508, 1986.

\bibitem{Arvanitaki:2022oby}
Asimina Arvanitaki and Savas Dimopoulos.
\newblock {Cosmic neutrino background on the surface of the Earth}.
\newblock {\em Phys. Rev. D}, 108(4):043517, 2023.

\bibitem{Gruzinov2024}
Andrei Gruzinov and Mehrdad Mirbabayi.
\newblock {The Density of Relic Neutrinos Near the Surface of Earth}.
\newblock 3 2024.

\bibitem{Buchmuller:1989rb}
W.~Buchmuller and F.~Hoogeveen.
\newblock {Coherent Production of Light Scalar Particles in Bragg Scattering}.
\newblock {\em Phys. Lett. B}, 237:278--283, 1990.

\bibitem{PASCHOS1994367}
E.A. Paschos and K.~Zioutas.
\newblock A proposal for solar axion detection via bragg scattering.
\newblock {\em Physics Letters B}, 323(3):367--372, 1994.

\bibitem{Creswick:1997pg}
R.~J. Creswick, F.~T. Avignone, III, H.~A. Farach, J.~I. Collar, A.~O. Gattone, S.~Nussinov, and K.~Zioutas.
\newblock {Theory for the direct detection of solar axions by coherent Primakoff conversion in germanium detectors}.
\newblock {\em Phys. Lett. B}, 427:235--240, 1998.

\bibitem{Sekiya2003}
Hiroyuki Sekiya, M.~Minowa, Y.~Shimizu, Y.~Inoue, and W.~Suganuma.
\newblock {Measurements of anisotropic scintillation efficiency for carbon recoils in a stilbene crystal for dark matter detection}.
\newblock {\em Phys. Lett. B}, 571:132--138, 2003.

\bibitem{creswick}
Richard~J. Creswick, Shmuel Nussinov, and Frank~T. {Avignone III}.
\newblock Direction dependence and diurnal modulation in dark matter detectors.
\newblock {\em Astroparticle Physics}, 35(2):62--66, 2011.

\bibitem{bozorgnia}
Nassim Bozorgnia, Graciela~B. Gelmini, and Paolo Gondolo.
\newblock Daily modulation due to channeling in direct dark matter crystalline detectors.
\newblock {\em Physical Review D}, 84(2), July 2011.

\bibitem{DAMA:2008jlt}
R.~Bernabei et~al.
\newblock {First results from DAMA/LIBRA and the combined results with DAMA/NaI}.
\newblock {\em Eur. Phys. J. C}, 56:333--355, 2008.

\bibitem{PhysRevLett.123.031302}
G.~Adhikari and P.~Adhikari et~al.
\newblock Search for a dark matter-induced annual modulation signal in nai(tl) with the cosine-100 experiment.
\newblock {\em Phys. Rev. Lett.}, 123:031302, Jul 2019.

\bibitem{bird}
Simeon Bird, Ilias Cholis, Julian~B. Muñoz, Yacine Ali-Haïmoud, Marc Kamionkowski, Ely~D. Kovetz, Alvise Raccanelli, and Adam~G. Riess.
\newblock Did ligo detect dark matter?
\newblock {\em Physical Review Letters}, 116(20), May 2016.

\bibitem{Barack:2018yly}
Leor Barack et~al.
\newblock {Black holes, gravitational waves and fundamental physics: a roadmap}.
\newblock {\em Class. Quant. Grav.}, 36(14):143001, 2019.

\bibitem{graham}
Peter~W. Graham, Surjeet Rajendran, and Jaime Varela.
\newblock Dark matter triggers of supernovae.
\newblock {\em Phys. Rev. D}, 92:063007, Sep 2015.

\bibitem{Khriplovich:2010hn}
I.~B. Khriplovich.
\newblock {Capture of dark matter by the Solar System. Simple estimates}.
\newblock {\em Int. J. Mod. Phys. D}, 20:17--22, 2011.

\bibitem{Damour}
Thibault Damour and Lawrence~M. Krauss.
\newblock New wimp population in the solar system and new signals for dark-matter detectors.
\newblock {\em Physical Review D}, 59(6), February 1999.

\bibitem{Dicke1965}
R.~H. Dicke, P.~J.~E. Peebles, P.~G. Roll, and D.~T. Wilkinson.
\newblock {Cosmic Black-Body Radiation}.
\newblock {\em Astrophys. J.}, 142:414--419, 1965.

\bibitem{Penzias:1965wn}
Arno~A. Penzias and Robert~Woodrow Wilson.
\newblock {A Measurement of excess antenna temperature at 4080-Mc/s}.
\newblock {\em Astrophys. J.}, 142:419--421, 1965.

\bibitem{Komargodski}
Zohar Komargodski and Adam Schwimmer.
\newblock {On Renormalization Group Flows in Four Dimensions}.
\newblock {\em JHEP}, 12:099, 2011.

\bibitem{agashe}
Kaustubh Agashe, Jae~Hyeok Chang, Steven~J. Clark, Bhaskar Dutta, Yuhsin Tsai, and Tao Xu.
\newblock Detecting axionlike particles with primordial black holes.
\newblock {\em Phys. Rev. D}, 108:023014, Jul 2023.

\bibitem{marcos}
Marcos~M. Flores and Alexander Kusenko.
\newblock Primordial black holes from long-range scalar forces and scalar radiative cooling.
\newblock {\em Phys. Rev. Lett.}, 126:041101, Jan 2021.

\bibitem{taylor}
Quinn Taylor, Glenn~D. Starkman, Michael Hinczewski, Deyan~P. Mihaylov, Joseph Silk, and Jose de~Freitas~Pacheco.
\newblock Extremal kerr black hole dark matter from hawking evaporation, 2024.

\bibitem{Alonso-Monsalve:2023brx}
Elba Alonso-Monsalve and David~I. Kaiser.
\newblock {Primordial Black Holes with QCD Color Charge}.
\newblock {\em Phys. Rev. Lett.}, 132(23):231402, 2024.

\bibitem{Loeb:2024tcc}
Abraham Loeb.
\newblock {Excluding Primordial Black Holes as Dark Matter Based on Solar System Ephemeris}.
\newblock {\em Res. Notes AAS}, 8(8):211, 2024.

\bibitem{Tran_2024}
Tung~X. Tran, Sarah~R. Geller, Benjamin~V. Lehmann, and David~I. Kaiser.
\newblock Close encounters of the primordial kind: A new observable for primordial black holes as dark matter.
\newblock {\em Physical Review D}, 110(6), September 2024.

\bibitem{Pospelov}
Maxim Pospelov, Adam Ritz, and Mikhail Voloshin.
\newblock Secluded wimp dark matter.
\newblock {\em Physics Letters B}, 662(1):53–61, April 2008.

\bibitem{Yang}
Yang Bai and Patrick~J. Fox.
\newblock Resonant dark matter.
\newblock {\em Journal of High Energy Physics}, 2009(11):052, nov 2009.

\bibitem{PTOLEMY:2019hkd}
M.~G. Betti et~al.
\newblock {Neutrino physics with the PTOLEMY project: active neutrino properties and the light sterile case}.
\newblock {\em JCAP}, 07:047, 2019.

\bibitem{Nussinov:2021zrj}
Shmuel Nussinov and Zohar Nussinov.
\newblock {Quantum induced broadening: A challenge for cosmic neutrino background discovery}.
\newblock {\em Phys. Rev. D}, 105(4):043502, 2022.

\bibitem{PhysRevC.91.035505}
L.~I. Bodine, D.~S. Parno, and R.~G.~H. Robertson.
\newblock Assessment of molecular effects on neutrino mass measurements from tritium $\ensuremath{\beta}$ decay.
\newblock {\em Phys. Rev. C}, 91:035505, Mar 2015.

\bibitem{cheipesh2021}
Yevheniia Cheipesh, Vadim Cheianov, and Alexey Boyarsky.
\newblock Navigating the pitfalls of relic neutrino detection, 2021.

\bibitem{Grossman_2019}
Yuval Grossman, Roni Harnik, Ofri Telem, and Yue Zhang.
\newblock Self-destructing dark matter.
\newblock {\em Journal of High Energy Physics}, 2019(7), July 2019.

\bibitem{Geller2}
Michael Geller and Ofri Telem.
\newblock Self-destructing atomic dark matter.
\newblock {\em Phys. Rev. D}, 104:035010, Aug 2021.

\bibitem{Giudice_2018}
Gian~F. Giudice, Doojin Kim, Jong-Chul Park, and Seodong Shin.
\newblock Inelastic boosted dark matter at direct detection experiments.
\newblock {\em Physics Letters B}, 780:543–552, May 2018.

\bibitem{Y_ksel_2007}
Hasan Yüksel, Shunsaku Horiuchi, John~F. Beacom, and Shin’ichiro Ando.
\newblock Neutrino constraints on the dark matter total annihilation cross section.
\newblock {\em Physical Review D}, 76(12), December 2007.

\bibitem{PhysRevD.104.103026}
Haipeng An, Haoming Nie, Maxim Pospelov, Josef Pradler, and Adam Ritz.
\newblock Solar reflection of dark matter.
\newblock {\em Phys. Rev. D}, 104:103026, Nov 2021.

\bibitem{pospelov2}
Maxim Pospelov and Harikrishnan Ramani.
\newblock Earth-bound millicharge relics.
\newblock {\em Phys. Rev. D}, 103:115031, Jun 2021.

\bibitem{Berlin:2022hmt}
Asher Berlin, Jeff~A. Dror, Xucheng Gan, and Joshua~T. Ruderman.
\newblock {Millicharged relics reveal massless dark photons}.
\newblock {\em JHEP}, 05:046, 2023.

\bibitem{Krnjaic2022}
G.~Krnjaic et~al.
\newblock {A Snowmass Whitepaper: Dark Matter Production at Intensity-Frontier Experiments}.
\newblock 7 2022.

\bibitem{Grabowska_2018}
Dorota~M. Grabowska, Tom Melia, and Surjeet Rajendran.
\newblock Detecting dark blobs.
\newblock {\em Physical Review D}, 98(11), December 2018.

\bibitem{nussinov2020dark}
Shmuel Nussinov and Yongchao Zhang.
\newblock Dark matter clusters and time correlations in direct detection experiments, 2020.

\bibitem{Arvanitaki:2019rax}
Asimina Arvanitaki, Savas Dimopoulos, Marios Galanis, Luis Lehner, Jedidiah~O. Thompson, and Ken Van~Tilburg.
\newblock {Large-misalignment mechanism for the formation of compact axion structures: Signatures from the QCD axion to fuzzy dark matter}.
\newblock {\em Phys. Rev. D}, 101(8):083014, 2020.

\bibitem{FrancoAbellan:2021hdb}
Guillermo Franco~Abell\'an, Zackaria Chacko, Abhish Dev, Peizhi Du, Vivian Poulin, and Yuhsin Tsai.
\newblock {Improved cosmological constraints on the neutrino mass and lifetime}.
\newblock {\em JHEP}, 08:076, 2022.

\bibitem{Serpico:2008zza}
Pasquale~D. Serpico.
\newblock {Neutrinos and cosmology: a lifetime relationship}.
\newblock {\em J. Phys. Conf. Ser.}, 173:012018, 2009.

\bibitem{Bulbul_2014}
Esra Bulbul, Maxim Markevitch, Adam Foster, Randall~K. Smith, Michael Loewenstein, and Scott~W. Randall.
\newblock Detection of an unidentified emission line in the stacked x-ray spectrum of galaxy clusters.
\newblock {\em The Astrophysical Journal}, 789(1):13, June 2014.

\bibitem{King:2013iva}
Stephen~F. King.
\newblock {Minimal predictive see-saw model with normal neutrino mass hierarchy}.
\newblock {\em JHEP}, 07:137, 2013.

\bibitem{Roach2022}
Brandon~M. Roach, Steven Rossland, Kenny C.~Y. Ng, Kerstin Perez, John~F. Beacom, Brian~W. Grefenstette, Shunsaku Horiuchi, Roman Krivonos, and Daniel~R. Wik.
\newblock {Long-exposure NuSTAR constraints on decaying dark matter in the Galactic halo}.
\newblock {\em Phys. Rev. D}, 107(2):023009, 2023.

\bibitem{Dodelson:1993je}
Scott Dodelson and Lawrence~M. Widrow.
\newblock {Sterile-neutrinos as dark matter}.
\newblock {\em Phys. Rev. Lett.}, 72:17--20, 1994.

\bibitem{DeGouvea:2019wpf}
Andr\'e De~Gouv\^ea, Manibrata Sen, Walter Tangarife, and Yue Zhang.
\newblock {Dodelson-Widrow Mechanism in the Presence of Self-Interacting Neutrinos}.
\newblock {\em Phys. Rev. Lett.}, 124(8):081802, 2020.

\bibitem{Abazajian2017}
Kevork~N. Abazajian.
\newblock {Sterile neutrinos in cosmology}.
\newblock {\em Phys. Rept.}, 711-712:1--28, 2017.

\bibitem{PhysRevLett.60.1793}
Georg Raffelt and David Seckel.
\newblock Bounds on exotic-particle interactions from sn1987a.
\newblock {\em Phys. Rev. Lett.}, 60:1793--1796, May 1988.

\bibitem{Sudakov:1954sw}
V.~V. Sudakov.
\newblock {Vertex parts at very high-energies in quantum electrodynamics}.
\newblock {\em Sov. Phys. JETP}, 3:65--71, 1956.

\bibitem{Friedland:2010sc}
Alexander Friedland.
\newblock {Self-refraction of supernova neutrinos: mixed spectra and three-flavor instabilities}.
\newblock {\em Phys. Rev. Lett.}, 104:191102, 2010.

\bibitem{Smirnov_2005}
A~Yu Smirnov.
\newblock The msw effect and matter effects in neutrino oscillations.
\newblock {\em Physica Scripta}, T121:57–64, January 2005.

\bibitem{Wolfenstein:1979ni}
L.~Wolfenstein.
\newblock {Neutrino Oscillations and Stellar Collapse}.
\newblock {\em Phys. Rev. D}, 20:2634--2635, 1979.

\bibitem{Frishman:1980dq}
Y.~Frishman, A.~Schwimmer, Tom Banks, and S.~Yankielowicz.
\newblock {The Axial Anomaly and the Bound State Spectrum in Confining Theories}.
\newblock {\em Nucl. Phys. B}, 177:157--171, 1981.

\bibitem{Ando:2005ka}
Shin'ichiro Ando, John~F. Beacom, and Hasan Yuksel.
\newblock {Detection of neutrinos from supernovae in nearby galaxies}.
\newblock {\em Phys. Rev. Lett.}, 95:171101, 2005.

\bibitem{Adler:1969er}
Stephen~L. Adler and William~A. Bardeen.
\newblock {Absence of higher order corrections in the anomalous axial vector divergence equation}.
\newblock {\em Phys. Rev.}, 182:1517--1536, 1969.

\bibitem{casher}
A.~Casher, H.~Neuberger, and S.~Nussinov.
\newblock model of particle production.
\newblock {\em Phys. Rev. D}, 20:179--188, Jul 1979.

\bibitem{COLEMAN1982205}
Sidney Coleman and Bernard Grossman.
\newblock 't hooft's consistency condition as a consequence of analyticity and unitarity.
\newblock {\em Nuclear Physics B}, 203(2):205--220, 1982.

\bibitem{20.gould}
Andrew Gould, {Bruce T.} Draine, {Roger W.} Romani, and Shmuel Nussinov.
\newblock Neuton stars: Graveyard of charged dark matter.
\newblock {\em Physics Letters B}, 238(2-4):337--343, April 1990.

\bibitem{20.derujula}
A.~{De Rújula}, S.L. Glashow, and Uri Sarid.
\newblock Charged dark matter.
\newblock {\em Nuclear Physics B}, 333(1):173--194, 1990.

\bibitem{20.goldman}
Itzhak Goldman and Shmuel Nussinov.
\newblock Weakly interacting massive particles and neutron stars.
\newblock {\em Phys. Rev. D}, 40:3221--3230, Nov 1989.

\bibitem{Blum2016}
Kfir Blum and Doron Kushnir.
\newblock {Neutrino Signal of Collapse-induced Thermonuclear Supernovae: the Case for Prompt Black Hole Formation in SN1987A}.
\newblock {\em Astrophys. J.}, 828(1):31, 2016.

\bibitem{Stodolsky:1987vd}
Leo Stodolsky.
\newblock {The Speed of Light and the Speed of Neutrinos}.
\newblock {\em Phys. Lett. B}, 201:353--354, 1988.

\bibitem{Longo:1987gc}
Michael~J. Longo.
\newblock {New Precision Tests of the Einstein Equivalence Principle From {SN1987A}}.
\newblock {\em Phys. Rev. Lett.}, 60:173, 1988.

\bibitem{Ellis2005sjy}
John~R. Ellis, Nick~E. Mavromatos, Dimitri~V. Nanopoulos, Alexander~S. Sakharov, and Edward K.~G. Sarkisyan.
\newblock {Robust limits on Lorentz violation from gamma-ray bursts}.
\newblock {\em Astropart. Phys.}, 25:402--411, 2006.
\newblock [Erratum: Astropart.Phys. 29, 158--159 (2008)].

\bibitem{Brzeminski2022rkf}
Dawid Brzeminski, Saurav Das, Anson Hook, and Clayton Ristow.
\newblock {Constraining Vector Dark Matter with neutrino experiments}.
\newblock {\em JHEP}, 08:181, 2023.

\bibitem{Chang:2022aas}
Po-Wen Chang, Ivan Esteban, John~F. Beacom, Todd~A. Thompson, and Christopher~M. Hirata.
\newblock {Toward Powerful Probes of Neutrino Self-Interactions in Supernovae}.
\newblock {\em Phys. Rev. Lett.}, 131(7):071002, 2023.

\bibitem{Barbiellini:1987zz}
G.~Barbiellini and G.~Cocconi.
\newblock {Electric Charge of the Neutrinos from SN1987A}.
\newblock {\em Nature}, 329:21--22, 1987.

\bibitem{Barbieri:1988av}
Riccardo Barbieri and Rabindra~N. Mohapatra.
\newblock {Limits on Right-handed Interactions From {SN1987A} Observations}.
\newblock {\em Phys. Rev. D}, 39:1229, 1989.

\bibitem{Goldman:1987fg}
I.~Goldman, Y.~Aharonov, G.~Alexander, and S.~Nussinov.
\newblock {Implications of the Supernova Sn1987a Neutrino Signals}.
\newblock {\em Phys. Rev. Lett.}, 60:1789, 1988.

\bibitem{Hidaka_2006}
Jun Hidaka and George~M. Fuller.
\newblock Dark matter sterile neutrinos in stellar collapse: Alteration of energy/lepton number transport, and a mechanism for supernova explosion enhancement.
\newblock {\em Physical Review D}, 74(12), December 2006.

\bibitem{Fiorillo:2022cdq}
Damiano F.~G. Fiorillo, Georg~G. Raffelt, and Edoardo Vitagliano.
\newblock {Strong Supernova 1987A Constraints on Bosons Decaying to Neutrinos}.
\newblock {\em Phys. Rev. Lett.}, 131(2):021001, 2023.

\bibitem{PhysRevLett.62.505}
Edward~L. Chupp, W.~Thomas Vestrand, and Claus Reppin.
\newblock Experimental limits on the radiative decay of sn 1987a neutrinos.
\newblock {\em Phys. Rev. Lett.}, 62:505--508, Jan 1989.

\bibitem{Dar:1987nq}
Arnon Dar and Shlomo Dado.
\newblock {Constraints on the Lifetime of Massive Neutrinos From Sn1987a}.
\newblock {\em Phys. Rev. Lett.}, 59:2368, 1987.

\bibitem{Grifols_1996}
J.~A. Grifols, E.~Massó, and R.~Toldrà.
\newblock Gamma rays from sn 1987a due to pseudoscalar conversion.
\newblock {\em Physical Review Letters}, 77(12):2372–2375, September 1996.

\bibitem{Dessert:2019dos}
Christopher Dessert, Joshua~W. Foster, and Benjamin~R. Safdi.
\newblock {Hard X-ray Excess from the Magnificent Seven Neutron Stars}.
\newblock {\em Astrophys. J.}, 904(1):42, 2020.

\bibitem{Caputo:2023cpv}
Andrea Caputo, Samuel~J. Witte, Alexander~A. Philippov, and Ted Jacobson.
\newblock {Pulsar Nulling and Vacuum Radio Emission from Axion Clouds}.
\newblock 11 2023.

\bibitem{McKeen_2021}
David McKeen, Maxim Pospelov, and Nirmal Raj.
\newblock Neutron star internal heating constraints on mirror matter.
\newblock {\em Physical Review Letters}, 127(6), August 2021.

\bibitem{Yakovlev_2004}
D.G. Yakovlev and C.J. Pethick.
\newblock Neutron star cooling.
\newblock {\em Annual Review of Astronomy and Astrophysics}, 42(1):169–210, September 2004.

\bibitem{Goldman:2022rth}
Itzhak Goldman, Rabindra~N. Mohapatra, Shmuel Nussinov, and Yongchao Zhang.
\newblock {Neutron\textendash{}Mirror-Neutron Oscillation and Neutron Star Cooling}.
\newblock {\em Phys. Rev. Lett.}, 129(6):061103, 2022.

\bibitem{Peccei:1977ur}
R.~D. Peccei and Helen~R. Quinn.
\newblock {Constraints Imposed by CP Conservation in the Presence of Instantons}.
\newblock {\em Phys. Rev. D}, 16:1791--1797, 1977.

\bibitem{PhysRevLett.40.223}
Steven Weinberg.
\newblock A new light boson?
\newblock {\em Phys. Rev. Lett.}, 40:223--226, Jan 1978.

\bibitem{PhysRevLett.40.279}
F.~Wilczek.
\newblock Problem of strong $p$ and $t$ invariance in the presence of instantons.
\newblock {\em Phys. Rev. Lett.}, 40:279--282, Jan 1978.

\bibitem{Witten:1979ey}
Edward Witten.
\newblock {Dyons of Charge e theta/2 pi}.
\newblock {\em Phys. Lett. B}, 86:283--287, 1979.

\bibitem{Graham:2019bfu}
Peter~W. Graham, David~E. Kaplan, and Surjeet Rajendran.
\newblock {Relaxation of the Cosmological Constant}.
\newblock {\em Phys. Rev. D}, 100(1):015048, 2019.

\bibitem{Dine:1981rt}
Michael Dine, Willy Fischler, and Mark Srednicki.
\newblock {A Simple Solution to the Strong CP Problem with a Harmless Axion}.
\newblock {\em Phys. Lett. B}, 104:199--202, 1981.

\bibitem{Zhitnitsky:1980tq}
A.~R. Zhitnitsky.
\newblock {On Possible Suppression of the Axion Hadron Interactions. (In Russian)}.
\newblock {\em Sov. J. Nucl. Phys.}, 31:260, 1980.

\bibitem{Shifman:1979if}
Mikhail~A. Shifman, A.~I. Vainshtein, and Valentin~I. Zakharov.
\newblock {Can Confinement Ensure Natural CP Invariance of Strong Interactions?}
\newblock {\em Nucl. Phys. B}, 166:493--506, 1980.

\bibitem{Raffelt:1987im}
Georg Raffelt and Leo Stodolsky.
\newblock {Mixing of the Photon with Low Mass Particles}.
\newblock {\em Phys. Rev. D}, 37:1237, 1988.

\bibitem{DiLuzio:2020wdo}
Luca Di~Luzio, Maurizio Giannotti, Enrico Nardi, and Luca Visinelli.
\newblock {The landscape of QCD axion models}.
\newblock {\em Phys. Rept.}, 870:1--117, 2020.

\bibitem{P.PDBook}
K.~{Hagiwara}, K.~{Hikasa}, K.~{Nakamura}, M.~{Tanabashi}, M.~{Aguilar-Benitez}, C.~{Amsler}, R.M. {Barnett}, P.R. {Burchat}, C.D. {Carone}, C.~{Caso}, G.~{Conforto}, O.~{Dahl}, M.~{Doser}, S.~{Eidelman}, J.L. {Feng}, L.~{Gibbons}, M.~{Goodman}, C.~{Grab}, D.E. {Groom}, A.~{Gurtu}, K.G. {Hayes}, J.J. {Hern\'andez-Rey}, K.~{Honscheid}, C.~{Kolda}, M.L. {Mangano}, D.M. {Manley}, A.V. {Manohar}, J.~{March-Russell}, A.~{Masoni}, R.~{Miquel}, K.~{M\"onig}, H.~{Murayama}, S.~{Navas}, K.A. {Olive}, L.~{Pape}, C.~{Patrignani}, A.~{Piepke}, M.~{Roos}, J.~{Terning}, N.A. {T\"ornqvist}, T.G. {Trippe}, P.~{Vogel}, C.G. {Wohl}, R.L. {Workman}, W.-M. {Yao}, B.~{Armstrong}, P.S. {Gee}, K.S. {Lugovsky}, S.B. {Lugovsky}, V.S. {Lugovsky}, M.~{Artuso}, D.~{Asner}, K.S. {Babu}, E.~{Barberio}, M.~{Battaglia}, H.~{Bichsel}, O.~{Biebel}, P.~{Bloch}, R.N. {Cahn}, A.~{Cattai}, R.S. {Chivukula}, R.D. {Cousins}, G.~{Cowan}, T.~{Damour}, K.~{Desler}, R.J. {Donahue}, D.A. {Edwards}, V.D. {Elvira}, J.~{Erler}, V.V. {Ezhela}, A.~{Fass\`o},
  W.~{Fetscher}, B.D. {Fields}, B.~{Foster}, D.~{Froidevaux}, M.~{Fukugita}, T.K. {Gaisser}, L.~{Garren}, H.-J. {Gerber}, F.J. {Gilman}, H.E. {Haber}, C.~{Hagmann}, J.~{Hewett}, I.~{Hinchliffe}, C.J. {Hogan}, G.~{H\"ohler}, P.~{Igo-Kemenes}, J.D. {Jackson}, K.F. {Johnson}, D.~{Karlen}, B.~{Kayser}, S.R. {Klein}, K.~{Kleinknecht}, I.G. {Knowles}, P.~{Kreitz}, Yu.V. {Kuyanov}, R.~{Landua}, P.~{Langacker}, L.~{Littenberg}, A.D. {Martin}, T.~{Nakada}, M.~{Narain}, P.~{Nason}, J.A. {Peacock}, H.R. {Quinn}, S.~{Raby}, G.~{Raffelt}, E.A. {Razuvaev}, B.~{Renk}, G.~{Rolandi}, M.T. {Ronan}, L.J. {Rosenberg}, C.T. {Sachrajda}, A.I. {Sanda}, S.~{Sarkar}, M.~{Schmitt}, O.~{Schneider}, D.~{Scott}, W.G. {Seligman}, M.H. {Shaevitz}, T.~{Sj\"ostrand}, G.F. {Smoot}, S.~{Spanier}, H.~{Spieler}, N.J.C. {Spooner}, M.~{Srednicki}, A.~{Stahl}, T.~{Stanev}, M.~{Suzuki}, N.P. {Tkachenko}, G.~{Valencia}, K.~{van Bibber}, M.G. {Vincter}, D.~{Ward}, B.R. {Webber}, M.~{Whalley}, L.~{Wolfenstein}, J.~{Womersley}, C.L. {Woody}, and O.V
  {Zenin}.
\newblock {Review of Particle Physics}.
\newblock {\em {Physical Review D}}, 66:010001+, 2002.

\bibitem{Bloch:2022kjm}
Itay~M. Bloch, Roy Shaham, Yonit Hochberg, Eric Kuflik, Tomer Volansky, and Or~Katz.
\newblock {Constraints on axion-like dark matter from a Serf Comagnetometer}.
\newblock {\em Nature Commun.}, 14(1):5784, 2023.

\bibitem{Creswick:2020amx}
Richard Creswick and Frank~T. Avignone.
\newblock {ALP search using precessing light in a magnetized Fabry Perot cavity}.
\newblock {\em JCAP}, 04(04):005, 2022.

\bibitem{Chelouche:2008ax}
Doron Chelouche and Eduardo~I. Guendelman.
\newblock {Cosmic Analogues of the Stern-Gerlach Experiment and the Detection of Light Bosons}.
\newblock {\em Astrophys. J. Lett.}, 699:L5--L8, 2009.

\bibitem{21B3_hook}
Anson Hook and Junwu Huang.
\newblock Probing axions with neutron star inspirals and other stellar processes.
\newblock {\em Journal of High Energy Physics}, 2018(6), June 2018.

\bibitem{Chang:1985mu}
D.~Chang, R.~N. Mohapatra, and S.~Nussinov.
\newblock {Could goldstone bosons generate an observable $1/R$ potential? }.
\newblock {\em Phys. Rev. Lett.}, 55:2835, 1985.

\bibitem{Creswick:2008as}
R.~J. Creswick, S~Nussinov, and F.~T. Avignone, III.
\newblock {Density Gradient and Absorption Effects in Gas-Filled Magnetic Axion Helioscopes}.
\newblock {\em Phys. Rev. D}, 78:017702, 2008.

\bibitem{kim2024oscillationsatomicenergylevels}
Hyungjin Kim and Gilad Perez.
\newblock Oscillations of atomic energy levels induced by qcd axion dark matter, 2024.

\bibitem{Gasser:1983yg}
J.~Gasser and H.~Leutwyler.
\newblock {Chiral Perturbation Theory to One Loop}.
\newblock {\em Annals Phys.}, 158:142, 1984.

\bibitem{Weinberg:1978kz}
Steven Weinberg.
\newblock {Phenomenological Lagrangians}.
\newblock {\em Physica A}, 96(1-2):327--340, 1979.

\bibitem{banerjee2023oscillating}
Abhishek Banerjee, Dmitry Budker, Melina Filzinger, Nils Huntemann, Gil Paz, Gilad Perez, Sergey Porsev, and Marianna Safronova.
\newblock Oscillating nuclear charge radii as sensors for ultralight dark matter, 2023.

\bibitem{Banerjee:2024bkp}
Abhishek Banerjee.
\newblock {Probing (Ultra-) Light Dark Matter Using Synchrotron Based M\"ossbauer Spectroscopy}.
\newblock 8 2024.

\bibitem{21b4_PRX}
Dmitry Budker, Peter~W. Graham, Harikrishnan Ramani, Ferdinand Schmidt-Kaler, Christian Smorra, and Stefan Ulmer.
\newblock Millicharged dark matter detection with ion traps.
\newblock {\em PRX Quantum}, 3:010330, Feb 2022.

\bibitem{21b4_SRF}
Asher~Berlin et~al.
\newblock Searches for new particles, dark matter, and gravitational waves with srf cavities, 2022.

\bibitem{caputo2024sensitivitynuclearclocksnew}
Andrea Caputo, Doron Gazit, Hans-Werner Hammer, Joachim Kopp, Gil Paz, Gilad Perez, and Konstantin Springmann.
\newblock On the sensitivity of nuclear clocks to new physics, 2024.

\bibitem{Agrawal2019lkr}
Prateek Agrawal, Anson Hook, and Junwu Huang.
\newblock {A CMB Millikan experiment with cosmic axiverse strings}.
\newblock {\em JHEP}, 07:138, 2020.

\bibitem{Dicke:1954zz}
R.~H. Dicke.
\newblock {Coherence in Spontaneous Radiation Processes}.
\newblock {\em Phys. Rev.}, 93:99--110, 1954.

\bibitem{East:2022rsi}
William~E. East and Junwu Huang.
\newblock {Dark photon vortex formation and dynamics}.
\newblock {\em JHEP}, 12:089, 2022.

\bibitem{Brzeminski:2024drp}
Dawid Brzeminski, Anson Hook, Junwu Huang, and Clayton Ristow.
\newblock {Searching for String Bosenovas with Gravitational Wave Detectors}.
\newblock 7 2024.

\bibitem{VanTilburg:2020jvl}
Ken Van~Tilburg.
\newblock {Stellar basins of gravitationally bound particles}.
\newblock {\em Phys. Rev. D}, 104(2):023019, 2021.

\bibitem{Langhoff2022bij}
Kevin Langhoff, Nadav~Joseph Outmezguine, and Nicholas~L. Rodd.
\newblock {Irreducible Axion Background}.
\newblock {\em Phys. Rev. Lett.}, 129(24):241101, 2022.

\bibitem{MAGIS100:2021}
Mahiro Abe et~al.
\newblock {Matter-wave Atomic Gradiometer Interferometric Sensor (MAGIS-100)}.
\newblock {\em Quantum Sci. Technol.}, 6(4):044003, 2021.

\bibitem{Daylan_2016}
Tansu Daylan, Douglas~P. Finkbeiner, Dan Hooper, Tim Linden, Stephen~K.N. Portillo, Nicholas~L. Rodd, and Tracy~R. Slatyer.
\newblock The characterization of the gamma-ray signal from the central milky way: A case for annihilating dark matter.
\newblock {\em Physics of the Dark Universe}, 12:1–23, June 2016.

\bibitem{Blum:2017qnn}
Kfir Blum, Kenny Chun~Yu Ng, Ryosuke Sato, and Masahiro Takimoto.
\newblock {Cosmic rays, antihelium, and an old navy spotlight}.
\newblock {\em Phys. Rev. D}, 96(10):103021, 2017.

\bibitem{Qin2023kkk}
Wenzer Qin, Julian~B. Munoz, Hongwan Liu, and Tracy~R. Slatyer.
\newblock {Birth of the first stars amidst decaying and annihilating dark matter}.
\newblock {\em Phys. Rev. D}, 109(10):103026, 2024.

\bibitem{Chou2016lxi}
John~Paul Chou, David Curtin, and H.~J. Lubatti.
\newblock {New Detectors to Explore the Lifetime Frontier}.
\newblock {\em Phys. Lett. B}, 767:29--36, 2017.

\bibitem{Alekhin2015by}
Sergey Alekhin et~al.
\newblock {A facility to Search for Hidden Particles at the CERN SPS: the SHiP physics case}.
\newblock {\em Rept. Prog. Phys.}, 79(12):124201, 2016.

\bibitem{nussinov2009}
Shmuel Nussinov.
\newblock Some comments on possible preferred directions for the seti search.
\newblock {\em astro-ph.EP}, 0903.1628, 2009.

\bibitem{Nussinov2014xua}
Shmuel Nussinov.
\newblock {Quirks and strings attached as the ultimate communication and acceleration devices}.
\newblock 10 2014.

\bibitem{25.Luty'sQuirksi}
Junhai Kang and Markus~A Luty.
\newblock Macroscopic strings and``quirks''at colliders.
\newblock {\em Journal of High Energy Physics}, 2009(11):065, 2009.

\bibitem{25.Junhai}
Junhai Kang, Markus~A. Luty, and Salah Nasri.
\newblock The relic abundance of long-lived heavy colored particles.
\newblock {\em Journal of High Energy Physics}, 2008(09):086, sep 2008.

\bibitem{Jacoby:2007nw}
Chen Jacoby and Shmuel Nussinov.
\newblock {The Relic Abundance of Massive Colored Particles after a Late Hadronic Annihilation Stage}.
\newblock 12 2007.

\bibitem{schwinger1951gauge}
Julian Schwinger.
\newblock On gauge invariance and vacuum polarization.
\newblock {\em Physical Review}, 82(5):664, 1951.

\bibitem{Harnik_2011}
Roni Harnik, Graham~D. Kribs, and Adam Martin.
\newblock Quirks at the tevatron and beyond.
\newblock {\em Physical Review D}, 84(3), August 2011.

\bibitem{25.feng}
Jonathan~L. Feng, Jinmian Li, Xufei Liao, Jian Ni, and Junle Pei.
\newblock Discovering quirks through timing at faser and future forward experiments at the lhc, 2024.

\bibitem{Khoury2001bz}
Justin Khoury, Burt~A. Ovrut, Nathan Seiberg, Paul~J. Steinhardt, and Neil Turok.
\newblock {From big crunch to big bang}.
\newblock {\em Phys. Rev. D}, 65:086007, 2002.

\bibitem{27.Aharonov}
Y.~Aharonov, A.~Casher, and S.~Nussinov.
\newblock {The Unitarity Puzzle and Planck Mass Stable Particles}.
\newblock {\em Phys. Lett. B}, 191:51, 1987.

\bibitem{27.bekenstein}
Jacob~D. Bekenstein.
\newblock Universal upper bound on the entropy-to-energy ratio for bounded systems.
\newblock {\em Phys. Rev. D}, 23:287--298, Jan 1981.

\bibitem{Casher1997rr}
Aharon Casher and Shmuel Nussinov.
\newblock {Is the Planck momentum attainable?}
\newblock 9 1997.

\bibitem{N.Harnik}
Roni Harnik, Graham~D. Kribs, and Gilad Perez.
\newblock A universe without weak interactions.
\newblock {\em Physical Review D}, 74(3), August 2006.

\bibitem{N.clavelli}
L.~Clavelli and R.~E.~White III.
\newblock Problems in a weakless universe.
\newblock hep-ph/0609050, 2006.

\bibitem{N.Fedrow}
Joseph~M. Fedrow and Kim Griest.
\newblock Anti-anthropic solutions to the cosmic coincidence problem.
\newblock {\em Journal of Cosmology and Astroparticle Physics}, 2014(01):004, jan 2014.

\bibitem{Milgrom:1983ca}
M.~Milgrom.
\newblock {A Modification of the Newtonian dynamics as a possible alternative to the hidden mass hypothesis}.
\newblock {\em Astrophys. J.}, 270:365--370, 1983.

\bibitem{Mannheim:1992tr}
Philip~D. Mannheim and Demosthenes Kazanas.
\newblock {Newtonian limit of conformal gravity and the lack of necessity of the second order Poisson equation}.
\newblock {\em Gen. Rel. Grav.}, 26:337--361, 1994.

\bibitem{Bekenstein:2004ne}
Jacob~D. Bekenstein.
\newblock {Relativistic gravitation theory for the MOND paradigm}.
\newblock {\em Phys. Rev. D}, 70:083509, 2004.
\newblock [Erratum: Phys.Rev.D 71, 069901 (2005)].

\bibitem{27.Milgrom}
Mordehai Milgrom.
\newblock {MOND vs. dark matter in light of historical parallels}.
\newblock {\em Stud. Hist. Phil. Sci. B}, 71:170--195, 2020.

\bibitem{27.Edmonds}
Doug Edmonds, Duncan Farrah, Chiu~Man Ho, Djordje Minic, Y.~Jack Ng, and Tatsu Takeuchi.
\newblock Testing mondian dark matter with galactic rotation curves.
\newblock {\em The Astrophysical Journal}, 793(1):41, sep 2014.

\bibitem{27.lisanti}
Mariangela Lisanti, Matthew Moschella, Nadav~Joseph Outmezguine, and Oren Slone.
\newblock Testing dark matter and modifications to gravity using local milky way observables.
\newblock {\em Phys. Rev. D}, 100:083009, Oct 2019.

\bibitem{NUSSINOV1995497}
S.~Nussinov.
\newblock Summary talk: “neutrino 94”.
\newblock {\em Nuclear Physics B - Proceedings Supplements}, 38(1):497--517, 1995.
\newblock Neutrino 94.

\bibitem{Chacko2018}
Zackaria Chacko, David Curtin, Michael Geller, and Yuhsin Tsai.
\newblock {Cosmological Signatures of a Mirror Twin Higgs}.
\newblock {\em JHEP}, 09:163, 2018.

\bibitem{BEREZHIANI_2004}
ZURAB BEREZHIANI.
\newblock Mirror world and its cosmological consequences.
\newblock {\em International Journal of Modern Physics A}, 19(23):3775–3806, September 2004.

\bibitem{Foot_2014}
R.~Foot.
\newblock Mirror dark matter: Cosmology, galaxy structure and direct detection.
\newblock {\em International Journal of Modern Physics A}, 29(11n12):1430013, April 2014.

\bibitem{Blum_2016}
Kfir Blum, Ryosuke Sato, and Tracy~R. Slatyer.
\newblock Self-consistent calculation of the sommerfeld enhancement.
\newblock {\em Journal of Cosmology and Astroparticle Physics}, 2016(06):021–021, June 2016.

\bibitem{Rubinstein1966zza}
H.~R. Rubinstein and H.~Stern.
\newblock {Nucleon - antinucleon annihilation in the quark model}.
\newblock {\em Phys. Lett.}, 21:447--449, 1966.

\bibitem{Slansky1981yr}
R.~Slansky.
\newblock {Group Theory for Unified Model Building}.
\newblock {\em Phys. Rept.}, 79:1--128, 1981.

\bibitem{Mohapatra2003ah}
R.~N. Mohapatra, S.~Nussinov, and Abdel Perez-Lorenzana.
\newblock {Large extra dimensions and decaying $K K$ recurrences}.
\newblock {\em Phys. Rev. D}, 68:116001, 2003.

\bibitem{Gell-Mann:1954yli}
Murray Gell-Mann and F.~E. Low.
\newblock {Quantum electrodynamics at small distances}.
\newblock {\em Phys. Rev.}, 95:1300--1312, 1954.

\bibitem{StueckelbergdeBreidenbach1952pwl}
Ernst Carl~Gerlach Stueckelberg~de Breidenbach and Andreas Petermann.
\newblock {Normalization of constants in the quanta theory}.
\newblock {\em Helv. Phys. Acta}, 26:499--520, 1953.

\bibitem{Machacek:1992yf}
M.~E. Machacek, E.~D. Carlson, and L.~J. Hall.
\newblock {Selfinteracting dark matter: An alternative scenario?}
\newblock In {\em {16th Texas Symposium on Relativistic Astrophysics and 3rd Particles, Strings, and Cosmology Symposium}}, pages 681--685, 12 1992.

\end{thebibliography}

\end{document}